\documentclass[english,12pt]{article}
\usepackage{graphicx}
\usepackage{amsmath}
\usepackage{amssymb}
\usepackage{bm}
\usepackage{slashed}
\usepackage{dcolumn}
\usepackage{epsfig}
\usepackage{times}
\usepackage{cite}
\usepackage{afterpage}
\usepackage{color}
\usepackage[T1]{fontenc}
\usepackage[latin9]{inputenc}
\usepackage{babel}
\usepackage{lipsum}
\usepackage{ulem} 
\usepackage{url}
\usepackage[breaklinks]{hyperref}
\usepackage{hyperref}
\hypersetup{
    colorlinks,
    citecolor=black,
    filecolor=black,
    linkcolor=black,
    urlcolor=black
}

\allowdisplaybreaks[1]

\newcommand{\et}{\right\rangle}
\newcommand{\br}{\left\langle}
\newcommand{\bea}{\begin{eqnarray}}
\newcommand{\eea}{\end{eqnarray}}
\newcommand{\beq}{\begin{equation}}
\newcommand{\eeq}{\end{equation}}
\newcommand \hmu {\hat{\mu}}

\def\lambdabar{\lambda\kern-1ex\raise0.65ex\hbox{-}}

\usepackage[margin=1in]{geometry}
\begin{document}
\hfill{\bf Date:} {\today}
\begin{center}
{\huge\bf Workshop on Excited Hyperons in QCD Thermodynamics at Freeze-Out \\
	(YSTAR2016) \\
	Mini-Proceedings} 
\end{center}
\hspace{0.1in}
\begin{center}
{\large 16th - 17th November, 2016}
{\large Thomas Jefferson National Accelerator Facility, Newport 
	News, VA, U.S.A.}
\end{center}
\begin{center}
P.~Alba,
M.~Amaryan,
V.~Begun,
R.~Bellwied,
S.~Borsanyi,
W.~Broniowski,
S.~Capstick,
E.~Chudakov,
V.~Crede,
B.~D\"onigus,
R.~G.~Edwards,
Z.~Fodor,
H.~Garcilazo,
J.~L.~Goity,
M.~I.~Gorenstein,
J.~G\"unther,
L.~Guo,
P.~Huovinen,
S.~Katz,
M.~Mai,
D.~M.~Manley,
V.~Mantovani~Sarti,
E.~Meg\'{\i}as,
F.~Myhrer,
J.~Noronha-Hostler,
H.~Noumi,
P.~Parotto,
A.~Pasztor,
I.~Portillo~Vazquez,
K.~Rajagopal,
C.~Ratti,
J.~Ritman,
E.~Ruiz~Arriola,
L.~L.~Salcedo,
I.~Strakovsky,
J.~Stroth,
A.~H.~Tang,
Y.~Tsuchikawa,
A.~Valcarce,
J.~Vijande, and
V.~Yu.~Vovchenko
\end{center}
\begin{center}
\textbf{Editors}: M.~Amaryan, E.~Chudakov, K.~Rajagopal, C.~Ratti, 
	J.~Ritman, and I.~Strakovsky
\end{center}

\noindent

\begin{center}{\large\bf Abstract}\end{center}

This Workshop brought top experts, researchers, postdocs, and 
students from high-energy heavy-ion interactions, lattice QCD and 
hadronic physics communities together.  YSTAR2016 discussed the 
impact of "missing" hyperon resonances on QCD thermodynamics, on 
freeze-out in heavy ion collisions, on the evolution of early 
universe, and on the spectroscopy of strange particles. Recent 
studies that compared lattice QCD predictions of thermodynamic 
properties of quark-gluon plasma at freeze-out with calculations 
based on statistical hadron resonance gas models as well as 
experimentally measured ratios between yields of different hadron 
species in heavy ion collisions provide indirect evidence for the 
presence of "missing" resonances in all of these contexts. The aim 
of the YSTAR2016 Workshop was to sharpen these comparisons and 
advance our understanding of the formation of strange hadrons from 
quarks and gluons microseconds after the Big Bang and in todays 
experiments at LHC and RHIC as well as at future facilities like 
FAIR, J-PARC and KL at JLab.  

It was concluded that the new initiative to create a secondary 
beam of neutral kaons at JLab will make a bridge between the 
hardron spectroscopy, heavy-ion experiments and lattice QCD 
studies addressing some major issues related to thermodynamics 
of the early universe and cosmology in general. 


\newpage
\tableofcontents

\newpage
\pagenumbering{arabic}
\setcounter{page}{1}
\section{Preface}
\halign{#\hfil&\quad#\hfil\cr
}

\begin{enumerate}
\item \textbf{Preface}

This volume presents the mini-proceedings of the workshop on "Excited
Hyperons in QCD Thermodynamics at Freeze-Out" which was held at the
Jefferson Laboratory (JLab) in Newport News, Virginia, USA, November
16-17, 2016. This workshop is a successor of the workshop "Physics
with Neutral Kaon Beam at JLab"~\cite{Albrow:2016ibsm} held at JLab,
February 1-3, 2016.  Both workshops were organized as a followup on 
the letter of intent, "Physics Opportunities with Secondary KL beam 
at JLab," (LOI-12-15-001), which was presented to PAC 43 at JLab in 
2015.  The aim of the current workshop was to discuss kaon-nucleon 
scattering on unpolarized and polarized targets with GlueX setup in 
Hall~D with the emphasis on hyperon spectroscopy and its connection 
to heavy-ion interactions. The physics impact  of the measurement of 
all missing hyperon states on  hadron spectroscopy and  thermodynamics 
of the early universe at freeze-out was the main focus of the workshop.

The unique feature of this series of workshops is a mixture of theory
and experiment. Significant progress made by lattice QCD calculations
and heavy-ion interactions over the last decade made a quest to unravel
missing baryon resonances even stronger due to their great impact on the
understanding of thermodynamic properties of the early universe around 
a microsecond after the Big Bang.

The workshop consisted of 25 talks and summary session. We had about
71 paricipants from 11 different countries with a large amount of 
younger people showing that this field is interesting and has growing 
importance.  A prominent representatives from different experimental 
collaborations from different laboratories including JLab, RHIC, LHC, 
J-PARC, and GSI presented current results and future plans related to 
hadron spectroscopy with experimental data strongly supporting evidence 
for  a transition from Quark-Gluon Plasma (QGP) to Hadron Resonance Gas 
(HRG) at critical temperature. The opening  talks given by M.~Amaryan 
and E.~Chudakov about the proposed facility of neutral long lived kaon
($KL$) beam at JLab and GlueX setup laid out a ground for further 
discussions.

Although this mini-proceedings summarizes all talks presented to the
workshop, the actual talks in pdf format may be found on the web page 
of the workshop: \\
https://www.jlab.org/conferences/YSTAR2016/ .

Finally we would like to take this opportunity to thank all speakers,
session chairs, secretaries, and all participants for making this 
workshop a real success.

\newpage
\item \textbf{Acknowledgments}

The workshop could't have been done without dedicated work of many
people. First, we would like to thank service group and the staff
of JLab for all their efforts. We would like to thank JLab management, 
especially Robert~McKeown for their help and encouragement to organize 
this workshop. Financial support was provided by the JLab, J\"ulich 
Forschungszentrum, The George Washington and Old Dominion 
universities.

\vskip 1cm
\leftline {Newport News, November 2016.}

\rightline {M.~Amaryan}
\rightline {E.~Chudakov}
\rightline {K.~Rajagopal}
\rightline {C.~Ratti}
\rightline {J.~Ritman}
\rightline {I.~Strakovsky}
\end{enumerate}


\newpage
\section{Summaries of Talks}

\subsection{The K$^0_L$ Beam Facility at JLab}
\addtocontents{toc}{\hspace{2cm}{\sl M.~Amaryan}\par}
\setcounter{figure}{0}
\setcounter{table}{0}
\setcounter{equation}{0}
\halign{#\hfil&\quad#\hfil\cr
\large{Moskov Amaryan}\cr
\textit{Department of Physics}\cr
\textit{Old Dominion University}\cr
\textit{Norfolk, VA 23529, U.S.A.}\cr}

\begin{abstract}
In this talk we discuss the possibility to create a secondary
$K_L^0$ beam to be used with GlueX detector in Hall-D at JLab
for spectroscopy of excited hyperons.
\end{abstract}

\rightline{ \it Nel mezzo del cammin di nostra vita}
\rightline  {\it mi ritrovai per una selva oscura,}
\rightline {\it che la diritta via era smarrita.}
\vskip 3mm
\rightline{Dante Alighieri}

\begin{enumerate}
\item \textbf{Introduction}

The constituent quark model  was astonishingly effective in
describing all observed  hadrons at a time and even made a
predictions that were experimentally confirmed, however even
after 50 years hundreds  of regular hadronic states made of $q\bar 
q$ and $qqq$ predicted by quark models are still missing.
Moreover it is fair  to say that the very  existence of exotic
multi-quark states and  QCD inspired hybrids and glueballs is
questionable and lucks solid experimental proof.

The reason for this is twofold: on one hand there were not enough 
dedicated efforts, on the other hand until recently there were not 
adequate high quality experimental facilities to carry out such 
studies.

Among predicted regular baryon states, hundreds of missing
hyperons attracted much interest due to the presence of heavy
strange quark in their wave functions, which makes their widths 
much smaller compared to solely light-quark made baryons and 
therefore experimentally easier to be measured. There is a wide 
experimental program to study hyperons at JLab, but as we 
understand not everything can be done with electromagnetic probe.

The best  way to map out strange baryons and mesons is to use a 
secondary beam of strange mesons, as it provides one of $s(\bar 
s)$-quark, which otherwise is absent in the initial state and 
will require production of $s\bar s$ when  hyperons are produced 
by other incoming particles.

The physics impact and necessity of data obtained by a secondary 
beam of $K_L$ mesons  created by high intensity electron beam of 
CEBAF at JLab was extensively discussed during KL2016 
workshop~\cite{Albrow:2016ibsm} held in February 2016. The current 
YSTAR2016 Workshop is organized to explore wider impact of hyperon 
spectroscopy on problems, which  are in the main focus of heavy-ion 
ineractions and their connections to  thermodynamics of the early 
universe at freeze-out, microseconds after the Big Bang.

In this talk, we describe conceptually the main  steps needed to 
produce intensive $K_L$ beam. We discuss momentum resolution of the 
beam using time-of-flight technique, as well as the ratio of $K_L$ 
over neutrons as a function of their momenta simulated based on well 
known production processes.  In some examples the quality of 
expected experimental data obtained by using GlueX setup in Hall-D 
will be demonstrated using results of Monte Carlo studies.

The experimental program to search for hybrid mesons in the GlueX 
experiment at JLab is underway. Over the last decade, significant 
progress in our understanding of baryons made of light $(u,d)$ quarks 
have been made in CLAS at JLab. However, systematic studies of excited 
hyperons are very much lacking with only decades old very scarce data 
filling the world database in different channels. In this experiment, 
we propose to fill this gap and study spectra of excited hyperons 
using the modern CEBAF facility with the aim to impinge produced 
$K^0_L$ beam on physics target of  GlueX experiment in Hall~D. The 
goal is to study $K_L-p$ and $K_L-d$ interactions and do the hyperon 
and possibly strange meson spectroscopy.

Unlike to cases with pion or photon beams, kaon beams are crucial to 
provide the data needed to identify and characterize the properties 
of hyperon resonances.

Our current experimental knowledge of strange resonances is far worse 
than our knowledge of $N$ and $\Delta$ resonances; however, within the 
quark model, they are no less fundamental. Clearly there is a need to 
learn about baryon resonances in the "strange sector" to have a 
complete understanding of three-quark bound states.

The masses and widths of the lowest mass baryons were determined with
kaon-beam experiments in the 1970s~\cite{Olive:2016xmwm}. First
determination of the pole positions, for instance for $\Lambda(1520)$,
were obtained only recently from analysis of Hall~A measurement at
JLab~\cite{Qiangm}. An intense kaon beam would open up a new window
of opportunity not only to locate missing resonances, but also to
establish their properties including their quantum numbers and 
branching ratios to different decay modes.

A comprehensive review of physics opportunities with meson beams is
presented in a recent paper~\cite{Briscoe:2015qiam}. Importance of 
baryon spectroscopy in strangeness sector was discussed 
in~\cite{Kronfeld:2013uoam}.

\item \textbf{Reactions that Could be Studied with $K_L^0$ Beam}

\begin{enumerate}
\item \textbf{Two-body Reactions Producing $S=-1$ Hyperons}

\begin{eqnarray}
        K_L^0p\to \pi^+\Lambda\\
        K_L^0p\to \pi^+\Sigma^0
\end{eqnarray}

\item \textbf{Three-body Reactions Producing $S=-1$ Hyperons}

\begin{eqnarray}
        K_L^0p\to \pi^+\pi^0\Lambda \\
        K_L^0p\to \pi^+\pi^0\Sigma^0 \\
        K_L^0p\to \pi^0\pi^0\Sigma^+  \\
        K_L^0p\to \pi^+\pi^- \Sigma^+  \\
        K_L^0p\to \pi^+\pi^- \Sigma^-
\end{eqnarray}

\item \textbf{Two- and Three-body Reactions Producing $S=-2$ Hyperons}

\begin{eqnarray}
        K_L^0p\to K^+\Xi^0 \\
        K_L^0p\to \pi^+K^+\Xi^-\\
        K_L^0p\to K^+\Xi^{0*} \\
        K_L^0p\to \pi^+K^+\Xi^{-*}
\end{eqnarray}

\item \textbf{Three-body Reactions Producing $S=-3$ Hyperons}

\begin{eqnarray}
        K_L^0p\to K^+K^+\Omega^-\\
        K_L^0p\to K^+K^+\Omega^{-\ast}
\end{eqnarray}
\end{enumerate}

\item \textbf{The $K_L^0$ Beam in HALL~D}

\begin{figure}[htb!]
\centerline{\includegraphics[width=450pt]{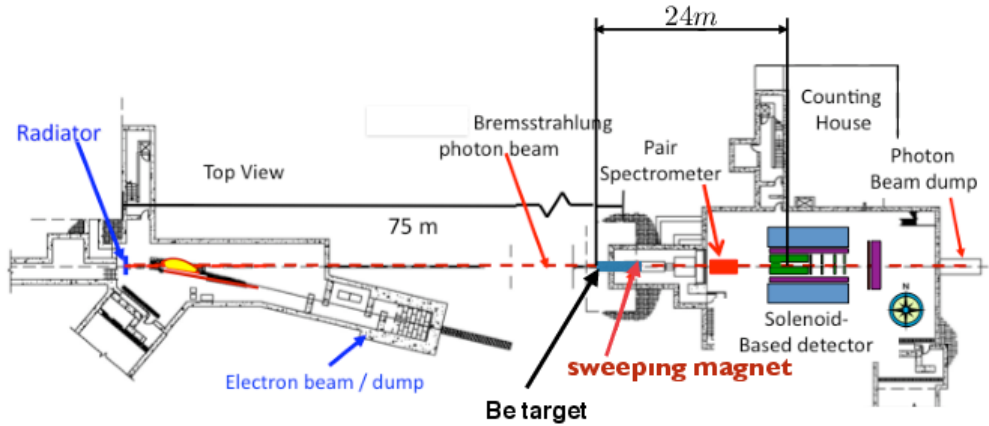}}

 \caption{Schematic view of Hall~D beamline. See a text for explanation.
        \label{fig:setup}}
\end{figure}
In this chapter, we describe photo-production of secondary $K_L^0$ 
beam in Hall~D.  There are few points that need to be decided. To 
produce intensive photon beam, one needs to increase radiation length 
of the radiator up to 10$\%$ radiation length. In a first scenario, 
$E_e=12$~GeV electrons produced at CEBAF will scatter in a radiator 
in the tagger vault,  generating  intensive beam of bremsstrahlung 
photons. This may will then require removal of all tagger counters 
and electronics and very careful design of radiation shielding, which 
is very hard to optimize and design. In a second scenario one may use 
Compact Photon Source design (for more details see a talk by 
Pavel~Degtiarenko in~\cite{Albrow:2016ibsm}) installed after the tagger 
magnet, which will produce bremsstrahlung photons and dump electron 
beam inside the source shielding the radiation inside. At the second 
stage, bremsstrahlung photons interact with Be target placed on a 
distance 16~m upstream of liquid hydrogen ($LH_2$) target of GlueX 
experiment in Hall~D producing $K_L^0$ beam. To stop photons a
30 radiation length lead absorber will be installed in the beamline
followed by a sweeping magnet to deflect the flow of charged particles.
The flux of $K_L$ on ($LH_2$) target of GlueX experiment in Hall~D will
be measured with pair spectrometer upstream the target. Details of this
part of the beamline (for a details see a talk by Ilya~Larin
in~\cite{Albrow:2016ibsm}). Momenta of $K_L$ particles will be
measured using the time-of-flight between RF signal of CEBAF and start
counters surrounding  $LH_2$ target. Schematic view of beamline is
presented in Fig.~\ref{fig:setup}. The bremsstrahlung photons, created
by electrons at a distance about 75~m upstream, hit the Be target and
produce $K_L^0$ mesons along with neutrons and charged particles. The
lead absorber of $\sim$30 radiation length is installed to absorb
photons  exiting Be target. The sweeping magnet deflects any remaining
charged particles (leptons or hadrons) remaining after the absorber.
The pair spectrometer will monitor the flux of $K_L^0$ through the
decay rate of kaons at given distance about 10~m from Be target. The
beam flux could also be monitored by installing nuclear foil in front
of pair spectrometer to measure a rate of $K^0_S$ due to regeneration
process $K_L +p \to K_S +p$ as it was done at NINA (for a details see
a talk my Mike~Albrow at this workshop).

Here we outline experimental conditions and simulated flux of $K_L^0$
based on GEANT4 and known cross sections of underlying 
subprocesses~\cite{Seraydaryan:2013ijam,Titov:2003bkm,Mcclellan:1971tkm}.

The expected flux of $K_L^0$ mesons integrated in the range of momenta
$P=0.3-10~GeV/c$ will be few times  $10^4$~$K_L^0/s$ on the physics
target of the GlueX setup under the following conditions:
\begin{itemize}
\item A thickness of the radiator 10$\%$.
\item The distance between Be and $LH_2$ targets in the range of 
	$(20-24)$~m.
\item The Be target with a length $L=40$~cm.
\item $LH_2$ target with $R=3$~cm and $L=30$~cm.
\end{itemize}

In addition, the  lower repetition rate of electron beam with 64~ns
spacing between bunches will be required to have enough time to measure
time-of-flight of the beam momenta and  to avoid an overlap of events
produced from alternating pulses. Low repetition rate was already
successfully used by G0 experiment in Hall~C at 
JLab~\cite{Androic:2011rham}.

The radiation length of the radiator needs further studies in order to
estimate the level of radiation and required shielding in the tagger
region. During this experiment all photon beam tagging detector systems
and electronics will be removed.

The final flux of $K_L^0$ is presented with 10$\%$ radiator, 
corresponding to maximal rate .

In the production of a beam of neutral kaons, an important factor is 
the rate of neutrons as background. As it is well known, the ratio
$R=N_n/N_{K_L^0}$ is on the order $10^3$ from primary proton
beams~\cite{Cleland:1975exm}, the same ratio with primary 
electromagnetic interactions is much lower. This is illustrated in 
Fig.~\ref{fig:ratio}, which  presents the rate of kaons and neutrons 
as a function of the momentum, which resembles similar behavior as it 
was measured at SLAC~\cite{Brandenburg:1972pmm}.
\begin{figure}[htb!]
\centerline{\includegraphics[width=3.5in]{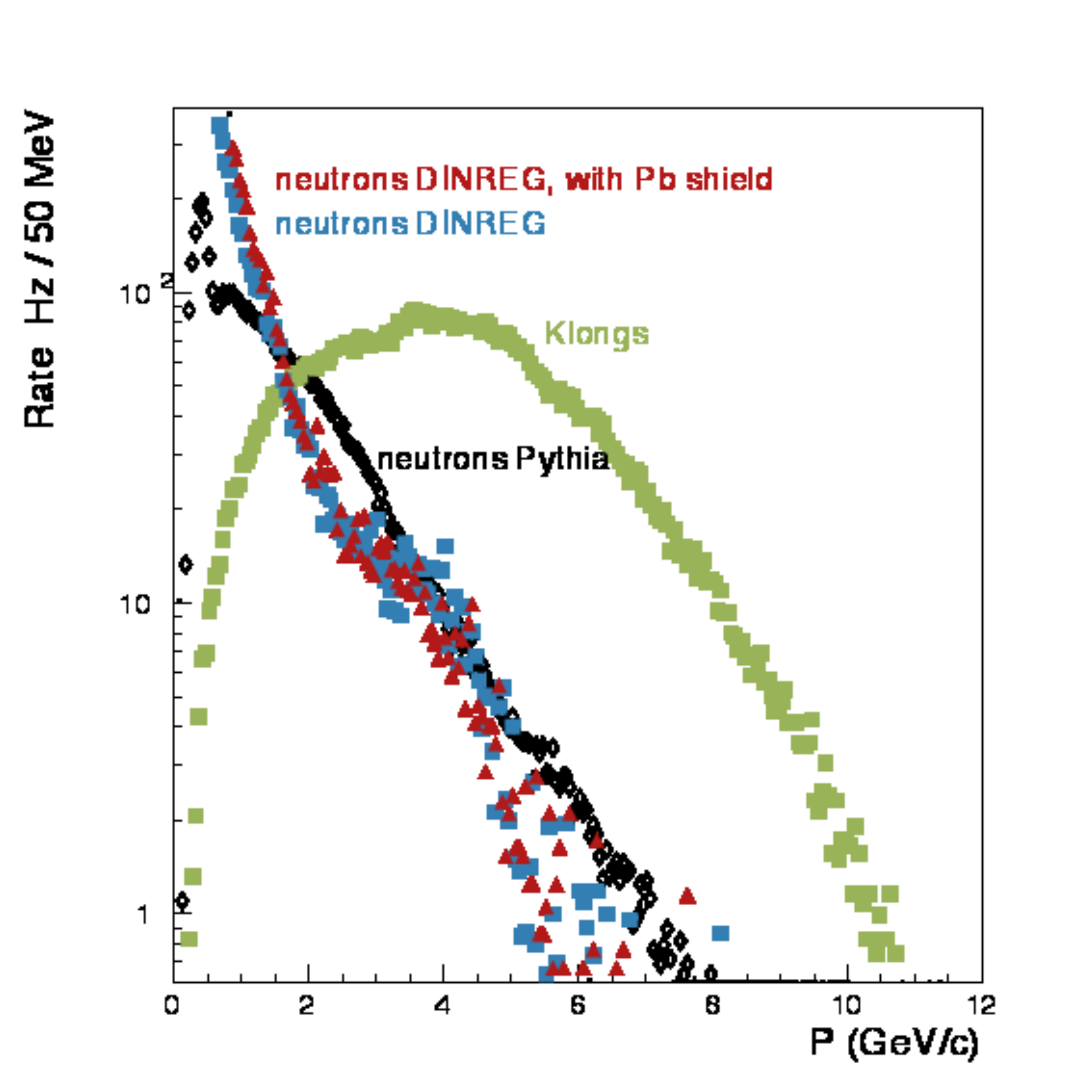}}

 \caption{\baselineskip 13pt The rate of neutrons (open symbols) and
	$K_L^0$ (full squares) on $LH_2$ target of Hall~D as a function
	of their momenta simulated with different MC generators with
        $10^4K^0_L$/sec. \label{fig:ratio}}
\end{figure}

Shielding of the low energy neutrons in the collimator cave and flux of
neutrons has been estimated to be affordable, however detailed 
simulations are under way to show the level of radiation along the 
beamline.

The response of GlueX setup, reconstruction efficiency and resolution 
are presented in a talk by Simon~Taylor in~\cite{Albrow:2016ibsm}.

\item \textbf{Expected Rates}

In this section, we discuss expected rates of events for some selected
reactions. The production of $\Xi$ hyperons has been measured only with
charged kaons with very low statistical precision and never with 
primary $K_L^0$ beam. In Fig.~\ref{fig:xi_prod} panel a) shows existing 
data for the octet ground state $\Xi$'s with theoretical model 
predictions for $W$ (the reaction center of mass energy) distribution, 
panel b) shows the same model prediction~\cite{Jackson:2015dvam} 
presented with expected experimental points and statistical error for 
10 days of running with our proposed setup with a beam intensity 
$2\times 10^3 K_L$/sec using missing mass of $K^+$ in the reaction 
$K_L^0+p\to K^+\Xi^0$ without detection of any of decay products of 
$\Xi^0$ (for more details on this topic see a talk by Kanzo~Nakayama 
in~\cite{Albrow:2016ibsm}).
\begin{figure}[htb!]
\begin{center}
\includegraphics[width=0.55\textwidth ]{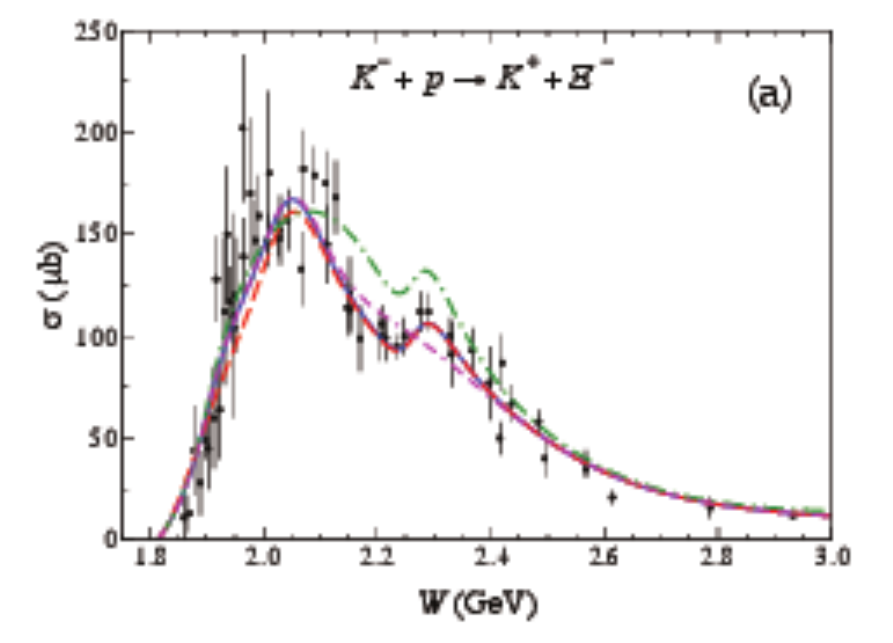} \\
\includegraphics[width=0.55\textwidth ]{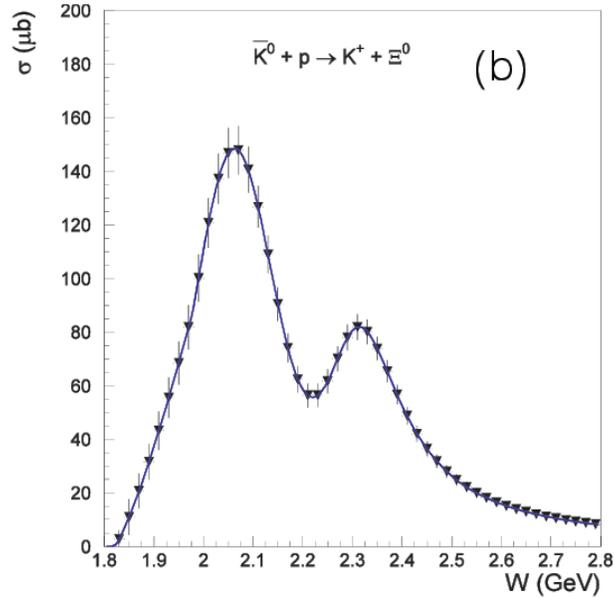}
\end{center}

 \caption{\baselineskip 13pt a) Cross section for existing 
	world data on $K^-+p\to K^+\Xi^-$ reaction with
	model predictions from~\protect\cite{Jackson:2015dvam};
	b) expected statistical precision for the reaction
        $K_L^0+p\to K^+\Xi^0$ in 10~days of running with a
        beam intensity $2\times 10^3 K_L$/sec overlaid on
        theoretical prediction~\protect\cite{Jackson:2015dvam}.
        \label{fig:xi_prod}}
\end{figure}

The physics of excited hyperons is not well explored, remaining
essentially at the pioneering stages of '70s-'80s. This is especially
true for $\Xi^\ast(S=-2)$  and $\Omega^\ast(S=-3)$ hyperons.
For example, the $SU(3)$ flavor symmetry allows as many $S=-2$ baryon
resonances, as there are $N$ and $\Delta$ resonances combined
($\approx 27$); however, until now only three [ground state
$\Xi(1382)1/2^+$, $\Xi(1538)3/2^+$, and $\Xi(1820)3/2^-$] have their
quantum numbers assigned and few more states have been
observed~\cite{Olive:2016xmwm}. The status of $\Xi$ baryons is
summarized in a table presented in Fig.~\ref{fig:table1} together
with the quark model predicted states~\cite{Chao:1980emm}.
\begin{figure}[htb!]
\begin{center}
\includegraphics[width=6in]{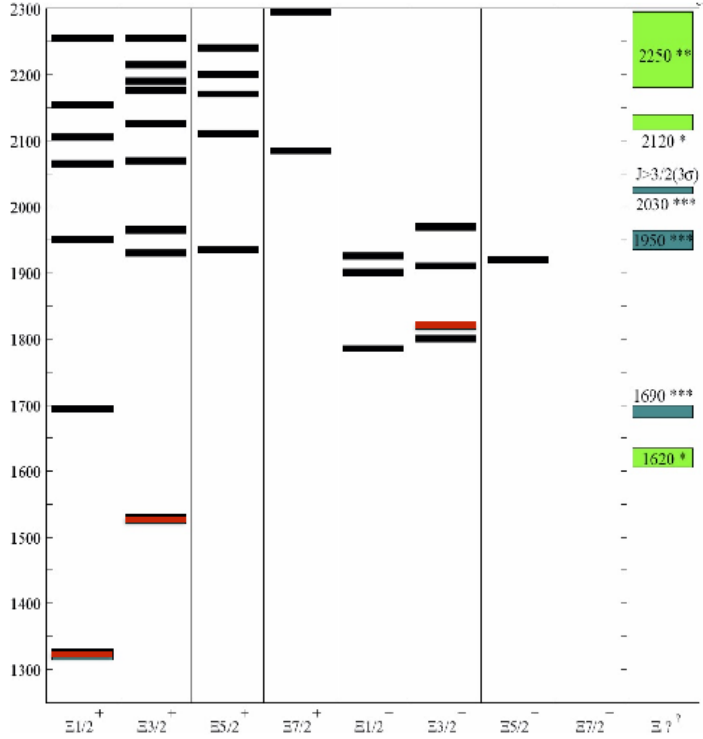}
\end{center}

 \caption{\baselineskip 13pt Black bars: Predicted
	$\Xi$ spectrum based on the quark model
	calculation~\protect\cite{Chao:1980emm}. Colored bars:
	Observed states. The two ground octet and decuplet states
	together with $\Xi(1820)$ in the column $J^P=3/2^-$ are 
	shown in red color. Other observed states with unidentified
	spin-parity are plotted in the rightest column.
        \label{fig:table1}}
\end{figure}

Historically the $\Xi^\ast$ states were intensively searched for mainly
in bubble chamber experiments using the $K^-p$ reaction in '60s-'70s.
The cross section was estimated to be on the order of 1-10~$\mu b$ at
the beam momenta up to $\sim$10~GeV/c. In '80s-'90s, the mass or width 
of ground and some of excited states were measured with a spectrometer 
in the CERN hyperon beam experiment. Few experiments have studied 
cascade baryons with the missing mass technique. In 1983, the production 
of $\Xi^\ast$ resonances up to 2.5~GeV were reported from $p(K^-,K^+)$
reaction from the measurement of the missing mass of
$K^+$~\cite{Jenkins:1983pmm}. The experimental situation with 
$\Omega^{-\ast}$'s is even worse than the $\Xi^\ast$ case, there are 
very few data for excited states. The main reason for such a scarce 
dataset in multi strange hyperon domain is mainly due to very low 
cross section in indirect production with pion or in particular- 
photon beams. Currently only ground state $\Omega^-$ quantum numbers 
are identified. Recently significant progress is made in lattice QCD 
calculations of excited baryon states~\cite{Engel:2013igm,Edwards:2012fxm} 
which poses a challenge to experiments to map out all predicted states  
(for more details see a talk by Robert~Edwards at this workshop). The 
advantage of baryons containing one or more strange quarks for
lattice calculations is that then number of open decay channels is in
general smaller than for baryons comprising only the light u and d
quarks. Moreover, lattice calculations show that there are many states
with strong gluonic content in positive parity sector for all baryons.
The reason why hybrid baryons have not attracted the same attention as
hybrid mesons is mainly due to the fact that they lack manifest
"exotic" character. Although it is difficult to distinguish hybrid
baryon states, there is significant theoretical insight to be gained
from studying spectra of excited baryons, particularly in a framework
that can simultaneously calculate properties of hybrid mesons.
Therefore this program will be very much complementary to the GlueX
physics program of hybrid mesons.

The proposed experiment  with a beam intensity $10^4 K_L$/sec will
result in about $2\times 10^5$~$\Xi^\ast$'s  and $4\times
10^3$~$\Omega^\ast$'s per month.

A similar program for KN scattering is under development at J-PARC
with charged kaon beams~\cite{bib17:2013m}. The current maximum
momentum of secondary beamline of 2~GeV/c is available at the K1.8
beamline. The beam momentum of 2~GeV/c corresponds to $\sqrt{s}$
=2.2~GeV in the $K^-p$ reaction which is not enough to generate 
even the first excited $\Xi^\ast$ state predicted in the quark 
model. However, there are plans to create high energy beamline in 
the momentum range 5-15~GeV/c to be used with the spectrometer 
commonly used with the J-PARC P50 experiment which will lead to 
expected yield of $(3-4)\times10^5$~$\Xi^\ast$'s and 
$10^3$~$\Omega^\ast$'s per month.

Statistical power of proposed experiment with $K_L$ beam at JLab will 
be of the same order as that in J-PARC with charged kaon beam.

An experimental program with kaon beams will be much richer and allows
to perform a complete experiment using polarized target and measuring
recoil polarization of hyperons. This studies are under way to find an
optimal solution for the GlueX setup.

\item \textbf{Summary}

In summary we intend to propose production of high intensity $K_L$ 
beam using photoproduction processes from a secondary Be target. A 
flux as high as $\sim 10^4 K_L$/sec could be achieved. Momenta of 
$K_L$ beam particles will be measured with time-of-flight. The flux 
of kaon beam will be measured through partial detection of $\pi^+
\pi^-$ decay products from their decay to $\pi^+\pi^-\pi^0$ by 
exploiting similar procedure used by LASS experiment at 
SLAC~\cite{Brandenburg:1972pmm}.  Besides using unpolarized liquid 
hydrogen target currently installed in GlueX experiment the 
unpolarized deuteron target may be installed.  Additional studies 
are needed to find an optimal choice of polarized targets. This 
proposal will allow to measure $KN$ scattering with different final 
states including production of strange and multi strange baryons 
with unprecedented statistical precision to test QCD in non 
perturbative domain. It has a potential to distinguish between
different quark models and test lattice QCD predictions for excited
baryon states with strong hybrid content. 

\newpage
\item \textbf{Acknowledgments}

This work is supported, in part, by the U.S. Department of Energy, 
Office of Science, Office of Nuclear Physics, under Award Number 
DE--FG02--96ER40960.
\end{enumerate}


\newpage
\subsection{Overview of Hall~D}
\addtocontents{toc}{\hspace{2cm}{\sl E.~Chudakov}\par}
\halign{#\hfil&\quad#\hfil\cr
\large{Eugene Chudakov}\cr
\textit{Thomas Jefferson National Accelerator Facility}\cr
\textit{Newport News, VA 23606, U.S.A.}\cr}

\begin{enumerate}
\item

Hall~D is a new experimental hall at Jefferson Lab, designed for 
experiments with a photon beam. The primary motivation for Hall~D 
is the GlueX experiment~\cite{Dugger:2012qra,AlekSejevs:2013mkl}, 
dedicated to meson spectroscopy. The Hall~D complex consists of:

\begin{itemize}
\item  An electron beam line used to extract the 5.5-pass electrons from
         the accelerator into the Tagger Hall. The designed beam energy
         is $E_{\circ}=$12~GeV.
\item The Tagger Hall, where the electron beam passes through a thin
         radiator ($\sim{}$0.01\% R.L.) and is deflected into the beam 
	dump. The electrons that lost $>$30\% of their energy in the 
	radiator are detected with scintillator hodoscopes
         providing a $\sim$0.1\% energy resolution for the tagged photons.
         Aligned diamond radiators allow to produce linearly polarized 
	photons via the Coherent Bremsstrahlung. The beam dump is limited 
	to 60~kW (5~$\mu$A at 12~GeV).
\item The Collimator Cave contains a collimator for the photon beam
         and dipole magnets downstream in order to remove charged particles.
         The 3.4~mm diameter collimator, located about 75~m downstream of the 
	radiator, selects the central cone of the photon beam increasing its 
	average linear polarization, up to $\sim$40\% in the coherent peak at 
	9~GeV.
\item Hall~D contains several elements of the photon beam line, and the main 
	spectrometer.
         A Pair Spectrometer consists of a thin converter, a dipole magnet, and
         a two-arm detector used to measure the energy spectrum of the photon beam.
         The main spectrometer is based on a 2-T superconducting solenoid, 4~m long 
	and 1.85~m
         bore diameter. The liquid hydrogen target is located in the front part the
         solenoid. The charged tracks are detected with a set of drift chambers;
         photons are detected with two electromagnetic calorimeters. There are also
         scintillator hodoscopes for triggering and time-of-flight measurements. The 
	spectrometer
         is nearly hermetic in an angular range of $1^{\circ}<\theta{}<120^{\circ}$. 
	The momentum
         resolution is $\sigma{}_P/p\sim{}1-3$\% depending on the polar angle $\theta$. 
	The energy resolution of the electromagnetic calorimeters is about 7\% at 1~GeV.
\end{itemize}

The main spectrometer is designed for photon beam rates below 100~MHz 
in the coherent peak. Such a rate can be provided by a 2.2~$\mu$A beam 
on a 0.02~mm = 0.0125\% R.L. diamond crystal. The 1-st stage of GlueX 
is planned to run at a lower rate of 10~MHz.

Hall D and the GlueX experiment had 3 commissioning runs in 2014-2016.
By April 2016 all the systems have been commissioned at some level
and most of them have reached the specifications. Preliminary results
of the 2014-2015 commissioning have been reported~\cite{Ghoul:2015ifw}.

In addition to the GlueX experiment, two other experiments (both
using Primakoff-type reactions) have been approved by the Program 
Advisory Committee (PAC). In total, about 500 days of effective running
have been approved by the PAC.

\item \textbf{Acknowledgments}

This material is based upon work supported by the U.S. Department of
Energy, Office of Science, Office of Nuclear Physics under contract
DE--AC05--06OR23177.
\end{enumerate}


\newpage
\subsection{Can K-Long Beams Find Missing Hyperon Resonances ?}
\addtocontents{toc}{\hspace{2cm}{\sl D.M.~Manley}\par}
\setcounter{figure}{0}
\setcounter{table}{0}
\setcounter{equation}{0}
\setcounter{footnote}{0}
\halign{#\hfil&\quad#\hfil\cr
\large{D.~Mark Manley}\cr
\textit{Department of Physics}\cr
\textit{Kent State University}\cr
\textit{Kent, OH 44242 U.S.A.}\cr}

\begin{abstract}
The problem of ``missing resonances'' is discussed with particular 
emphasis on missing hyperon resonances.  It is shown that $K_L^0p$ 
reactions would be an ideal means to search for $\Sigma^\ast$ 
resonances due to their isospin-1 selectivity. Production reactions 
will likely be needed in order to search for missing $\Lambda^\ast$ 
and $\Sigma^\ast$ resonances that couple weakly to $K_L^0p$.  
Suggestions are given for specific production reactions that would 
allow searches for missing $\Lambda^\ast$, $\Sigma^\ast$, 
$\Xi^\ast$, and $\Omega^\ast$ resonances.
\end{abstract}

\begin{enumerate}
\item \textbf{Introduction}

It is convenient to describe baryons with a ``quark shell model" in 
which each quark moves in a mean field generated mainly by the 
gluons in the hadron.  Lattice gauge calculations have shown that 
the predicted spectrum of excited states is more-or-less consistent 
with what is expected from SU(6) symmetry.  In such models, baryons 
are grouped into three possible SU(6) multiplets consisting of 
flavor octets, decuplets, or singlets with total quark intrinsic 
spin of either 1/2 (spin doublets) or 3/2 (spin quartets):
\begin{eqnarray*}
	{\bf 56}_S &=& {^2}{8} + {^4}{10}, \\
	{\bf 70}_M &=& {^2}{8} + {^4}{8} + {^2}{10} + {^2}{1}, \\
	{\bf 20}_A &=& {^2}{8} + {^4}{1}. \\
\end{eqnarray*}

The $uds$ basis, which takes into account the heavier mass of the 
strange quark, provides a better physical description of the 
hyperon resonances, but the SU(6) basis is simpler for counting 
the number of states.  In a harmonic-oscillator shell-model 
description, the SU(6) multiplets contained in the first three 
oscillator bands have the following structure~\cite{Faiman1968}:
\begin{eqnarray*}
	N=0 \qquad \psi({\bf 56},0^+) &=& (1s)^3,\\
	N=1 \qquad \psi({\bf 70},1^-) &=& (1s)^2(1p),\\
	N=2 \qquad \psi({\bf 56},0^+) &=& \sqrt{\frac{2}{3}}(1s)^2(2s) + \sqrt{\frac{1}{3}}(1s)(1p)^2, \\
	\psi({\bf 70},0^+) &=& \sqrt{\frac{1}{3}}(1s)^2(2s) + \sqrt{\frac{2}{3}}(1s)(1p)^2, \\
	\psi({\bf 56},2^+) &=& \sqrt{\frac{2}{3}}(1s)^2(1d) - \sqrt{\frac{1}{3}}(1s)(1p)^2, \\
	\psi({\bf 70},2^+) &=& \sqrt{\frac{1}{3}}(1s)^2(1d) - \sqrt{\frac{2}{3}}(1s)(1p)^2, \\
	\psi({\bf 20},1^+) &=& (1s)(1p)^2. \\
\end{eqnarray*}

The eight SU(6) multiplets that are allowed in the $N=3$ band are:
$({\bf 56},1^-)$, 
$({\bf 70},1^-)$, 
$({\bf 70},1^-)$, 
$({\bf 20},1^-)$, 
$({\bf 70},2^-)$, 
$({\bf 56},3^-)$, 
$({\bf 70},3^-)$, and 
$({\bf 20},3^-)$ and the allowed shell-model configurations are listed 
in Table~1.  Here $L$ is the total orbital angular momentum of the 
constituent quarks and the total parity is shown as a superscript.
\begin{table}[htb!]
\begin{center}
\begin{tabular}{lll} \hline\hline
$(1s)^2(2p)$ & \qquad\qquad\qquad & $L=1$ \\
$(1s)^2(1f)$ & & $L=3$ \\ \hline
$(1s)(1p)(2s)$ & & $L=1$ \\
$(1s)(1p)(1d)$ & & $L=1,2,3$ \\ \hline
$(1p)^3$ & & $L=1,3$ \\ \hline\hline
\end{tabular}
\centerline{\parbox{0.80\textwidth}{
\caption[] {\protect Shell-model configurations allowed for baryons 
	in the $N=3$ oscillator band.}}}
\end{center}
\end{table}

Pure hyperon states in the $N=2$ (${\bf 20}$,$1^+$) multiplet cannot 
couple to $\overline{K}N$ via a single-quark transition operator. 
They will not be considered further.  Similarly, pure hyperon states 
in the $N=3$ (${\bf 20}$,$1^-$), (${\bf 70}$,$2^-$), and (${\bf 
20}$,$3^-$) multiplets cannot couple to $\overline{K}N$ via a 
single-quark transition operator.  They will also not be considered 
further.  All baryons predicted to belong to these multiplets are
``missing'' and likely to stay that way.

\item \textbf{Missing Resonances}

\begin{table}[htb!]
\begin{center}
\begin{tabular}{lllllll} \hline\hline
$^2$8 & $N$(939) & **** & \qquad\qquad\qquad & $^{4}$10 & $\Delta$(1232) & **** \\
1/2$^+$ & $\Lambda$(1116) & **** & & 3/2$^+$   & $\Sigma$(1385) & **** \\
& $\Sigma$(1193)   & **** & & & $\Xi$(1530)    & **** \\
& $\Xi$(1322)      & **** & & & $\Omega$(1672) & **** \\ \hline\hline
\end{tabular}
\centerline{\parbox{0.80\textwidth}{
\caption[] {\protect The $N=0$ $({\bf 56}$,0$^+$) ground-state baryons.
        All are well-established 4-star states and none are missing.}}}
\end{center}
\end{table}
\begin{table}[htb!]
\begin{center}
\begin{tabular}{lllllllllllllll} \hline\hline
$^2$8 & $N$(1535) & **** & \qquad & $^{2}$8 & $N$(1520) & **** \\
1/2$^-$ & $\Lambda$(1670) & **** & & 3/2$^-$ & $\Lambda$(1690) & **** \\
& $\Sigma$(1620)& * & & & $\Sigma$(1670) & **** \\
& $\Xi$(1690) & *** & & & $\Xi$(1820) & *** \\ \hline
$^4$8 & $N$(1650) & **** & \qquad & $^{4}$8 & $N$(1700) & *** & \qquad
& $^{4}$8 & $N$(1675) & **** \\
1/2$^-$ & $\Lambda$(1800) & *** & & 3/2$^-$ & $\Lambda$ & & & 5/2$^-$
& $\Lambda$(1830) & ****\\
& $\Sigma$(1750)& *** & & & $\Sigma$ & & & & $\Sigma$(1775)& **** \\
& $\Xi$(1950) & *** & & & $\Xi$ & & & & $\Xi$(2030) & *** \\ \hline\
$^2$10 & $\Delta$(1620) & **** & \qquad & $^{2}$10 & $\Delta$(1700)
& **** & \qquad & $^{2}$1 &
$\Lambda$(1405) & **** & \qquad & $^{2}$1 & $\Lambda$(1520) & ****\\
1/2$^-$ & $\Sigma$ &  & & 3/2$^-$ & $\Sigma$(1940) & *** & & 1/2$^-$
&  &  & & 3/2$^-$ &  & \\
& $\Xi$ &  & & & $\Xi$ & & & &  &  & &  &  & \\
& $\Omega$ &  & & & $\Omega$ & & & & &  & &  & & \\ \hline\hline
\end{tabular}
\centerline{\parbox{0.80\textwidth}{
\caption[] {\protect The $N=1$ $({\bf 70}$,1$^-$) negative-parity 
	excited states. The spin-parity quantum numbers of 
	$\Xi$(1690), $\Xi$(1950), and $\Xi$(2030) are unmeasured.}}}
\end{center}
\end{table}
\begin{table}[htb!]
\begin{center}
\begin{tabular}{lllllll} \hline\hline
$^2$8 & $N$(1440) & **** & \qquad\qquad\qquad & $^{4}$10 & $\Delta$(1600) & *** \\
1/2$^+$ & $\Lambda$(1600) & *** & & 3/2$^+$ & $\Sigma$ &  \\
& $\Sigma$(1660)& *** & & & $\Xi$ &  \\
& $\Xi$ &  & & & $\Omega$ &  \\ \hline\hline
\end{tabular}
\centerline{\parbox{0.80\textwidth}{
\caption[] {\protect The $N=2$ $({\bf 56}$,0$^+$) positive-parity
        excited states.}}}
\end{center}
\end{table}

Tables~2 to 7 compare experimental observations with predictions 
for low-lying states in the first three harmonic-oscillator bands.  
Assignments for some states are educated guesses. Star ratings are 
from the {\it Review of Particle Physics}~\cite{RPP}.  Table~8
gives a summary of the number of missing states for each oscillator 
band.  For counting purposes, 1-star states are included in the 
count for missing states, but 2-star states are not. As one might 
expect, most predicted $\Xi^\ast$ and $\Omega^\ast$ resonances are 
missing. Candidate states have been found for most of the predicted 
$N^\ast$ and $\Delta^\ast$ resonances, if one ignores states in the 
(${\bf 20}$,$1^+$) multiplet. However, a large number of predicted 
$\Sigma^\ast$ resonances are also missing.  Many or most of these 
states might be found using measurements of $K_L^0p$ scattering, 
which is isospin-1 selective and would therefore be an ideal means 
to study $\Sigma^\ast$ resonances.  Further details are provided
in the next section.
\begin{table}[htb!]
\begin{center}
\begin{tabular}{lllllllllllllll} \hline\hline
$^2$8 & $N$(1720) & **** & \qquad & $^{2}$8 & $N$(1680) & **** \\
3/2$^+$ & $\Lambda$(1890) & **** & & 5/2$^+$ & $\Lambda$(1820) & **** \\
& $\Sigma$ &  & & & $\Sigma$(1915) & **** \\
& $\Xi$ &  & & & $\Xi$ &  \\ \hline
$^4$10 & $\Delta$(1910) & **** & \qquad & $^{4}$10 & $\Delta$(1920) 
& *** & \qquad & $^{4}$10 & $\Delta$(1905)
& *** & \qquad & $^{4}$10 & $\Delta$(1905) & ***\\
1/2$^+$ & $\Sigma$ &  & & 3/2$^+$ & $\Sigma$ & & & 5/2$^+$ & $\Sigma$ 
&  & & 7/2$^+$ & $\Sigma$(2030) & ****\\
& $\Xi$ &  & & & $\Xi$ & & & & $\Xi$ &  & &  & $\Xi$ & \\
& $\Omega$ &  & & & $\Omega$ & & & & $\Omega$ &  & &  & $\Omega$ 
& \\ \hline\hline
\end{tabular}
\centerline{\parbox{0.80\textwidth}{
\caption[] {\protect The $N=2$ $({\bf 56}$,2$^+$) positive-parity 
	excited states.}}}
\end{center}
\end{table}
\begin{table}[htb!]
\begin{center}
\begin{tabular}{lllllllllllllll} \hline\hline
$^2$8 & $N$(1710) & *** & \qquad & $^{4}$8 & $N$ \qquad\qquad &  & \qquad & $^{2}$10 & $\Delta$(1750)
& * & \qquad & $^{2}$1 & $\Lambda$(1710) & *\\
1/2$^+$ & $\Lambda$(1810) & *** & & 3/2$^+$ & $\Lambda$ & & & 1/2$^+$ & $\Sigma$ &  & & 1/2$^+$ &  & \\
& $\Sigma$(1880) & ** & & & $\Sigma$ & & & & $\Xi$ &  & &  &  & \\
& $\Xi$ &  & & & $\Xi$ & & & & $\Omega$ &  & &  &  & \\ \hline\hline
\end{tabular}
\centerline{\parbox{0.80\textwidth}{
\caption[] {\protect The $N=2$ $({\bf 70}$,0$^+$) positive-parity excited 
	states.}}}
\end{center}
\end{table}
\begin{table}[htb!]
\begin{center}
\begin{tabular}{lllllllllllllll} \hline\hline
$^2$8 & $N$ &  & \qquad & $^{2}$8 & $N$(1860) & * \\
3/2$^+$ & $\Lambda$ &  & & 5/2$^+$ & $\Lambda$ &  \\
& $\Sigma$ &  & & & $\Sigma$ &  \\
& $\Xi$ &  & & & $\Xi$ &  \\ \hline
$^4$8 & $N$(1880) & ** & \qquad & $^{4}$8 & $N$(1900) & *** & \qquad & $^{4}$8 & $N$(2000) & ** &
\qquad & $^{4}$8 & $N$(1990) & **\\
1/2$^+$ & $\Lambda$ &  & & 3/2$^+$ & $\Lambda$ & & & 5/2$^+$ & $\Lambda$ &  & & 7/2$^+$ & $\Lambda$(2020) & *\\
& $\Sigma$ &  & & & $\Sigma$ & & & & $\Sigma$ &  & &  & $\Sigma$ & \\
& $\Xi$ &  & & & $\Xi$ & & & & $\Xi$ &  & &  & $\Xi$ & \\ \hline
$^2$10 & $\Delta$ &  & \qquad & $^{2}$10 & $\Delta$(2000) & ** & \qquad & $^{2}$1 & $\Lambda$ &  &
\qquad & $^{2}$1 & $\Lambda$(2110) & ***\\
3/2$^+$ & $\Sigma$ &  & & 5/2$^+$ & $\Sigma$ &  & & 3/2$^+$ &  &  & & 5/2$^+$ &  & \\
& $\Xi$ &  & & & $\Xi$ & & & &  &  & &  &  & \\
& $\Omega$ &  & & & $\Omega$ & & & & &  & &  & & \\ \hline\hline
\end{tabular}
\centerline{\parbox{0.80\textwidth}{
\caption[] {\protect The $N=2$ $({\bf 70}$,2$^+$) positive-parity 
	excited states.}}}
\end{center}
\end{table}
\begin{table}[htb!]
\begin{center}
\begin{tabular}{ccccccc}\hline\hline
Baryon & \qquad\qquad  & $N=0$ & \qquad\qquad & $N=1$ & \qquad\qquad & $N=2$ \\ \hline
$N$       & & 0 & & 0 & &  2 \\
$\Delta$  & & 0 & & 0 & &  2 \\
$\Lambda$ & & 0 & & 1 & &  9 \\
$\Sigma$  & & 0 & & 3 & & 15 \\
$\Xi$     & & 0 & & 3 & & 19 \\
$\Omega$  & & 0 & & 2 & &  8 \\ \hline\hline
\end{tabular}
\centerline{\parbox{0.80\textwidth}{
\caption[] {\protect Summary of numbers of low-lying missing baryon 
	resonances. For counting purposes, 1-star states are included 
	as ``missing" but 2-star states are not.  All predicted states 
	in the $N=2$ (${\bf 20}$,$1^+$) multiplet are missing but are 
	not included in these totals.}}}
\end{center}
\end{table}

\item \textbf{PWA Formalism}

Here, we summarize some of the physics issues involved with 
$K^0_L p$ scattering. The differential cross section and 
polarization for $K^0_L p$ scattering are given by 
$$\frac{d\sigma}{d\Omega} = \lambdabar^2 (|f|^2 + |g|^2),$$
$$P \frac{d\sigma}{d\Omega} = 2\lambdabar^2 {\rm Im}(fg^\ast),$$
where $\lambdabar = \hbar/k$, with $k$ the magnitude of c.m.\ 
momentum for the incoming meson.  Here $f = f(W,\theta)$ and 
$g = g(W,\theta)$ are the usual spin-nonflip and spin-flip 
amplitudes at c.m.\ energy $W$ and meson c.m.\ scattering 
angle $\theta$.

In terms of partial waves, $f$ and $g$ can be expanded as
$$
	f(W,\theta) = \sum_{l=0}^\infty [(l+1)T_{l+} + 
	lT_{l-}]P_l(\cos\theta),
$$
$$
	g(W,\theta) = \sum_{l=1}^\infty [T_{l+} - T_{l-}]P_l^1(\cos\theta).
$$
Here $l$ is the initial orbital angular momentum, $P_l(\cos\theta)$ 
is a Legendre polynomial, and $P_l^1(\cos\theta) = \sin\theta \times 
dP_l(\cos\theta)/d(\cos\theta)$ is an associated Legendre function.
The total angular momentum for $T_{l+}$ is $J=l+\frac{1}{2}$, while 
that for $T_{l-}$ is $J=l-\frac{1}{2}$.

We may ignore small CP-violating terms and write
$$
	K_L^0 = \frac{1}{\sqrt{2}} (K^0 - \overline{K^0}) \qquad {\rm and} \qquad
	K_S^0 = \frac{1}{\sqrt{2}} (K^0 + \overline{K^0}).
$$

We have both $I=0$ and $I=1$ amplitudes for $KN$ and $\overline{K}N$ 
scattering, so that amplitudes $T_{l\pm}$ can be expanded in isospin 
amplitudes as 
$$
	T_{l\pm} = C_0 T^0_{l\pm} + C_1 T^1_{l\pm},
$$
where $T_{l\pm}^I$ are partial-wave amplitudes with isospin $I$ and 
total angular momentum $J = l \pm \frac{1}{2}$, with $C_I$ the 
appropriate isospin Clebsch-Gordon coefficients. The amplitudes for 
$K_L^0 p$ reactions leading to two-body final states are:
\begin{eqnarray*}
	T(K_L^0p\to K_S^0 p) &=& \frac{1}{2} \left (\frac{1}{2}T^1(KN \to KN)
	+ \frac{1}{2}T^0(KN \to KN) \right )\\ &-& \frac{1}{2}T^1({\overline K}N \to {\overline K}N), \\
	T(K^0_Lp\to \pi^+ \Lambda) &=& -\frac{1}{\sqrt{2}}T^1({\overline K}N \to \pi\Lambda),\\
	T(K^0_Lp\to \pi^+ \Sigma^0) &=& -\frac{1}{2}T^1({\overline K}N \to \pi\Sigma),\\
	T(K^0_Lp\to \pi^0 \Sigma^+) &=& \frac{1}{2}T^1({\overline K}N \to \pi\Sigma),\\
	T(K^0_Lp\to K^+\Xi^0) &=& -\frac{1}{\sqrt{2}}T^1({\overline K}N \to K\Xi).\\
\end{eqnarray*}

These amplitudes show that only $\Sigma^\ast$ resonances are formed as 
$s$-channel intermediate states in $K_L^0 p$ reactions. The reaction 
$K_L^0 p \to K_S^0 p$ is not ideal for finding missing $\Sigma^\ast$ 
states that couple weakly to $\overline{K}N$ because of nonresonant 
$KN$ background and because the amplitude involves $\overline{K}N$ in 
both initial and final states.  The inelastic two-body reactions 
listed above would be better probes for finding missing $\Sigma^\ast$ 
states due to isospin selectivity, absence of nonresonant $KN$ 
background, and the fact that their amplitudes only involve 
$\overline{K}N$ coupling in the initial state.

To search for missing $\Sigma^\ast$ states that couple weakly to 
$\overline{K}N$, one should also use production reactions such as 
$K_L^0 p \to \pi^+ \Sigma^{0*}$, with $\Sigma^{0*}\to \pi^0 \Lambda$, 
or use $K_L^0 p \to \pi^0  \Sigma^{+*}$, with $\Sigma^{+*} \to \pi^+
\Lambda$. Note that the $\pi \Lambda$ decays establish $\Sigma^\ast$ 
states ($I=1$) uniquely. In principle, one could search for 
$\Lambda^\ast$ resonances using $K_L^0 n$ reactions, but doing so 
presents additional problems such as Fermi motion smearing and the 
need to identify quasi-free $K_L^0 n$ reactions from measurements on 
a deuterium target. To search for missing $\Lambda^\ast$ states that 
couple weakly to $\overline{K}N$, one should instead use production 
reactions such as $K_L^0 p \to \pi^+ \Lambda^\ast$, with $\Lambda^\ast
\to\pi^+\Sigma^-$ , $\Lambda^\ast\to\pi^-\Sigma^+$, or $\Lambda^\ast
\to\pi^0\Sigma^0$. Note that the $\pi^0 \Sigma^0$ decays establish 
$\Lambda^\ast$ states ($I=0$) uniquely. Narrow states could be 
identified using missing-mass techniques, but the determination
of spin-parity quantum numbers would need a partial-wave analysis 
analogous to those used in meson spectroscopy studies. To search 
for missing $\Xi^\ast$ or $\Omega^\ast$ states, one needs production 
reactions such as $K^0_Lp\to K^+\Xi^{0*}$, $K^0_Lp\to\pi^+K^+
\Xi^{-*}$, and $K^0_L p\to K^+K^+\Omega^{-*}$.

\item \textbf{Summary and Conclusions}

New data for inelastic $K_L^0 p$ scattering would significantly 
improve our knowledge of $\Sigma^\ast$ resonances.  To search for 
missing hyperon resonances, we need measurements of production 
reactions:

\begin{tabular}{ll}
	$\Sigma^\ast$:  & $K_L^0p\to\pi\Sigma^\ast\to\pi\pi\Lambda$, \\
	$\Lambda^\ast$: & $K_L^0p\to\pi\Lambda^\ast\to\pi\pi\Sigma$, \\
	$\Xi^\ast$:     & $K_L^0p\to K\Xi^\ast$, $\pi K\Xi^\ast$, \\
	$\Omega^\ast$:  & $K_L^0p\to K^+K^+\Omega^\ast$. \\
\end{tabular}

If such measurements can be performed with good energy and angle 
coverage and good statistics, then it is very likely that 
measurements with $K_L^0$ beams would find several missing hyperon 
resonances.

\item \textbf{Acknowledgments}

The author thanks Dr.~Igor~Strakovsky and the other organizers 
for inviting him to speak at the workshop. This material is based 
upon work supported by the U.S. Department of Energy, Office of 
Science, Office of Medium Energy Nuclear Physics, under Award No. 
DE--SC0014323.
\end{enumerate}


\newpage
\subsection{The Quark Model and the Missing Hyperons}
\addtocontents{toc}{\hspace{2cm}{\sl F.~Myhrer}\par}
\setcounter{figure}{0}
\setcounter{table}{0}
\setcounter{equation}{0}
\setcounter{footnote}{0}
\halign{#\hfil&\quad#\hfil\cr
\large{Fred Myhrer}\cr
\textit{Department of Physics and Astronomy}\cr
\textit{University of South Carolina}\cr
\textit{Columbia, SC 29208, U.S.A.}\cr}

\begin{abstract}
The  mass spectrum of the first excited negative parity $Y^\ast$ states 
is presently not well established. The missing hyperon states will 
provide important clues about the dominant dynamics among the light 
quarks ($u$, $d$, and $s$) including their confinement properties. 
Accurate information of the partial decays of the first excited 
negative parity $\Lambda^\ast$ and $\Sigma^\ast$ states is extremely 
valuable in determining the main features of the interactions among 
the quarks. These quark-model  interactions are due to gluon-  and 
pseudo-meson exchanges. The latter is required by  chiral symmetry. 
As will be presented, these interactions will strongly affect the 
partial decay widths of these hyperons.
\end{abstract}

\begin{enumerate}
\item \textbf{The Quark Models}

The interactions among the quarks are guided by the dynamics and 
symmetry properties of QCD. For high four-momentum hadronic processes 
where the strong coupling, $\alpha_s$, is small, perturbative QCD is 
applicable and has been successful in describing these processes.
However, at low energy $\alpha_s$ is not small and a model approach 
is necessary in order to study the ground- and excited states of 
baryons.  In order to build a reasonable model we need to investigate 
which dynamical forces dominate among the ``light" quarks, $u$, $d$ 
and $s$.  The $\Lambda^\ast$ and $\Sigma^\ast$ states have one $s$ 
quark and two almost massless $u$ and $d$ quarks. Could this extra 
feature of having one quark with mass $m_s \gg m_u, m_d$ give us 
some extra insight into QCD beyond what the $N^\ast$ and 
$\Delta^\ast$ states can provide?  These quarks have current masses 
smaller than the QCD scale $\Lambda_{QCD}$, and it appears reasonable 
to assume that chiral symmetry considerations will be of importance 
for baryons involving these three light quarks.  As we will discuss, 
chiral symmetry requirements introduce quark-interactions which 
affect the structure and the decays of the excited baryons.

A quark model for the hyperons includes quark confinements and 
effective quark-quark interactions. One presumes that the exchange 
of multiple soft gluons including self-interactions among the gluons 
will give raise to quark confinement. However, quark confinement is 
not really understood at present.  The non-relativistic quark model 
(NRQM) of Isgur and Karl and their coworkers~\cite{IsgurKarlf} uses 
the harmonic oscillator potential to simulate confinement and the 
effective one-gluon exchange (OGE) to describe the spin-dependent 
quark interactions. This NRQM is reasonably successful in reproducing 
the measured baryon mass-spectrum and the decays of the excited 
baryons, see also the presentation by Capstick~\cite{Capstickf}. The 
NRQM of Isgur and Karl was the main focus of an earlier conference
presentation~\cite{K0L-myhrerf}.

In this talk, I will concentrate on the MIT bag model where a 
confinement condition is imposed on the relativistic $u$, $d$ and $s$ 
quarks. In the bag model chiral symmetry is implemented 
phenomenologically by requiring the axial current to be continuous. 
This means that outside the confinement region the axial current is 
carried by a pseudo-scalar meson cloud surrounding the three-quark 
bag (the ``bare" baryon). In the chiral (cloudy) bag model these 
pseudo-scalar mesons mimic $\bar{q} q$ $0^-$ excitations, and 
effectively introduces spin-dependent interactions among the quarks 
with a coupling strength determined by the axial current.  In 
addition, an effective interaction due to one-gluon exchange among 
the confined, relativistic quarks is included. This chiral (cloudy) 
bag model will predict the decay channel widths of excited baryons 
to ground state baryons plus a meson ($\pi$, $K$ and/or $\eta$) or 
their electromagnetic transition to a lower baryon state.  In this 
paper, I will concentrate only on the  \underline{first} excited 
hyperons, the negative parity $\Lambda^\ast$ and $\Sigma^\ast$  
states, and make some observation on what we could learn about the 
effective quark dynamics in a quark model. We will assume that these 
hyperons are mainly composed of three valence quarks and present how 
the quark interactions will change the hyperon states' compositions 
and their partial decays. Some comments regarding the $\Lambda(1405)$ 
will be made.

According to the quark model there are several states among the first 
excited hyperon states which are missing or are not established. A 
natural question is: Do QCD require that these missing states should 
exist, or does QCD require further model restrictions not implemented 
in todays quark models? At this conference, we learned that in order 
to understand measurements in heavy-ion collisions, the number of 
hyperon states with masses below about 2.5~GeV is very important,
e.g., see talk by Rene~Bellwied~\cite{Bellwiedf}.  As presented, it 
appears that the heavy-ion data require more hyperon states than
what have been listed by PDG~\cite{PDGf}.

The generic non-relativistic baryon wave function has the following 
structure
\begin{eqnarray}
	\Psi =  \Psi_{color}
	\;\Psi_{flavor}\; \Psi_{spin}\;\Psi_{space}\; .
	\label{eq:baryonwf}
\end{eqnarray}
We assume that isospin is a good symmetry, {\it i.e.}, in the bag model
the $u$ and $d$ quarks are massless, {\it i.e.,} $m_u$ = $m_d$ = $m_q$ 
=$0$. Since the $s$ quark has a mass $m_s > m_q$, SU$_F$(3) is a broken 
symmetry and we therefore adopt the $uds$ basis when  the baryon wave 
functions are determined. Other quark model assumptions are:
\begin{itemize}
\item All hadrons are $SU(3)$-color singlets, {\it i.e.},  $\Psi_{color}$ 
	is a totally anti-symmetric wave function under the interchange 
	of any two quarks.
\item Confinement of quarks is universal condition, the same for all 
	quark flavors.
\item The Pauli principle requires that two identical quarks must have 
	a totally anti-symmetric  wave function. Since $\Psi_{color}$ 
	is anti-symmetric the product of the other components in 
	Eq.(\ref{eq:baryonwf}) must be symmetric under the interchange 
	of any two quarks.
\item Non-relativistic quarks interact via an effective one-gluon-exchange 
	a la De Rujula {\it et al.}~\cite{DeRujulaf}.  This effective 
	gluon exchange generates a \underline{spin-dependent interaction}
	among the quarks and makes the {\it decuplet}
	baryons \underline{heavier} than the {\it octet} baryons.
\end{itemize}
The non-relativistic effective  OGE between quarks $i$ and $j$ is:
\begin{eqnarray}
	H_{hyp}^{ij} &=& A_{ij} \left\{
	\frac{8\pi}{3} \vec{S}_i\cdot \vec{S}_j\delta^3(\vec{r}_{ij})
	+\frac{1}{r_{ij}^3}\left(
	\frac{ 3 (\vec{S}_i\cdot \vec{r}_{ij})(\vec{S}_j\cdot\vec{r}_{ij})}{r_{ij}^2}
	- \vec{S}_i\cdot \vec{S}_j
	\right) \right\} \, ,
	\label{eq:hyperfine}
\end{eqnarray}
where $A_{ij}$ is a constant which depends on the constituent quark 
masses~\cite{DeRujulaf}.  Note that  this non-relativistic reduction of 
the effective OGE quark-quark interaction neglects the spin-orbit force.
Isgur and Karl argue that the spin-orbit force should be small. This 
argument is confirmed in the cloudy bag model evaluation where the 
relativistic quark P-state with j=3/2 has a lower energy than  j=1/2, 
{\it i.e.}, the bag model's confinement condition introduces an 
effective spin-orbit splitting of quark states. Fortunately, in the 
chiral bag the  relativistic OGE introduces a spin-orbit force of 
opposite sign and  basically cancel the one from confinement.
{\it Effectively what remains are the spin-spin and tensor interactions
due to OGE and  the pseudo-scalar meson (e.g., pion, K) cloud surrounding 
the quark core}, see for example Refs.~\cite{MyhrerWroldsenf,um91f}.

The  spin-spin and the tensor quark-quark interactions in 
Eq.(\ref{eq:hyperfine}) are closely related, and the tensor component 
will produce a spatial D-state quark wave function. These interactions 
are  similar to the meson exchange forces between nucleons in $^3H$ 
and $^3He$, and analogously we expect the three valence quark baryon 
ground states to have similarly mixed three-quark spatial wave 
functions, {\it i.e.,} a mixture of  S, S$^\prime$ and D quark states.
For example, due to $H_{hyp}^{ij}$ Isgur and Karl~\cite{IsgurKarlf} 
find  that the nucleon  has the  structure:
\begin{eqnarray}
	| N \rangle &\simeq & 0.90 | ^2S_S\rangle -0.34 | ^2S_S^\prime\rangle
	-0.27 | ^2S_M\rangle -0.06 | ^2D_M\rangle \, ,
	\label{eq:N}
\end{eqnarray}
where $ |S^\prime \rangle $ and $|D\rangle$ are the excited S- and D- 
quark states of the harmonic oscillator. The subscripts $S$ and $M$ 
denote symmetric and mixed symmetry spatial states, respectively.
Similarly, Isgur and Karl find the $\Lambda(1116)$ state to be:
\begin{eqnarray}
	| \Lambda \rangle &\simeq & 0.93 | ^2S_S\rangle -0.30 | ^2S_S^\prime\rangle
	-0.20 | ^2S_M\rangle -0.03 | ^4D_M\rangle -0.05 |\mathbf{1}, ^2S_M\rangle .
	\label{eq:L}
\end{eqnarray}
The effective pseudo-scalar meson cloud of the chiral bag model 
surrounding the quark core of the baryons will also contribute to the 
spatial mixture of states. The spatial admixtures in the baryon ground 
states beyond the completely symmetric $| ^2S_S\rangle$ state strongly 
affect some excited hyperon partial decay widths. In the next section, 
we will focus on the cloudy bag model evaluation of the spin-flavor 
admixture of the excited hyperon states.

\item \textbf{Some Excited Hyperon states and Their Decays}

The main decay channels of the some of the negative parity $\Lambda^\ast$ 
and the $\Sigma(1775)$ states are tabulated in Table~1.  The partial 
widths of the $\Lambda^\ast$ states to the open meson baryon ground 
states and the $\Sigma(1775)$ decay to $\bar{K} N$ are from 
Ref.~\cite{PDGf}. Except for the two lowest $\Lambda^\ast$ mass states, 
the table makes clear that most of the partial widths are not well 
determined.  The masses of these first excited hyperon states are shown 
in Fig.~\ref{fig:masses} where the uncertainties in the masses are 
indicated by the heights of the boxes. Also included are the two states 
given three stars in the PDG classification. In Fig.~\ref{fig:masses} 
the boxes with the question marks have two or one star PDG 
classification~\cite{PDGf}. Two states ``predicted" by the quark model, 
one $\Lambda^\ast$ and one $\Sigma^\ast$, are completely missing in 
Ref.~\cite{PDGf}, {\it i.e.,} they have not been observed. The mass 
spectrum and the decay widths into specific final states can be 
strongly influenced by the spin-flavor admixtures as well as the 
spatial admixtures of the ground state baryons illustrated in 
Eqs.~(\ref{eq:N}) and (\ref{eq:L}).
\newpage
 \begin{center}
  Table~1 : \parbox[t]{5.3in}{ Ref.~\protect\cite{PDGf} gives the 
	hadronic \underline{decay branching ratios} for  some 
	$\Lambda^\ast$ states and $\Sigma(1775)$.}
\end{center}
  $$
  \begin{array}{|| l ||r ||r ||c ||c ||}
\hline\hline
  \Lambda^\ast - j^P & {\rm Decay} \;\;  {\rm Product} &  \% & {\rm Comment} \\
\hline \hline
 \Lambda(1405) - \frac{1}{2}^-\; \; \; &  \pi\Sigma& 100 & K^- p\;\; {\rm bound}\;{\rm  state}? \\ 
\hline
\Lambda(1520) - \frac{3}{2}^-\; & \bar{K} N& 45\pm 1 & - \\ 
\hline
 \; \; \; & \pi\Sigma & 42\pm 1  &  - \\ 
\hline
\; \; \; & \pi\pi\Lambda &10\pm 1 &  {\rm via}\;\; \pi\Sigma(1385) \; {\rm mainly}? \\ 
\hline
\; & \pi\pi\Sigma  & 0.9 \pm 0.1 & \gamma\Lambda\sim 0.85\% \\ 
\hline
\Lambda(1670) - \frac{1}{2}^- \; &  \bar{K} N & 20 - 30 & - \\ 
\hline
\; & \pi\Sigma  &  25 - 55 & - \\ 
\hline
\; \; \; & \eta\Lambda   &  10 - 25 &  - \\ 
\hline
 \Lambda(1690) - \frac{3}{2}^- \; & \bar{K} N & 20 - 30 & - \\ 
\hline
  \; & \pi\Sigma & 20 - 40 & - \\ 
\hline
   \; & \pi\pi\Lambda & \sim 25 & - \\ 
\hline
    \; & \pi\pi\Sigma & \sim 20 & - \\ 
\hline
     \; & \eta\Lambda & ? & - \\ 
\hline
 {\bf  \Lambda(1850) - \frac{5}{2}^-} & {\bf \bar{K} N} & 3 - 10 & - \\ 
\hline
  \; & \pi\Sigma & 35 - 75 & - \\ 
\hline
   \; & \eta\Lambda & ? & - \\ 
\hline
  {\bf  \Sigma(1775) - \frac{5}{2}^-} & {\bf \bar{K} N} & 10 - 40 & - \\ 
\hline\hline
    \label{tabl:one}
\end{array}
$$

This section will highlight results obtained in the chiral (cloudy) 
bag model. In a bag model evaluation the three valence quarks move 
relativistically. We assume that one valence quark is in one of the 
two possible $P$-states ($P_{1/2}$ or $P_{3/2}$). The two others are 
in the $S$ state. We further assume that since the three-quark 
center-of-mass (c.m.) operator is a symmetric operator acting on 
the dominant symmetric three-quark ground state, the symmetric 
spatial wave function of the excited hyperons will describe their 
c.m. motion. The first excited negative parity $\Lambda^\ast$ and 
$\Sigma^\ast$ states therefore have a wave function, $\Psi_{space}$, 
which has mixed spatially symmetry. This will place restrictions on 
the possible flavor and spin wave function given in 
Eq.~(\ref{eq:baryonwf}). Note, for example, that the $\Lambda(1850)$ 
or the $\Sigma(1775)$ $\frac{5}{2}^-$- states are pure $|^48\rangle$ 
states, {\it i.e.}, their spin wave function, $\Psi_{spin}$, is 
completely symmetric.
\begin{figure}
\begin{center}
 \includegraphics[width=6in]{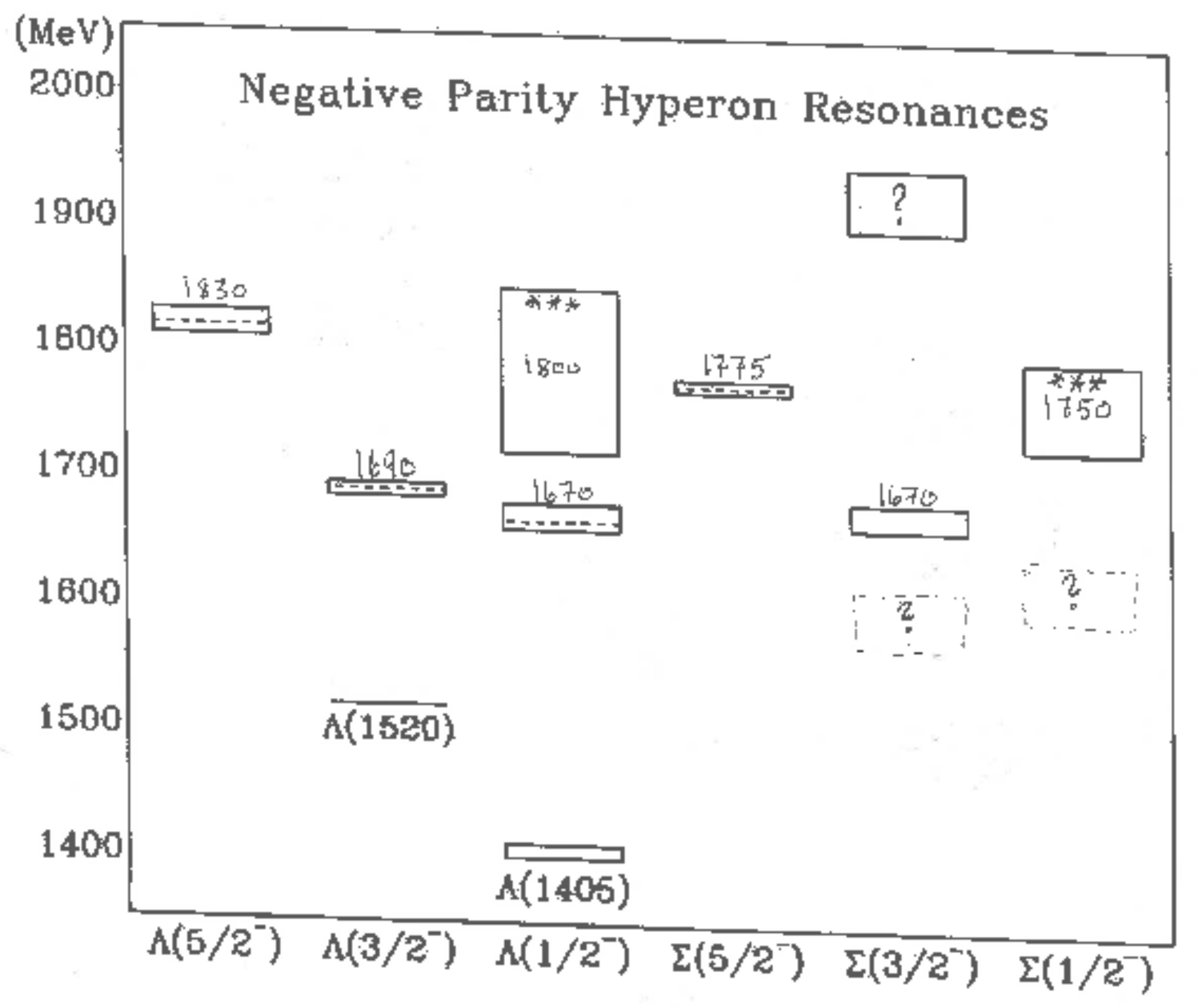}
\end{center}

\caption{The confirmed mass spectrum of the first excited negative
        parity $\Lambda^\ast$ and $\Sigma^\ast$ states where we
        have included  three-star states in Ref.~\protect\cite{PDGf}.
        The \underline{height} of the boxes illustrate the
        \underline{mass} \underline{uncertainties} of the
        ``established" states.  The boxes with the question marks
        are states which are controversial. According to quark
        models there are two completely missing three-quark states
        in this figure.} \label{fig:masses}
\end{figure}

In this bag model it is the $P$-state quark which couples to the 
outgoing meson ($\pi$, $K$, or $\eta$) or the emitted photon in 
the decay of this excited state. The quark model predicts three 
possible $\Lambda^\ast$ $\frac{3}{2}^-$-states and three possible 
$\Lambda^\ast$ $\frac{1}{2}^-$-states, and  gives the following 
general decomposition in terms of spin and SU$_F$(3) multiplets:
\begin{eqnarray}
	| \Lambda^\ast \rangle & = & a | ^2\mathbf{1}\rangle +
	b | ^4\mathbf{8}\rangle +
	c | ^2\mathbf{8}\rangle \; .
	\label{eq:Lsymm}
\end{eqnarray}
The corresponding $\Sigma^\ast$-states have the general structure:
\begin{eqnarray}
	| \Sigma^\ast\rangle &\simeq & \tilde{a} | ^2\mathbf{10}\rangle 
	+\tilde{b} | ^4\mathbf{8}\rangle +
	\tilde{c} | ^2\mathbf{8}\rangle .
	\label{eq:Ssymm}
\end{eqnarray}
Different quark models give different values for the sets of the 
three coefficients. Using the harmonic oscillator of the NRQM of 
Isgur and Karl find $a=0.92 $, $b= -0.04  $ and $c= 
0.39$~\cite{IsgurKarlf}, which result in the following values for 
the two electromagnetic widths: $\Gamma [\Lambda(1520)\to\Lambda 
\gamma] = 96$~keV and $\Gamma [\Lambda(1520)\to \Sigma^0 \gamma] 
= 74$~keV~\cite{dhk83f}. The cloudy bag model calculation produces 
the set of values $a= 0.95$, $b= -0.09 $ and $c= 0.30 $, which 
gives for the same photon decay widths:
$\Gamma [\Lambda(1520)\to \Lambda \gamma] = 32$~keV,
$\Gamma [\Lambda(1520)\to \Sigma^0 \gamma] = 49$~keV~\cite{um91f}.
In other words, a small change in the spin-flavor mixture of the 
$\Lambda^\ast$ states will strongly influence the decay widths of 
these states.

In order to further dissect  these results, we list in Table~2 
the cloudy bag model intermediate results of Ref.~\cite{um91f} 
for the electromagnetic decay widths from each of the spin-flavor 
components of two  excited $\Lambda^\ast$ states.  Further it is 
assumed that both $\Lambda(1116)$ and  $\Sigma^0$ ground states 
have a totally symmetric spatial $|^2S_S\rangle$ state\footnote{
Including the configuration mixing in, e.g., $\Lambda(1116)$, 
Eq.~(\ref{eq:L}) may change $\Gamma_\gamma$ by 50\% or more,
see, e.g., the discussion in Ref.~\cite{K0L-myhrerf}.}.
(Note that in the cloudy bag model the photon is also emitted 
from the virtual meson cloud.) Comparing these results one  
infers that these decay rates are very sensitive to the 
coefficients $a$, $b$, and $c$ in Eq.(\ref{eq:Lsymm}).
\begin{center}
  Table~2 : \parbox[t]{5.3in}{
	The pure spin-flavor radiative decay widths of $\Lambda(1520)$ 
	and $\Lambda(1405)$ in keV to ground state hyperons are 
	extracted from evaluation-notes of Ref.~\protect\cite{um91f}.
	Each column assumes the excited $\Lambda^\ast$ is only in the 
	indicated pure spin-flavor state.}
\end{center}
$$
  \begin{array}{|l|r | |r|r|r|c|r|}
\hline\hline
 {\rm Transition } \; \; \; \,  &|^21\rangle 
 & |^48\rangle & |^28\rangle \\  
\hline\hline
 \Lambda(1520)\to\gamma\Lambda  & 24 & 21   & 23 \\ 
\hline
\Lambda(1520)\to\gamma\Sigma^0  & 91 & 58   & 93 \\ 
\hline
\Lambda(1405)\to\gamma\Lambda   & 42 & 0.2  & 33 \\ 
\hline
\Lambda(1405)\to\gamma\Sigma^0  & 98 & 0.03 & 91 \\ 
\hline\hline
\end{array}
$$
\begin{figure}
\begin{center}
 \includegraphics[width=6in]{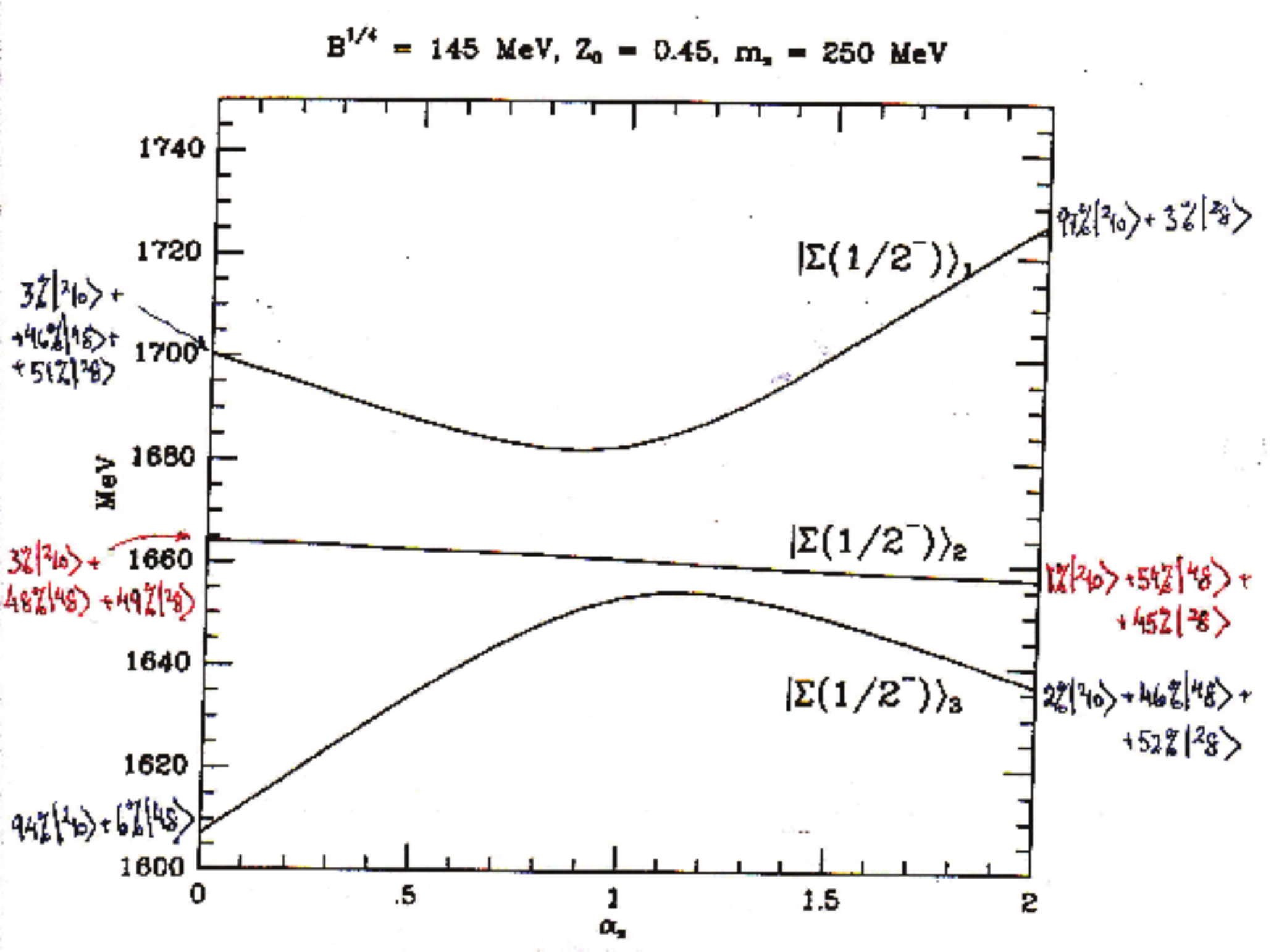}

\caption{ The intermediate evaluated masses of the three
	$\Sigma^\ast$ $\frac{1}{2}^-$ states as a function 
	of the strength $\alpha_s$ of the effective OGE.
	The spin-flavor multiplet contents of the states 
	are listed for $\alpha_s=0$ and $\alpha_s=2.0$ and
	indicate how the spin-flavor of the states changes 
	with $\alpha_s$. The pseudo-scalar cloud interaction 
	with the quarks also affects the flavor-spin 
	multiplet content of the states. } \label{fig:S12}
\end{center}
\end{figure}

The admixture of the spin-flavor multiplets in excited hyperon 
states changes with the interactions among quarks. This is 
illustrated in Fig.~\ref{fig:S12} where we plot and list the 
masses of the three $\Sigma^\ast$ $\frac{1}{2}^-$ states versus 
the strength $\alpha_s$ of the effective OGE interaction taken 
from the intermediate evaluations of Ref.~\cite{um91f}. As a 
function of the strength of the meson-cloud interaction one 
observes similar behaviors. (As stated earlier, the coupling of 
the meson cloud is fixed by the continuity requirement of the 
axial current.) The graphs in Fig.~\ref{fig:S12} give the 
intermediate hyperon masses for bag model parameters, 
$B^{1/4}=145$~MeV, ``zero-point" energy parameter $Z_0=0.45$ and 
strange quark mass $m_s=250$~MeV, see Ref.~\cite{um91f}. Listed 
are the spin flavor compositions of the three states for 
$\alpha_s=0$ and $\alpha_s=2$. The figure shows how  rapidly the 
spin flavor composition can change with increasing strength of 
the OGE interaction, $\alpha_s$. Ref.~\cite{um91f} found a 
reasonable hyperon mass spectrum (apart from the lowest 
$\Lambda^\ast$ $\frac{1}{2}^-$ state mass) for $\alpha_s=1.5$.

\newpage
\item \textbf{An Illuminating Hadronic Decay Rate Argument}

Following Ref.~\cite{IsgurKarlf}, the spatial wave function of 
three quarks is given by the two relative coordinates between 
the three quarks
\begin{eqnarray}
	\vec{\rho} &=& \left(\vec{r}_1 -\vec{r}_2\right)/\sqrt{2} \; \; 
	{\rm and} \; \;
	\vec{\lambda} = \left(\vec{r}_1 +\vec{r}_2 -
	2\vec{r}_3\right)/\sqrt{6} .
	\label{eq:lambda}
\end{eqnarray}
Here quarks 1 and 2 are the $u$ and $d$ quarks, and quark 3 is the 
$s$-quark. In the NRQM, the corresponding reduced constituent quark 
masses are $m_\rho=m_q^{con}$ and $m_\lambda= 3m_q^{con} 
m_s^{con}/(2m_q^{con}+m_s^{con})$. As presented by Capstick~\cite{Capstickf},
the harmonic oscillator potential gives a difference in $\rho$ and 
$\lambda$ oscillator frequencies, $\omega_\rho$ and $\omega_\lambda$ 
due to the difference in the two reduced quark masses~\cite{IsgurKarlf} 
resulting in a hyperon mass difference of,
\begin{eqnarray}
	\hbar (\omega_\rho-\omega_\lambda) = \hbar \omega_\rho \left[ 1-
	\left(\frac{2(m_q^{con}/m_s^{con})+1}{3}\right)^{1/2}\right]  
	\simeq 75 \ {\rm MeV} \, .
	\label{eq:frequencies}
\end{eqnarray}
As shown by Capstick~\cite{Capstickf} and also in a talk at the KL2016 
Workshop~\cite{K0L-myhrerf}, the mass splitting between the 
$\Lambda^\ast(5/2^-)$ and the $\Sigma^\ast(5/2^-)$ states, shown in 
Fig.~1, can easily be understood by this difference in frequencies.
($H_{hyp}$ will modify the mass difference between the two 
$(5/2)^-$ states~\cite{IsgurKarlf}.)

The difference in the spatial decomposition of the two $J^P= 5/2^-$ 
states' wave functions has the following decay implications:  The spin 
wave function, $\Psi_{spin}$, of both $J^P= 5/2^-$ states is completely 
symmetric. The $\Lambda^\ast(1830)$ is an isospin singlet and therefore 
$\Psi_{space}$ of Eq.~(\ref{eq:baryonwf}) must have the anti-symmetric  
$\vec{\rho}$-dependence, {\it i.e.,} it decouples from  $\bar{K} N$ 
since the nucleon spatial wave function is symmetric under the 
interchange $1\leftrightarrow 2$.  The $\Sigma^\ast(1775)$ is symmetric 
in isospin and therefore $\Psi_{space}$ must be symmetric, {\it i.e.,} 
it has a symmetric $\vec{\lambda}$-dependence and couples easily to 
$\bar{K} N$. When  modifications due to $H_{hyp}$ is included, the 
difference in the  two observed decays widths are explained~\cite{IsgurKarlf} 
as follows.  

The expressions in Eqs.~(\ref{eq:N}) and (\ref{eq:L}) derived using 
$H^{ij}_{hyp}$ tell us that the ground state baryons are not pure 
symmetric $|^2S_S\rangle$ states.  By including the mixed symmetric 
component $|^2S_M\rangle$ of the nucleon state, Ref.~\cite{IsgurKarlf} 
finds the following ratio of decay amplitudes, 
\begin{eqnarray}
	\frac{{\rm A}(\Lambda^\ast(1830)\to \bar{K}N)}{{\rm A}
	(\Sigma^\ast(1775)\to \bar{K}N)} \simeq -0.28 \, .  
	\label{eq:BR} 
\end{eqnarray} 

This example illustrates the close relations between the internal 
structure of the initial and final baryon and the  magnitude of the 
corresponding decay width. This ratio should be compared with the 
ratio of decay widths extracted from the Ref.~\cite{PDGf} and listed 
in Table~1.

The cloudy (chiral) bag model where the coupling to the outgoing 
meson is fixed by the required continuation of the axial current, 
the hadronic decay widths are determined. However, as shown, the 
spin-flavor mixing of the excited states (and the ground states)
due to the quark interactions are critical in evaluating the 
various hadronic (and electromagnetic) partial decay widths.
A more precise determination of these partial decay widths will 
allows us to better describe the effective interactions among the 
light confined quarks.

\item \textbf{Comments Regarding $\Lambda(1405)$}

A common puzzle within quark models, which has not been resolved 
satisfactory, is: Why is $\Lambda(1405)$ about 100~MeV below 
$\Lambda(1520)$ in mass? If we assume that the $\Lambda^\ast$ 
states are mainly three-quark states, quark models have serious 
problems generating this large observed spin-orbit-like mass 
splitting between these two $\Lambda^\ast$ states.  At present 
it is very difficult within quark models to generate a large 
mass difference between these two states since, as shown, the 
effective spin-orbit interaction among quarks appears to be
negligibly small. In the cloudy bag model, the P-state quark 
with $j=3/2$ has a lower energy than the $j=1/2$ state. This 
means the bag confinement condition imposes an effective 
spin-orbit splitting of the quark P-state which is mostly 
cancelled by the effective OGE interactions. This strengthen 
the original argument of Isgur and Karl~\cite{IsgurKarlf} that 
the effective spin-orbit interaction is negligible in quark 
models. Effectively what remains are the spin-spin and tensor 
interactions among the quarks. Could a strong non-linear 
coupling of the lowest $J^P = \frac{1}{2}^-$  three-quark state
to the meson-baryon decay channels (beyond how this is presently 
treated in cloudy bag models) explain this mass-splitting?

Historically, Dalitz and Tuan~\cite{Dalitzf} proposed that 
$\Lambda(1405)$ is a $K^-p$ bound state (a ``quark molecule").
Could it have a large multi-quark (pentaquark) state component?
What are the differences between a pentaquark state and a 
``quark molecule"? A renewed and better treatment of the meson 
cloud in the chiral bag model could possibly generate a strong 
enough meson-baryon interaction in order to generate a $K^-p$ 
bound state.  Apart from recent measurements of the $K^-p$ 
atom~\cite{KpAtomf}, most $\bar{K} p$ scattering data are very 
old. We urgently need more and better data to settle the numerous 
theoretical discussions regarding the nature of $\Lambda(1405)$.

\item \textbf{Summary}

It is imperative that we can experimentally establish the mass 
spectrum of the lowest excited negative parity $\Lambda^\ast$ and 
$\Sigma^\ast$ states in order to enhance our understanding of how 
QCD operates among the three confined "light" quarks $u$, $d$ and 
$s$.  The forthcoming JLab proposal on a $K_L^0$ beam is promising.
$K_L^0$ scattering off a hydrogen target can access $\Sigma^\ast$ 
states and hopefully could firmly establish some of the missing
$\Sigma^\ast$ states in Fig.~1. The $\Sigma^\ast$ states decay to 
$\bar{K} N$ or $\pi \Sigma$ or possibly both, as well as $\pi 
\Lambda$ according to theory estimates. However, some $\Sigma^\ast$ 
states might not couple strongly to the $\bar{K} N$. In order to 
explore the $\Lambda^\ast$ states, the reaction $\gamma +p \to 
K^+ \Lambda^\ast$ appears to be a possible $\Lambda^\ast$ direct 
production channel.

As is evident from Table~1 and this presentation, more precise
measurements of the branching ratios of these excited hyperon 
decays will enhance our understanding of these states, and give 
better insight of the behavior of the interactions of the confined 
light quarks.

\item \textbf{Acknowledgments}

The author is grateful for generous travel support to the YSTAR 
conference from Jefferson Lab. and the USC's Physics and Astronomy 
Department.
\end{enumerate}


\newpage
\subsection{Ambiguities of the Antikaon-Nucleon Scattering Amplitude}
\addtocontents{toc}{\hspace{2cm}{\sl M.~Mai}\par}
\setcounter{figure}{0}
\setcounter{table}{0}
\setcounter{equation}{0}
\setcounter{footnote}{0}
\halign{#\hfil&\quad#\hfil\cr
\large{Maxim Mai}\cr
\textit{Institute for Nuclear Studies and Department of Physics}\cr
\textit{The George Washington University}\cr
\textit{Washington, DC 20052, U.S.A.}\cr}

\begin{abstract}
In this talk we discuss the modern approaches of the theory of
antikaon-nucleon scattering based on unitarization of the interaction 
potential derived from Chiral Perturbation Theory. Such approaches 
rely on very old experimental data, when fitting their free parameters. 
Thus, ambiguities arise the description of the data. We demonstrate 
these ambiguities on the example of one specific framework and discuss 
several possibilities to overcome these.
\end{abstract}

\begin{enumerate}
\item \textbf{Introduction}

Antikaon-nucleon scattering is one of the most discussed reactions among 
the meson-baryon interaction channels from the corresponding ground state 
octets. At energies below the chiral symmetry breaking scale, Chiral 
Perturbation Theory (ChPT) allows one to investigate the meson-baryon 
scattering amplitudes systematically, see Ref.~\cite{Bernard:2007zu} for 
a recent review on baryon ChPT. The full calculation of the scattering 
length up to next-to-next-to-leading chiral order yields $a_{\bar KN}^{I=0}
=+0.53+0.97-(0.40-0.22i)=+1.11+0.22i$~fm, see Ref.~\cite{Mai:2009ce}. The 
convergence of the series is rather slow, which is presumably due to the 
large kaon mass as well as large difference of the coupled channels 
thresholds. Further, the net result disagrees with the experimental one
($a_{\bar KN}^{I=0}\approx-0.53+0.77i$~fm) even in the sign. This result 
is derived from from the measurement of the energy shift and width of 
kaonic hydrogen in the SIDDHARTA experiment at 
DA$\Phi$NE~\cite{Bazzi:2011zj}, using the Deser-type formula from 
Ref.~\cite{Meissner:2004jr}. The main reason for this behavior is the 
presence of a sub-threshold resonance, the so-called $\Lambda(1405)$ in 
this channel, which was already indicated in the early studies of total 
cross section~\cite{Dalitz:1959dq}.  Therefore, a perturbative treatment 
inevitably breaks down and non-perturbative techniques are required.

The chiral unitary approach (UChPT) is considered to be the best 
non-perturbative framework to address the SU(3) dynamics in such a type 
of systems. Many studies have been performed in the last two decades 
using different versions of this framework, see, e.g., 
Refs.~\cite{Borasoy:2006sr,Oset:1997it,Jido:2003cb,Hyodo:2011ur,
Kaiser:1995eg,Mai:2014xna,Mai:2012dt,Oller:2000fj}. Exploring the complex 
energy plane, it has been observed in Ref.~\cite{Oller:2000fj} for the
first time that the resonant behavior in the $\bar K N$ channel is 
associated with a two-pole structure, see Ref.~\cite{Hyodo:2011ur} for 
a review. However, the available $K^-p \to MB$ total cross section data 
alone do not allow to pin down the positions of both poles with good 
precision, see, e.g., Ref.~\cite{Borasoy:2006sr,Mai:2014xna}. Further, 
in a direct comparison~\cite{Cieply:2016jby} of the most recent 
approaches on antikaon-nucleon scattering~\cite{Mai:2014xna,Ikeda:2012au,
Guo:2012vv,Cieply:2011nq} it was found that the predictions of different 
approaches disagree strongly on the scattering amplitude in the Isospin 
$I=1$ sector already in the close proximity of the $\bar KN$ threshold. 
Further, it was found that in all analyzed models not only the position, 
but also the origin of the wider pole varies between different approaches 
and even within different fitting scenarios of each single approach.

The main purpose of this work is to study this ambiguity as well as to 
suggest ways to reduce it. To perform this study in a systematic manner, 
we will restrict ourselves here to one single, but the most general 
framework derived in Refs.~\cite{Bruns:2010sv,Mai:2014xna}. We will show 
that at least several solutions for the antikaon-nucleon scattering 
amplitude agree with the experimental scattering data. However, including
data, e.g, from photoproduction experiment at CLAS~\cite{Moriya:2014kpv}, 
some of these solutions can be disregarded as unphysical. Further, using 
synthetic data we discuss the possible impact of the new measurements of 
cross sections, which might become available in the proposed 
$K_{\rm long}$-beam experiment at Jefferson Lab~\cite{Amaryan:2015swp}.

\item \textbf{Chiral Unitary Approach}

\begin{enumerate}
\item \textbf{Model}

The present analysis is based on the amplitude constructed and described 
in detail in Refs.~\cite{Mai:2014xna,Mai:2012wy,Bruns:2010sv}, to which 
we refer the reader for the conceptual details. We start from the chiral 
Lagrangian of the leading (LO) and next-to-leading (NLO) chiral order, 
see Refs.~\cite{Krause:1990xc,Frink:2004ic}.  For the reasons presented 
in Refs.~\cite{Mai:2014xna,Mai:2012wy,Bruns:2010sv}, the $s$- and 
$u$-channel one-baryon exchange diagrams are neglected, leaving us with
the following chiral potential
\begin{align}\label{eqn:potential}
	V(\slashed{q}_2, \slashed{q}_1; p)=&A_{WT}(\slashed{q_1}+\slashed{q_2})
	+A_{14}(q_1\cdot q_2)+A_{57}[\slashed{q_1},\slashed{q_2}] +A_{M}
        \nonumber \\ 
	+A_{811}\Big(\slashed{q_2}(q_1\cdot p)+\slashed{q_1}(q_2\cdot p)\Big)\,,
\end{align}
where the incoming and outgoing meson four-momenta are denoted 
by $q_1$ and $q_2$, whereas the overall four-momentum of the meson-baryon 
system is denoted by $p$. The symbols $A_{WT}$, $A_{14}$, $A_{57}$, $A_{M}$ 
and $A_{811}$ denote the 10-dimensional matrices which encode the coupling 
strengths between all 10 channels of the meson-baryon system for strangeness 
$S=-1$, i.e., $\{K^-p$, $\bar K^0 n$, $\pi^0\Lambda$, $\pi^0\Sigma^0$, 
$\pi^+\Sigma^-$, $\pi^-\Sigma^+$, $\eta\Lambda$, $\eta \Sigma^0$, 
$K^+\Xi^-$, $K^0\Xi^0\}$. These matrices  depend on the meson decay 
constants, the baryon mass in the chiral limit, the quark masses as 
well as 14 low-energy constants (LECs) as specified in the original
publication~\cite{Mai:2014xna}.

Due to the appearance of the $\Lambda(1405)$ resonance just below the 
$\bar K N $ threshold and large momentum transfer, the strict chiral 
expansion is not applicable for the present system. Instead, the above 
potential is used as a driving term of the coupled-channel Bethe-Salpeter 
equation (BSE), for NLO approaches see, e.g., Ref.~\cite{Mai:2012dt,
Borasoy:2005ie,Oller:2006jw,Ikeda:2011pi,Ikeda:2012au,Guo:2012vv,
Borasoy:2006sr}. For the meson-baryon scattering amplitude
$T(\slashed{q}_2, \slashed{q}_1; p)$ the integral equation to be solved 
reads
\begin{align}\label{eqn:BSE}
	T(\slashed{q}_2, \slashed{q}_1; p)=V(\slashed{q}_2, \slashed{q}_1;p)
	+i\int\frac{d^d l}{(2\pi)^d}V(\slashed{q}_2, \slashed{l}; p)
	S(\slashed{p}-\slashed{l})\Delta(l)T(\slashed{l}, \slashed{q}_1; p)\,,
\end{align}
where $S$ and $\Delta$ represent the baryon (of mass $m$) and the meson 
(of mass $M$) propagator, respectively, and are given by $iS(\slashed{p}) 
= {i}/({\slashed{p}-m+i\epsilon})$ and 
$i\Delta(k) = {i}/({k^2 - M^2 + i\epsilon})$. Moreover, $T$, $V$, $S$, and 
$\Delta$ in the last expression are matrices in the channel space. The 
loop diagrams appearing above are treated using dimensional regularization 
and applying the usual $\overline{\rm MS}$ subtraction scheme in the spirit 
of our previous work~\cite{Bruns:2010sv}. Note that the modified loop 
integrals are still scale-dependent. This scale $\mu$ reflects the 
influence of the higher-order terms not included in our potential. It is 
used as a fit parameter of our approach. To be precise, we have 6 such 
parameters in the isospin basis.

The above equation can be solved analytically, if the kernel contains 
contact terms only, see Ref.~\cite{Mai:2012wy} for the corresponding 
solution. Using this solution for the strangeness $S=-1$ system, we 
have shown in Ref.~\cite{Mai:2012dt} that once the full off-shell 
amplitude is constructed, one can easily reduce it to the on-shell 
solution, i.e., setting all tadpole integrals to zero. It appears that 
the double pole structure of the $\Lambda(1405)$ is preserved by this 
reduction and that the position of the two poles are changing only by 
about $20$~MeV in the imaginary part. On the other hand, the use 
of the on-shell approximation of the Eq.~\eqref{eqn:BSE} reduces the 
computational time roughly by a factor of 30. Therefore, since we wish
to explore the parameter space in more detail, it seems to be safe and 
also quite meaningful to start from the solution of the BSE~\eqref{eqn:BSE} 
with the chiral potential \eqref{eqn:potential} on the mass-shell. Once 
the parameter space is explored well enough we can slowly turn on the 
tadpole integrals obtaining the full off-shell solution. Such a solution 
will become a part of a more sophisticated two-meson photoproduction 
amplitude in a future publication.

\begin{figure}[t]
\begin{center}
 \includegraphics[width=0.9\linewidth]{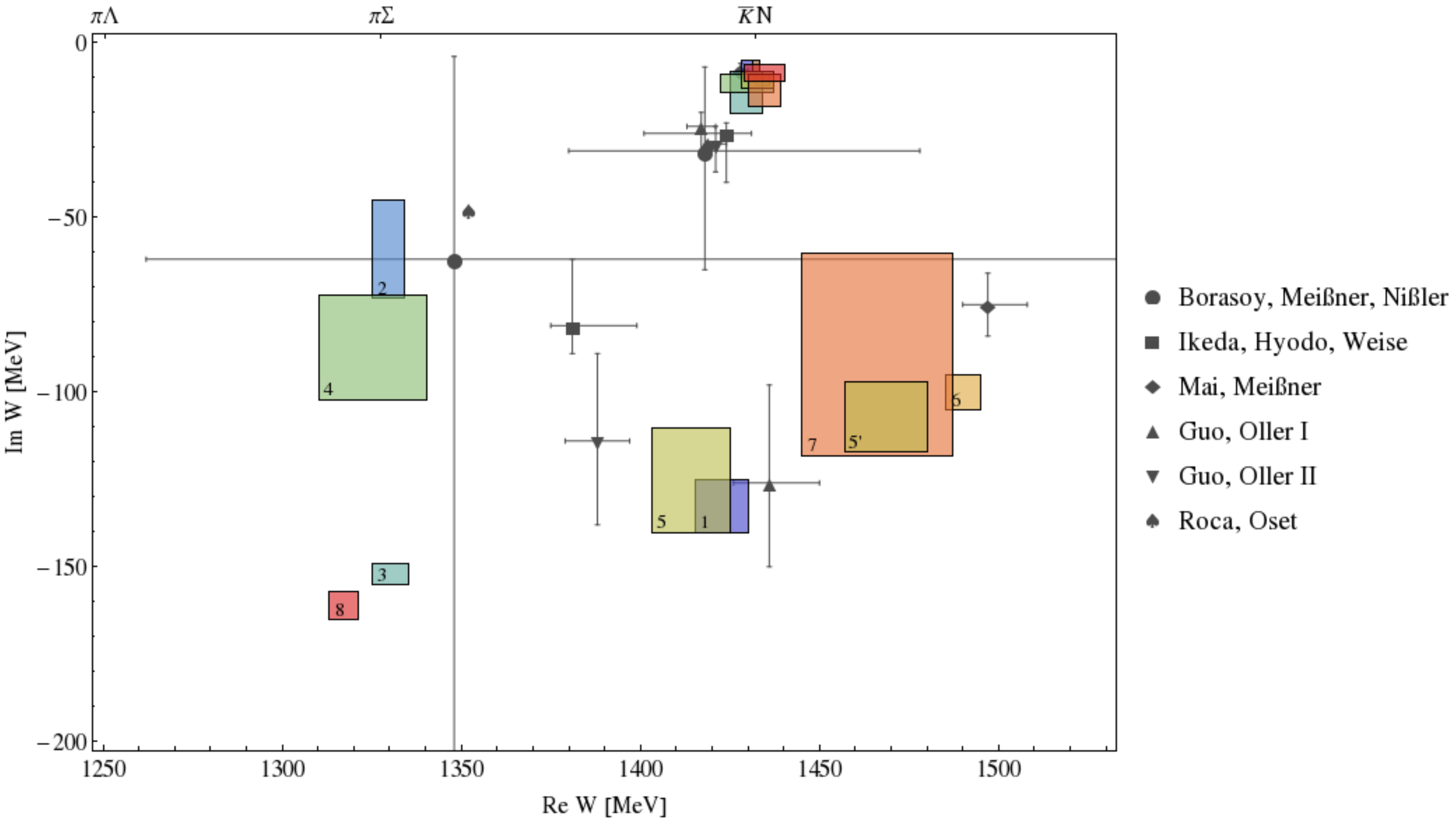}

\caption{Double pole structure of the $\Lambda(1405)$ in the complex
        energy plane for the eight solutions that describe the
        scattering and the SIDDHARTA data. For easier reading, we
        have labeled the second pole of these solutions by the
        corresponding fit \#, where $5$ and $5'$ denote the second
        pole on the second Riemann sheet, connected to the real axis
        between the $\pi\Sigma-\bar K N$ and $\bar K N-\eta\Lambda$
        thresholds, respectively. For comparison, various results
        from the literature are also shown, see
        Refs.~\protect\cite{Borasoy:2006sr,Guo:2012vv,Ikeda:2012au,
        Mai:2012dt,Roca:2013av}. } \label{fig:poles1}
\end{center}
\end{figure}

\item \textbf{Fit Procedure and Results}

The free parameters of the present model, the low-energy constants as well 
as the regularization scales $\mu$ are adjusted  to reproduce all known 
experimental data in the meson-baryon sector. The main bulk of this data 
consists of the cross sections for the processes $K^-p\to K^-p$, 
$K^-p\to\bar K^0n$, $K^-p\to\pi^0\Lambda$, $K^-p\to\pi^+\Sigma^-$, 
$K^-p\to\pi^0\Sigma^0$, $K^-p\to\pi^-\Sigma^+$ for laboratory momentum 
$P_{\rm lab}<300$~MeV from Refs.~\cite{Ciborowski:1982et,Humphrey:1962zz,
Sakitt:1965kh,Watson:1963zz}. Electromagnetic effects are not included 
in the analysis and assumed to be negligible at the measured values of
$P_{\rm lab}$. Additionally, at the antikaon-nucleon threshold, the 
following decay ratios from Refs.~\cite{Tovee:1971ga,Nowak:1978au} as 
well as the energy shift and width of kaonic hydrogen in the 1s state, 
i.e., $\Delta E -i\Gamma/2=(283\pm42)-i(271\pm55)$~eV from the SIDDHARTA 
experiment at DA$\Phi$NE~\cite{Bazzi:2011zj}. The latter two values are 
related to the $K^-p$ scattering length via the modified Deser-type 
formula~\cite{Meissner:2004jr}.

The fit procedure was performed as follows: First, for randomly chosen 
starting values of the free parameters (in a natural range) the fit was 
performed to all threshold values and cross section data. Repeating this 
procedure several thousand times, we ended with several dozen of parameter 
sets that describe the data equally well. For each of these sets the 
amplitudes were analytically continued to the positive and negative 
complex energy plane. Thereafter, every unphysical solution, e.g., those
with poles on the first Riemann sheet for ${\rm Im}(W)<200$~MeV 
($W:=\sqrt{p^2}$), was sorted out. Eight best solutions were obtained by 
this procedure, see Tab.~\ref{tab:photo}, whereas the next best 
$\chi_{\rm d.o.f.}^2$ are at least one order of magnitude larger. Although 
the fit results look very promising, we would like to point out that there 
are quite a few (20) free parameters in the model. The latter are assumed 
to be of natural size, but not restricted otherwise. Thus, we can not 
exclude that there might be more solutions which describe the assumed 
experimental data equally well.

The data are described equally well by all eight solutions, showing, 
however, different functional behavior of the cross sections as a 
function of $P_{\rm lab}$. These differences are even more pronounced 
for the scattering amplitude $f_{0+}$, which is fixed model independently 
only in the ${K^-p}$ channel at the threshold by the scattering length 
$a_{K^-p}$. Similar observation was made in the comparison of this 
approach with other most recently used UChPT models in 
Ref.~\cite{Cieply:2016jby}.

When continued analytically to the complex $W$ plane, all eight 
solutions confirm the double pole structure of the $\Lambda(1405)$, 
see Fig.~\ref{fig:poles1}. The scattering amplitude is restricted 
around the $\bar K N$ threshold by the SIDDHARTA measurement quite 
strongly. Therefore, in the complex $W$ plane we observe a very 
stable behavior of the amplitude at this energy, i.e., the position 
of the narrow pole agrees among all solutions within the $1\sigma$ 
parameter errors, see Fig.~\ref{fig:poles1}. This is in line with 
the findings of other groups~\cite{Ikeda:2012au,Borasoy:2006sr,
Guo:2012vv}, i.e., one observes stability of the position of the 
narrow pole. Quantitatively, the first pole found in these models
is located at somewhat lower energies and is slightly broader than 
those of our model.  In view of the stability of the pole position, 
we trace this shift to the different treatment of the Born term 
contributions to the chiral potential utilized in 
Refs.~\cite{Ikeda:2012au,Borasoy:2006sr,Guo:2012vv}.

The position of the second pole is, on the other hand, less restricted.
To be more precise, for the real part we find three clusters of these 
poles: around the $\pi\Sigma$ threshold, around the $\bar K N$ threshold 
as well as around $1470$~MeV. For several solutions there is some 
agreement in the positions of the second pole between the present 
analysis and the one of Ref.~\cite{Guo:2012vv} and of our previous 
work~\cite{Mai:2012dt}. However, as the experimental data is described 
similarly well by all fit solutions, one  can not reject any of them. 
Thus, the distribution of poles represents the systematic uncertainty 
of the present approach.  It  appears to be quite large, but is still
significantly smaller than in the older analysis of 
Ref.~\cite{Borasoy:2006sr} based on scattering data only.
\end{enumerate}

\item \textbf{Reduction of {\bf the} Model Ambiguities}

\begin{figure}[t]
\begin{center}
 \includegraphics[width=\linewidth]{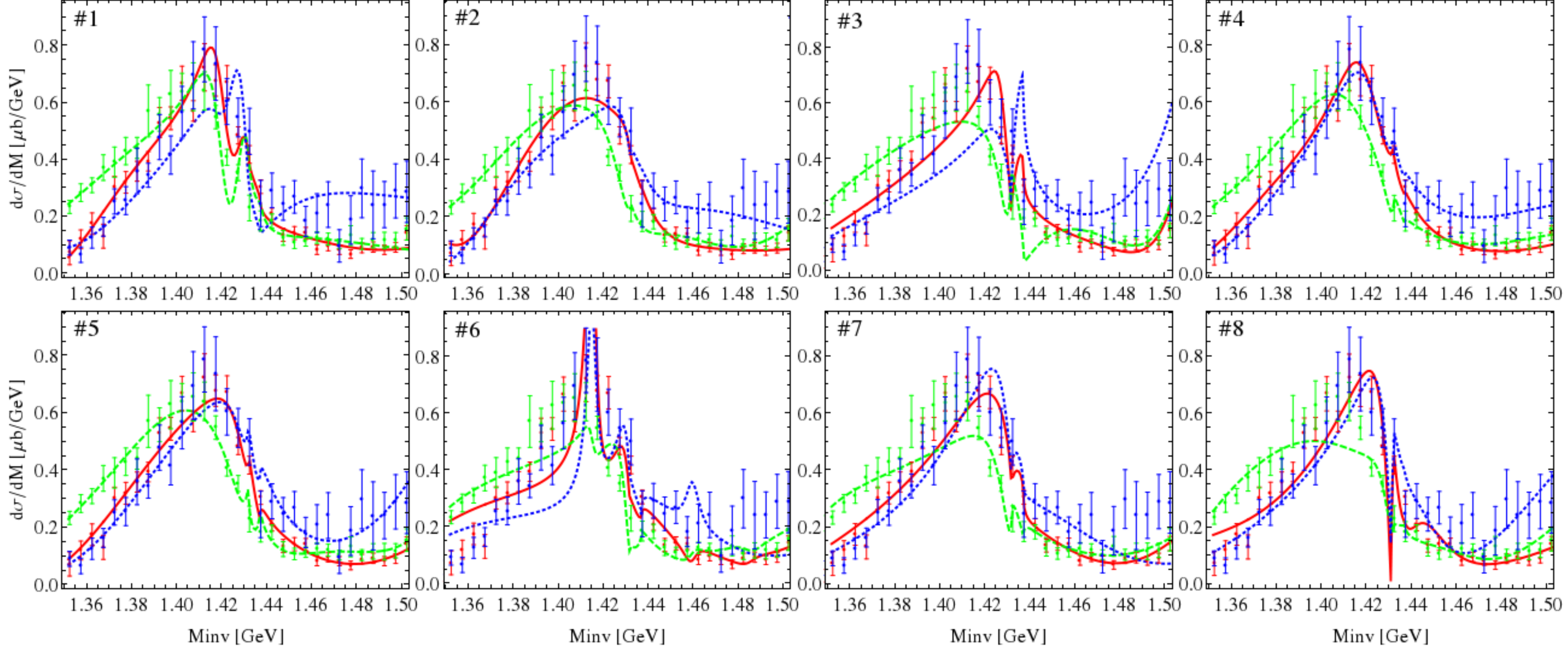}

\caption{Comparison of all solutions describing the $\pi 
	\Sigma$ mass distribution at $\tilde W=2.5$~GeV in all 
	three channels $\pi^+\Sigma^-$ (green, dashed), 
	$\pi^-\Sigma^+$ (full, red) and  $\pi^0\Sigma^0$ (blue, 
	dotted).} \label{fig:aus}
\end{center}
\end{figure}

In the previous section we have exemplified that the old scattering 
data~\cite{Ciborowski:1982et,Humphrey:1962zz,Sakitt:1965kh,Watson:1963zz} 
together with the recent and very important measurement of the kaonic 
hydrogen characteristics~\cite{Bazzi:2011zj} alone do not fix the 
antikaon-nucleon scattering amplitude well enough. There are, however, 
several (proposed) measurements which may lead to a reduction of this 
ambiguity. In the following we wish to present what is in our view most 
promising ways to do so.

\renewcommand{\baselinestretch}{1.25}
\begin{table*}[t]
\begin{center}
\begin{tabular}{|c|cccccccc|}
\hline
~~~~~~~~Fit \#~~~~~~~~ &~~~~1~~~~~&~~~~2~~~~&~~~~3~~~~&~~~~4~~~~&~~~~5~~~~&~~~~6~~~~&~~~~7~~~~&~~~~8~~~~\\
\hline
~~$\chi_{\rm d.o.f.}^2$ (hadronic data)~~ &1.35 &1.14&0.99 &0.96 &1.06 &1.02
&1.15 &0.90\\
\hline
$\chi_{\rm p.p.}^2$   (CLAS data)~~ &3.18&1.94&2.56&1.77&1.90&6.11&2.93&3.14\\
\hline
\end{tabular}
\caption{Quality of the various fits in the description of the hadronic
        and the photoproduction data from CLAS. For the definition of
        $\chi_{\rm p.p.}^2$, see the text. \label{tab:photo}}
\end{center}
\end{table*}

\begin{enumerate}
\item \textbf{Photoproduction data}

Recently, very sophisticated measurements of the reaction $\gamma p\to 
K^+\Sigma \pi$ were performed by the CLAS collaboration at JLab, see 
Ref.~\cite{Moriya:2013eb}. There, the invariant mass distribution of 
all three $\pi\Sigma$ channels was determined in a broad energy range 
and with high resolution. Finally, from these data the spin-parity 
analysis of the $\Lambda(1405)$ was performed in 
Ref.~\cite{Moriya:2014kpv}.

First theoretical analyses have been performed on the basis of a 
chiral unitary approach in Refs.~\cite{Roca:2013av,Roca:2013cca,
Nakamura:2013boa,Mai:2014xna}. In the following, we present the 
results of the analysis of Ref.~\cite{Mai:2014xna}, which relies on 
the hadronic scattering amplitude described in the previous section. 
In this study, we assumed the simplest ansatz for the photoproduction 
amplitude
\begin{equation}\label{eq:photo}
	\mathcal M^j(\tilde W,M_{\rm inv})
	= \sum_{i=1}^{10} C^i(\tilde W)\,G^i(M_{\rm inv})\,f_{0+}^{i,j}(M_{\rm inv})\,,
\end{equation}
where $\tilde W$ and $M_{\rm inv}$ denote the total energy of the system 
and the invariant mass of the $\pi\Sigma$ sub-system, respectively. For 
a specific meson-baryon channel $i$, the Greens function is denoted by 
$G^i(M_{\rm inv})$, and the energy-dependent (and in general complex 
valued) constants  $C^i(\tilde W)$ describe the reaction mechanism of 
$\gamma p\to K^+M_iB_i$. The hadronic final-state interaction is captured 
by the standard H\"ohler partial waves $f_{0+}$, derived from 
Eq.~\eqref{eqn:BSE}. For more details, i.e., the explicit form of the 
Greens function we refer the reader to the original 
publication~\cite{Mai:2014xna}.

Clearly, a more sophisticated ansatz is required to address scattering 
and photoproduction data simultaneously, while fulfilling in the same 
time the gauge invariance. Such an approach can be developed along the 
techniques used for the analysis of the single meson photoproduction, 
see Refs.~\cite{Borasoy:2007ku,Mai:2012wy}. The question we wish to 
address here is, however, different. Namely, whether all obtained 
hadronic solutions allow for a good description of the photoproduction 
data using such a flexible ansatz. Thus, without altering the parameters
of the hadronic part (8 solutions) in the photoproduction amplitude 
Eq.~\eqref{eq:photo}, we fit the unknown constants $C^i(\tilde W)$ to 
reproduce the CLAS data in all three measured final states 
($\pi^+\Sigma^-$, $\pi^0 \Sigma^0$, $\pi^-\Sigma^+$) and for all 9 
measured total energy values ($\tilde W=2.0, 2.1, .., 2.8$~GeV).

The resulting values of $\chi^2$ per data point for these fits are 
collected in the third row of the Table~\ref{tab:photo}, whereas for 
the further details and error analysis we again refer to the original 
publication~\cite{Mai:2014xna}. They show that even within such a  
flexible ansatz the solutions~\#1, \#3, \#6, \#7 and \#8 of the eight 
hadronic solutions do not allow for a decent description of the 
high-quality CLAS data. The failure of these solutions becomes
even more evident when comparing the $\pi\Sigma$ mass distribution 
of all eight solutions at one (typical) energy $\tilde W=2.5$~GeV, 
see Fig.~\ref{fig:aus}. Consequently, only three of originally eight 
solutions can be considered as physical with respect to the CLAS data, 
which indeed reduces the ambiguity of the antikaon-nucleon scattering 
amplitude substantially.

\item \textbf{Kaonic Deuterium}

The $K^-p$ scattering amplitude at the threshold is fixed very well 
by the strong energy shift and width of the kaonic hydrogen, measured 
in the SIDDHARTA experiment~\cite{Bazzi:2011zj}. However, this does 
not fix the full $\bar KN$ scattering amplitude at the threshold, 
which has two complex valued components, i.e., for isospin $I=0$ and 
$I=1$. To fix both components one requires another independent 
measurement, such as the energy shift and width of the kaonic
deuterium. Such a measurement is proposed at LNF~\cite{LNF} and 
J-PARC~\cite{JPARK}. Ultimately, this quantity can be related to 
the antikaon-deuteron scattering length by the well known Deser-type 
relations (see, e.g., Refs.~\cite{Deser:1954vq,Meissner:2004jr,
Meissner:2006gx,physrep}) and then to the antikaon-nucleon scattering 
length, using an effective field theory framework, see, e.g.,
Ref.~\cite{Mai:2014uma}.

\item \textbf{Additional Scattering Data}

The scattering data~\cite{Ciborowski:1982et,Humphrey:1962zz,
Sakitt:1965kh,Watson:1963zz} used to fix the $\bar KN$ scattering 
amplitude as described in the last section stem from very old 
bubble chamber experiments. From the theoretical point of view, 
improvement of these data would be the best way to reduce the 
ambiguity of the theoretical predictions. The proposed measurement 
of the two-body interaction of $K_{\rm long}$-beam and the proton
target at JLab~\cite{Amaryan:2015swp} can potentially lead to such 
an improvement of the data.

In the following, we wish to quantify the above statement, using the 
already obtained solutions of the hadronic model of the last section. 
We do so by generating a set of new synthetic data on total cross 
sections in the same channels as before, using our best solution 
\#4 in the momentum interval $P_{\rm lab}=100-300$~MeV. We assume 
the energy binning to be fixed, and randomize the synthetic data by 
Gaussian distribution with a standard deviation of $\Delta\sigma$ 
for the charged and $2\Delta\sigma$ for the neutral final state 
channels. The latter is assumed to account for the fact that neutral 
channels are usually more intricate to measure.

We have tested different scenarios - considering different energy 
binning ($\Delta P$), measurement accuracy ($\Delta\sigma$) and whether 
the new synthetic data complements or replaces the old data. Without 
further fitting, we have compared the new $\chi^2_{\rm d.o.f.}$ of all 
solutions, obtained in the previous section. We found that at least 
four of the obtained eight solutions are not compatible with the 
updated data for $\Delta \sigma \le 5$~MeV and $\Delta P\sim 10$~MeV. 
This procedure appears to be more sensitive to the measurement accuracy 
than on the energy binning - for $\Delta\sigma\ge10$~MeV none of the 
solutions could be sorted out for any of chosen values of $\Delta P$. 
A complete replacement of the old by the new (synthetic) data does not 
change our findings qualitatively, but increases the differences 
between new $\chi^2_{\rm d.o.f.}$ values slightly. In summary, this 
preliminary and simplistic analysis underlines the importance of the
re-measurement of the cross section data on $\bar K N$ scattering in 
a modern experimental setup, such as the one proposed in 
Ref.~\cite{Amaryan:2015swp}.
\end{enumerate}

\item \textbf{Acknowledgments}

The speaker is grateful to the organizers of the workshop for the 
invitation. The vast part of the work presented at the YSTAR2016 
workshop was performed in the HISKP at the University of Bonn in 
collaboration with Ulf-G.~Mei\ss ner, whom he thanks for a careful 
reading of the manuscript. The speaker is grateful for multiple 
interesting discussions with M.~D\"oring, A.~Cieply, P.~Bruns and 
D.~Sadasivan on this matter. Currently, the speaker is supported 
by the German Research Foundation (DFG) under the fellowship MA 
7156/1--1.
\end{enumerate}


\newpage
\subsection{Bound States of $N$'s and $\Xi$'s}
\addtocontents{toc}{\hspace{2cm}{\sl V.~Crede and S.~Capstick}\par}
\setcounter{figure}{0}  
\setcounter{table}{0}   
\setcounter{equation}{0}
\setcounter{footnote}{0}
\halign{#\hfil&\quad#\hfil\cr
\large{Volker~Crede}\cr
\textit{Department of Physics}\cr
\textit{Florida State University}\cr
\textit{Tallahassee, FL 32306-4350, U.S.A.}\cr\cr
\large{Simon~Capstick}\cr
\textit{Department of Physics}\cr
\textit{Florida State University}\cr  
\textit{Tallahassee, FL 32306-4350, U.S.A.}\cr}
 
\begin{abstract}
Experimental information on the spectrum, structure, and decays of strangeness
$-2$ Cascade baryons is sparse compared to non-strange and strangeness~$-1$
baryons. It is argued that an experimental program at Jefferson Lab using the 
photo-production and the GlueX detector to study the physics of Cascades is   
of considerable interest, since it is likely that the lightest Cascade baryons of
a given spin and parity are relatively narrow. If this is verified in an experiment,
this would confirm the flavor independence of the confining interaction that is
assumed in models. These narrow widths may also make it possible to measure
the isospin-symmetry violating mass splittings in a spatially-excited baryon for
the first time. Copious data for excited strangeness $-1$ baryons will be
collected along with the data for Cascade baryons in such an experimental 
program. Photo-production reactions which can be used to study excited Cascade
baryons are described, and simulations made to understand the production of a  
ground-state and an excited-state Cascade baryon in the GlueX experiment are  
discussed, along with possible sources of background.
\end{abstract}

\begin{enumerate}
\item \textbf{Introduction}

The spectrum of multi-strange hyperons is poorly known, with only a few
resonances whose existence is well established. Among the doubly-strange 
states, the two ground-state Cascades, the octet member~$\Xi$ and the decuplet
member $\Xi^\ast(1530)$, have four-star status in the RPP~\cite{Agashe:2014kdaw},  
with only four other three-star candidates. On the other hand, more than 
20~$N^\ast$ and $\Delta^\ast$ resonances are rated with at least three stars 
by the Particle Data Group (PDG). Of the six $\Xi$~states that have at least 
three-star ratings, only two are listed with weak experimental evidence for 
their spin-parity $(J^P)$ quantum numbers: $\Xi(1530)\,
\frac{3}{2}^+$~\cite{Aubert:2008tyw}, $\Xi(1820)\,
\frac{3}{2}^-$~\cite{Biagi:1986vsw}. All other $J^P$~assignments are based on 
quark-model predictions. 

Flavor~SU(3) symmetry predicts as many $\Xi$~resonances as $N^\ast$ and
$\Delta^\ast$~states combined, suggesting that many more Cascade resonances
remain undiscovered. The three lightest quarks, $u$, $d$, and $s$, have 27
possible flavor combinations: $3\otimes 3\otimes 3 = 1\oplus 8\oplus
8\,^{\prime}\oplus 10$ and each multiplet is identified by its spin and
parity,~$J^P$. Flavor SU(3) symmetry implies that the members of the
multiplets differ only in their quark makeup, and that the basic properties
of the baryons should be similar, although the symmetry is known to be
broken by the strange-light quark mass difference. The octets consist of
$N^*$, $\Lambda^*$, $\Sigma^*$, and $\Xi^*$~states. We thus expect that
for every $N^*$ state, there should be a corresponding $\Xi^*$~state \emph{with
similar properties}. Additionally, since the decuplets consist of $\Delta^*$,
$\Sigma^*$, $\Xi^*$, and  $\Omega^*$~states, we also expect for every
$\Delta^*$ state to find a decuplet~$\Xi^*$ with similar properties. In a
simple quark model picture, the strange states will fit into multiplets which
correspond to those of the $u,\ d$~sector. However, it could be that the
dynamics of the excited baryons differ from those of the lower-lying
states; for example, the pattern of their decays may be systematically different.
Parity doublets may appear in some sectors with increasing mass. Should we expect
doubly-strange baryons with properties similar to those of the $\Lambda(1405)$
with $J^P = 1/2^-$ and the Roper $N^\ast$ with $J^P = 1/2^+$, which do not fit
easily the conventional picture of three quarks in the baryon? The dependence of
the physics of these unusual states on the number of strange quarks is of crucial
importance to our understanding of them, which motivates the collection of a 
significant database on multi-strange baryons.

\item \textbf{$\Xi$ Spectrum and Decays}

The $\Xi$~hyperons have the doubly-strange quark content $|ssu\,\rangle$ and
$|ssd\,\rangle$. An interesting feature of the $\Xi$~spectrum is that there are
fewer degeneracies than in the light-quark baryon spectrum. If the confining
potential is independent of quark flavor, the energy of spatial excitations of a
given pair of quarks will be inversely proportional to their reduced mass.
If all three quark masses are the same, the excitation energy of either of the
two relative coordinates will be the same, which will lead to degeneracies in the
excitation spectrum. However, with two strange quarks and one light quark, the
excitation energy of the relative coordinate of the strange quark pair is smaller. This
means that the lightest excitations in each partial wave are between the two strange
quarks, and that the degeneracy between excitations of the two relative coordinates
is lifted. The spectrum of $\Xi$~baryons calculated using the relativized quark
model~\cite{Capstick:1986bmw} along with information about $\Xi$~states extracted
from experiment is shown in Figure~\ref{Figure:Spectrum}. A comparison of results
from this model for the masses of non-strange and strangeness $-1$ baryons with
those extracted from experimental data makes it likely that the lowest-mass 
positive-parity excited $\Xi^\ast$~states are lower than shown in 
Fig.~\ref{Figure:Spectrum}, and that the spectrum of negative-parity excited states 
should have larger splittings.

In the absence of configuration mixing and in a spectator decay model, $\Xi$~states
with the relative coordinate of the strange-quark pair excited cannot decay to the
ground state $\Xi$ and a pion, because of orthogonality of the part of the spatial wave
function between the two strange quarks in the initial excited state and in the final
ground state. Having instead to decay to final states that include Kaons rules out the
decay channel with the largest phase space for the lightest states in each partial wave,
substantially reducing their widths~\cite{Chao:1980emw}. This selection rule
is modified by (configuration) mixing in the wave function; however, color-magnetic
hyperfine mixing is weaker in $\Xi$~states because this interaction is smaller between
quarks of larger masses. The flavor-spin [SU(6)] coupling constants at the decay
vertices for $N,\ \Delta \to N\pi,\ \Delta\pi$ are significantly larger than those for
$\Xi,\ \Xi^\ast \to \Xi\pi,\ \Xi^\ast\pi$ decays~\cite{Zenczykowski:1985uhw}, which also
reduces these widths. The result is that the well known lower-mass resonances have
widths $\Gamma_{\Xi^\ast}$ of about 10 - 20~MeV, which is 5 - 30 times narrower
than is typical for $N^\ast$, $\Delta$, $\Lambda$, and $\Sigma$~states.
\begin{figure}[t!]
\begin{center}
\includegraphics[width=0.7\textwidth]{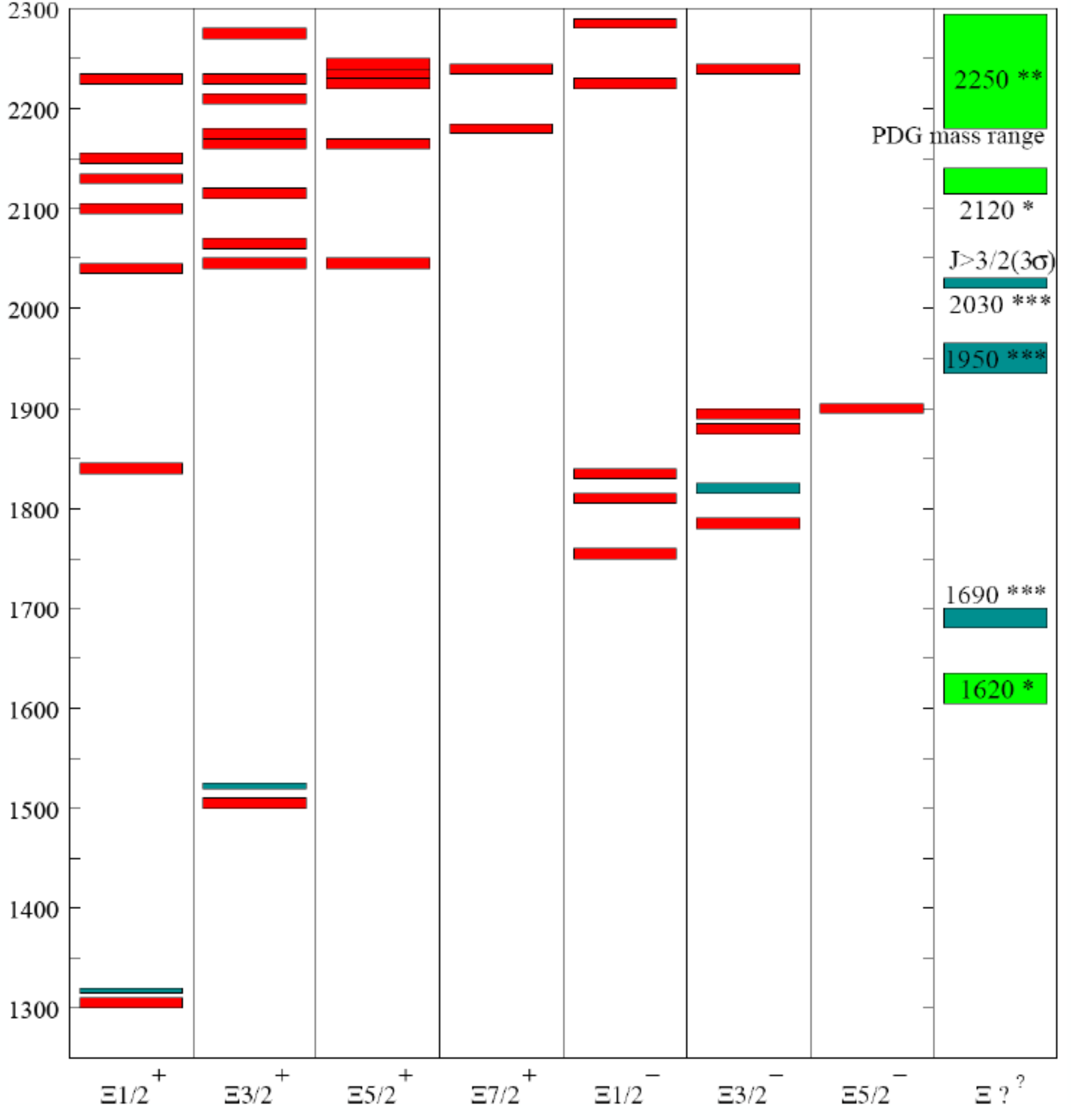}

\caption{\label{Figure:Spectrum} Relativized model spectrum of
  $\Xi$~baryons~\protect\cite{Capstick:1986bmw} below 2300~MeV (red bars)
  compared to masses extracted from experiment and their uncertainties
  (dark green bars for well known states, light green bars for
  tentative states). Experimental states with undetermined spins
  and/or parities are placed in the last column on the right.}
\end{center}
\end{figure}

The first excited state with nucleon quantum numbers, the Roper
resonance at 1440~MeV, is interesting because its low mass is hard to
explain in models containing only three quarks.  It is likely that
this is because the pole position of this resonance is shifted because
of strong coupling to the $N\pi\pi$ channel, which also contributes to
its large width of about 350~MeV. Extraction of its properties from
partial-wave analysis of $\pi N$ scattering and photo-production data
is made difficult by a complicated pole structure and by the presence
of a second nucleon resonance with $J^P={1\over 2}^+$ at 1710~MeV with
a width of roughly 100~MeV. The equivalent $\Xi^\ast$~state should be
quite narrow and so be relatively easy to separate from the next
$\Xi^\ast\,\frac{1}{2}^+$, which itself should be relatively narrow,
and from the lightest negative-parity excitations, which are also
relatively narrow.

These features render possible a wide-ranging program to study
the physics of the Cascade hyperon and its excited states. The study
of these hyperons has focused until recently on their production in
$K^- p$~reactions, although some $\Xi^\ast$~states were found using high-energy
hyperon beams. Photo-production appears to be a very promising
alternative. Results from earlier Kaon-beam experiments indicate that
it is possible to produce the $\Xi$~ground state through the decay of
high-mass $Y^\ast$~states~\cite{Tripp:1967kjw,Burgun:1969eew,Litchfield:1971riw}.
It is therefore possible to produce Cascade resonances through
$t$-channel photo-production of hyperon resonances using the
photo-production reaction $\gamma p\to K K\,\Xi^{(\ast)}$. The
CLAS collaboration investigated this reaction~\cite{Guo:2007dww},
but no signicant signal for an excited Cascade state was observed, other 
than that of
the decuplet ground state $\Xi(1530)$. The absence of higher-mass signals
is very likely due to the low photon energies available to these experiments 
and the limited
acceptance of the CLAS detector. Equipped with a Kaon-identification
system, the GlueX experiment will be well suited to search for and
study excited $\Xi$~resonances.

To summarize, it would be interesting to see in a Cascade physics program
at Jefferson Lab the lightest excited $\Xi^\ast$~states of a given spin and
parity~$J^P$ decoupling from the $\Xi\pi$~channel, confirming the
flavor independence of confinement. Measurements of the isospin-symmetry
violating mass splittings ($\Xi^{\ast -}-\Xi^{\ast 0}$) in spatially 
excited Cascade
states are also possible, for the first time in a spatially-excited hadron. 
Currently,
mass splittings like $n-p$ or $\Delta^0 - \Delta^{++}$ are only available
for the octet and decuplet ground states, but are hard to measure in
excited $N,~\Delta$ and $\Sigma,~\Sigma^\ast$~states, which are
broad. The lightest Cascade baryons are expected to be narrower, and
measuring the $\Xi^- - \Xi^0$ splitting of spatially-excited $\Xi$~states
remains a strong possibility. Such measurements would allow an
interesting probe of excited-hadron structure, and would provide
important input for quark models which explain the isospin-symmetry violating
mass splittings by the effects of the difference of the $u$- and $d$-quark masses
and of the electromagnetic interactions between the quarks.

\item \textbf{$\Xi$ Searches using the GlueX Experiment}

The Cascade octet ground states $(\Xi^0,\,\Xi^-)$ can be studied in
the GlueX experiment {\it via} exclusive $t$-channel (meson exchange)
processes in the reactions
\begin{eqnarray}
	\label{Equation:GroundOctet}
	\gamma p\,\to\, K\,Y^\ast\,\to\,K^+\,(\,\Xi^-\,K^+\,),~K^+\,(\,\Xi^0\,K^0\,),
	~K^0\,(\,\Xi^0\,K^+\,)\,.
\end{eqnarray}
The production of such two-body systems involving a $\Xi$~particle
also allows the study of highly-excited $\Lambda^\ast$ and $\Sigma^\ast$
states. Initially, the $\Xi$~octet ground states ($\Xi^0$ and $\Xi^-$) will be
challenging to study via exclusive $t$-channel (meson exchange) production.
The typical final states have kinematics for which the baseline GlueX detector
has very low acceptance due to the high-momentum forward-going Kaon and
the relatively low-momentum pions produced in the $\Xi$~decay. However,
the production of the $\Xi$~decuplet ground state, $\Xi(1530)$, and other
$\Xi^\ast$ states decaying to $\Xi\pi$ results in a lower momentum Kaon at
the upper vertex, and heavier $\Xi$~states produce higher momentum
pions in their decays.

The Cascade decuplet ground state, $\Xi(1530)$, and other excited Cascades
can be searched for and studied in the reactions
\begin{eqnarray}
	\label{Equation:GroundDecuplet}
	\gamma p\,\to\, K\,Y^\ast\,\to\,K^+\,(\,\Xi\,\pi\,)\,K^0,~K^+\,
	(\,\Xi\,\pi\,)\,K^+,~K^0\,(\,\$
\end{eqnarray}
The lightest excited $\Xi$~states of a given spin and parity $J^P$ are expected 
to decouple from $\Xi\pi$ and can be searched for and studied in the reactions
\begin{eqnarray}
	\label{Equation:Excited}
	\gamma p\,\to\,K\,Y^\ast\,\to\,K^+\,(\,K\Lambda\,)_{\Xi^{-\ast}}\,
	K^+,~K^+\,(\,K\Lambda\,)_{\Xi^{0\ast}}\,K^0,~
	K^0\,(\,K\Lambda\,)_{\Xi^{0\ast}}\,K^+\,,\\[1ex]
	\gamma p\,\to\, K\,Y^\ast\,\to\,K^+\,(\,K\Sigma\,)_{\Xi^{-\ast}}\,
	K^+,~K^+\,(\,K\Sigma\,)_{\Xi^{0\ast}}\,K^0,~
	K^0\,(\,K\Sigma\,)_{\Xi^{0\ast}}\,K^+\,.
\end{eqnarray}
\begin{figure}[t!]
\begin{center}
\begin{tabular}{cc}
\includegraphics[width=0.5\textwidth]{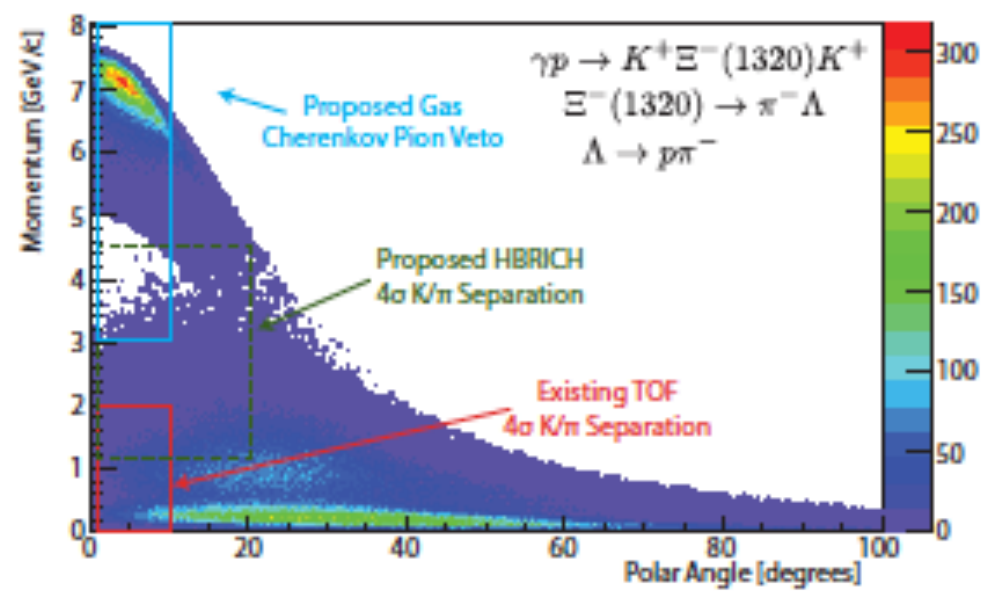}
\includegraphics[width=0.5\textwidth]{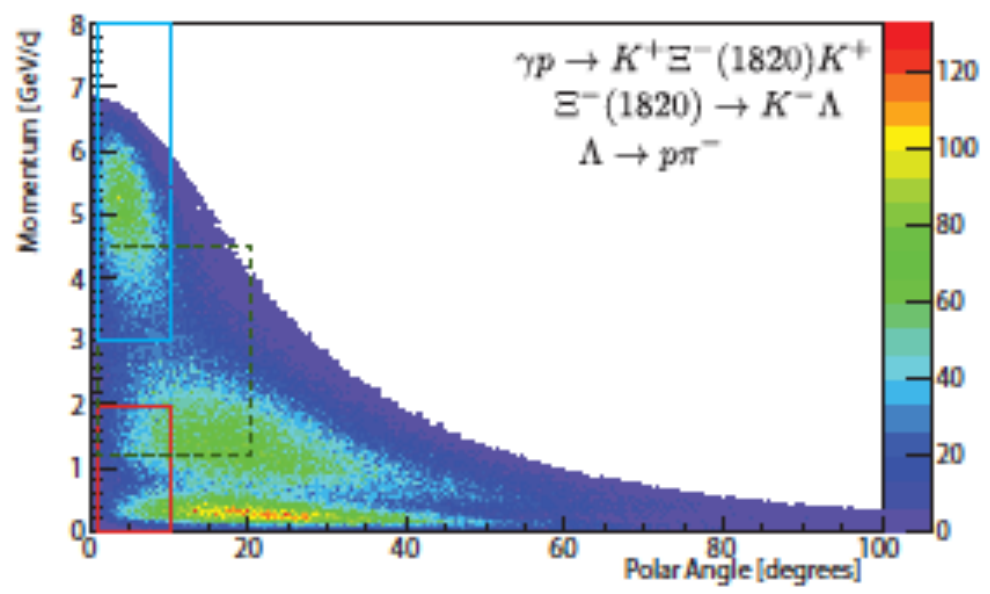}
\end{tabular}

\caption{\label{Figure:Production} Generated momentum versus polar
    angle for all tracks in the simulated reactions (a) $\gamma p\to
    K^+\,\Xi^-(1320) K^+$ and (b) $\gamma p\to K^+\,\Xi^-(1820)
    K^+$. The decay of the Cascade resonance and assigments of
    final-state particles to observed high-density regions in phase
    space are described in the plots.}
\end{center}
\end{figure}

Our simulations and application of state-of-the-art analysis 
tools~\cite{AlekSejevs:2013mklw}
show that background from competing hadronic reactions will be much reduced 
because of the unique signature provided by the two associated Kaons in
photo-production, in combination with the additional information found by 
analyzing
the weak-decay secondary vertices of ground-state strangeness $-1$ hyperons 
in the final state.

However, decays such as $Y^\ast\to\phi\Lambda,~\phi\Sigma$ might
contribute to the background for certain final states.
Larger contributions to the background will more likely come from events with
pions misidentified as Kaons, as well as other reconstruction and detector
inefficiencies. To extract small Cascade signals at masses above the
$\Xi(1530)$, it will therefore be important to reduce the background
by kinematically reconstructing complete final states. A full exclusive
reconstruction also enhances the possibility of being able to measure
the $J^P$ of these states.

We have simulated the production of the $\Xi^-(1320)$ and
$\Xi^-(1820)$~resonances to better understand the kinematics of these
reactions. The photo-production of the $\Xi^-(1320)$ decaying to
$\pi^-\Lambda$ and of the $\Xi^-(1820)$ decaying to $\Lambda K^-$ is
shown in Fig.~\ref{Figure:Production}. These reactions results in the
$K^+K^+\pi^-\pi^-p$ and $K^+K^+K^-\pi^-p$ final states,
respectively. Reactions involving excited Cascades have ``softer''
forward-going Kaons, and there is more energy available on average to
the Cascade's decay products. Both plots show three regions of high
density. The upper momentum region ($>4$~GeV/$c$) consists of
forward-going $K^+$ tracks from the associated production of an
excited hyperon. The middle momentum regions (1-2~GeV/$c$) are a
mixture of $K^-$, $K^+$, and proton tracks, while the lower region
(below 1 GeV/$c$) contains mostly $\pi^-$~tracks. The high-momentum
Kaon tracks with momenta larger than about 2.5~GeV/$c$ cannot be
identified with the current GlueX PID system. Shown in 
Fig.~\ref{Figure:Production}
in solid (red), dashed (green), and dotted (blue) are the regions of
phase space where the existing time-of-flight (TOF) detector, the proposed
Hadron Blind RICH (HBRICH) detector, and proposed gas Cherenkov detector
provide pion/Kaon discrimination at the four standard deviation 
level~\cite{Dugger:2014xaaw}.

\item \textbf{Acknowledgments}

The authors thank the technical staff at Jefferson Lab and at all 
the participating institutions for their invaluable contributions 
to the success of the experiment. This material is based upon work 
supported by the U.S. Department of Energy, Office of Science, Office 
of Nuclear Physics, under Contract No. DE--AC05--06OR23177. The GlueX 
simulations discussed here comprise part of the PhD dissertation work 
of FSU graduate student Nathan Sparks. Florida State University 
acknowledges support under grant DE--FG02--92ER40735.
\end{enumerate}


\newpage
\subsection{Hyperon Photoproduction at Jefferson Lab}
\addtocontents{toc}{\hspace{2cm}{\sl L.~Guo}\par}
\setcounter{figure}{0}
\setcounter{table}{0}
\setcounter{equation}{0}
\setcounter{footnote}{0}
\halign{#\hfil&\quad#\hfil\cr
\large{Lei Guo (for CLAS Collaboration)}\cr
\textit{Physics Department}\cr
\textit{Florida International University}\cr
\textit{Miami, FL 33199, U.S.A}\cr}

\begin{abstract}
Compared to the many recent experimental progress made in the nucleon 
resonances, the advances in hyperon spectroscopy have been scarce. 
The large amount of photoproduction data that have been collected in 
the past decade by the CLAS collaboration, and the next generation of 
experiments to be erson Lab, will make it possible to investigate the 
production mechanisms of all three sectors of hyperon states. It could
also become possible to discover the missing hyperon states as expected 
by various quark model predictions and Lattice QCD calculations.
\end{abstract}

\begin{enumerate}
\item \textbf{Introduction}

The strange quark plays an important role in understanding the strong 
interaction of nucleons. Even though photo- and electroproduction of
strangeness has been carried out since the 1950s, there is still no 
unambiguous and comprehensive model describing the reaction mechanism
of baryon and hyperon ($S=-1, -2$ and $-3$) resonances. This is due, 
in part, to the difficulties encountered in modeling the strong 
interaction in the non-perturbative regime.  As such, the problem 
has been approached through the use of effective field 
theories~\cite{haber,MB,Puente,maxwell}, Regge models~\cite{Guidal2003} 
and hybrid Regge-plus-resonance (RPR) models~\cite{corthals,
tom-vrancx}, and more recently, through coupled-channel 
analyses~\cite{Anisovich2012,Penner,Doring,Kamano}. All of these 
methods require large and precise data sets in order to constrain 
fitting parameters.  In addition to the crucial study of the 
strange-quark production mechanism itself, an important part of 
these efforts is the identification of nucleon and hyperon resonances 
predicted by various QCD-based models~\cite{Koniuk80,Capstick98,Isgur} 
but not previously observed. Recent progress in the hyperon sectors 
in lattice calculations~\cite{Edwards} has also made it more urgent 
to obtain experimental data. Compared with the nucleon resonances
sectors, the status of hyperon spectroscopy leaves much to be
desired. In the search for missing nucleon resonances, a major
difficulty arises from many overlapping broad states. Cascade
resonances are typically much narrower and comparatively easier to 
identify. The status $S=-1$ hyperon states lie somewhere between the 
nucleon resonances and the cascade sector. In particular, more than 
30 excited states are predicted to lie between $2$~GeV and 
$2.3$~GeV~\cite{Capstick98}. Currently, there are only five $S=-1$ 
states considered to be established, with three or four star rating 
in the PDG~\cite{PDG} in that region. On the other hand,  the $S=-3$ 
$\Omega^-$ state has never been observed in photoproduction. The 
main issue with the multi-strangeness sector can be attributed to 
the lack of experimental data as a result of small cross sections.
With the existing high statistics data that have been collected
in the past decade at Jefferson Lab, and the future experiment
at CLAS12 and GlueX, there is a significant opportunity for making
progress at all three sectors of the hyperon spectroscopy.

\item \textbf{Photoproduction Mechanism for Hyperons}

With a high energy photon beam, it is generally understood that
hyperons can be produced via a series of kaon exchanges. Near
threshold, the contributions from intermediate nucleon resonances
that decay to strange particles, such as $K\Lambda$ and $K\Sigma$,
are expected to be significant. In fact, recent CLAS
data~\cite{Bradford,McC} on the polarization observables for
$\Lambda$ and $\Sigma$ photoproduction have played an important
role in establishing nucleon resonances. However, the 
non-resonance contributions of hyperon photoproduction remain 
not entirely understood. For example, a recent study~\cite{Freese} 
showed that $\Lambda$ photoproduction on a proton target, seems to 
be consistent with only $K$ exchanges. $\Sigma^0$ photoproduction
on a proton target, on the other hand, has comparable
contributions from both $K$ and $K^\ast$ exchanges. Decay
angular distributions for excited hyperons such as $\Lambda(1520)$, 
have been investigated in the past to probe the exchange mechanism 
in both photoproduction~\cite{Barber} and 
electroproduction~\cite{Steve}. However, higher statistics was
needed. Recent CLAS data on $\Lambda(1520)$, $\Lambda(1405)$ and
$\Sigma(1385)$ photoproduction~\cite{Moriya} are also suggestive
of the contributions of intermediate nucleon resonances. The
non-resonance contributions for these states can be further
constrained by investigating these states at higher energies,
expected to be feasible at both CLAS12 and GlueX.

\item \textbf{Beam Helicity Asymmetry in $K^+K^-$ Photoproduction}

Excited $S=-1$ hyperons could be produced in reactions such as
$\gamma p\rightarrow p K^+K^-$. However, such a reaction also
has significant contribution from intermediate meson resonances
that decay to $K\bar{K}$. This would be even more complicated
in the analogous reaction in the nucleon sector, $\gamma p
\rightarrow p\pi^+\pi^-$, where the two pion photoproduction
is believed to be important for identifying the missing nucleon
states. In that reaction, both pions can also resonance with
the nucleons, in addition to them being the decay products of
intermediate meson states. The $K^+K^-$ photoproduciton, on the
hand, does have the advantage due to the lack of $NK$ resonances.
In the end, however, the complete understanding of two 
pesudoscalar meson photoproduction typically still rely on models 
using effective lagrangian approach. In particular, it is 
important to point out that polarization observables such as the 
beam helicity asymmetry $I^\odot$ is expected to be sensitive to 
the interference of the various competing 
mechanisms~\cite{RobertsKK}, and essential to extract the 
information of the intermediate resonances.
\begin{figure}[tbh]
\begin{center}
 \includegraphics[width=92mm, height=74mm]{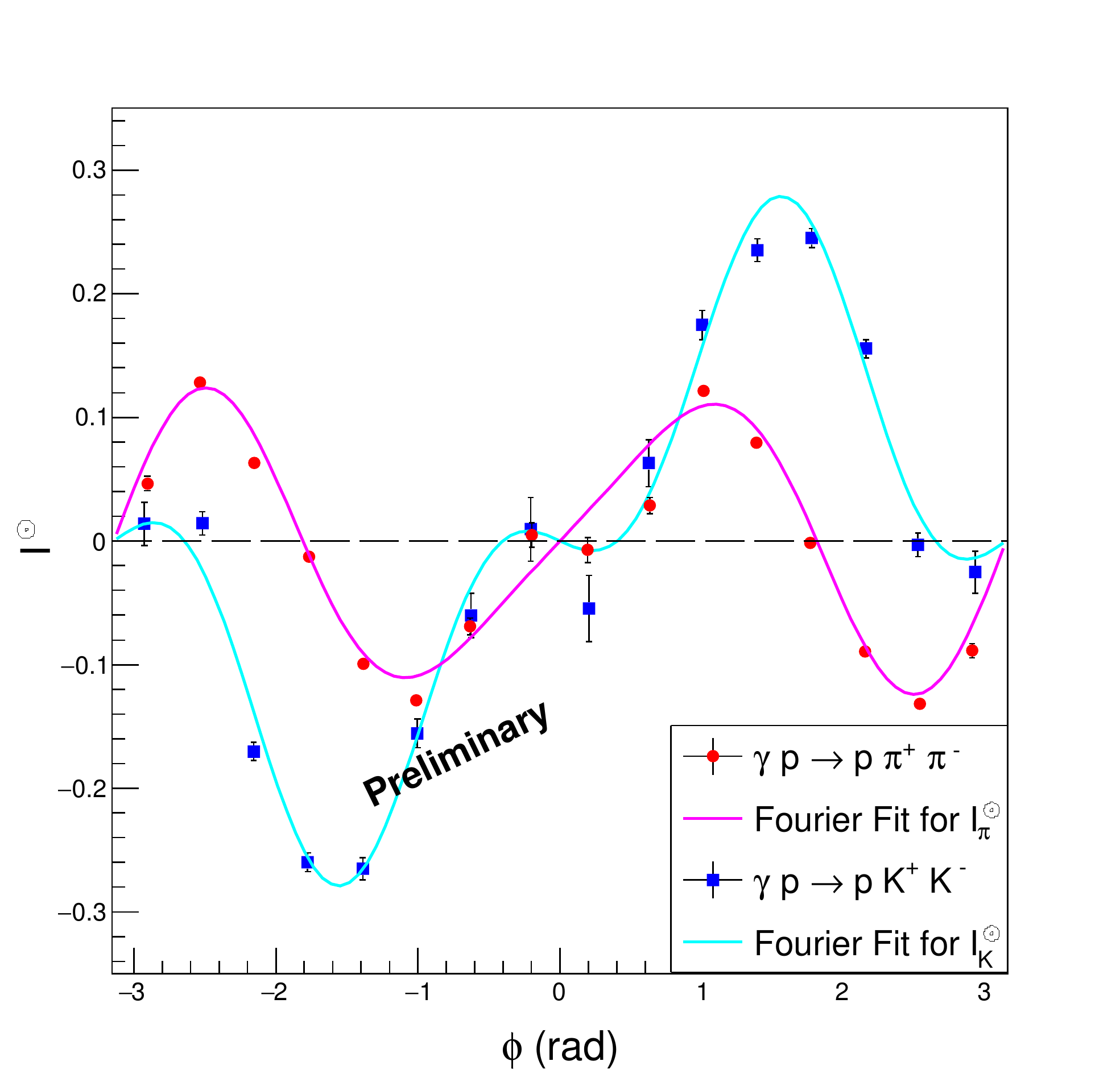}

  \caption{The beam helicity asymmetries, as a function of $\phi$,
	for the reaction of $\gamma p\rightarrow p\pi^+\pi^-$(red 
	solid circles), and $\gamma p\rightarrow p K^+K^-$ (blue 
	solid squares), from the g12 data set. $\phi$ is the angle 
	between the production plane and the two-meson plane. The 
	results are integrated over for the $E_{\gamma}$ range of 
	$2.8-5.4$~GeV and other kinematic variables.} 
	\label{KK_pipi}
\end{center}
\end{figure}

The beam helicity asymmetry $I^\odot(\tau)$, is defined by
\begin{equation}
\begin{centering}
	I^\odot(\tau)=\frac{1}{P_{\gamma}(\tau)}\frac{\sigma^+(\tau)
	-\sigma^-(\tau)}{\sigma^+(\tau)+\sigma^-(\tau)} ,
\end{centering}
\end{equation}
where $\tau$ is a kinematic bin, and $\sigma^{\pm}$ is cross
sections for photons in a $\pm$ helicity state. $I^\odot$ is
typically measured as a function of the angle $\phi$ between
different planes, such as the $K^+K^-$ plane and the production
plane in the center-of-momentum frame. Although the beam helicity
asymmetries have been measured in reactions such as  $\gamma p
\rightarrow p\pi^+\pi^-$~\cite{Strauch}, no data exist for the
two-kaon counterparts. The CLAS experiment E04-005 
(g12)~\cite{g12}, using a photon beam with energies up to 
$5.4$~GeV, and circular polarization up to $70\%$, has made it 
possible to perform these measurements for the two-kaon 
photoproduction on a proton target~\cite{Badui}. In 
Fig.~\ref{KK_pipi}, the asymmetries between the $K^+K^+$ and 
$\pi^-\pi^+$ are shown together, for events with $E_{\gamma}>
2.8$~GeV, as a function of the angle between the two-meson plane 
and the production plane. The features of the two reactions are 
strikingly different, with the two-pion channel showing a 
dominant $sin(2\phi)$ behavior, while the two-kaon data is 
mostly changing as a function of $sin(\phi)$. This could be due 
to the fact of that two-kaon photproduction is not expected to 
have contributions of $pK^+$ resonances, while $p\pi^{+/-}$ 
resonances are certainly not forbidden. However, in order to 
further probe the underlying contributions of various 
intermediate resonances, the beam helicity asymmetry must be 
measured as a function of various kinematic variables, such as 
$w$ (Fig.~\ref{ciw}) and $t$.

The angular dependence of the asymmetries, can be fitted to a
Fourier series. The sensitivities of these Fourier coefficients
to kinematic variables such as $pK^-$ invariant mass
(Fig.~\ref{cipk}), as well as the definition of the planes, are
indicative of the possible contributions of various intermediate
hyperon resonances. However, it is important to point out that
these results must be combined with cross section measurements,
in order to provide meaningful constraints for the production
models of two-kaon photoproduction. These efforts are in 
progress, and could further our understanding of the intermediate 
$S=-1$ hyperon resonances.
\begin{figure}[!tbh]
\begin{center}
 \includegraphics[width=92mm, height=74mm]{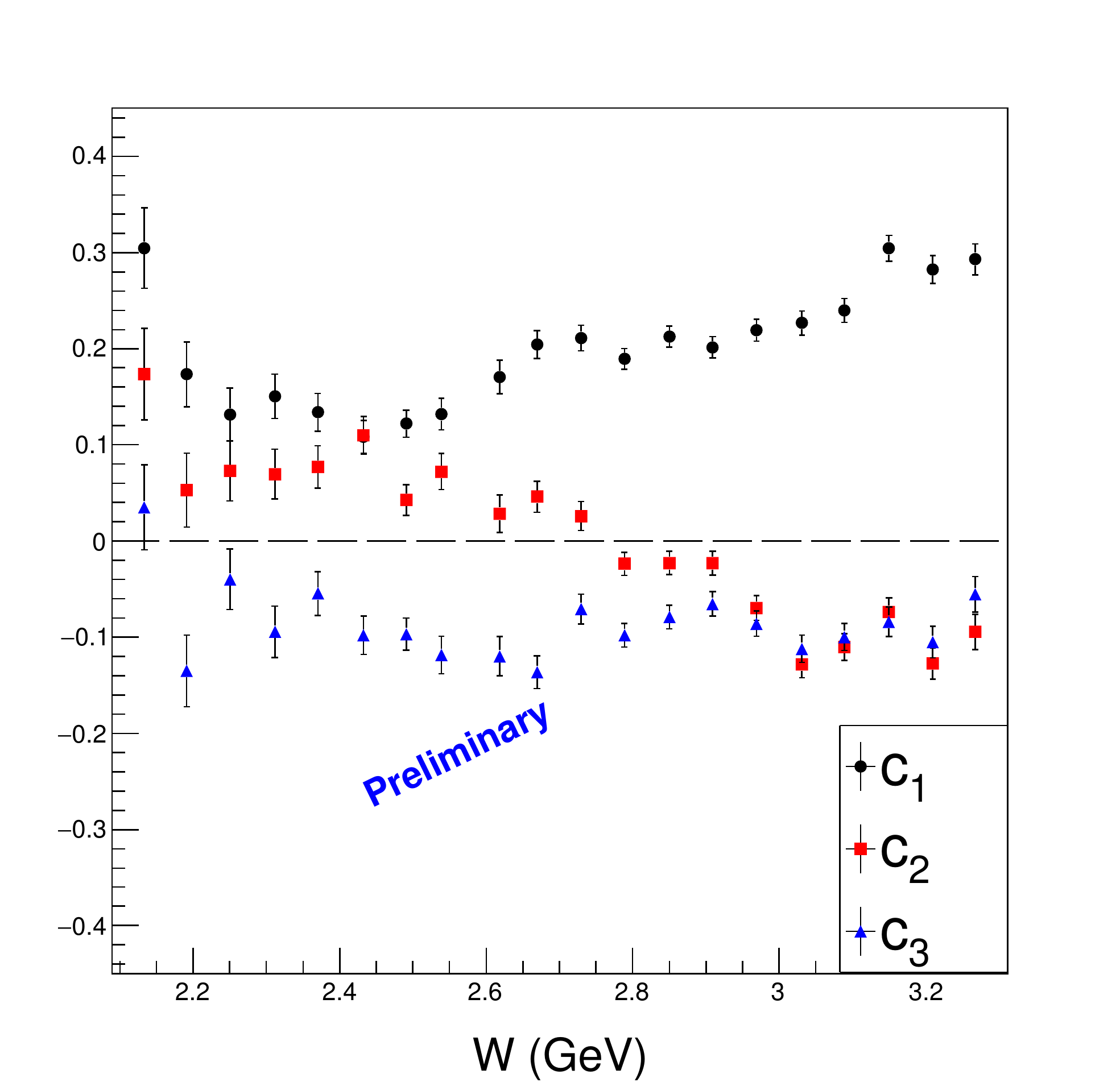}

  \caption{The beam helicity asymmetries as a function of $\phi$,
	are fitted to a Fourier series up to $sin(3\phi)$. The 
	coefficients are plotted as a function of $W$.}
	\label{ciw}
\end{center}
\end{figure}
\begin{figure}[!tbh]
\begin{center}
 \includegraphics[width=92mm, height=74mm]{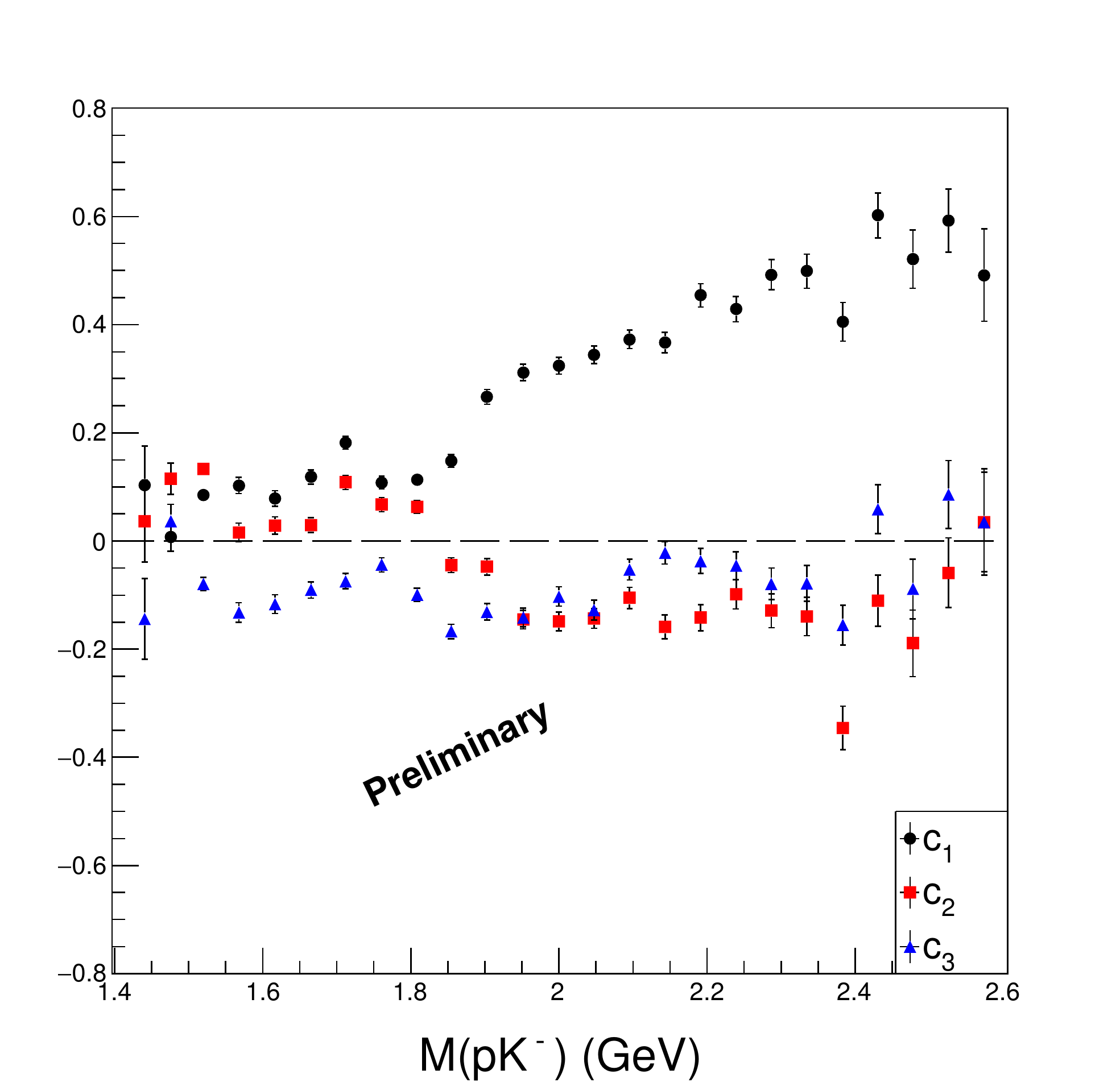}

  \caption{The beam helicity asymmetries as a function of $\phi$,
	are fitted to a Fourier series up to $sin(3\phi)$. The 
	coefficients are plotted as a function of $pK^-$ invariant 
	mass} \label{cipk}
\end{center}
\end{figure}

\item \textbf{Cascade Polarization in Photoproduction}

Recent CLAS data established that the lowest excited cascades,  
such as $\Xi(1320)$ and $\Xi(1530)$, can be produced copiously 
using a photon beam and a thick target for high 
luminosity~\cite{Guo}. Cascade production is also intimately 
related to excited hyperons~\cite{Nakayama, Man}. The $S=-1$ 
hpyeron states above $2$~GeV can be studied in unique channels 
such as $Y^\ast\rightarrow\Xi^-K^+$, as well as the  typical 
decay mode of $Y^\ast\rightarrow N\bar{K}^{(\ast)}$. Similar to 
the important roles of the $\Lambda$ and $\Sigma$ polarizations 
played in extracting the information of the intermediate nucleon 
resonances, the $\Xi^-$ polarization has also been expected to 
be essential in constraining the contributions of various 
intermediate high-mass $S=-1$ hyperons states. Due to the 
self-analyzing nature of the $\Xi(1320)$ weak decay,  its 
polarization can be measured in various photo-nucleon reactions. 
In reactions such as  $\gamma p\rightarrow K^+K^+\Xi^-(1320)$, 
with two pseudoscalar mesons ($K^+$) and one $J=\frac{1}{2}$ 
baryon ($\Xi^-(1320)$) in the final state, the expectation for 
$\Xi^-$ polarization could be very different from that of 
$\Lambda$ in $\gamma p\rightarrow K^+\Lambda$.  In fact, 
recent results on the polarization observables in reactions 
such as $\gamma p\rightarrow\pi\pi N$, using a more realistic 
three-body framework~\cite{Roberts}, suggests that the 
$\Xi^-(1320)$ polarization in all three directions could be 
non-zero. In addition, quasi-two-body models for $\Xi^-$
photoproduction also suggest non-zero $\Xi^-$ transferred
polarization ($C_{z}$)~\cite{Nakayama}. Therefore, the 
measurement of $\Xi^-$ polarization is an important tool to 
reveal the production mechanism, for which differential cross 
section measurements alone are not sufficient.

Recent CLAS data collected by the experiment E04-005 has made 
the measurement of $\Xi^-$ polarization in photoproduction 
possible for the first time~\cite{Bono}. In reaction $\gamma p 
\rightarrow K^+K^+\pi^-(\Lambda)$, the $\Xi^-$ was constrained 
from both the $K^+K^+$ missing mass and $\pi^-\Lambda$ invariant 
mass.  The $\Lambda$ is identified using the missing mass 
technique. The $\Xi^-$ decay is then fully reconstructed. The 
circular polarization of the beam is a function of the beam 
energy and the electron beam polarization, allowing the 
determination of the transferred polarizations $C_x$ and $C_z$ 
as well. The $z$-axis is along the beam, and the $y$-axis is 
along the norm of the production plane, defined by the beam, 
target and $\Xi^-$ vectors in the center-of-momentum frame. 
Preliminary results of the measured polarizations are shown 
as a function of  $\Xi^-$ center-of-momenum angle, in 
Fig.~\ref{g12pol}. These results are compared with calculations 
using parameters from Ref.~\cite{Nakayama}, and Ref.~\cite{Man}, 
which includes contributions from intermediate hyperons with 
$J>\frac{3}{2}$. It is important to point out that the lack of 
statistics in the existing data does not provide any 
differentiating power for distinguish the models. Furture 
experiments, discussed in the next section, will certainly be 
able to take these measurements to the necessary levels.
\begin{figure}[tbh]
\begin{center}
 \includegraphics[width=150mm, height=80mm]{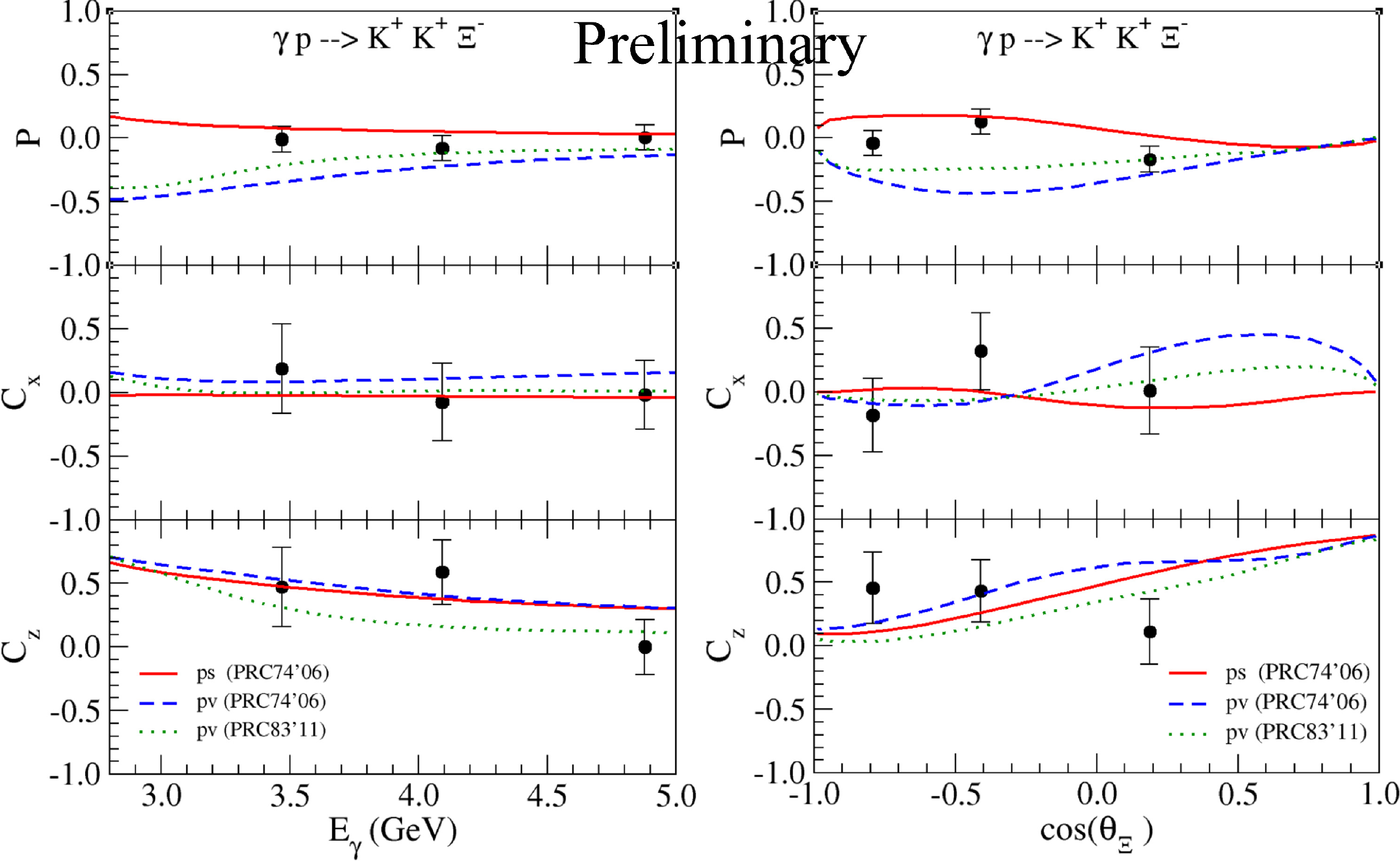}

  \caption{$\Xi^-$ polarization observables (Top: $P$, Middle:
	$C_x$, Bottom: $C_z$) as a function of $E_{\gamma}$ (Left) 
	and $\Xi^-$ angles in the center-of-momentum frame(Right). 
	Only the statistical uncertainties, which are dominant, 
	are shown. The red solid curve is based on the model from  
	Ref.~\protect\cite{Nakayama}, using pseudo-vector coupling. 
	The blue dash-dotted curve uses pseudo-scalar coupling. 
	The green dashed curve includes contributions from 
	intermediate hyperons with $J>\frac{3}{2}$~\protect\cite{Man}. 
	The theoretical curves are for $E_{\gamma}=4$~GeV.}
	\label{g12pol}
\end{center}
\end{figure}

\item \textbf{$\Omega^-$ Photoproduction and the Very Strange 
	Experiment at CLAS12}

The prediction and discovery of the $\Omega^-$ baryon certainly
was one of the great triumphs of the quark model. However, half
a century later, there has been little new information about the 
$\Omega$ and $\Xi$ baryons. In fact, only two $\Omega$ states 
and six $\Xi$ states are considered to be well-established, with 
at least three-star ratings in the PDG~\cite{PDG}.   Production 
of doubly- or triply-strange baryons by means of a photon beam 
(as with CLAS currently and with CLAS12 and GlueX in the future) 
is expected to shed light on the genesis of these states which 
involves the production of multiple $s\bar{s}$ pairs from the 
vacuum.  This significant change in baryon strangeness number 
from initial ($S=0$) to final state ($S=-3, -2$) could result 
from direct production via vector-meson dominance or from a 
sequence of intermediate transitions. Inference on the production 
mechanisms of these states in $\gamma p$ collisions can be 
obtained from precision measurements of the cross section and the 
invariant mass of these states.

The CLAS12 Very Strange Experiment, E12-11-005a~\cite{strange},
is expected to yield valuable data in the physics of $\Omega$
and cascade states. This experiment takes advantage of the
large luminosity after the $12$~GeV upgrade at the Thomas
Jefferson Laboratory. Using the CLAS12 spectrometer and the
Forward Tagger (FT) for the quasi-real photon beams, E12-11-005a
is expected to yield unprecedented statistics in the production
of Cascade and Omega baryons. Excited cascades  can be investigated
in reactions such as $\gamma p\rightarrow K^+K^+\Xi^-(1820)$,
$\Xi^-(1820) \rightarrow K^-\Lambda$, $\Lambda\rightarrow p\pi^-$. 
This would allow the determination of the spin and parity of the 
observed excited cascade states~\cite{DMA}. The polarization of 
ground state $\Xi$ can also be measured with uncertainties
sufficiently small to constrain production models that currently
can not separate the contributions of various intermediate hyperon
resonances.  As for the $\Omega^-$ photoproduction, detailed
differential cross section measurements can be performed,
necessary for the understanding the production mechanism,
differentiating various models such as vector meson dominance
and a sequential decay of intermediate states.

The expected statistics of various reactions for E12-11-005a,
is summarized in the Table~1:
\begin{table}[tbh]
\begin{center}
\label{t1}
\begin{centering}
\begin{tabular}{llll}
\hline
&Detected Particles & Measured Decays & Total Detected\\
$\Omega^- $& $K^+K^+K^0$ & &$ \sim7$k \\
$\Omega^-$ & $K^+K^+K^0 K^-$ &$\Omega^-$ &$\sim1$k \\
$\Xi^-$ & $K^+K^+\pi^-$ &$\Xi^-$ &$\sim0.9$M \\
$\Xi^-(1530)$ & $K^+K^+\pi^-$ &$\Xi^-(1530)$ &$\sim270$k \\
$\Xi^-(1820)$ & $K^+K^+K^-p$ &$\Xi^-(1820),\Lambda$ &$\sim12$k \\
\hline
\end{tabular}
\end{centering}
\caption{Expected Particle Rate for the CLAS12 Very Strange 
	Experiment (E12-11-005a), based on the simulation 
	including detection efficiency and branching ratios. 
	80 beam days were assumed.}
\end{center}
\end{table}

\item \textbf{Summary}

With the effort of the Jefferson Lab $12$~GeV upgrade ongoing, and 
the next generation of high statistics photoproduction experiments 
expected to yield unprecedented amount of data for hyperon states, 
it is an exciting time for hyperon spectroscopy. The existing data 
from the CLAS collaboration have already demonstrated the feasibility 
and importance of the various polarization observables in hyperon 
production, and its relation with the intermediate resonances. 
Future data from both CLAS12 and GlueX will no doubt provide more 
detailed measurements to understand the production mechanisms of 
the various hyperon states.

\item \textbf{Acknowledgments}

The authors would like to thank the Jefferson Lab Hall B staff, the 
very strange collaboration, and the CLAS g12 group. This work was 
supported by DOE award DE-SC0013620.
\end{enumerate}


\newpage
\subsection{The Role of Hadron Resonances in Hot Hadronic Matter}
\addtocontents{toc}{\hspace{2cm}{\sl J.L.~Goity}\par}
\setcounter{figure}{0}
\setcounter{table}{0}
\setcounter{footnote}{0}
\setcounter{equation}{0}
\halign{#\hfil&\quad#\hfil\cr
\large{Jos\'e~L.~Goity}\cr
\textit{Department of Physics}\cr
\textit{Hampton University}\cr
\textit{Hampton, VA 23668, U.S.A. \&}\cr
\textit{Thomas Jefferson National Accelerator Facility}\cr
\textit{Newport News, VA 23606, U.S.A.}\cr}

\begin{abstract}
Hadron resonances can play a significant role in hot hadronic matter. 
Of particular interest for this workshop are the contributions of 
hyperon resonances. The question about how to quantify the effects 
of resonances is here addressed. In the framework of the hadron 
resonance gas, the chemically equilibrated case, relevant in the 
context of lattice QCD calculations, and the chemically frozen case 
relevant in heavy ion collisions are discussed.
\end{abstract}

\begin{enumerate}
\item \textbf{Introduction}

Lattice QCD (LQCD) and high energy heavy collisions (RHICs) give access 
to QCD thermodynamics in the limit of low or vanishing conserved charges 
(Baryon number B, strangeness S or electric charge Q). While LQCD 
addresses the case of chemically equilibrated hot matter, which 
corresponds ot the early universe, HICs produces a fireball which 
expands too fast for chemical equilibrium to be maintained giving rise 
to a hot hadronic system which at kinetic freeze out is well off chemical 
equilibrium. At temperatures between 0.15 -- 0.17~GeV a cross over 
transition occurs from a quark-gluon to a hadronic phase. The rigorous 
description of the hadronic phase is in principle possible with a full 
knowledge of the S-matrix~\cite{Dashenj}. Absent that knowledge, one needs 
to consider models. The simplest model, known as the hadron resonance gas 
(HRG) turns out to provide a remarkably good  description of thermodynamic 
observables.
The HRG  is to a first approximation an ideal gas of hadrons, consisting 
of mesons and baryons and their resonances. The HRG gas is then determined
simply by the hadron spectrum. The hadron spectrum is however
incompletely known, in particular for baryons, and thus one question is
how important the role of such "missing" states may be in the HRG; this
issue is the main focus of this note. Several indications of missing
states exist, namely the known hadron spectrum has very few complete
$SU(3)$ multiplets, and recent LQCD calculations show the existence of
yet unobserved states, albeit at larger quark masses for
baryons~\cite{Edwardsj}. A first estimation of the missing states is based
on SU(3) and the PDG listed baryons is given by the number of different
strangeness isospin multiplets, namely: $\#\Sigma=\# \Xi=\# N+\#\Delta$
(PDG- 26; 12; 49), $\#\Omega=\#\Delta$ (4; 22), and $\#\Lambda=\#N+ \#
\text{ singlets}$ (18; 29). Thus on this count alone we are missing the
following isospin multiplets: 23 $\Sigma$, 11 $\Lambda$, 37 $\Xi$ and 18
$\Omega$. One expects even more missing states according to the quark
model, LQCD, and/or  the $SU(6)\times O(3)$ organization of multiplets.
The question is therefore how sensitive is the HRG to those missing
states, which consist in particular of a large number of hyperons.

The HRG is determined by the pressure, where the contribution to the
partial pressure by a given iso-multiplet $i$ is:
\beq
	p_i=T\frac{\partial}{\partial V}\log Z_i=T^2\, m_i^2 \,d_i \,
	\frac{1}{2\pi^2}\sum_{k=1}^{\infty}\frac{(-1)^{(1+k)B_i}}{k^2} \;
	K_2(k\frac{m_i}T)\;e^{k\mu_i/T} ,
\eeq
where $B_i$ the baryon number, $d_i=(2I_i+1)(2 J_i+1)$, and $\mu_i$
is a chemical potential, and $K_2$ is the modified Bessel function.
For our purposes where $T<0.16 \text{ GeV}$, keeping only the first
term in the sum is sufficient for all hadrons except the $\pi$, $K$
and $\eta$ mesons. Here we use the meson resonances listed in the PDG,
and for baryons we choose to use $SU(6)\times O(3)$ multiplets with
the mass formulas provided in Ref.~\cite{GMj}, where we will include
the $\bf{56}$ and $\bf{70}$ multiplets with $\ell=0,\cdots 4$.

\item \textbf{HRG in Chemical Equilibrium: LQCD}

We use here the results for QCD thermodynamics obtained in LQCD, and
the results are those of Ref.~\cite{Bazavov1j,Bazavov2j,Bazavov3j}. Above
the cross over transition there is a slow evolution towards the ideal
quark-gluon gas. Below the transition the HRG gives a remarkably good
description of the thermodynamic observables, as shown in Fig.~1. The
figure also shows the effect of excluding  baryon resonances; those
effects for the pressure $p$ and the entropy $s$ are modest  for
$T<0.15$~GeV, and the effects of the hyperon resonances become almost
insignificant.  Figs.~2 and 3 show the effects of resonances in particle
densities. Clearly it is not possible to disentangle the baryon resonance
effects through the global thermodynamic observables of the hadron gas
vis-\`a-vis the LQCD results. In chemical equilibrium, resonances rapidly
disappear with the falling temperature and so do their effects on total
thermodynamic observables.
It is therefore necessary to have  more sensitive observables  in order
to find the composition of the hadron gas:  for hyperons one needs  to
filter strangeness. This is achieved via the study of
correlations. In particular the susceptibilities (see for instance
Refs.~\cite{Rattij,corrBellwiedj,ERAj})  provide a useful tool. They are
defined by:
\beq
	\chi_2^{QQ'}\equiv \frac {1}{T^2}\frac{\partial^2 p}{\partial
	\mu_Q\partial \mu_{Q'}},
\eeq
where $Q$ and $Q'$ are conserved charges. If we consider only baryons,
we have that $\chi_2^{BB}\sim (n_B+n_{\overline B})/T^3$ and $\chi_2^{BS}
\sim (n_Y+n_{\overline Y})/T^3$: the HRG gives a very simple relation of
the susceptibilities to the particle number densities, which can be 
tested with the LQCD~\cite{Bazavov2}, as shown in Fig.~4. The agreement 
is reasonably good, in particular for $\chi_2^{BS}$, which provides
perhaps the best indication of the role of excited hyperons for
$T>0.13$~GeV. For more extensive discussions of fluctuations and LQCD
results see~\cite{Rattij,Bellwied1j,ERAj}.

Although one expects resonances to have contributions whose magnitude
is similar to the ones estimated with the HRG, it is also true that
deviations from the approximation of the HRG may be of similar
significance, and thus the LQCD results do not seem to permit for a
definite estimate of what we are missing in terms of baryon resonance
states.
\begin{figure}
\begin{center}
\includegraphics[width=0.35\linewidth]{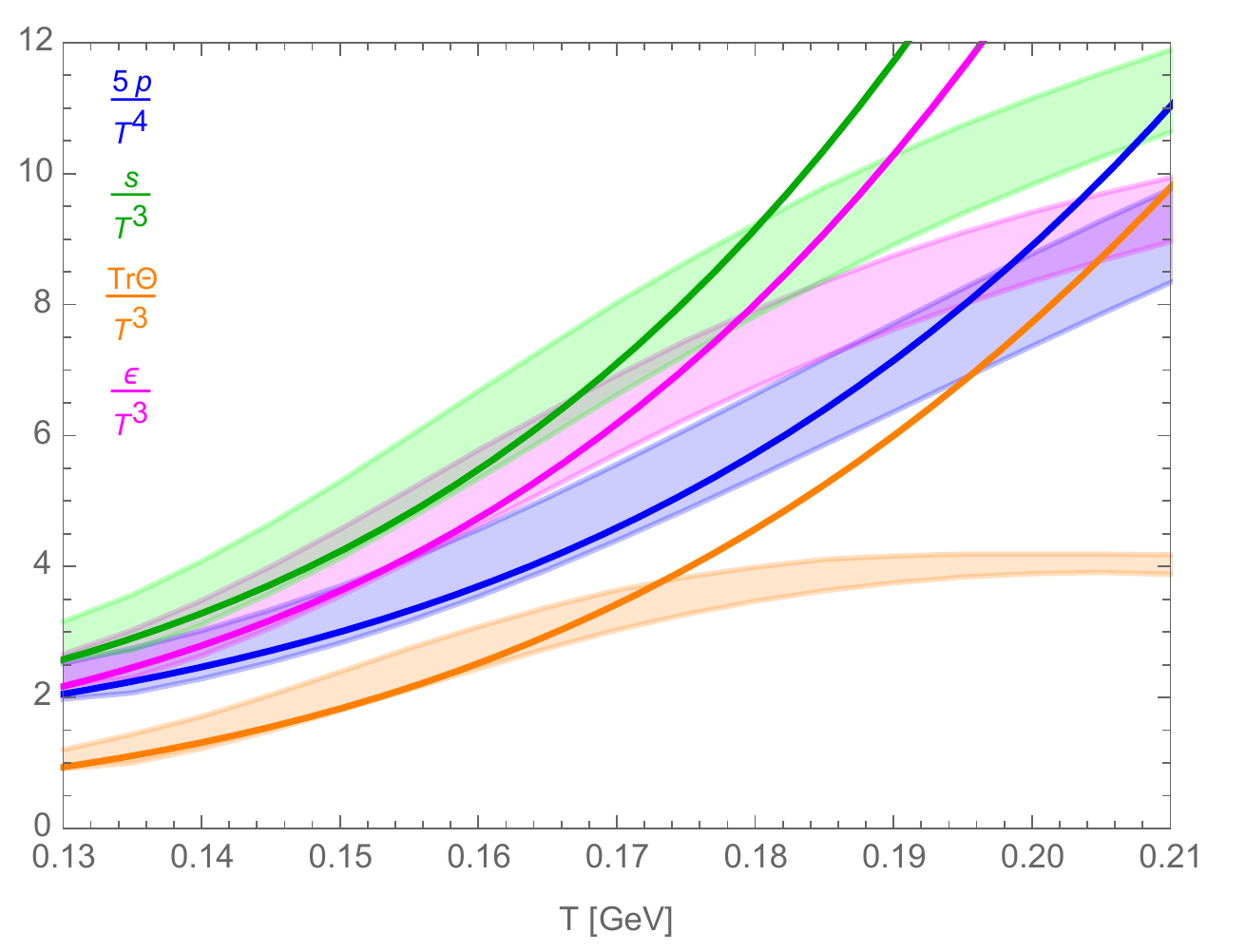}\\
\hspace*{-4mm}
\includegraphics[width=0.35\linewidth]{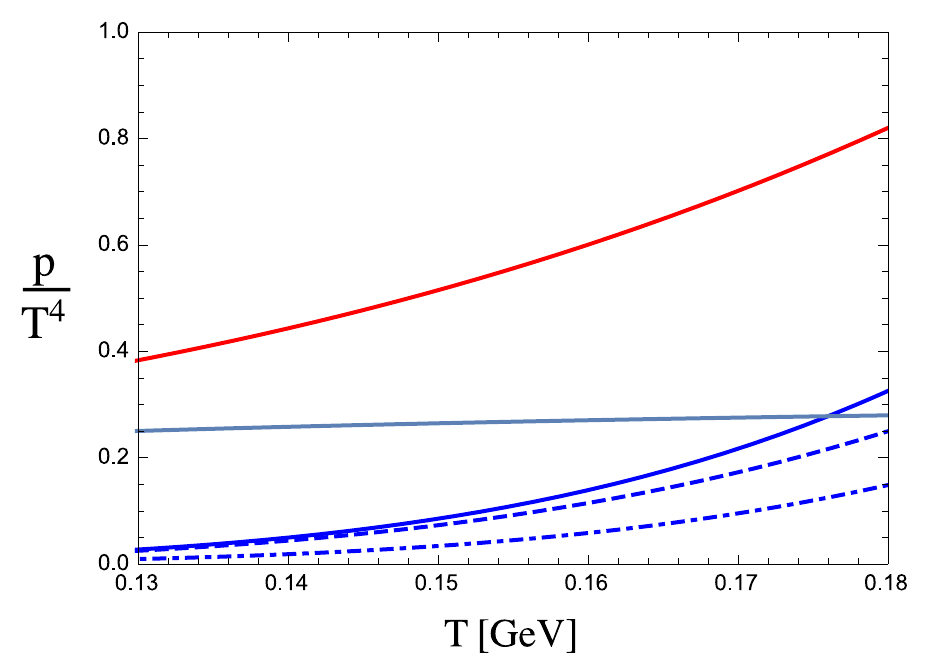}
\includegraphics[width=0.35\linewidth]{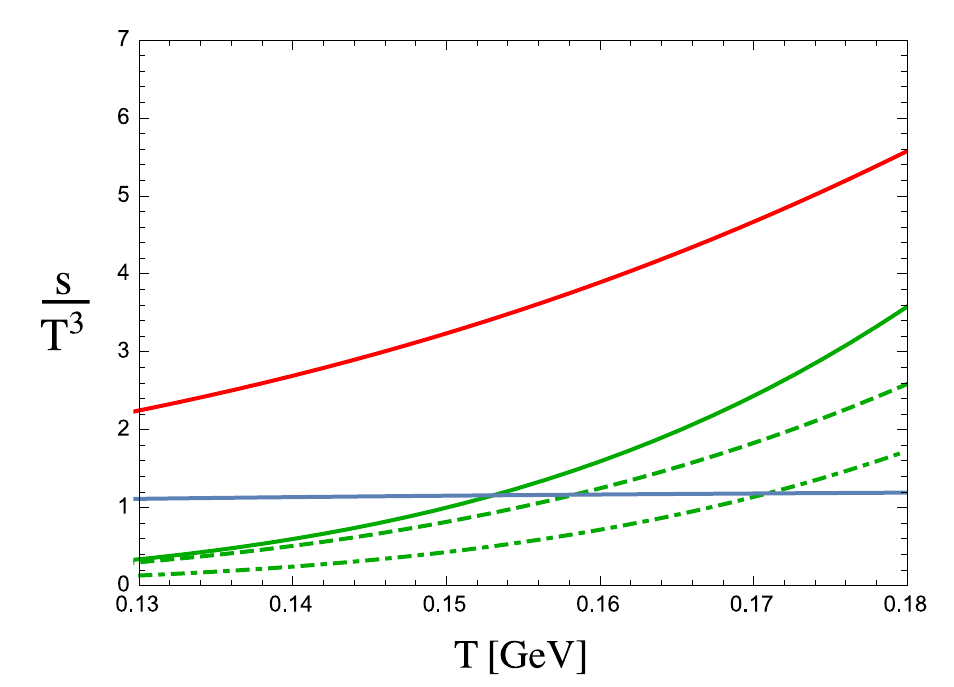}
\vspace*{-.8cm}
\end{center}

\caption{Thermodynamic observables at vanishing chemical potentials.
	Upper panel: bands from LQCD calculations~\protect\cite{Bazavov1j}, 
	solid lines from HRG. $s$ is the entropy density, $\epsilon$ is 
	the internal energy, and Tr$\Theta$ is the trace of the energy 
	momentum tensor. Lower panels show meson contributions (red), 
	pion contributions (gray), baryon contributions (solid), baryons 
	with hyperon resonances removed (dashed) and only hyperons 
	(dotdash). }
\end{figure}
\begin{figure}
\begin{center}
\includegraphics[width=1.005\linewidth]{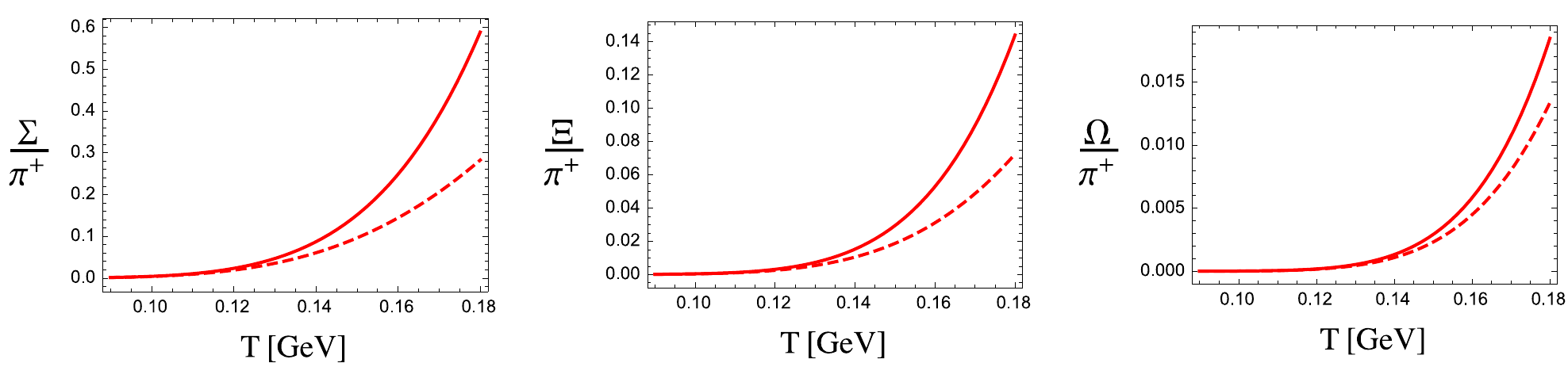}
\end{center}
\vspace*{-.8cm}

\caption{HRG hyperon number ratios with respect to $\pi^+$.
	Solid (dashed) line includes (excludes) baryon resonances.  }
\end{figure}
\begin{figure}
\begin{center}
\includegraphics[width=0.5\linewidth]{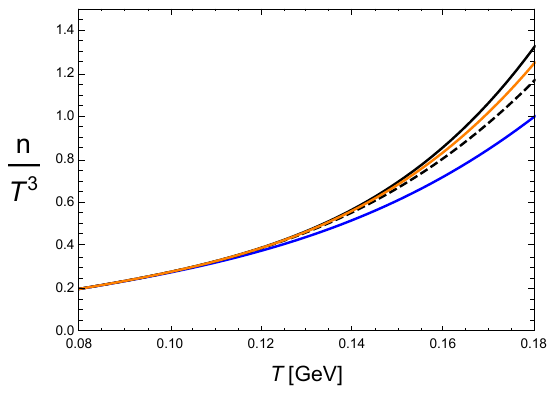}\\
\end{center}
\vspace*{-0.3cm}

\caption{Total particle number density at vanishing chemical
	potentials. All hadrons (black), all hadrons except hyperon
	resonances (orange), all hadron except all baryon resonances
	(dashed), and mesons only (blue).}
\end{figure}
\begin{figure}
\vspace*{-0.29cm}
\begin{center}
\includegraphics[width=0.895\linewidth]{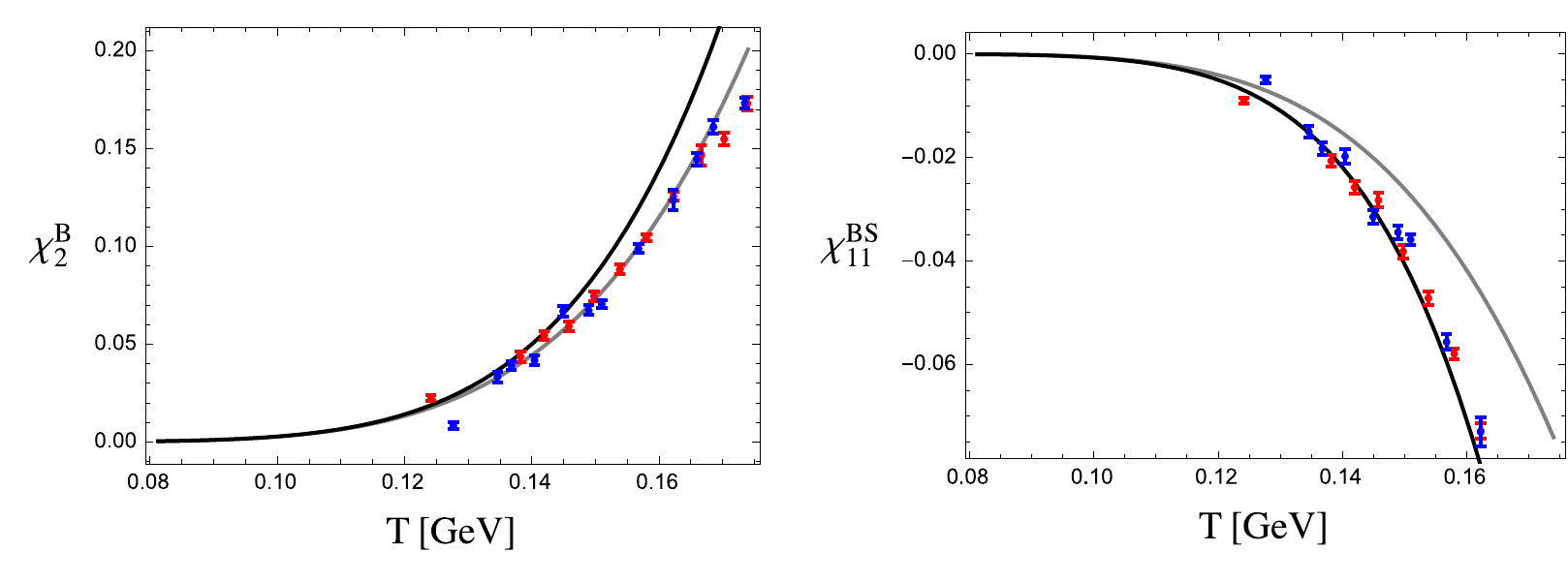}\\
\end{center}
\vspace*{-0.cm}

\caption{Baryon-Baryon number density susceptibility. In black the
	HRG, in gray  the HRG with hyperon resonances removed, and 
	the LQCD data is from Ref.~\protect\cite{Bazavov2j}
}
\end{figure}

\item \textbf{HRG off Chemical Equilibrium: RHICS}

The hot hadronic system produced in high energy HICs is for most of
its brief expansion off chemical equilibrium. In the HRG description,
this requires the inclusion of chemical potentials  to account for
the overabundance of the different hadrons. The presentation here is
basic, ignoring possible effects of hydrodynamics, and corresponds to
describing the thermodynamics of the HRG in the local co-moving frame;
it should be reasonably good for discussing particle yields. In the
absence of net B, S and Q, we associate to each isospin multiplet a
chemical potential, equal to that of the corresponding antiparticles.
Processes which remain in equilibrium  give relations between chemical
potentials, e.g., if $A+B\leftrightarrow C+D$ is in equilibrium, then
$\mu_A+\mu_B=\mu_C+\mu_D$, or $A+B\leftrightarrow C^\ast$ implies 
$\mu_{C^\ast}=\mu_A+\mu_B$. The existence of different reaction channels   
requires information about partial rates. Such information is extremely 
poorly known for most resonances, in particular baryons, and therefore 
one needs to resort to models. For the purpose of our discussion we 
adopt a very simple model, which assumes: 
i) resonances are in chemical equilibrium with respect to their decay 
	products, 
ii) non-strange meson resonances decay only into pions, 
iii) strange meson resonances decay into one Kaon and pions, 
iv) all baryon resonances have 2-body decays into the ground state 
	octet and decuplet, 
v) all non-strange baryon resonances decay only via pion emission, 
vi) $\Sigma$, $\Lambda$ and $\Xi$ resonances decay with different rates 
emitting pions and K or $\overline K$. Resonance chemical potentials are 
then given by:
\bea
	\mu_i^{M^\ast}=\sum_{j=\pi,K} \nu_{ij}^{M^\ast} \mu_j~~,&~~~&
	\mu_i^{B^\ast}=\sum_{j=N,\Sigma,\Lambda,\Xi,\Omega}
	\nu_{ij}^{B^\ast}(\mu_j
	+\delta_{S_i\,S_j}\,\mu_\pi+\delta_{S_i\,(S_j\pm1)}\,\mu_K),
\eea
where the decay rates are encoded in:
\bea
	\nu_{ij}^{M^\ast} &=&\delta_{S_i0}\;\delta_{j\pi} \;\overline{\eta}_i
	+\delta_{S_i\;\pm 1}\,(\delta_{jK}+(\overline{\eta}_i-1)\delta_{j\pi})
	\nonumber\\
	\nu_{ij}^{B^\ast}&=&\delta_{S_i0}\;\delta_{S_j 0} +\delta_{|S_i| 3}\;
	\delta_{|S_j| 2}+\sum_{S=1,2} \delta_{S\; |S_i|} \,(r_i \,\delta_{S_i\;
	S_j}+\frac{1-r_i}{2}\,\delta_{S_i\,(S_j\pm 1)})~~,~~~
	\sum_j  \nu_{ij}^{B^\ast}=1,
\eea
where $\overline{\eta}_i$ is the average particle multiplicity in
the decay of the meson resonance $i$, and $r_i$ is the branching fraction
of pion emission in the decay of the baryon resonance $i$. With this one
can define effective particle number densities:
\bea
	\overline{n}^M_i= n^M_i+\sum_j \nu^{M^\ast}_{ji} n_j^{M^\ast}~~,&~~~&
	\overline{n}^B_i= n^B_i+\sum_j \nu^{B^\ast}_{ji} n_j^{B^\ast},
\eea
where $i$ indicates a stable meson or baryon, and $n$ are the number
densities obtained with the corresponding chemical potentials. For
simplicity we have neglected the baryon resonance decay contribution to
$\overline{n}^M_i$. The particle number densities  $\overline n_i=\partial
p/\partial\mu_i$ are the ones observed after kinetic freeze out.

Here the discussion is focused on the possible effects of resonances in
the observed particle yields in RHICs. One can consider ratios of yields,
and also fluctuations. The ratios remain constant after chemical freeze
out. One can consider more detailed chemical freeze out for different
hadrons~\cite{Bellwied:2016kpjj}, but for brevity we take a simple one,
namely freeze out at about $T=0.15$~GeV for all stable hadrons. Due to our
lack of knowledge and for simplicity we take $r_i=r$ for all $i$, and
check the dependencies on $r$. Using the yield ratios with respect to
pions from ALICE~\cite{ALICEj}, Fig.~5, we choose $\mu_\pi=0$ at
$T=0.15$~GeV (this choice is  arbitrary here, and in particular the
value of $\mu_\pi$ at kinetic freeze out  is sensitive to it);
determinations of the initial $\mu_\pi$ can be improved via knowledge of
it at kinetic freeze out. Fitting to the ALICE yield ratios one fixes the
chemical potentials $\mu_i$ at the that initial $T$. To see how the HRG
evolves one uses the approximation~\cite{BGGLj} that the ratios of the
densities of conserved particle numbers to total entropy remain
approximately constant, namely $\overline{n}_i/s\sim\text{const}$.
This gives the chemical potentials in Figs.~5 and 6. One notices that,
as expected, the baryon resonances play a very marginal a role in the
evolution of the meson component of the HRG (mostly because of the
assumption in Eqn.~(5)); for nucleons the effects are also marginal
except at higher $T$. The main sensitivity to resonances is on the
hyperons,  and it depends  very much on the value of $r$. The effects
of baryon resonances on the effective chemical potentials defined by the
relation $n_i(T,\overline{\mu}_i)=\overline{n}_i(T,\{\mu_j\})$ are 
also relatively small. Clearly the effects of the baryon resonances 
are below other effects, such as the initial conditions right after 
hadronization and/or the uncertainties in the resonances' partial decay 
fractions. Thus, it is necessary to use more sensitive observables to 
acquire more sensitivity. For that purpose, one can consider 
susceptibilities such as the ones mentioned earlier, now adapted to a 
HRG off chemical equilibrium. They are simply defined by:
\beq
	\chi_2^{ij}=\frac{1}{T^2}\,\frac{\partial^2 p}{\partial \mu_i 
	\partial \mu_j},
\eeq
where the experimentally accessible quantities are ratios of those
susceptibilities, namely:
\beq
  R^{ij}_{kl}=\frac{ \chi_2^{ij}}{ \chi_2^{kl}}.
\eeq
Using the HRG off chemical equilibrium one easily calculates the ratios.
The most relevant ones involving hyperons are shown in Fig.~7. The ratios
are sensitive to the value of $r$: while for the case $r=1$ one cannot
distinguish effects of resonances, those effects are  significant for
$r=0.5$. In particular one needs the resonances to have a non-vanishing
$R^{\Omega\Xi}_{NN}$. Thus, the analysis of particle number fluctuations
and correlations are the most sensitive tool to extract information on
resonance effects from RHICs data.  Those effects come however modulated
by resonance partial decay rates which are little known. For further 
discussion  involving higher order correlations/fluctuations see  
Ref.~\cite{BellwiedProcj}.
\begin{figure}
\begin{center}
\includegraphics[width=0.8\linewidth]{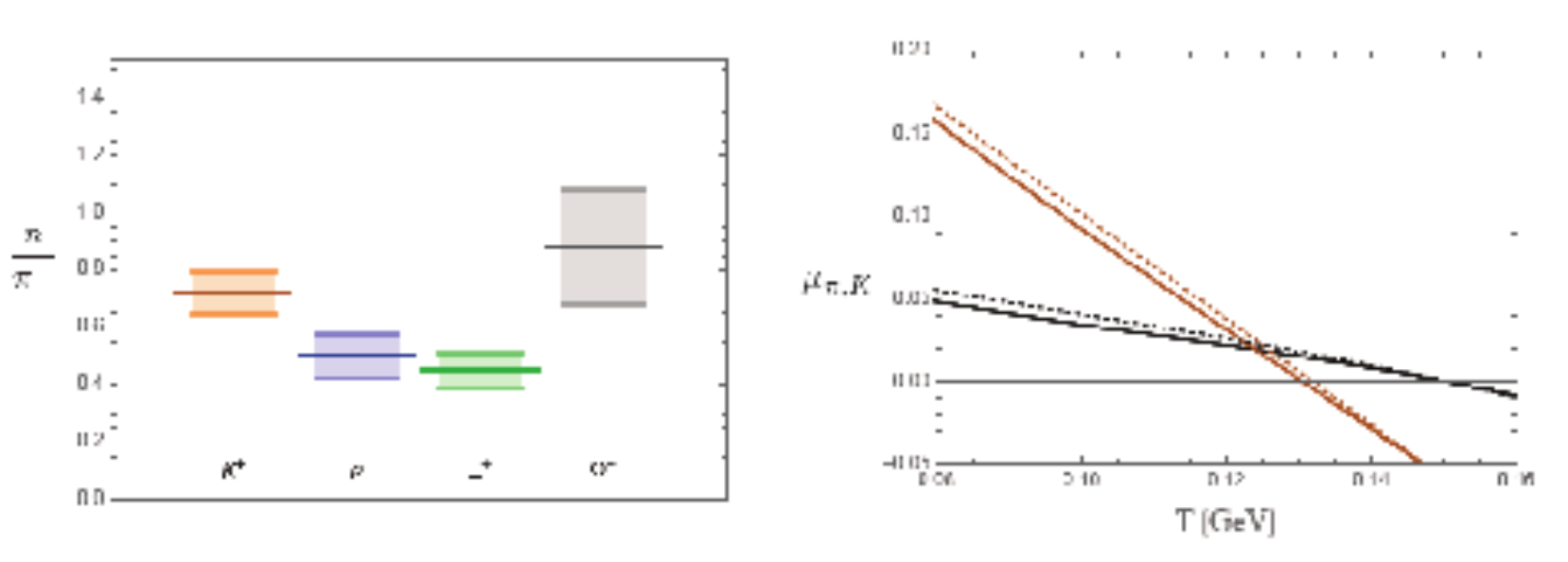}\\
\includegraphics[width=0.45\linewidth]{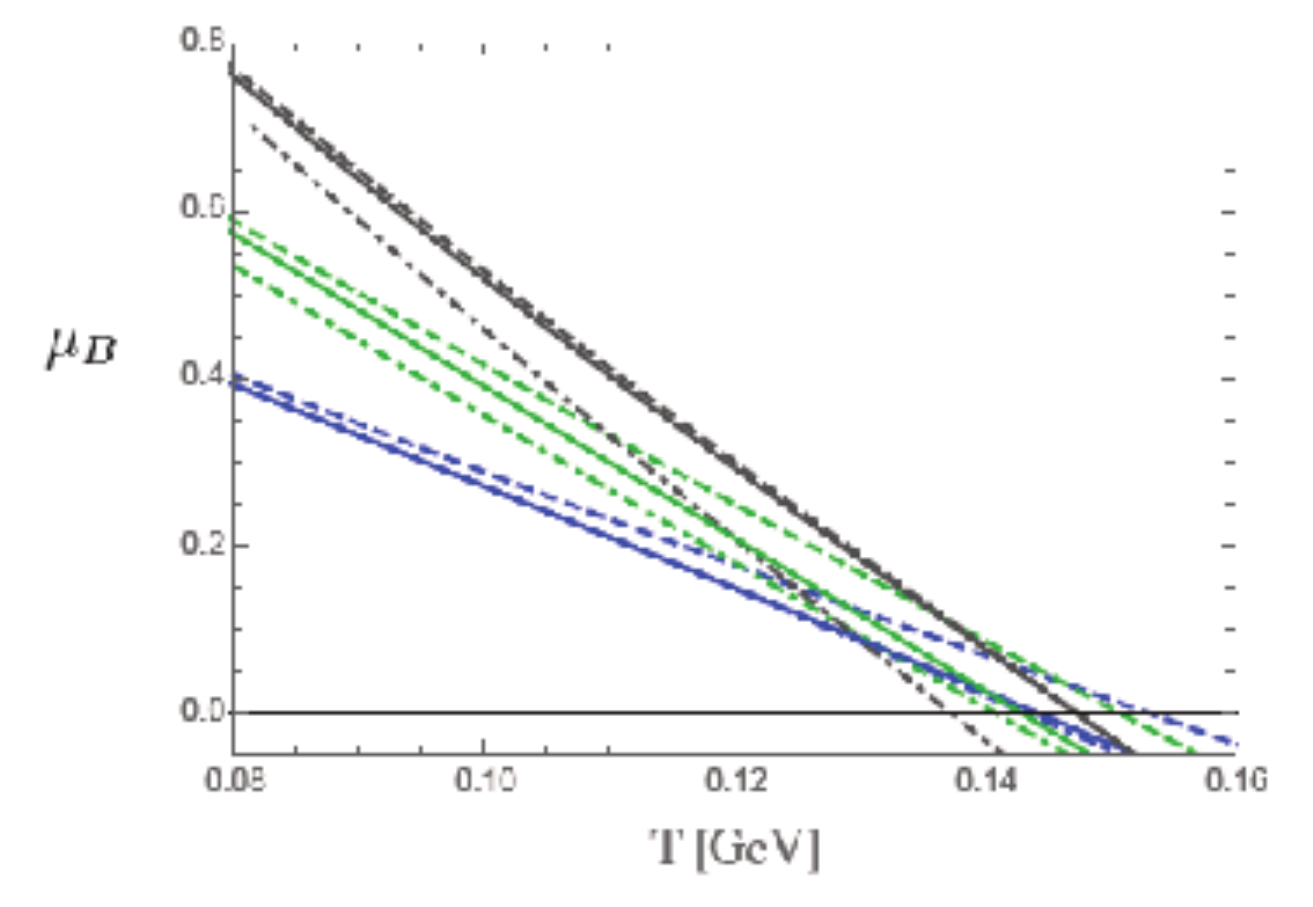}
\end{center}

\caption{Upper left: particle number density ratios with respect to
	$\pi^+$ from ALICE Pb-Pb collisions at 2.76~TeV~\protect\cite{ALICEj}, 
	scaled according to: $5\times K^+$, $10\times p$, $100\times\Xi^+$ 
	and $10^3\times \Omega^-$. They are inputs to fit initial values of 
	chemical potentials at $T=0.15$~GeV. Upper right: evolution of 
	$\mu_\pi$ (black) and $\mu_K$ (brown) with $T$: full HRG (solid) 
	and HRG with no baryon resonances (dotted). Lower panel: baryon 
	chemical potentials, $N$ (blue), $\Xi$ (green) and $\Omega^-$ 
	(gray); full HRG with $r=1$ (solid) and $r=0.5$ (dotdash), and HRG 
	without resonances (dashed).}
\end{figure}
\begin{figure}
\begin{center}
\includegraphics[width=0.8\linewidth]{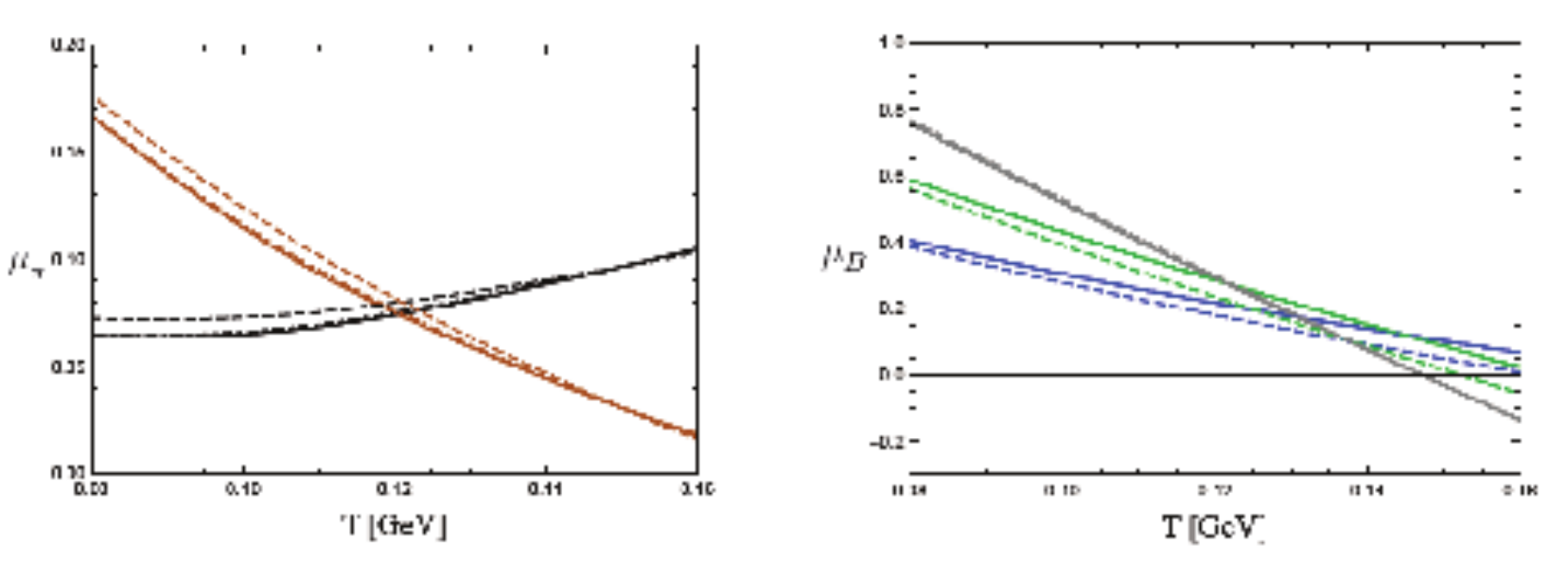}
\end{center}

\caption{Left: pion effective chemical potential in GeV, and  right:
	baryon effective chemical potentials. Same conditions as in 
	Fig.~5.}
\end{figure}
\begin{figure}
\begin{center}
\includegraphics[width=1\linewidth]{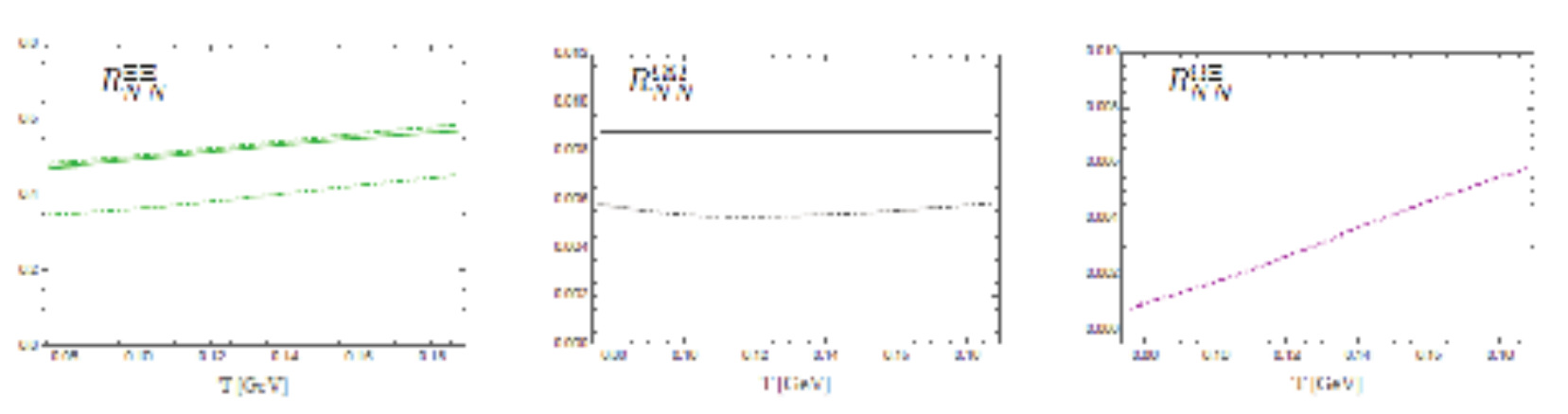}
\end{center}

\caption{Correlation ratios: HRG with $r=1$ (solid), with $r=0.5$ 
	(dotdash) and with no resonances (dashed).}
\end{figure}

\vspace{1cm}
\item \textbf{Comments}

In principle, it is possible to obtain indications of hadron resonance
effects, in particular baryons, in the hadronic phase of hot QCD. Studies
with LQCD and RHICs are the sources of relevant information. Using the
HRG one can then quantify those effects within the model, and draw
conclusions as discussed in this note. For the case of LQCD, the main
sensitivity to resonances resides in fluctuations/correlations  such as
susceptibilities, which strongly indicate the importance of hyperon
resonances for $T>0.13$~GeV. In the case of RHICs there is less
certainty on the conclusions due to lack of knowledge of the initial
thermodynamic state of the hadronic  fireball and  of  the partial
decay rates of resonances which affect the  expansion off chemical
equilibrium. There is however interesting information encoded in
ratios of correlations, which may definitely require the effect of
resonances to be explained.

\newpage
\item \textbf{Acknowledgments}

This work was supported in part by DOE Contract No. DE--AC05--06OR23177 
under which JSA operates the Thomas Jefferson National Accelerator 
Facility and by the National Science Foundation through grants 
PHY--1307413 and PHY--1613951.
\end{enumerate}


\newpage
\subsection{Bound States of $N$'s and $\Xi$'s}
\addtocontents{toc}{\hspace{2cm}{\sl H.~Garcilazo, A.~Valcarce, and 
	J.~Vijande}\par}
\setcounter{figure}{0}
\setcounter{table}{0}
\setcounter{equation}{0}
\setcounter{footnote}{0}
\halign{#\hfil&\quad#\hfil\cr
\large{Humberto~Garcilazo}\cr
\textit{Escuela Superior de F\' \i sica y Matem\'aticas}\cr
\textit{Instituto Polit\'ecnico Nacional}\cr
\textit{07738 M\'exico D.F., Mexico}\cr\cr
\large{Alfredo~Valcarce}\cr
\textit{Departamento de F\'\i sica Fundamental and IUFFyM}\cr
\textit{Universidad de Salamanca}\cr
\textit{E-37008 Salamanca, Spain}\cr\cr
\large{Javier~Vijande}\cr
\textit{Departamento de F\'{\i}sica At\'{o}mica, Molecular y Nuclear}\cr
\textit{Universidad de Valencia \&}\cr
\textit{IFIC (UV-CSIC)}\cr
\textit{E-46100 Valencia, Spain}\cr}

\begin{abstract}
Using local central Yukawa-type Malfliet-Tjon interactions reproducing
the low-energy parameters and phase shifs of the $NN$ system and the
latest updates of the $N\Xi$ and $\Xi\Xi$ Nijmegen ESC08c potentials, 
we point out that these interactions predict bound states of 2, 3, and 
4 particles consisting of nucleons and cascades so that very likely 
there will be a full periodic table of $\Xi$ hypernuclei. We also 
discuss a resonant solution of the coupled $\Xi NN$-$\Lambda\Lambda N$ 
system.
\end{abstract}

\begin{enumerate}
\item \textbf{S-wave Two-body Interactions}

We have studied the bound-state problem of few-body systems composed
of $N$'s and $\Xi$'s using two-body interactions of Malfliet-Tjon
type~\cite{Mal69} that fit the low-energy data and the phase shift
of a given two-body channel. These interactions are of the form
\begin{equation}
	V^{ij}(r)=-A\frac{e^{-\mu_Ar}}{r}+B\frac{e^{-\mu_Br}}{r} \, ,
\label{eq1}
\end{equation}
that is, an attractive and a repulsive Yukawa terms. We give in Table~1
the parameters of these interactions for the $NN$, $N\Lambda$, $N\Xi$,
and $\Xi\Xi$ systems. As shown in Ref.~\cite{Mal69}, the $NN$ 
interactions given in Table~1 predict a tritium binding energy of 
8.3~MeV which is not far from the experimental value of 8.48~MeV. Also, 
if we use those $NN$ interactions together witn the $N\Lambda$ 
interactions given in Table~1, we obtain a $\Lambda$-hypertriton 
separation energy of 144~keV which is quite consistent with the 
experimental value of 130 $\pm$ 50~keV. Thus, our approach appears to 
be reasonable.
\begin{table}[t]
\centerline{\parbox{0.80\textwidth}{
 \caption{Parameters of the local potentials given by Eq.~(\ref{eq1}) 
	for the $NN$, $N\Lambda$, $N\Xi$, and $\Xi\Xi$ subsystems.}} }
\vspace{0.3cm}
\begin{center}
\begin{tabular}{cccccc}
\hline \hline
& $(i,j)$ &  $A$(MeV fm) & $\mu_A({\rm fm}^{-1}$) & $B$(MeV fm) 
& $\mu_B({\rm fm}^{-1})$  \\
\hline
$NN$ & $(0,1)$  & $626.885$ & $1.55$  & $1438.72$ & $3.11$ \\
     & $(1,0)$  & $513.968$ & $1.55$  & $1438.72$ & $3.11$ \\ 
\hline
$\Lambda N$  & $(1/2,0)$ & $280$ & $2.00$ & $655$ & $3.55$ \\
             & $(1/2,1)$ & $170$ & $1.95$ & $670$ & $4.60$ \\ 
\hline
$\Xi N$      & $(0,0)$ & $120$  & $1.30$  & $510$ & $2.30$ \\
             & $(0,1)$ & $434$  & $2.68$  & $980$ & $6.61$ \\
             & $(1,0)$ & $290$  & $3.05$  & $155$ & $1.60$ \\
             & $(1,1)$ & $568$  & $4.56$  & $425$ & $6.73$ \\ 
\hline
$\Xi \Xi$    & $(0,1)$ &  $210$ & $1.60$  & $560$ & $2.05$ \\
             & $(1,0)$ &  $155$ & $1.75$  & $490$ & $5.60$ \\
\hline
\hline
\end{tabular}
\end{center}
\end{table}
The interactions of the $N\Xi$ and $\Xi\Xi$ subsystems are based in 
the latest updates of the Nijmegen ESC08c potentials~\cite{Nag15,Rij13}. 
In the $N\Xi$ case the singlet interactions are repulsive and the 
triplet ones are attractive, in particular the (1,1) interaction has 
a bound state of about 1.6~MeV the so-called $D^\ast$~\cite{Nag15}. 
For the $\Xi\Xi$ subsystem, on the other hand, the singlet channel 
is attractive and the triplet is repulsive.

\newpage
\item \textbf{Three-body Equations}

Restricting ourselves to configurations where all three particles 
are in $S$-wave states the Faddeev equations for the bound-state 
problem in the case of three baryons with total isospin $I$ and 
total spin $J$ are 
\begin{eqnarray}
	T_{i;IJ}^{i_ij_i}(p_iq_i) = &&\sum_{j\ne i}\sum_{i_jj_j}
	h_{ij;IJ}^{i_ij_i;i_jj_j}\frac{1}{2}\int_0^\infty q_j^2dq_j \nonumber
	\int_{-1}^1d{\rm cos}\theta\,
	t_{i;i_ij_i}(p_i,p_i^\prime;E-q_i^2/2\nu_i)
	\nonumber \\ &&
	\times\frac{1}{E-p_j^2/2\mu_j-q_j^2/2\nu_j}\;
	T_{j;IJ}^{i_jj_j}(p_jq_j),
	\label{eq2}
\end{eqnarray}
$t_{i;i_ij_i}$ is  the amplitude of the pair $jk$ with isospin $i_i$ and 
spin $j_i$ while $h_{ij;IJ}^{i_ij_i;i_jj_j}$ are the spin-isospin 
recoupling coefficients. $p_i$ is the momentum of the pair $jk$ (with $ijk$ 
an even permutation of 123) and $q_i$ the momentum of particle $i$ with 
respect to the pair $jk$ while $\mu_i$ and $\nu_i$ are the corresponding 
reduced masses.

\item \textbf{The Four-body Problem}

The four-body problem has been addressed by means of a generalized 
variational method. The nonrelativistic hamiltonian will be given by,
\begin{equation}
	H=\sum_{i=1}^4\frac{\vec p_{i}^{\,2}}{2m_{i}}+\sum_{i<j=1}^4V(\vec 
	r_{ij}) \, ,
	\label{ham}
\end{equation}
where the potentials $V(\vec r_{ij})$ have been discussed in the 
previous section. For each channel $s$, the variational wave function 
will be the tensor product of a spin ($\left|S_{s_1}\right>$), isospin 
($\left|I_{s_2}\right>$), and radial ($\left|R_{s_3}\right>$) component.
Once the spin and isospin parts are integrated out, the coefficients of 
the radial wave function are obtained by solving the system of linear 
equations,
\begin{equation}
\label{funci1g}
	\sum_{s'\,s} \sum_{i} \beta_{s_3}^{(i)}
	\, [\langle R_{s_3'}^{(j)}|\,H\,|R_{s_3}^{(i)}
	\rangle - E\,\langle
	R_{s_3'}^{(j)}|R_{s_3}^{(i)}\rangle \delta_{s,s'} ] = 0
	\qquad \qquad \forall \, j\, ,
\end{equation}
where the eigenvalues are obtained by a minimization procedure.

\item \textbf{Strong Decay of $\Xi$-Hypernuclei}

$\Xi$-hypernuclei can undergo a strong decay due to the process
\begin{equation}
	N+\Xi\to\Lambda+\Lambda,
	\label{eq7}
\end{equation}
which takes place in the $(i,j)=(0,0)$ two-body channel when only
S-wave configurations are taken into account. Thus, in that case
a $\Xi$-hypernucleus will be stable if the $\Xi N$ two-body channels
do not include the (0,0) channel. More generally, since the strong
decay process (\ref{eq7}) can only take place in isospin 0 two-body
channels, a $\Xi$-hypernucleus where all the two-body channels are 
in isospin 1 states will be stable under the strong interaction. We 
will consider examples of these two cases in the next sections.

\item \textbf{The $NN\Xi$ $(I,J^P)=(\frac{1}{2},\frac{3}{2}^+)$ State}

In this three-body state the spins of the three baryons are parallel
so that the two-body subsystems must all be in spin-triplet channels.
They are the $NN$ $^3S_1(I=0)$ channel which contains the deuteron
and the $N\Xi$ $^3S_1(I=0)$ and $^3S_1(I=1)$ channels. As we have
mentioned before, these two $N\Xi$ channels are attractive and 
moreover the last one contains the $D^\ast$ bound state. Therefore, 
one expects that the three-body state will be bound with a large 
binding energy and indeed that is the case since we obtain for the 
binding energy
\begin{equation}
	B=17.2\,\,\, {\rm MeV}.
	\label{eq8}
\end{equation}
Notice that the $\Xi N$ $^1S_0(I=0)$ (0,0) channel does not contributes
so that in this pure S-wave configuration the bound state is stable. Of 
course there will be a contribution to the width by the decay
$N+\Xi\to\Lambda+\Lambda$ when $N\Xi$ is in a P-wave and the spectator
nucleon is also in a P-wave but due to the angular momentum barriers
one expects that this width will be very small. In our previous
works~\cite{Gar1,Gar2} we have reported a smaller value for this
energy since we erroneously quoted the separation energy instead of
the binding energy as well as the effect of a small numerical error.

\item \textbf{Maximal Isospin States}

The decay process $N+\Xi\to\Lambda+\Lambda$ can not take place if the
$\Xi$-hypernucleus is in a state of maximal isospin since in that case
all the $\Xi N$ and $\Xi\Xi$ two-body subsystems must necessarily be 
in an isospin 1 state~\cite{Gar3}. This will be the case for systems 
consisting only of neutrons and negative $\Xi$'s or composed only of 
protons and neutral $\Xi$'s.

We show in Table~2 the first four maximal-isospin $\Xi$-hypernuclei. 
Since both $N$ and $\Xi$ have isospin 1/2 the total isospin of these 
hypernuclei is equal to the number of baryons divided by two. Since 
the space and spin parts of these $\Xi$-hypernuclei is the same as 
those of the corresponding ordinary nuclei they could be formed by 
simply replacing the protons of the ordinary nuclei by negative
$\Xi$'s or the neutrons by neutral $\Xi$'s. Very likely there will be 
a full periodic table of stable maximal-isospin $\Xi$-hypernuclei.
\begin{table}[t]
\centerline{\parbox{0.80\textwidth}{
\caption{\label{tab:table1} Binding energies of the first four 
	$\Xi$-hypernuclei.} }}
\vspace{0.3cm}
\begin{center}
\begin{tabular}{ccccc}
\hline
\hline
& hypernucleus   & $(I,J^P)$ & B (MeV) & \\
\hline
& $^2_\Xi H$ & $(1,1^+)$ & 1.6  & \\
& $^3_\Xi H$ & $(\frac{3}{2},\frac{1}{2}^+)$ & 2.9  & \\
& $^3_{\Xi\Xi} He$ & $(\frac{3}{2},\frac{1}{2}^+)$ & 4.5  & \\
& $^4_{\Xi\Xi} He$ & $(2,0^+)$ & 7.4  & \\
\hline
\hline
\end{tabular}
\end{center}
\end{table}

\newpage
\item \textbf{The KISO Event}

The only $\Xi$-hypernucleus observed up to now, the so-called 
KISO event~\cite{KISO}, takes place through the process
\begin{equation}
	\Xi^- + ^{14}N\to ^{10}_\Lambda Be +^5_\Lambda He,
	\label{eq9}
\end{equation}
where the bound state, once formed, immediately decays due to 
the strong process $N+\Xi\to\Lambda+\Lambda$ giving rise to the 
two fragments observed, each one being a $\Lambda$-hypernucleus.
The separation energy of the $\Xi$-hypernucleus is 4.4~MeV which 
is of the same order of magnitude as the binding energies of 
Table~2.

\item \textbf{The $\Xi NN$---$\Lambda\Lambda N$
	$(I,J^P)=(\frac{1}{2},\frac{1}{2}^+)$ Resonance}

The $\Xi NN$ system is coupled to the $\Lambda\Lambda N$ system in 
the $(I,J^P)=(\frac{1}{2},\frac{1}{2}^+)$ state since the $N\Xi$ 
(0,0) channel which is responsible for the strong decay is allowed 
to participate. Thus, in order to look for three-body resonances 
in this system one has to solve the integral equations of the 
coupled three-body system in the complex plane. This has been done 
using separable potentials~\cite{Gar4} taking the $\Xi$ mass as 
the average of the $\Xi^0$ and $\Xi^-$ and the nucleon mass as the 
average of the proton and neutron. We used the low-energy 
parameters of the Nigmegen ESC08 baryon-baryon interactions for the 
systems with strangeness 0, -1, and -2 to construct a separable 
potential model of the coupled $\Lambda\Lambda N$---$\Xi NN$ system 
in order to study the position and width of the three-body $(I,J^P)
=(\frac{1}{2},\frac{1}{2}^+)$ resonance.

We found that the three-body resonance lies at~\cite{Gar4}
\begin{equation}
	E_0=23.408-i0.045\,\,\,\, {\rm MeV}
	\label{eqapp}
\end{equation}
measured with respect to the $\Lambda\Lambda N$ threshold, {\it i.e.}, 
just 0.012~MeV below the $\Xi d$ threshold. Thus, this is a loosely 
bound state of a $\Xi$ and a deuteron with a small decay width into 
$\Lambda\Lambda N$.

The result (\ref{eqapp}) is somewhat intriguing, in particular
the very small width, since the $\Lambda\Lambda N$ threshold is open.
In order to understand that result we disconected the $N\Xi$ (0,0) 
channel and got an eigenvalue of $E_0=23.386$~MeV. Thus, the effect 
of the $\Lambda\Lambda-N\Xi$ (0,0) channel in the energy eigenvalue 
is negligible near the $\Xi d$ threshold. Adding the rest masses to 
the result (\ref{eqapp}) we get that  the three-body $(\frac{1}{2},
\frac{1}{2})$ resonance lies at $W_0 = 3194$~MeV and has a very small 
width of $\Gamma=0.09$~MeV.

The negligible effect of the absorption channel is a consequence of
the very small binding energy which implies that the main component 
of the wave function is a $\Xi$ orbiting around the deuteron at a 
very long distance. Using first order perturbation theory one can 
estimate the effect of the absorptive channel as 
$\delta E=\br\Psi\mid H_{ABS}\mid\Psi\et$, where 
$H_{ABS}=V_{N\Xi\to\Lambda\Lambda}G_0 V_{\Lambda\Lambda\to N\Xi}$. 
So that $H_{ABS}$ is a short-range operator while $\Psi$ is long 
ranged so that the overlap between $\Psi$ and $H_{ABS}$ is very 
small and consequently $\delta E$ is negligible.

Finally, since the effect of the absorptive channel is negligible,
we calculated the $\Xi NN$ bound state using the $NN$ and $N\Xi$ 
Malfliet-Tjon potentials of Table~1 without the $N\Xi$ (0,0) channel 
and found that the system is unbound by about 0.01~MeV which means 
that in this case the state will appear as a very sharp resonance 
just above the $\Xi d$ threshold.

Thus, the $(I,J^P)=(\frac{1}{2},\frac{1}{2}^+)$ state will appear 
as a very sharp resonance either just above or just below the 
$\Xi d$ threshold.

\item \textbf{Acknowledgments}

This work has been partially funded by COFAA-IPN (M\'exico),
by Ministerio de Educaci\'on y Ciencia and EU FEDER under
Contracts No. FPA2013-47443 and FPA2015-69714-REDT,
by Junta de Castilla y Le\'on under Contract No. SA041U16,
and by Generalitat Valenciana PrometeoII/2014/066.
H.~G. wishes to thank the organizers of YSTAR2016 for some 
finantial support during the workshop.
A.~V. is thankful for financial support from the
Programa Propio XIII of the University of Salamanca.
\end{enumerate}


\newpage
\subsection{Status and Plans of HADES}
\addtocontents{toc}{\hspace{2cm}{\sl J.~Stroth}\par}
\setcounter{figure}{0}
\setcounter{table}{0}
\setcounter{footnote}{0}
\setcounter{equation}{0}
\halign{#\hfil&\quad#\hfil\cr
\large{Joachim Stroth (for HADES Collaboration)}\cr
\textit{Institut f\"ur Kernphysik}\cr
\textit{Goethe-Universit\"at Frankfurt}\cr
\textit{D-60438 Frankfurt am Main, Germany}\cr}

\begin{abstract}
HADES is operated at the SIS18 and investigates the microscopic properties
of resonance matter formed in heavy-ion collision in the 1~A~GeV energy
regime. Important topics of the research program are the mechanisms of
strangeness production, the emissivity of resonance matter and the role
of baryonic resonances herein. The latter topic is addressed also by
investigation of exclusive channels in proton and pion beam induced
reaction, both for hadronic and semi-leptonic final states. To optimize
the performance of the spectrometer for the FAIR Phase-0 campaign,
several upgrade project are underway.
\end{abstract}

\begin{enumerate}
\item \textbf{Properties of Hadron Resonances and Resonance Matter}

In central collisions of heavy ions at energies around a few A~GeV ,
strongly interacting matter is substantially compressed and collective
kinetic energy dissipated into intrinsic degrees of freedom. As a result,
baryonic resonances are formed and in the final state of the reaction
increasing abundances of mesonic states are observed as the beam energy
rises. A scientific challenge is to understand the microscopic properties
of the matter formed in the early stages of such collisions. Indeed, from
studies using microscopic transport models, it is observed that the density
in the interior of the collision zone may reach up to three times nuclear
ground state density already at a beam energy of 1~A~GeV.  In such an
environment hadronic states will likely change their properties compared
to those observed in vacuum.  Several experiments have tried to search
for modifications of vector mesons states using their decay into lepton
pairs. Although the results are not conclusive yet, a general trend
examined is a strong broadening of vector meson states in the medium
already for cold nuclear matter. A crucial question of QCD is under
which conditions of density and temperature hadrons will ultimately
lose their hadronic character and the system in such collisions will
change from hadron gas into exotic states of strongly interacting matter?
The situation at colliders (RHIC, LHC) is qualitatively very different.
Due to the high gamma factors, the initial state at mid-rapidity is
characterized by nearly instantaneous parton parton interactions which
create a state of extreme energy density. The system then evolves through

Figure~\ref{fig:phasediagram} depicts a sketch of the phase diagram of
strongly interacting matter. The experimental information is depicted
by symbols localizing the chemical freeze-out region which depend on
system size and on beam energy. They remarkably line up along a narrow
corridor spanning from LHC ($T\simeq 160$~MeV, $\mu_B\simeq 0$~MeV)
down to SIS18 energies ($T\simeq 50$~MeV, $\mu_B\simeq 800$~MeV) and
are determined with the help of statistical hadronization models from
final state particle abundances measured in these collisions. Also shown
is the expectation value of  the chiral condensate calculated in a
Polyakov--Quark--Meson Model approach as a function of baryo-chemical
potential ($\mu$b) and temperature~\cite{Schaefer:2004ens}. The systematics
of these freeze-out points suggests that a thermal spectrum of hadrons is
observed even at low energies, where a quark gluon plasma is not formed.
\begin{figure}[tbh]
\begin{center}
 \includegraphics[width=4in]{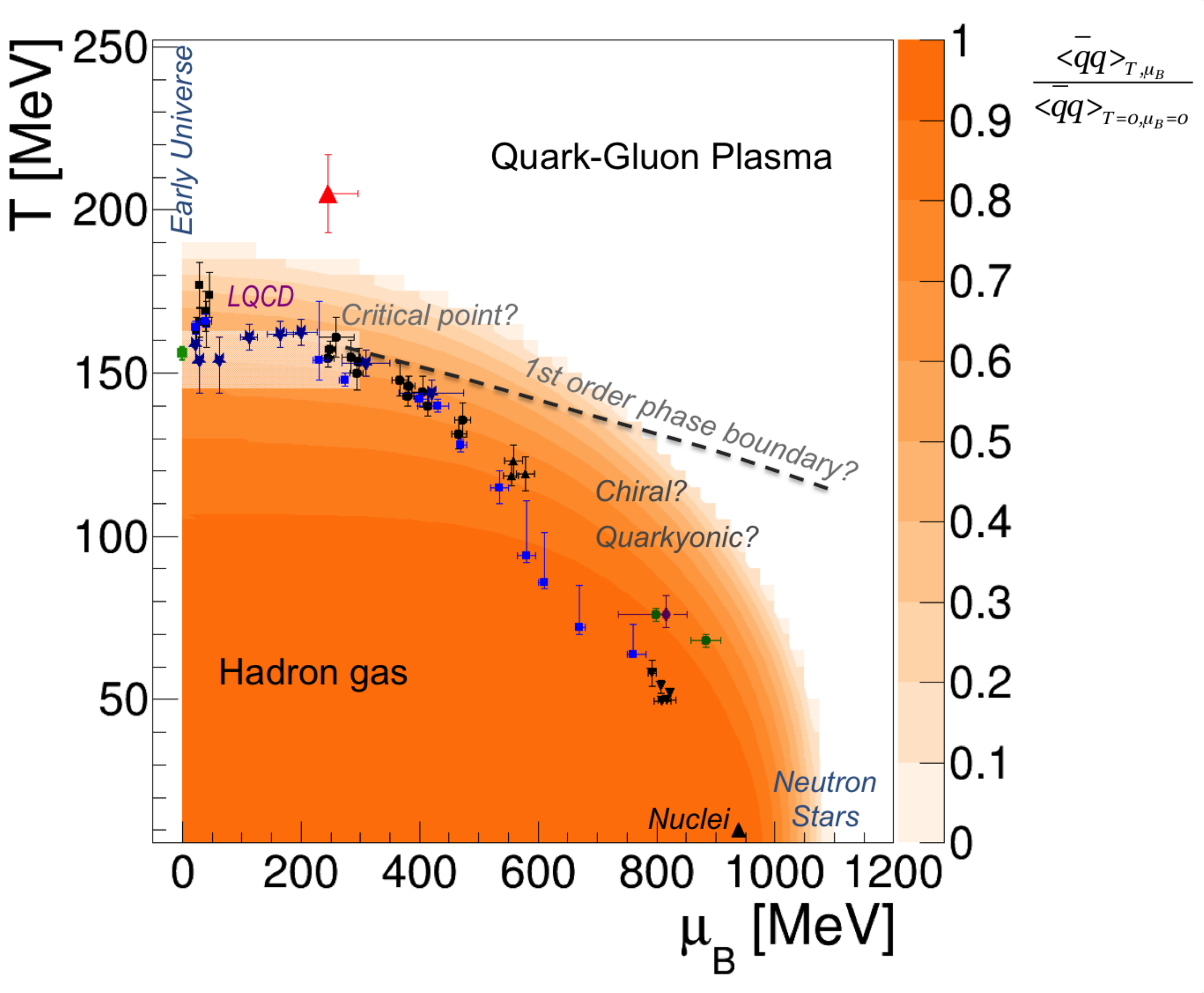}
\end{center}

   \caption{Sketch of the phase diagram of strongly interacting matter. 
	The shaded area reflects the expectation value of the chiral 
	condensate relative to its vacuum value. The data points are 
	freeze-out configurations obtained by analyzing hadron final 
	states in the framework of statistical hadronization models.}
	\label{fig:phasediagram}
\end{figure}

To learn more about the microscopic structure of matter in the region of
high baryo-chemical potential HADES pursues a strategy, which relies on
systematic measurements of strangeness production and virtual photon
emission in heavy-ion collisions. The latter observable has the advantage,
that this radiation is emitted through out the collision and hence is very
sensitive to the properties of the dense and hot system. The disadvantage
however is its small branching ratio and the fact that the spectrometer
integrates the radiation over time. Consequently, contributions from the
late stage of the collision have to be identified and subtracted before
conclusions about the radiation from the dense phase can be drawn. Hadrons
carrying strangeness are similarly interesting as their production threshold
is high compared to the available energy in an $NN$ collision at these beam
energies.  Consequently, their production requires a certain degree of
collectivity, like it is the case in multi-particle processes or in deep
off-shell production.

A central part of the physics program of HADES are measurements of rare
meson production in heavy-ion collisions and in respective elementary
collisions. The focus of the latter are the decay of baryonic resonances
into intermediate $\rho$ mesons and into final states with open strangeness
thus addressing the observables utilized for the study of dense matter. 
The program includes in particular the reconstruction of exclusive hadronic 
and semi-leptonic final states. The paper is organized in the following 
way: In the next Chapter, selected results of experiments addressing 
virtual photons are presented. The examples address the concept of a 
reference spectrum in comparison to which medium-effects are addressed. 
In Chapter~3 few results on strangeness are shown emphazising the 
connection to virtual photon production and the role of baryonic 
resonance it. Chapter~4 gives a short status of the HADES campaign 
2012--2014 and Chapter~5 provides an outlook on the upcoming physics 
program of HADES anticipated for the years 2018 to 2021 and the 
spectrometer upgrades planned for this campaign.

\item \textbf{Dilepton Emission from Hadronic Systems}

An important reference for the study of dilepton emission (here and in the
following $e^+e^-$ pairs) off strongly interacting matter is the production
of dileptons in $p+p$ and $n+p$ collisions at energies below the threshold
for $\eta$ production. Expected contributions are then solely by conventional
bremsstrahlung and due to Delta Dalitz-decay. HADES has observed a
surprisingly large isospin dependence comparing  $p+p$ with $n+p$
collisions at $T_{beam} = 1.25 (\sqrt{s} = 
2.4)$~GeV~\cite{Agakishiev:2011vfs}. This strong deviation appears in the
spectral distribution ($dP/dM_{e+e-}$) of the lepton pairs towards the 
kinematic limit. While conventional bremsstrahlung can not account for the 
effect, although there is a difference in $n+p$ and $p+p$ due to the 
absence of a electric dipole moment in the latter case, calculations 
including emission from the intrinsic charged pion line in an 
one-boson-exchange model for the $n+p$ case, and using a VDM form factor 
for the pion photon vertex, could reproduce the trend observed in the 
data. Charged pion transfer is only possible in $n+p$ reactions due to 
isospin restrictions. In a strict VDM picture such a process can be
interpreted as the formation of a deep off-shell $\rho$-meson in the 
overlapping cloud of the by-passing baryons. To what degree such a 
production would rather proceed through an intermediate baryon resonance 
state, such as a $\Delta(1232)$ resonance, remains an open question. 
Indeed, calculations of the $\Delta(1232)\rightarrow Ne^+e^-$
electromagnetic transitiofrom factor  in the kinematic region 
discussed above with the Spectator Quark Model~\cite{Ramalho:2015qnas} 
and a Two-component Model~\cite{Wan:2005dss} show only little or moderate 
deviations from solutions assuming pure QED (point-like) transitions, 
respectively.
\begin{figure}[tbh]
\begin{center}
 \includegraphics[width=4in]{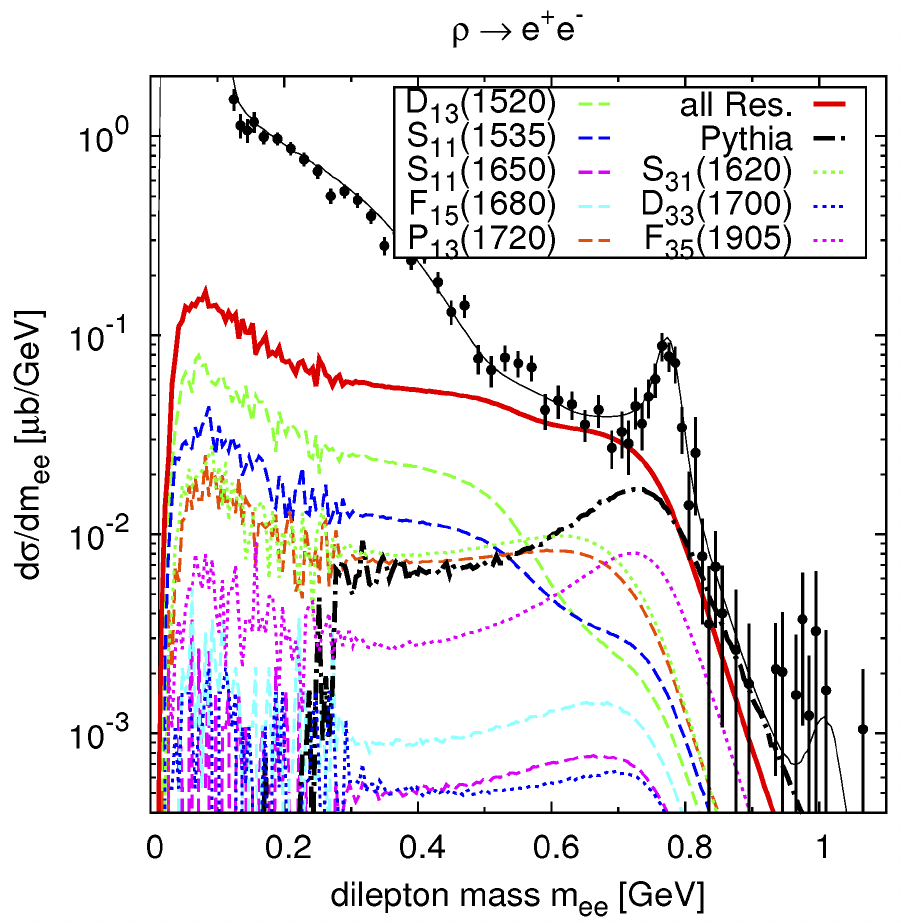}
\end{center}

   \caption{Dilepton invariant mass spectrum from $p+p$ collisions at 
	3.5~GeV (HADES data) compared to a calculation using a
	resonance model as implemented in microscopic transport codes. 
	The cross sections for the excitation of the different baryonic 
	resonances are obtained from a fit to data on meson production. 
	All baryonic resonances, except the $\Delta(1232)$, produce 
	lepton pairs via explicit decay to an intermediate $\rho$ meson 
	and subsequent annihilation into a pair of electrons.
	Figure taken from Ref.~\protect\cite{Weil:2012jis}.}
	\label{fig:Janus-pp}
\end{figure}

While at a proton beam energy of around  1~A~GeV essentially 
$\Delta(1232)$ resonances are active in electron-pair production, the 
situation changes substantially at an energy of 3.5~A~GeV .  
Fig.~\ref{fig:Janus-pp} shows the inclusive dilepton yield obtained by 
HADES in the channel $p+p \rightarrow  e^+e^-X$ as a function of the 
dilepton invariant mass. The cocktail shown has been calculated with a 
microscopic transport model and represents only prompt dileptons while 
the data contains also contributions from Dalitz- and direct decays of 
long-lived mesons (dominantly $\pi^0$, $\eta$ and $\omega$). In the 
model, prompt dileptons stem solely from (off-shell) $\rho$ meson 
production produced in baryonic resonance decays. The treatment of 
off-shell $\rho$--meson propagation leads to a modified mass distribution  
of the mesons governed by phase space constraints. The higher the 
resonance mass, the more prominently the $\rho$ pole emerges, yet, the 
resulting spectral distribution of $\rho$ mesons from baryonic resonance 
decays (all res.) significantly differs from a distribution obtained if  
prompt, non-resonant production is assumed (Phytia). The thin solid curve 
going through the data is the sum of mesonic and prompt dileptons. 
Although the model has a number of parameters which have to be adjusted, 
it illustrates how dilepton production might proceed through strict 
vector meson dominance and how important a detailed understanding of 
resonant $\rho$--nucleon coupling is for the theoretical description of 
dilepton production in heavy-ion collisions.

Indeed, the preparation of a reference spectrum based on elementary 
inclusive cross sections provided a conventional explanation for the 
inclusive electron pair spectra observed in $C+C$ collisions at 1 and 
2~A~GeV by HADES~\cite{Agakishiev:2007tss,Agakichiev:2006tgs,
Agakishiev:2009yfs} and earlier by the DLS 
Collaboration~\cite{Porter:1997rcs}, which could not be reproduced in 
microscopic model calculations at the time the DLS data was taken
(DLS puzzle). It later turned out that within $\simeq20$\% uncertainty, 
$C+C$ collisions at these energies essentially appear as mere 
superposition of individual $N+N$ collisions, while a true excess 
radiation could only be observed in the heavier $Ar+KCl$ ($Ca+Ca$ in 
case of DLS) reactions at $1.76$~A~GeV .

A good description of this excess radiation can be obtained if emission 
out of thermalized system is assumed~\cite{Gale:1987kis}. Under this 
assumption, the emissivity of matter at not too high temperatures is 
given by the thermal average of the in-medium $\rho$ propagator for 
which $\rho$--baryon couplings are fundamentally 
important~\cite{Rapp:1997fss}. The spectra observed in experiments is 
then given as a four-volume integral over the emissivity, \textit{weighted}
with the time-dependent temperature and density profile of the hot and 
dense system. The latter can, e.g., be derived by coarse graining a 
microscopic transport code.  Fig.~\ref{fig:Ar_excess} shows the dilepton 
excess radiation, obtained from the  $Ar+KCl$ data by subtracting all 
known sources from long-lived mesons decay, compared to a calculation 
assuming thermal emission and coarse graining 
UrQMD~\cite{Galatyuk:2015pkqs,Endres:2015fnas}. The calculations agree 
well with data and the spectra show  nearly exponential shapes, although 
the calculations are entirely based on VDM and hence dominated by large 
by off-shell $\rho$ meson.
\begin{figure}[tbh]
\begin{center}
 \includegraphics[width=4in]{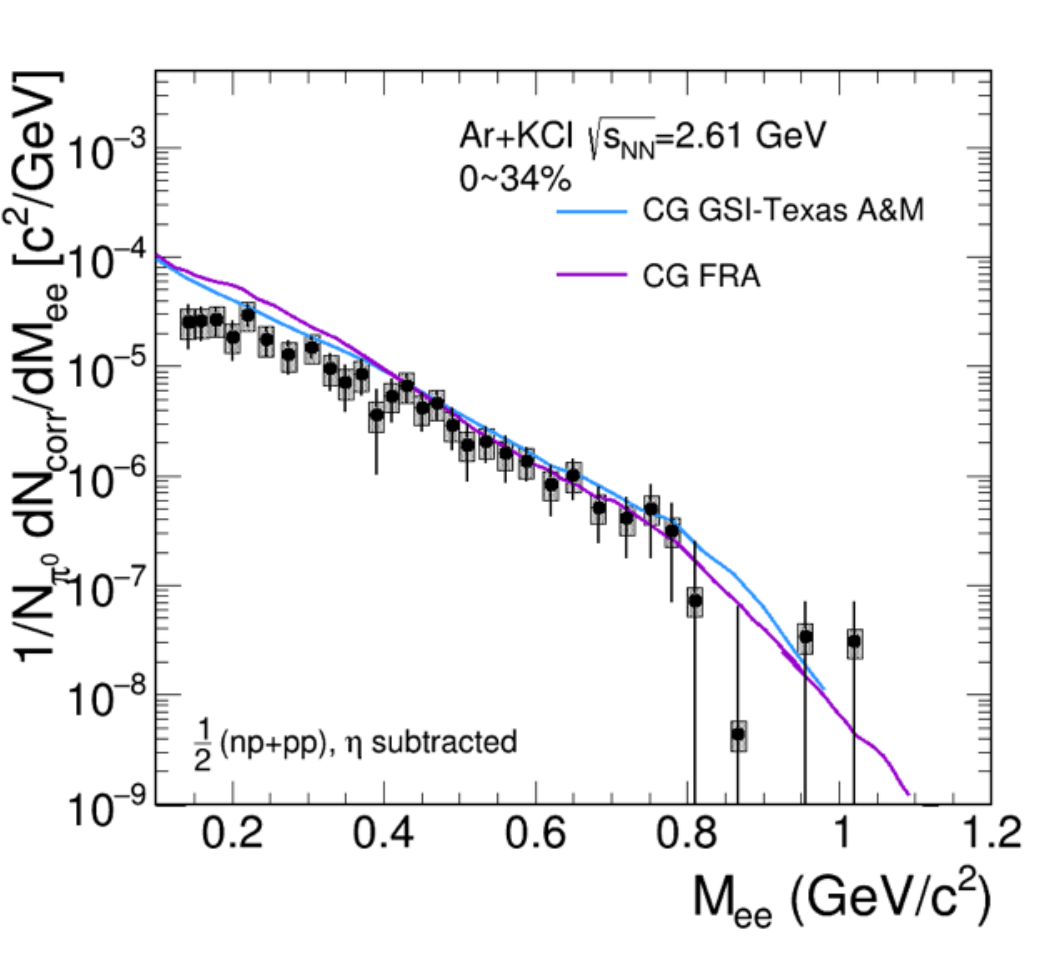}
\end{center}

   \caption{Dilepton excess radiation observed in $Ar+KCl$. Contributions 
	from first chance $NN$ collisions and late $\eta$--Dalitz
	decays have been subtracted from the data. The yield is normalized 
	to the number of produced $\pi^0$ (to remove trivial $A_{part}$ 
	dependences). The colored lines show the results of two versions 
	of corse grained UrQMD calculations using different concepts for 
	obtaining the thermal parameters. The data has been analyzed for 
	the 34\% most central events.}
	\label{fig:Ar_excess}
\end{figure}

\item \textbf{Selected Results on Strangeness Production}

Similarly important are baryonic resonances for the description of
strangeness production in heavy-ion collisions, in particular at beam
energies where the total energy in a single $N+N$ collision is not
sufficient to produce a final state with strangeness. If instead binary
collision occur, which include  secondary (or even higher generation)
baryonic resonance states, the kinematic constraints can be lifted
beyond the effect of fermi motion. A theoretical description of secondary
interactions beyond the low-density approximation is difficult but 
desirable as the collision frequencies are as high that asymptotic 
states are not easily reached between subsequent collisions.

A particular role in the production and propagation of strangeness is
played by the $\Lambda(1405)$ resonance. The existence of this resonance
influences the interaction of antikaons propagating through a baryonic
medium since this resonance can be understood as a dynamically produced
antikaon-nucleon quasi-bound state~\cite{Tolos:2007sts}. Such a picture 
is also supported by lattice data if the strange magnetic form factor 
of the resonance is investigated~\cite{Kamleh:2016dwes}. The 
$\Lambda(1405)$ has been observed by HADES in $p+p$ collisions at 
$3.5$~GeV collisions~\ref{fig:lambda1405}. The broad structure is 
attributed to the formation of the $\Lambda(1405)$ state which appears 
a little bit shifted, a hint that the state is a molecule and its 
appearance depending on the reaction dynamics.
\begin{figure}[tbh]
\begin{center}
 \includegraphics[width=4in]{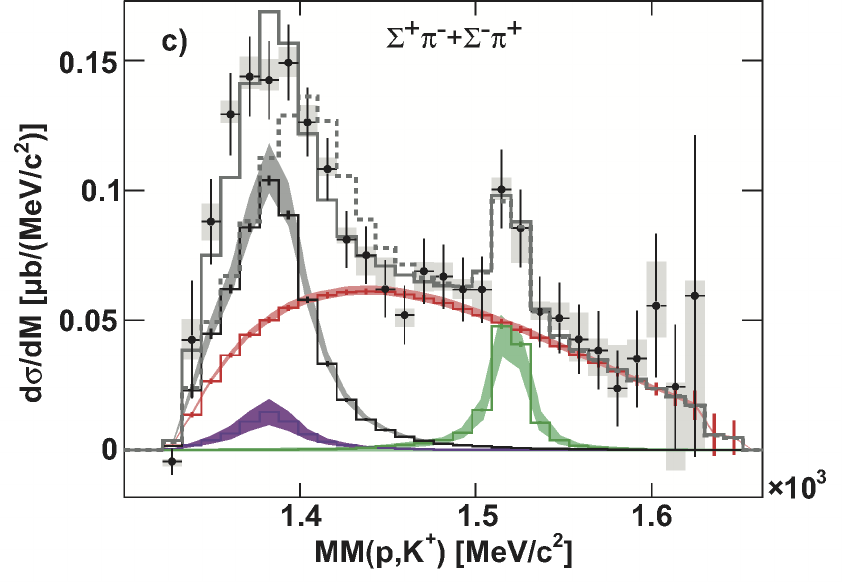}
\end{center}

   \caption{Missing mass distribution to the $p K^+$ subsystem of the 
	exclusive final state $p K^+ n \pi^+ \pi^-$ reconstructed in 
	$p+p$ reactions at $3.5$~GeV. The distribution is interpreted 
	in terms of contributions from $\Lambda(1405)$ and $\Lambda(1520)$, 
	histograms shaded in grey and
	green, respectively, $\Sigma(1385)$ (purple) and channels not 
	involving excited hyperons (red). The pole mass for the simulated 
	$\Lambda(1405)$ contribution was fitted to the experimental 
	distribution and favors a downward shift by $20$~MeV/$c^2$ .}
	\label{fig:lambda1405}
\end{figure}

The resonant scattering, as well as the different $NN$ thresholds for
associated ($K\Lambda$) and direct ($K\bar{K}$) kaon production, suggests
a rather complex structure of kaon production in heavy-ion collisions at
SIS18 (''threshold-'') energies. Yet, the yields observed in $Ar+KCl$
collisions at $1.76$~A~GeV  shows a very simple behaviour. The
multiplicities of all observed hadrons an be explained assuming a
break-up (freeze-out) of a thermalized hadronic system if strangeness
is treated canonically by implementing a \textit{strangeness correlation
radius} $R_c$. The latter leads to  suppression of hadronic states with
open strangeness but not for states with hidden strangeness like the
$\phi$. Fig.~\ref{fig:HADES_ArKCl_thermal} shows the reconstructed
hadron multiplicities in comparison with a fit using the THERMUS code.
All multiplicities including states involving strange quarks are
reproduced assuming a temperature of $\simeq76$~MeV  and a baryo-chemical
potential of $\simeq800$ MeV, except for the double strange baryon
$\Xi$~\cite{Agakishiev:2009rrs}. Note that $\phi$ is produced abundantly
as expected by the statistical model and holds up for $\simeq30$\% of
the produced antikaons~\cite{Agakishiev:2009ars}.

These observations, as well as those in dilepton production are accordance
with a picture describing the fireball as \textit{strongly} interacting
resonance gas. The observed hadron multiplicities resemble thermalization
although, e.g., the ow pattern of protons does not. That raises the 
question if the equilibration is driven by binary collisions of the 
(hadronic) constituents, as it is implied by the low-density approximation 
forming the basis of the model calculations used to describe heavy-ion 
collisions and medium-effects at these energies? Likely other mechanisms 
drive equilibration, like, e.g., a strong quantum mechanical 
entanglement~\cite{Stroth:2011zzs}.  To further scrutinize these 
observations, HADES focused on studying a heavy collision system at 
maximum SIS18 beam energy and on exploiting pion induced reactions.
\begin{figure}[tbh]
\begin{center}
        \includegraphics[width=3in]{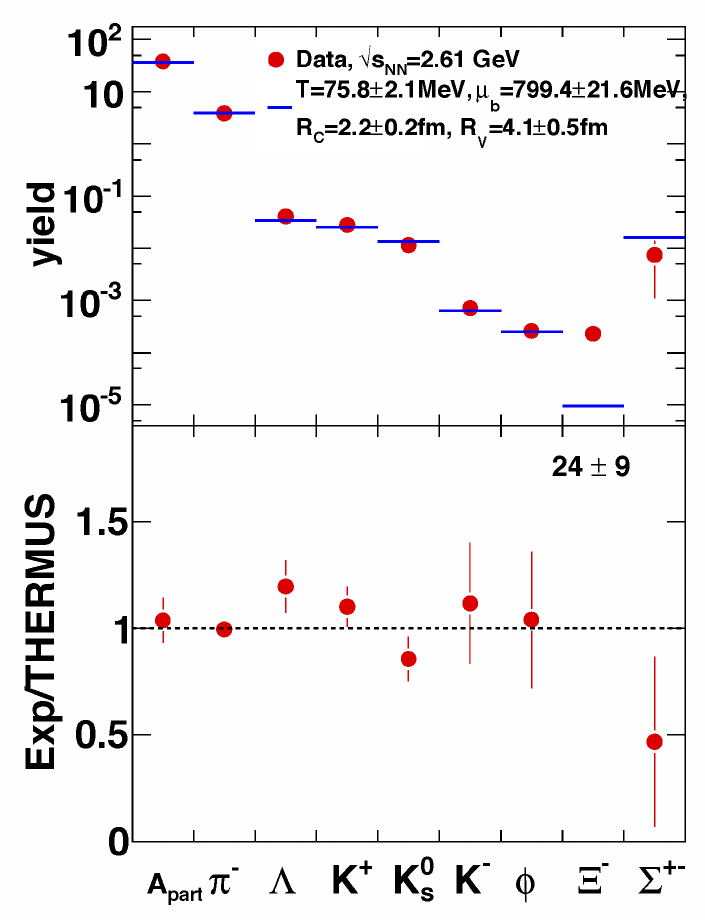}
\end{center}

   \caption{Hadron multiplicities in full phase space for 38\% most 
	central $Ar+KCl$ collisions at $1.76$~A~GeV. The data is 
	compared to the expectation of a Statistical Hadronization 
	Model (THERMUS). The extracted paramters for temperature and 
	baryochemical potential fit weel in the systematics shown
	in~Fig.~\protect\ref{fig:phasediagram}.}
	\label{fig:HADES_ArKCl_thermal}
\end{figure}

\item \textbf{The HADES campaign 2012--2014}

During the first measurement campaign (2002-2007), HADES was operating 
with a 18-fold segmented inner time-of-flight detector system. This low 
granularity made runs with heavy collision systems impossible. In the 
shutdown period for SIS18, the HADES collaboration replaced this system 
by a RPC system with more than 1000 individual cells. Also replaced was 
the innermost layer of drift chambers.

In April 2012 HADES took $7.2\times 10^9$ events during 30 days run for 
the collision system $Au+Au$ at $1.23$~A~GeV.  For the first time, a 
nearly complete set of strange hadrons ($\phi, K^+,K^-,K^0$) has been 
reconstructed at such low beam energies ($\sqrt{s}$ =2.4~A~GeV). Again a 
large $\phi$ meson abundancy, which accounts for about 25\% of the 
total $K^-$ production, could be observed - note that a $NN$ collisions 
at this beam energy is $\simeq500$~MeV below the $NN\phi$ threshold. 
Ongoing analysis work focuses on a further reduction of systematic 
errors and on a comprehensive comparison to results from transport 
calculations.

As expected already, the yield of dielectrons reconstructed in these
collisions shows a strong excess above a reference spectrum. Two different
analysis strategies have been implemented, one focusing on highest purity,
the other on highest efficiency for electron positron reconstruction. The
results nicely agree, though the one with higher statistics comes with the
price of an higher uncertainty in the overall normalization.  The spectrum
again drops off nearly exponentially.

A first experiment with the GSI pion beam has been realized in 2014 with
two weeks of beam on target in total. The first run was dedicated to
strangeness production in cold matter in pion-induced reactions on light
(12C) and heavy (74W) nuclei at a beam momentum of 1.7~GeV/c.  The goal
of the second run was to measure both double pion and dilepton production
in $\pi^-+p$ reactions around the pole of $N(1520)$ resonance using
polyethylene and carbon targets.  Data at four different pion beam momenta
(0.656, 0.69, 0.748 and 0.8~GeV/c) were collected with the largest
statistics in the case of 0.69~GeV/c momentum, aimed for dilepton
production studies.

The two-pion production data samples have been included in the
multichannel Partial Wave Analysis (PWA) developed by the Bonn-Gatchina
group, which allows for the extraction of the various $2 \pi  N$ channels
($\Delta \pi, \rho N, \cdots$). The total (resonant+non-resonant) $\rho$
is derived  and converted into a $e^+e^-$ contribution using the Vector
Dominance Model (VDM) assumption. Adding to this $\rho$ contribution a
cocktail of point-like baryonic sources, the exclusive dilepton production
can be described quite nicely. This demonstrates the consistency between
the dilepton and double pion production channels and suggests that the
involved time like baryonic electromagnetic transitions can be described
by VDM. The $\rho N$ couplings of the $N(1520)$ are also of high interest
due to the connection with the expected medium modification of the $\rho$
meson spectral function. Most of the new results will be presented on the
upcoming Quark Matter Conference in February 2017.

\item \textbf{HADES at FAIR Phase--0}

With the start of FAIR Phase 0 in 2018, most of the detector systems will
reach an age of more than 15 years. Hence, we have started an upgrade
program replacing the UV detector of the RICH by a detector based on
MAPMTs. The development of this detector is a joint initiative between
HADES and CBM, the same modules will later also be used in the CBM RICH.
The pre-shower detector (polar angle coverage from 18 to 45 degree), which
augmented electron/positron identification and also served as additional
tracking  detector in front of the low-granularity time-of-flight system
during the first experimental campaign, will be replaced by an
electromagnetic calorimeter (ECAL). This new detector is based on
recycled lead glass crystal from the former OPAL calorimeter.  For the
first time, this will enable studies of radiative baryonic resonance
decays with real photons. Moreover, the success of the experimental
program addressing elementary reactions calls for an instrumentation
of the acceptance region between polar angles of 2 to 8 degree in order
to enlarge the phase space acceptance. This solid angle will be equipped
with tracking stations and a new RPC-based time-of-flight wall. The
tracking system will be based on developments for the PANDA tracker
which will be composed of by 5 mm straw tubes. Last but not least a
moderate upgrade of the DAQ system will increase the band width such
that interaction rates of up to 100 kHz are in reach for elementary
reactions and around 20 kHz for heavy-ion reactions.

Further experiments on baryonic-resonance and cold-matter physics will
be realized, taking advantage of foreseen improvement of the extraction
and increase of the space charge limit. In addition, with the ECAL, the
capacity of HADES to provide high precision data for PWA for baryonic
resonances with masses up to 2~GeV/$c^2$ will be extended to exit channels
with neutral mesons.  HADES at the moment represents the only facility
world-wide, which combines a pion beam with dilepton spectrometry and
acts as a precursor in view of existing plans for meson beam facilities
for baryon spectroscopy.  Emphasis will be on the electromagnetic
transition form factors of baryonic resonances, the coupling of vector
mesons and kaons to baryons and medium effects in cold nuclear matter.

Measurements of hadron production off cold nuclear matter provide an
important reference for understanding heavy-ion collisions. They also
provide an ideal test bed for microscopic transport theory. HADES has
already studied $p+Nb$ collisions at 3.5~GeV in 2007. Yet, many of
the the interesting observation made suffer from statistical significance
or call for multi-differential analysis. Moreover, particle identification
in the 2007 run was limited since no fast start  detector had been available
at this time.  Meanwhile, HADES has developed such detectors based on
mono-crystal CVD diamond which combine high rate capability with excellent
time resolution for minimum ionizing particles. A high statistics run would
enable very important studies like, $\phi$  production and propagation both
in the hadronic and leptonic final state, in-medium w-meson, multi-strange
baryon production, short range correlation (SCA) and two particle correlation
studies aiming at determining, e.g., the $\lambda p$ phase shifts.

The combined measurement of dielectrons and strangeness performed by HADES 
in $Ar+KCl$ and Au+Au collisions have provided new intriguing results which
call for further systematic investigations. We propose an experiment which 
focuses on measuring a medium-heavy collision system at the maximum energy 
at SIS18, to increase the NN center-of-mass energy in favour of an enhanced
strangeness production. Silver, e.g., can be stripped to charge state $45^+$ 
behind the UNILAC and would then allow acceleration in SIS18 to 1.67~A~GeV.

The main goals are to search for multi-strange baryons in connection to our
finding in $p+Nb$ and $Ar+KCl$, and to extend the dielectron and strangeness
excitation functions. Indeed, the unexpected high cascade yield is by now not
explained theoretically and call for more data (see also above). In the
dielectrons channel, we will focus on studies of properties of the low-mass
excess as well as on vector meson ($\rho,\omega,\phi$) spectroscopy. The
strangeness study will include $\phi$ meson production via the $K^+K^?$
decay channel, kaon production characteristics, $\Lambda(1115)$,
$\Sigma(1385)$ and $\Xi(1321)$ strange baryon production, and HBT
correlations.

The motivation for operating HADES as part of the Compressed Baryonic
Matter program at SIS100 is two-fold: First, HADES can bridge the gap
from the SIS18 energies to the SIS100 region were CBM acceptance is
favourable. It enables the option to measure dilepton spectra in heavy-ion
collisions with two different spectrometers to so minimize systematic
uncertainties. Second, HADES can serve as ideal spectrometer to continue
reference measurements focusing on cold matter studies.  As medium-effects
show a strong momentum dependence,  the relevant phase space for meson
production and propagation is near to the target rapidity.

Finally, HADES can also be operated with a liquid hydrogen target, which
is not possible in CBM. Any reference measurements, as well as a
continuation of the physics program of HADES at SIS18 is possible at the
new experimental site.

\item \textbf{Acknowledgments}

The HADES collaboration gratefully acknowledges the support by the grants 
LIP Coimbra, Coimbra (Portugal) PTDC/FIS/113339/2009, UJ Krak\.ow (Poland) 
NCN 2013/10/M/ST2/00042, TU M\.unchen, Garching (Germany) MLL M\.unchen: 
DFG EClust 153, VH--NG--330 BMBF 06MT9156 TP5 GSI TMKrue 1012 NPI AS CR, 
Rez, Rez (Czech Republic) GACR 13--06759S, NPI AS CR, Rez, USC - S.~de 
Compostela, Santiago de Compostela (Spain) CPAN: CSD2007-00042, G\"othe 
University, Frankfurt (Germany): HA216/EMMI HIC for FAIR (LOEWE) 
BMBF:06FY9100I GSI F\&E, IN2P3/CNRS (France).
\end{enumerate}


\newpage
\subsection{Hadron Physics with High-Momentum Hadron Beams at J-PARC}
\addtocontents{toc}{\hspace{2cm}{\sl H.~Noumi}\par}
\setcounter{figure}{0}
\setcounter{table}{0}
\setcounter{footnote}{0}
\setcounter{equation}{0}
\halign{#\hfil&\quad#\hfil\cr
\large{Hiroyuki Noumi (for E50 Collaboration)}\cr
\textit{Research Center for Nuclear Physics (RCNP)}\cr
\textit{Osaka University}\cr
\textit{Osaka, 567-0047, Japan}\cr}

\begin{abstract}
Baryon spectroscopy with heavy flavors provides unique opportunities to
study internal motions of ingredients in baryons,
from which we can learn the effective degrees of freedom to describe
hadrons. We proposed an experiment on charmed baryon spectroscopy via
the $(\pi^-,D^{\ast -})$ reaction at the J-PARC high-momentum beam line.
Mass spectrum from the ground state to highly-excited states of the
excitation energy up to more than 1 GeV will be observed by means of a
missing mass technique. Production cross sections and decay branching
ratios of these states will be measured. It is worthy to compare how
nature of baryons with different flavors changes. In particular,
double-strangeness baryons, $\Xi$, are of interest. A neutral kaon
beam at J-Lab is unique to produce $\Xi$ baryons. Hadron beams at
J-PARC play  complimentary roles to the neutral kaon beam at J-Lab.
\end{abstract}

\begin{enumerate}
\item \textbf{Baryon Spectroscopy with Heavy Flavors}

"How hadrons are formed from quarks?" is a fundamental question in hadron
physics. We know that the quantum chromo-dynamics (QCD) is a fundamental
theory to describe dynamics of quarks and gluons.
However, it is still very hard to describe hadrons by solving the QCD
equation in low energy because of its non-perturbative nature of the
strong interaction, where the coupling constant becomes very large when
the energy scale is close to the scale parameter $\Lambda_{\rm QCD}$.
Quarks  drastically change their nature below $\Lambda_{\rm QCD}$.
Then, constituent quarks as effective degree of freedom to describe
hadrons seem to work rather well. Actually, the constituent quark model
well describes properties of hadrons in the ground state, such as masses,
spin-flavor classifications, magnetic moment of octet baryons, and so no.
However, it sometimes fails in excited states.
In particular, not only recent reports on so-called exotic hadrons,
such as $X$, $Y$, $Z$, and pentaquark states in heavy sector but also a
long-standing puzzle in $\Lambda$(1405)
indicate that we need a new aspect in describing hadrons.
Internal correlations among ingredients of hadrons such as diquarks and
hadron clusters are expected to play an important role. Since they are
confined in a hadron, hadron spectroscopy to look into more details of
internal structure or motions of the composites in hadrons is necessary.

Since the color magnetic interaction between quarks is proportional to
the inverse of the quark mass, spin-dependent interactions to a heavy
quark vanish in the heavy quark mass limit.
In the heavy quark mass limit, a heavy quark spin and the spin of the
other system become good quantum numbers. This is the so-called heavy
quark symmetry of QCD.
Let us consider a baryon with a heavy quark.
A relative motion between two light quarks ($\rho$ mode) and a collective
motion of the light quark pair ($\lambda$ mode) are separated in excited
states, as illustrated in Fig.~\ref{fig:rholam}. This is known as the
so-called isotope shift. These states are further split due to
spin-dependent interactions between quarks~\cite{yoshidat}. The
spin-correlation between light quarks becomes stronger than those to a
light quark and a heavy quark due to the heavy quark spin symmetry.
Internal structure of the baryon with a heavy quark is characterized by
the two light-quark (diquark) correlation. Nature of these baryons is
reflected in mass, decay width (branching ratio), and production rate.

Therefore, we proposed an experimental study of charmed baryons
via the ($\pi^-,D^{\ast -}$) reaction on hydrogen~\cite{e50t}
at the J-PARC high-momentum beam line. In the reaction,
we reconstruct charmed baryons by means of missing mass
technique.
An excitation spectrum of charmed baryon states can be measured
independent of their decay final states. We could also identify
decay modes with detecting a decay particle together with scattered
$D^{\ast -}$ and identifying a daughter particle in a missing mass. A
branching ratio (partial decay width) of the decay mode can be obtained
rather easily. Branching ratios provide information on diquark motions
of excited charmed baryons, as described later.
This is an advantage of the missing mass method.
\begin{figure}
\centerline{
 \includegraphics[width=8cm]{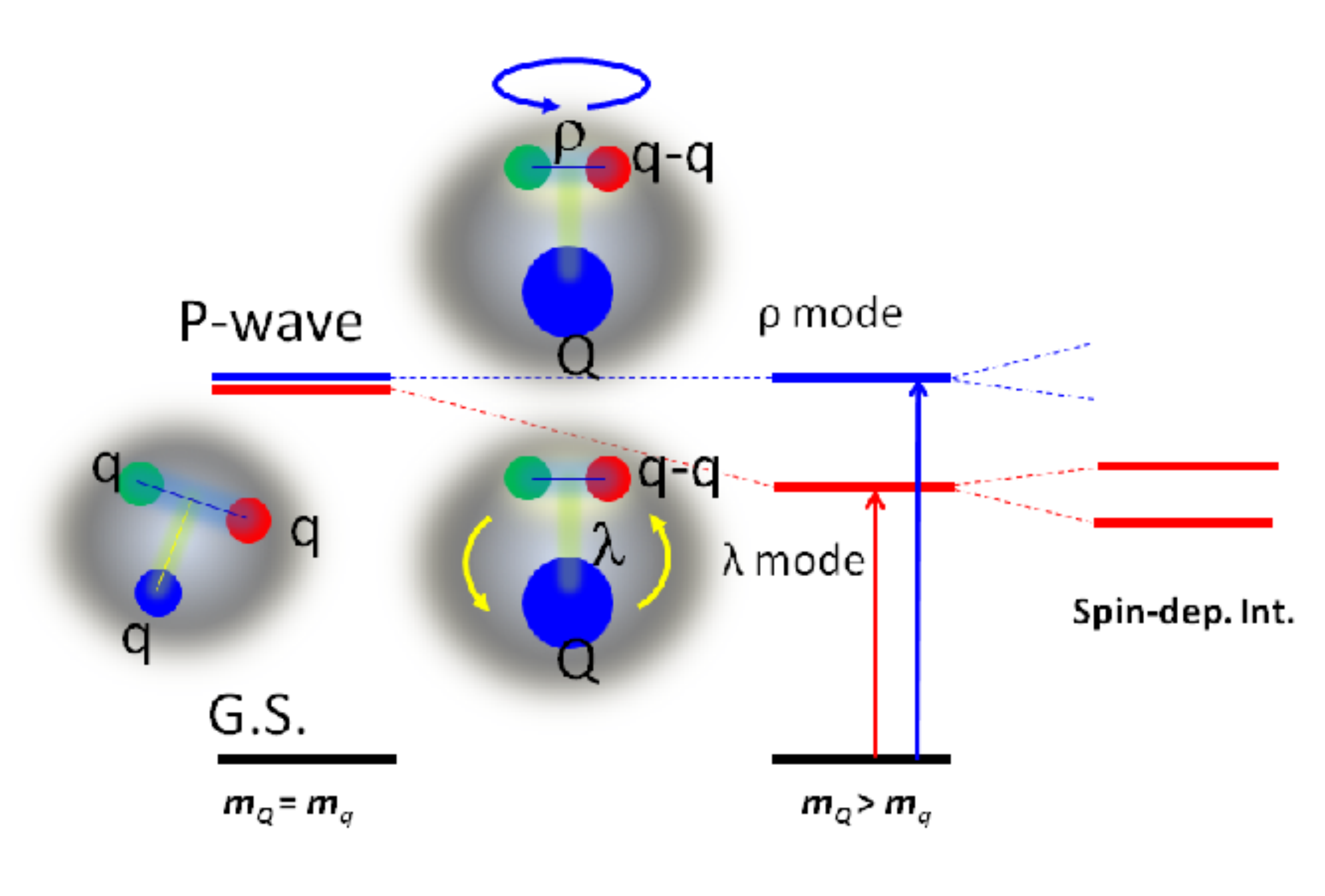}}%

\caption{\label{fig:rholam}Schematic level structure of excited baryons
	with a heavy quark}
\end{figure}
\begin{figure*}
\centerline{
 \includegraphics[width=15cm]{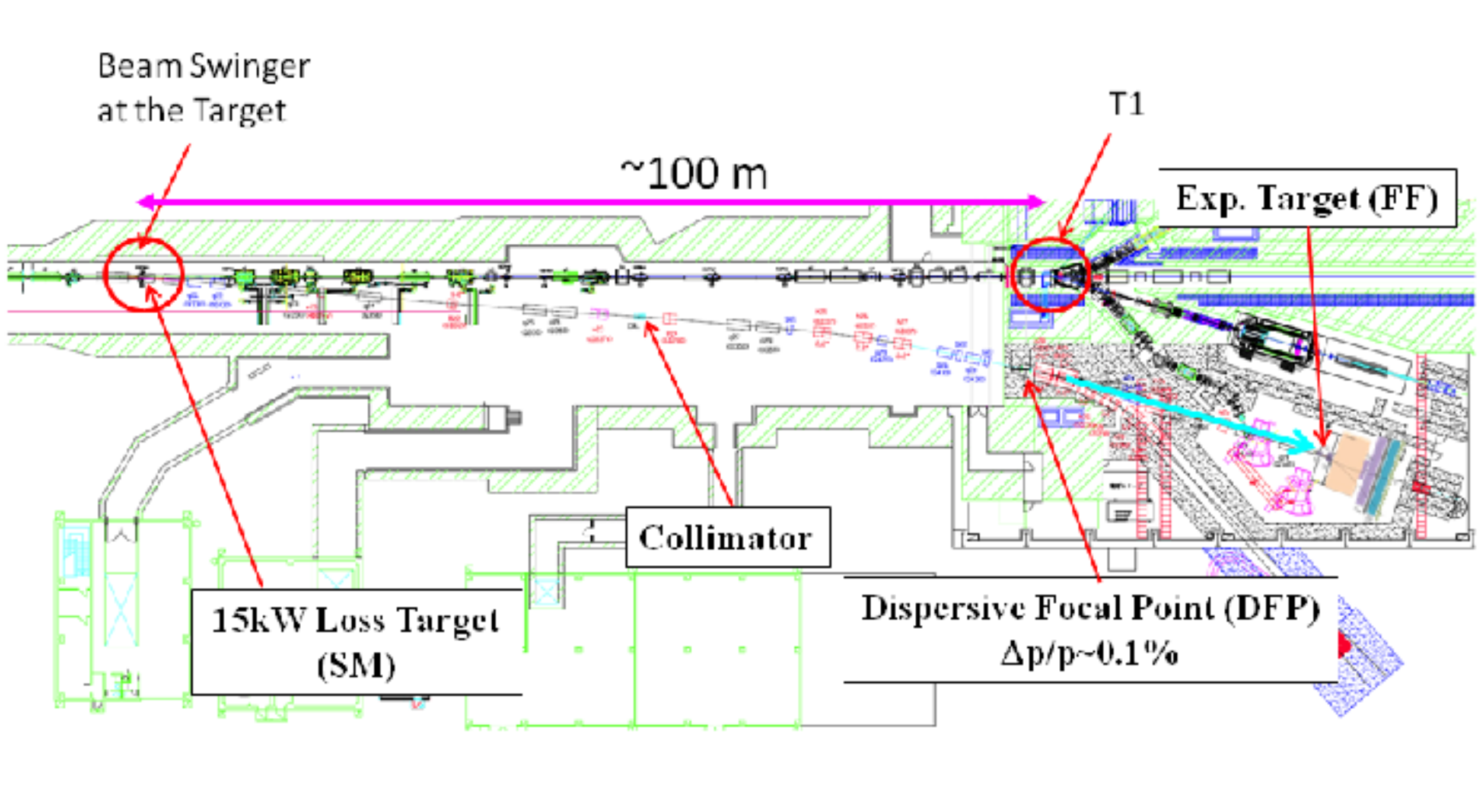}}%

\caption{\label{fig:layout}High-momentum beam line at the 
	J-PARC Hadron Experimental Facility}
\end{figure*}

\item \textbf{Beam Line}

A new beam line, called high-momentum beam line, is being constructed in
the Hadron Experimental Facility of the Japan Proton Accelerator Research
Complex (J-PARC).
It is branched from the slow extraction primary beam line in the switch yard.
A small fraction of the primary beam is transported to the experimental
area, located about 130 m downstream from the branch point, for the E16
experiment which aims at measuring spectral changes of vector mesons in
nuclei~\cite{e16t}.

The high-momentum beam line can deliver secondary beams if we install a
production target at the branching point. The layout of the beam line
magnets are arranged so as to transport secondary beams up to 20 GeV/$c$,
as shown in Fig.~\ref{fig:layout}. The beam line is carefully designed to
realize a dispersive beam at the dispersive focal point, where a momentum
and a horizontal position of the secondary particles are strongly correlated.
Fig.~\ref{fig:dispersion}-top demonstrates the correlation calculated by
the DECAY TURTLE~\cite{turtlet}. We place 3 sextupole magnets to reduce
second order aberrations and to sharpen the correlation. The calculation
tells that a momentum resolution of 0.12\% is expected by measuring a beam
position with a spatial resolution of 1 mm, as illustrated in
Fig.~\ref{fig:dispersion}-bottom. The momentum resolution is mostly
determined by a beam size at the production target (actually, its image
at the dispersive focal point). We assume the beam size of 1 mm in $\sigma$
in the horizontal direction at the production target.
\begin{figure}
\centerline{
 \includegraphics[width=8cm]{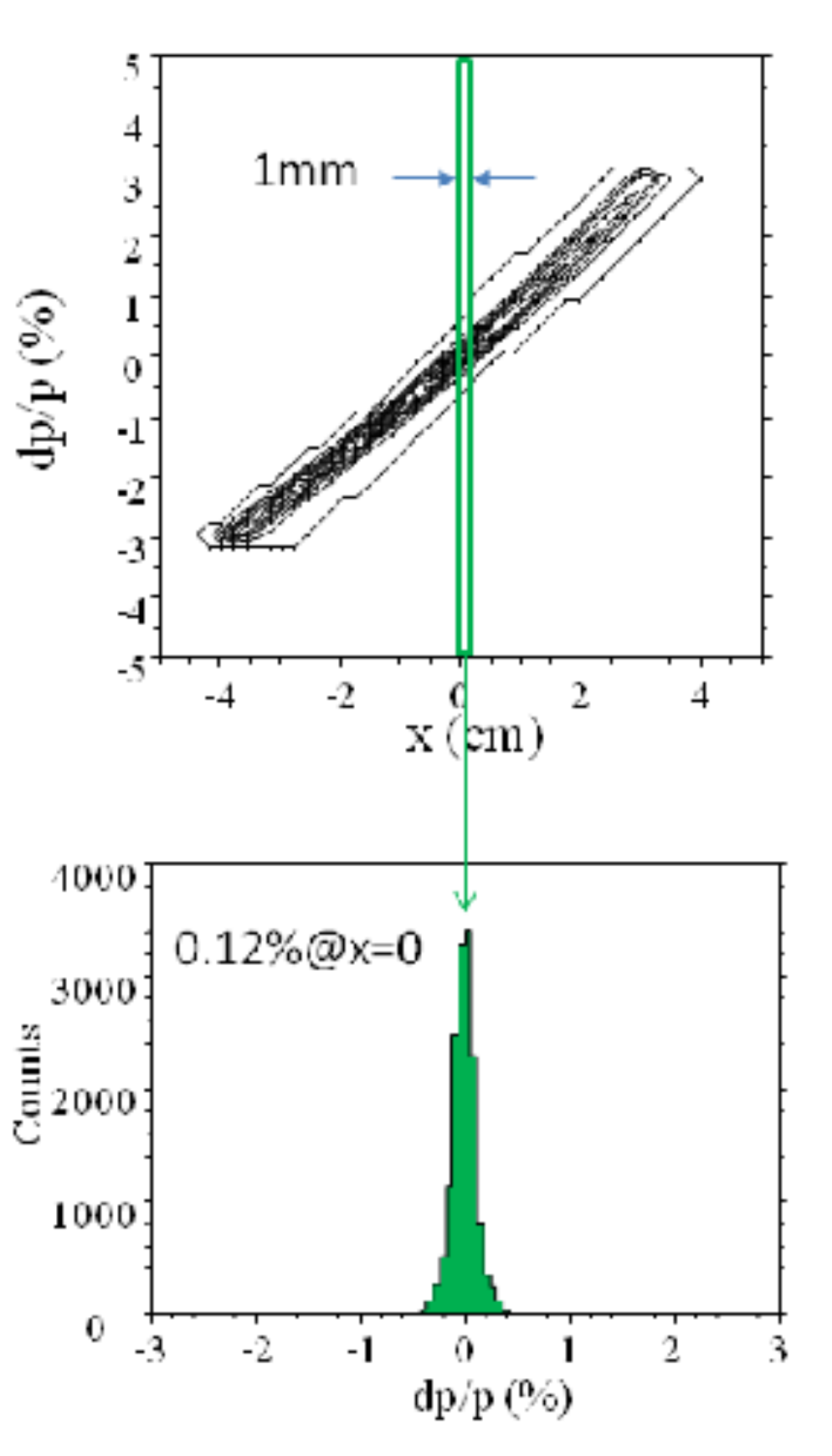}}%

\caption{\label{fig:dispersion}Left: correlation between beam momentum and
	position at the dispersive focal point. Right: momentum distribution
	within a 1-mm space in horizontal at the beam center.}
\end{figure}

The beam line length and acceptance are 133 m and 1.5~msr$\cdot$\%.
We estimated intensities of secondary particles by the so-called
Sanford-Wang formula~\cite{swt}, assuming that a 6-cm thick platinum
target is irradiated by a 30-GeV proton beam of 30~kW (15-kW beam loss
at the target), as shown in Fig.~\ref{fig:intensity}.
Here, a production angle for negative and positive particles are assumed
to be 0$^\circ$ and 3.9$^\circ$. We expect that the negative pion beam
intensity is more than 10$^7$ per second at 20~GeV/$c$.
\begin{figure}
\centerline{
 \includegraphics[width=8cm]{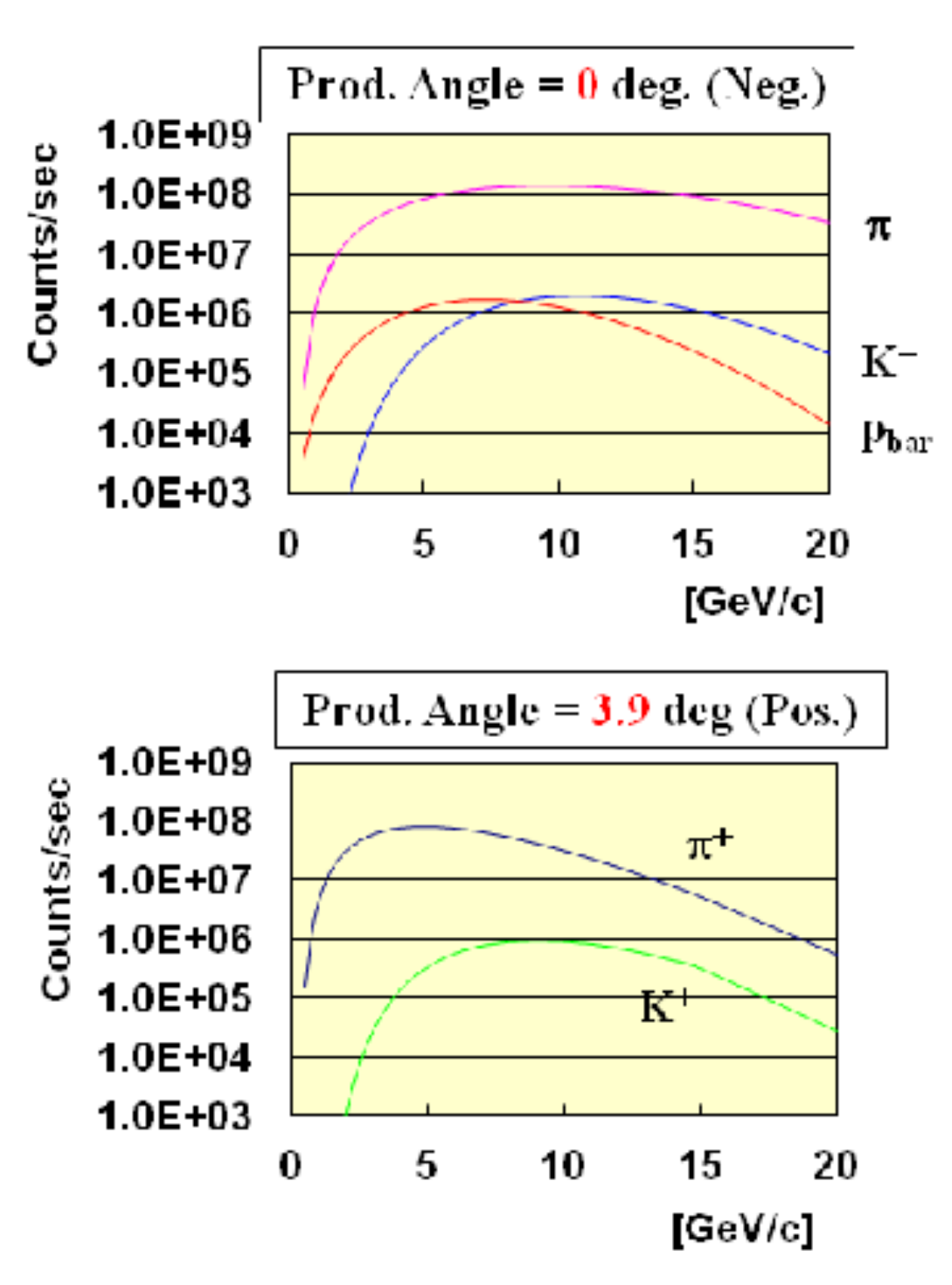}}%

\caption{\label{fig:intensity}Intensities of secondary beams calculated
	by Sanform-Wang's formula~\protect\cite{swt}. See text for assumed 
	conditions.}
\end{figure}

\item \textbf{Spectrometer}

An incident pion momentum will be measured at a resolution as good as
$\sim$0.1\% in the high-momentum beam line. We designed a spectrometer
system to reconstruct
scattered $D^{\ast -}$ from its decay chain of $D^{\ast -}\rightarrow
\bar{D}^0\pi^-$, $\bar{D}^0\rightarrow K^+\pi^-$, as shown in
Fig.~\ref{fig:spectrometer}. The spectrometer is based on a single
dipole magnet with a circular pole of 2.1~m in diameter and a gap of
1~m. A rigidity of the magnet is 2.3~Tm. A typical momentum resolution
is expected to be $\sim$0.5\% at 5~GeV/$c$. A liquid hydrogen target of
57~cm in length (4~g/cm$^2$ in thickness) will be placed close to the
entrance face of the magnet. Fiber trackers with 1 mm scintillating fiber
will be placed just after the target. A set of drift chambers will be
placed surrounding the pole and after the magnet to detect scattered
particles with lower and higher momenta, respectively. A ring image
cherenkov counter (RICH) with dual radiators of aerogel with a reflection
index of 1.04 and a C$_4$F$_{10}$ gas with an index of 1.00137 will be
used for identifying pion, kaon, and proton in a wide momentum range from
2 to 16~GeV/$c$. Time of flight counters will be placed to identify
scattered particles with lower momenta. The spectrometer covers about
60\% of solid angle for scattered $D^{\ast -}$ and about 80\% for decay
pions from produced charmed baryons.
\begin{figure}
\centerline{
 \includegraphics[width=6cm]{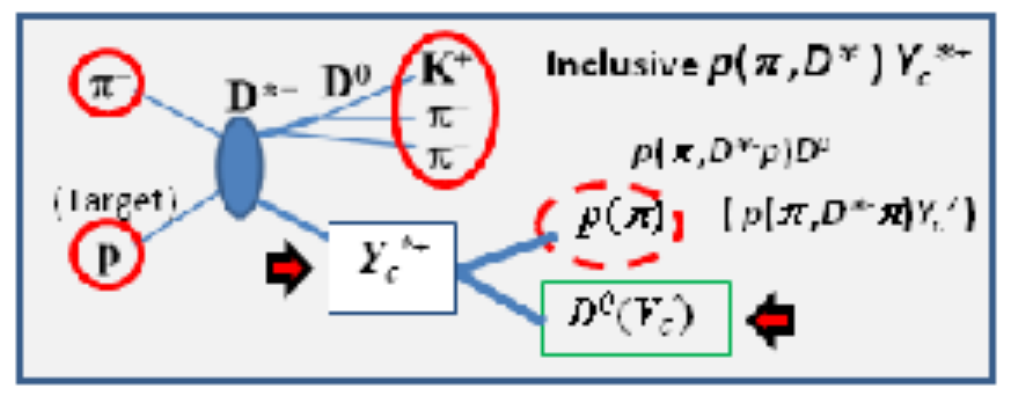}\\
 \vspace{0.5cm}
 \includegraphics[width=8cm]{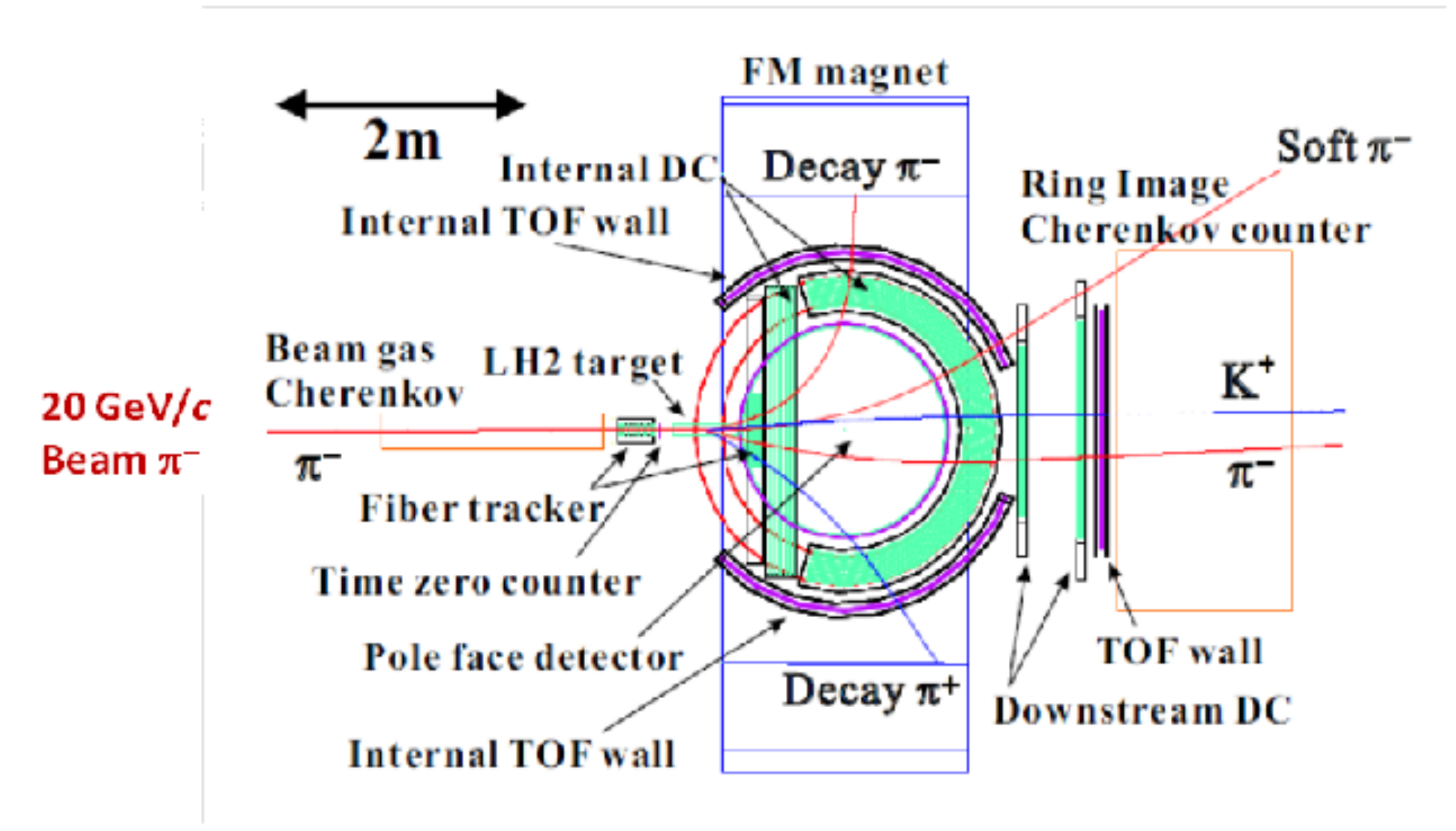}}%

\caption{\label{fig:spectrometer}Top: reaction scheme to identify charmed
	baryons and their decays by means of a missing mass technique.
	Bottom:designed spectrometer layout}
\end{figure}

Identifying two charmed mesons from the decay final state,
$K^+\pi^-\pi^-$, we could reduce huge background events of
$K^+\pi^-\pi^-$ productions by a factor of $\sim$10$^6$. Expected
charmed baryon spectrum is demonstrated by a Monte Carlo simulation
in Fig.~\ref{fig:spectrum}. Here, the states reported by the Particle
Data Group~\cite{pdgt} are taken into account. One sees that a series
of charmed baryons from the ground state to highly excited states with
higher spins are clearly observed.
\begin{figure}
\centerline{
 \includegraphics[width=10cm]{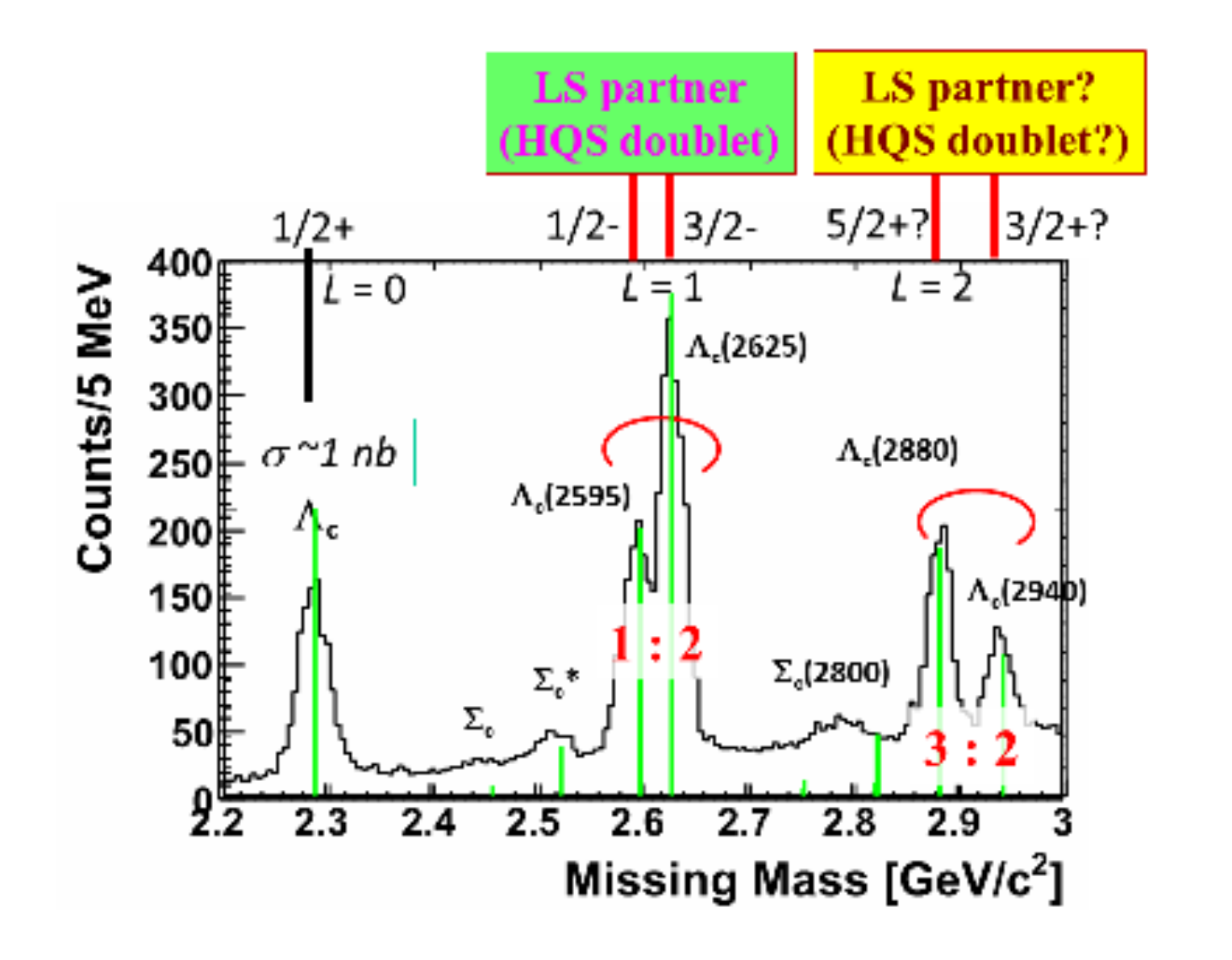}}%

\caption{\label{fig:spectrum}Expected missing mass spectrum in the
	($\pi^-,D^{\ast -}$) reaction on hydrogen.}
\end{figure}

We found that the production cross sections of the excited states
relative to that of the ground state do not go down. This is an
important feature of the $(\pi^-,D^{*-}$) reaction. We estimated the
production rates of the excited charmed baryons in the framework of a
$D^*$ exchange in the $t$-channel
at a very forward scattering angle~\cite{shkimt}.
Employing harmonic oscillator wave functions of constituent quarks in
the initial and final baryons, we estimate that the production rate, R,
is expressed as
\begin{eqnarray}
	R&\sim&<\phi_f|\sqrt{2}\sigma_-\exp(i{\bf q_{eff}r})|\phi_i>\\
	&\sim&(q_{eff}/A)^L\exp({-q_{eff}^2/2A^2)},
\end{eqnarray}
where $\phi_i$ and $\phi_f$ represent initial and final states of baryons.
An effective momentum transfer $q_{eff}$, taking a recoil effect of the
residual $ud$ diquark, is as large as 1.4~GeV/$c$ in the $p(\pi^-,D^{\ast -})$
reaction at the pion beam momentum of 20~GeV/$c$.
An oscillator parameter $A$ is taken to be $\sim$0.4~GeV,
which corresponds to the inverse of a typical baryon size.
In this reaction, a $u$ quark in a proton is converted to a $c$ quark in
the final charmed baryon in the reaction.
The $ud$ diquark behaves as a spectator.
Thus, this reaction well populates $\lambda$-mode excited states,
where an angular momentum $L$ is introduced between a $c$ quark and a
diquark.
Due to a large factor of $(q_{eff}/A)$, an absolute value of the production
cross section is reduced very much.
On the other hand, the ratio of $R$ for an excited state with $L$ to that
for the ground state ($L=0$) is $\sim(q_{eff}/A)^L$, which does not go down
even for $L>0$.
The $(\pi^-,D^{\ast -}$) reaction is suitable to populate higher spin states.

The $\lambda$-mode $\Lambda_c$ baryons with $L>0$ has two spin states coupled
to $L\pm1/2$.
These states are LS partners.
Therefore, we find that the production ratio of the two states should be
$L$:$L$+1.
Reversely, we can determine the spin-parity of the $\lambda$-mode
$\Lambda_c$ baryons by measuring the cross sections.

So far, a production cross section of $\Lambda_c$ in the p($\pi^-,D^{*-})$
reaction has not been measured.
Only upper limit, 7~nb, was reported in 1985~\cite{christensent}.
We estimate the production cross section to be a few nano barn at incident
pion momentum of 20~GeV/$c$~\cite{shkim2t} by employing a framework of reggeon
exchange model, which describes binary peripheral reactions well at high
energy.
We expect to observe 1000~events of the ground state $\Lambda_c$ production
for 100~days.

Decay branching ratios carry information on internal structure of a baryon.
The ratio of decay into a heavy meson and a light baryon to that into a
light meson and a heavy baryon is of particular interesting. The former is
expected to be dominant, if it is energetically allowed, in $\lambda$-mode
excited baryon, which is an orbital excitation
between a heavy quark and a light diquark.
The situation is to be opposite in $\rho$-mode.
One can find a suggestion in the case of $\Lambda$(1520), which is a P-wave
hyperon with spin-parity of 3/2$^-$. In $\Lambda$(1520) dominantly decays
into a kaon and a nucleon, while a Q-value in the decay is smaller than
that in the decay into a pion and a $\Sigma$ hyperon. The $\Lambda$(1520)
hyperon can be classified as a $\lambda$-mode hyperon although $\rho/\lambda$
mode classification has yet to be established in any baryon excited states.
Systematic measurements of decay branching ratios for the excited charmed
baryons are of particular importance.

\item \textbf{Baryon Spectroscopy with Different Flavors}

The above-mentioned discussion on internal structure of baryons with a
single heavy quark can be extended to baryons with double heavy quarks.
In the case of double heavy-quark baryons, the order of the excitation
energy for $\lambda$ and $\rho$ modes  interchanges. Here, the $\lambda$
mode is a motion of the light quark to the heavy-quark pair, and the $\rho$
mode is a relative motion between two heavy quarks.
One expects that a $\lambda$ mode excited state favors a decay into a light
meson and a double heavy-quark baryon. A $\rho$ mode state may dominantly
decay into a single-heavy meson and a single-heavy baryon.

A several states of cascade hyperons $\Xi$ are listed~\cite{pdgt}. Little is
known about their spin-parities and decay branching ratios. The
$(\bar{K},K^+)$ reaction is one of promising reactions to produce
cascade hyperons.
Since the $(\bar{K},K^+)$ reaction has no single-meson exchange process
in $t$ channel, $\Xi$ productions at backward angles are expected to play
a principal role.
$\rho$-mode $\Xi$ hyperons may be populated well through $u$-channel process.
It is quite worthy to measure production rates and decay branching ratios
of $\Xi$ hyperons.

\item \textbf{Concluding Remark}

\begin{itemize}
\item Masses, decay branching ratios, and production rates of baryons with
	heavy flavors provide information on internal motions of ingredients, 
	such as diquark correlation.
\item We proposed an experiment on charmed baryon spectroscopy via the
	$(\pi^-, D^{\ast -})$ reaction at the J-PARC high-momentum beam 
	line. We will measure a mass spectrum of charmed baryons from the 
	ground state to highly excited states in an excitation energy range 
	of more than 1~GeV by means of a missing mass technique. Production 
	cross sections and decay branching ratios of produced charmed baryons 
	will be measured.
\item The present argument on baryon spectroscopy with a charm quark should
	be extended to those with different flavors. In particular, $\Xi$ 
	baryons are of interest as double-heavy quark system that can be 
	accessible in experiment. Neutral kaon beam at Jlab is unique in 
	hadron spectroscopy and plays a complimentary role to the J-PARC. 
	Constructive collaboration to integrate efforts to realize hadron 
	spectroscopy with hadron beams in JLab and J-PARC is desired.
\end{itemize}
\end{enumerate}


\newpage
\subsection{$K^0 \Lambda$ Photoproduction on the Neutron at $E_\gamma <$
        1.2~GeV}
\addtocontents{toc}{\hspace{2cm}{\sl Y.~Tsuchikawa}\par}
\setcounter{figure}{0}
\setcounter{table}{0}
\setcounter{equation}{0}
\setcounter{footnote}{0}
\halign{#\hfil&\quad#\hfil\cr
\large{Yusuke Tsuchikawa (for FOREST Collaboration)}\cr
\textit{Department of Physics}\cr
\textit{Nagoya University}\cr
\textit{Nagoya 464-8602, Japan}\cr}

\begin{abstract}
Photoproduction of $K^0$ and $\Lambda$ has been studied at the
Research Center for Electron Photon Science (ELPH), Sendai, Japan.
The $\gamma d\to K^0\Lambda p$ events are detected with the
electromagnetic calorimeter FOREST near the threshold region. We
present the first measurement of $K^0\Lambda$ photoproduction
cross sections on the neutron.
\end{abstract}

\begin{enumerate}
\item \textbf{Introduction}

The simultaneous $K^0$ and $\Lambda$ photoproduction is a good probe to
study highly excited baryons which hardly couple to the well known
channels ({\it e.g.,} $\pi N$ and $\eta N$). The $\gamma n\to K^0\Lambda$
reaction is expected to be sensitive to not only nucleon but also hyperon
resonances. Because there is no charged $K$ exchange as the $t$-channel
contribution in the Born terms, it is expected there are relatively large
contributions from the $s$- and $u$-channels in the $K^0\Lambda$
photoproduction on the neutron in comparison to the $K^+\Lambda$
photoproduction on the proton.

In addition, the prominent structure at $W \sim 1.67$~GeV which is seen
in the $\eta$ photoproduction cross section on the neutron as a narrow
peak structure~\cite{etan_ishikawa,etan_MAINZ} is one of the special
interests in this study.  While many experiments observed the structure,
the origin of the structure is still an open question. The $K^0\Lambda$
system is similar to the $\eta n$ system from the following points;
having 1/2-isospin, produced by the $\gamma n$ initial state, having
an $s\bar{s}$ component. From the viewpoint, photoproduction of $K^0
\Lambda$ may be a good channel to reveal the origin of the prominent
structure.

Cross sections of the $\gamma d \to K^0\Lambda p$ reaction was reported
by Tsukada {\it et al.}~\cite{NKSK0L}. The differential cross sections
were measured for $\cos\theta_K^{Lab} = [0.9 , 1.0]$ and $E_\gamma
=[0.9,1.1]$~GeV, where $\cos\theta_K^{Lab}$ stands for the $K^0$
emission angle in the laboratory frame and $E_\gamma$ stands for the
incident photon energy.

This contribution presents the first measurement of $K^0$ angular
distribution for the $\gamma d\to K^0\Lambda p$ reaction in the
center-of-mass frame by using a $4\pi$ calorimeter named FOREST which
is introduced in the next section.

\item \textbf{Experiment}

Meson photoproduction experiments are carried out at ELPH, Sendai.
The bremsstrahlung photon beam~\cite{photonbeam} is generated by
inserting a 11~$\mu$m thick carbon fiber to the circulating electrons
in the 1.2~GeV synchrotron~\cite{STBring}. Liquid hydrogen or deuterium
target~\cite{target} is bombarded with the 0.5--1.2~GeV tagged photons.
A four pi calorimeter FOREST~\cite{FORESTNIM} detects the emitted
particles produced by the photo-induced reactions. FOREST consists of
three calorimeters made of 192 pure CsI crystals, 252 lead/scintillating
fiber modules, and 62 lead glasses. The energy resolutions of the
calorimeters are 3\%~\cite{SCISSORS}, 7\%~\cite{BG1,BG2}, and
5\%~\cite{Rafflesia} for 1~GeV photons, respectively. Three hodoscopes
are situated in front of the calorimeters for charge identification.
The hodoscopes are made of 72, 18, and 12 pieces of 5~mm thick plastic
scintillators.

\item \textbf{Analysis, Results, and Discussion}

The $\gamma n \to K^0 \Lambda$ reaction is identified by focussing on
the reaction chains; $K^0\to\pi^0\pi^0\to\gamma\gamma\gamma\gamma$ and
$\Lambda\to p\pi^-$. Event selection for the $K^0 \Lambda$ photoproduction
is performed by detecting all of the below particles by the calorimeters.
Hence, four neutral and two charged particles are required to be
detected as clusters in the calorimeters. Four neutral clusters within
2~ns in the FOREST detectors are required for four photons from $K^0$.
Two charged particles detected at later timing from the four photons
are required as $p$ and $\pi^-$. The proton and charged pion are
identified by using timing difference between the two charged particles.
Measured energies, polar angles, and azimuthal angles of the six
particles and incident photon energy are kinematically fitted to
satisfy four constraints which is related the $\gamma n\to K^0\Lambda$
reaction kinematics with assuming the target neutron is at rest. 
Figure~1 shows $\gamma \gamma$ invariant masses
of selected photons and $\Delta E$-$E$ plots of the selected proton and
that of the charged pions.
\begin{figure}
\begin{center}
    \includegraphics[scale=0.7]{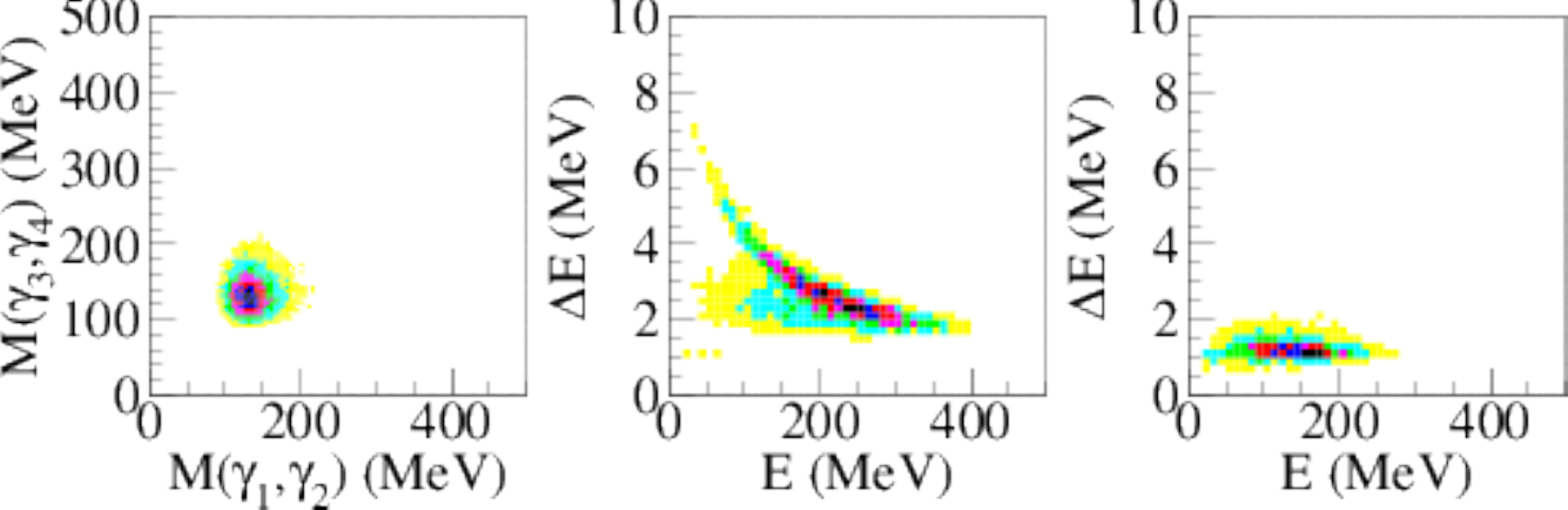}

  \caption{The $\gamma \gamma $ invariant masses (left), $\Delta
        E$-$E$ plots for proton (center) and charged pion (right).}
\end{center}
\label{fig:selectedevents}
\end{figure}

It is clearly shown that the particle selection for four photons, proton,
and charged pion works well. The four photon invariant mass distribution
shows a clear peak structure around the $K^0$ rest mass (Fig.~2). The 
$K^0$ yield as a function of $\cos\theta_K^{CM}$ for different energy 
bins are estimated by fitting linear combination of the gaussian and 
background distribution to the measured distributions. The background 
shape is estimated by using GEANT4 with $\gamma n\to\pi^0\pi^0\pi^-p$ 
reaction. Figure~2 shows the simulated distributions well reproduce the 
measured background shapes for all energy bins.
\begin{figure}
        \begin{center}
        \includegraphics[scale=0.5]{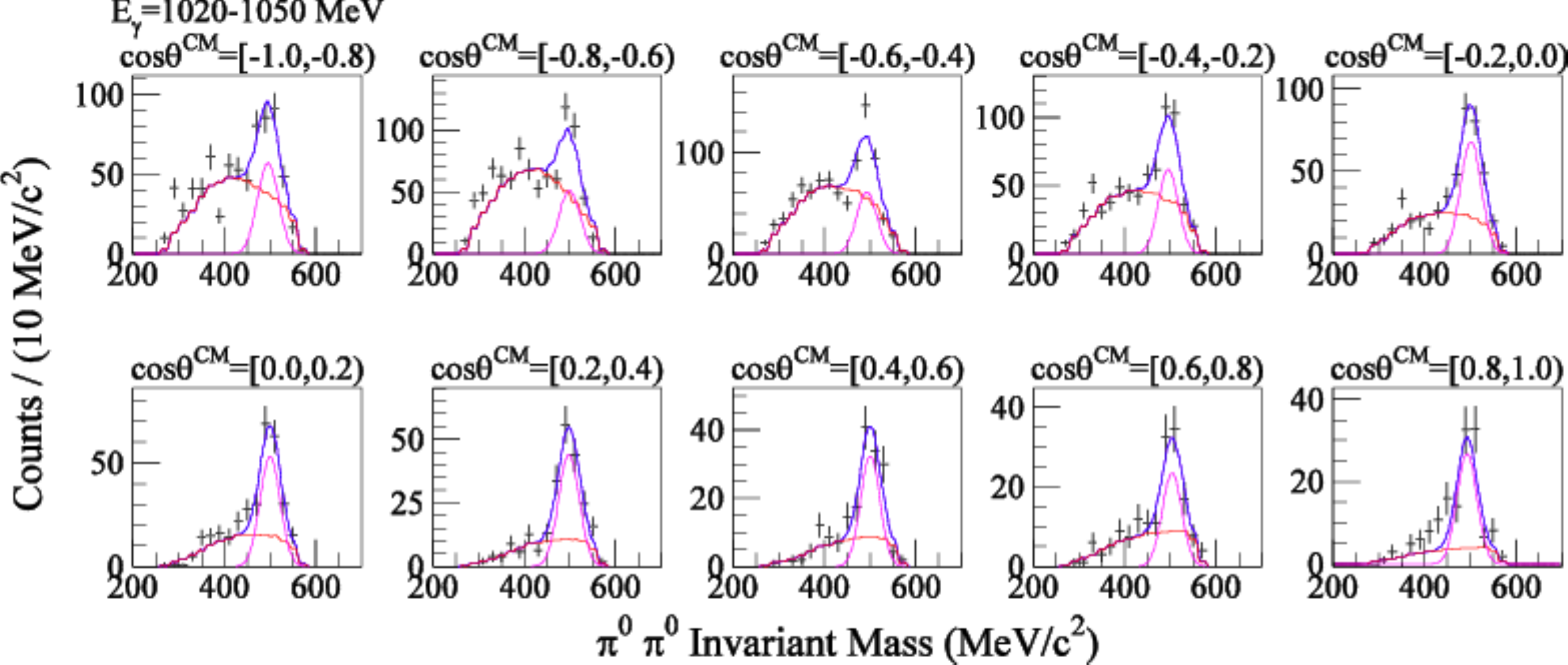}

        \caption{The four photon invariant mass distributions (black
        ponits) for different $\cos_K^{CM}$. The magenta line shows
        the fitted gaussian and red line shows simulated distribution.
        The blue line shows the combination of the above two
        functions.}
        \end{center}
        \label{fig:4gminvs}
\end{figure}

Differential cross sections are derived from the estimated yields of
$K^0$ signals.  Figure~3 shows the differential cross
sections as a function of $\cos\theta_K^{CM}$ for different incident
photon energy $E_\gamma$ where $\cos\theta_K^{CM}$ stands for the
$K^0$ emission angle in the center-of-mass (CM) frame and $E_\gamma$
stands for the incident photon energy.
\begin{figure}
        \begin{center}
        \includegraphics[scale=0.45]{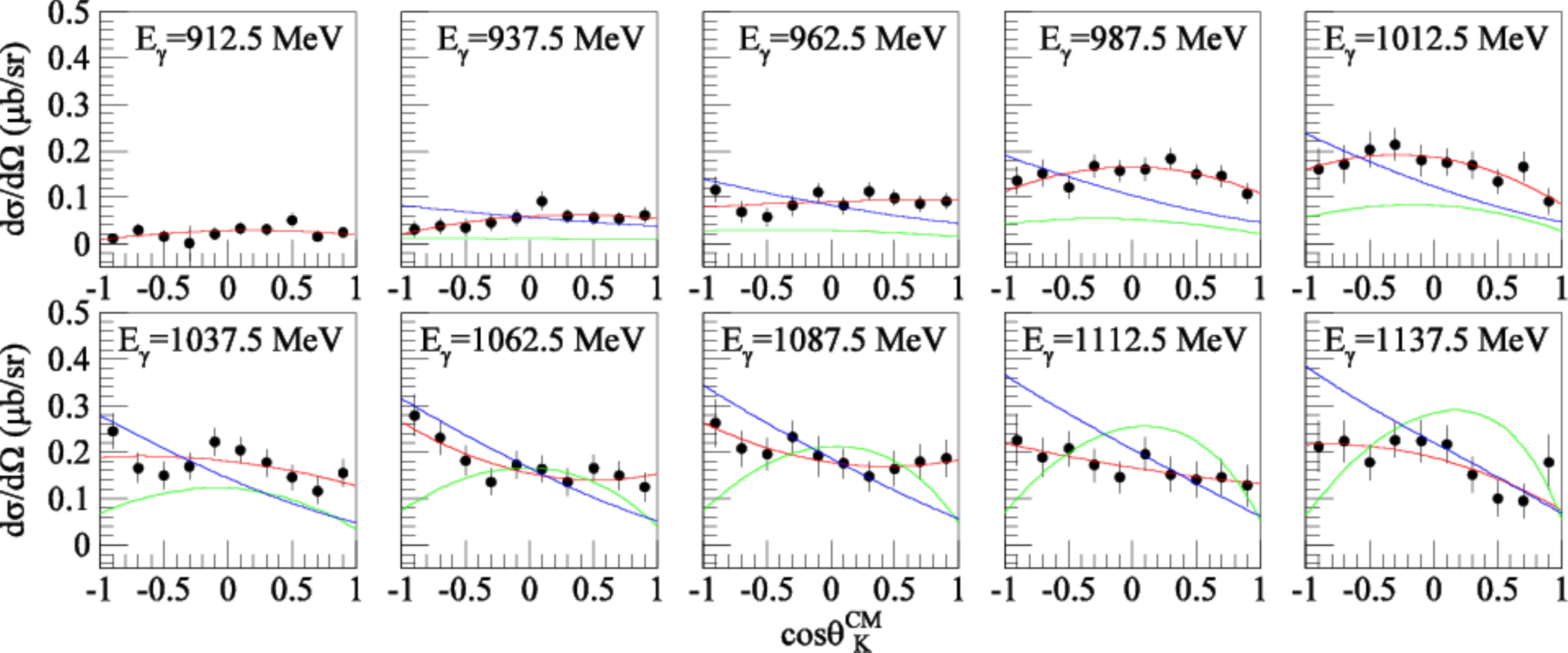}

        \caption{Differential cross sections of $\gamma n\to K^0
        \Lambda$ reaction as a function of $K^0$ emission angle in
        the CM frame $\cos \theta_K^{CM}$ for different incident
        photon energy $E_\gamma$ (black points). The green (blue) 
        line shows the theoretical curve given by
        Kaon-MAID~\protect\cite{KM1,KM2,KM3}
        (Saclay-Lyon~A~\protect\cite{SLA}).
        The red line shows fitted Legendre polynomials.}
        \label{fig:diffcs}
        \end{center}
\end{figure}

The first measurement of angular distribution for the $K^0\Lambda$
photoproduction in the CM frame shows flat shapes near the reaction
threshold but shows backward enhancement in the higher energy parts
in this energy region. Comparisons with the theoretical curves given
by the Kaon-MAID~\cite{KM1,KM2,KM3} and Saclay-Lyon~A~\cite{SLA}
models are performed. The measured shapes of the differential cross
sections favor the latter one. It indicates that hyperon resonances
play important roles in the $K^0~\Lambda$ photoproduction since the
difference between two models; the Kaon-MAID includes various nucleon
resonances, while the Saclay-Lyon~A includes various hyperon 
resonances.

\item \textbf{Summary}

The $K^0\Lambda$ photoproduction on the neutron is studied at ELPH
with the $4\pi$ electromagnetic calorimeter FOREST via $K^0\to\pi^0\pi^0
\to\gamma\gamma\gamma\gamma$ and $\Lambda\to p\pi^-$ decay modes.
The $K^0$ signal is clearly observed in the four photon invariant mass
distribution as a narrow peak. The first measurement of the differential
cross section as a function of $K^0$ emission angle in the CM frame
reveals the angular distribution of this reaction shows backward
enhancement in higher energy region, while it shows flat distribution
near the reaction threshold.  By comparing the measured distributions
with two theoretical curves given by Kaon-MAID and Saclay-Lyon~A models,
the measured distribution indicates that hyperon resonances play
important roles in the $\gamma n\to K^0\Lambda$ reaction. 

\item \textbf{Acknowledgments}

This work was supported in part by Grants-in-Aid for Scientific
Research (B) (17340063), for Specially promoted Research (19002003),
for Scientific Research (A) (24244022), and for Scientific Research
(C) (26400287).
\end{enumerate}


\newpage
\subsection{Excited Hyperon Possibilities in Relativistic Heavy 
	Ion Collisions}
\addtocontents{toc}{\hspace{2cm}{\sl R.~Bellwied}\par}
\setcounter{figure}{0}
\setcounter{table}{0}
\setcounter{equation}{0}
\setcounter{footnote}{0}
\halign{#\hfil&\quad#\hfil\cr
\large{Rene Bellwied}\cr
\textit{Physics Department}\cr
\textit{University of Houston}\cr
\textit{617 SR1 Building}\cr
\textit{Houston, TX 77204, U.S.A.}\cr}

\begin{abstract}
I will review the latest experimental measurements from RHIC and the 
LHC in the context of predictions by lattice QCD and statistical 
hadronization models with regard to ground state, resonance and exotic 
particle production in the strange quark sector. I will propose possible 
future measurements at YSTAR and the heavy ion experiments at RHIC and 
the LHC which might shed light on the question of hadronization in the 
non-perturbative regime and the production of, and interactions in, 
specic hadronic bound states.
\end{abstract}

\begin{enumerate}
\item \textbf{Introduction}

\begin{figure}[tbh]
\begin{center}
\includegraphics[width=3.5in]{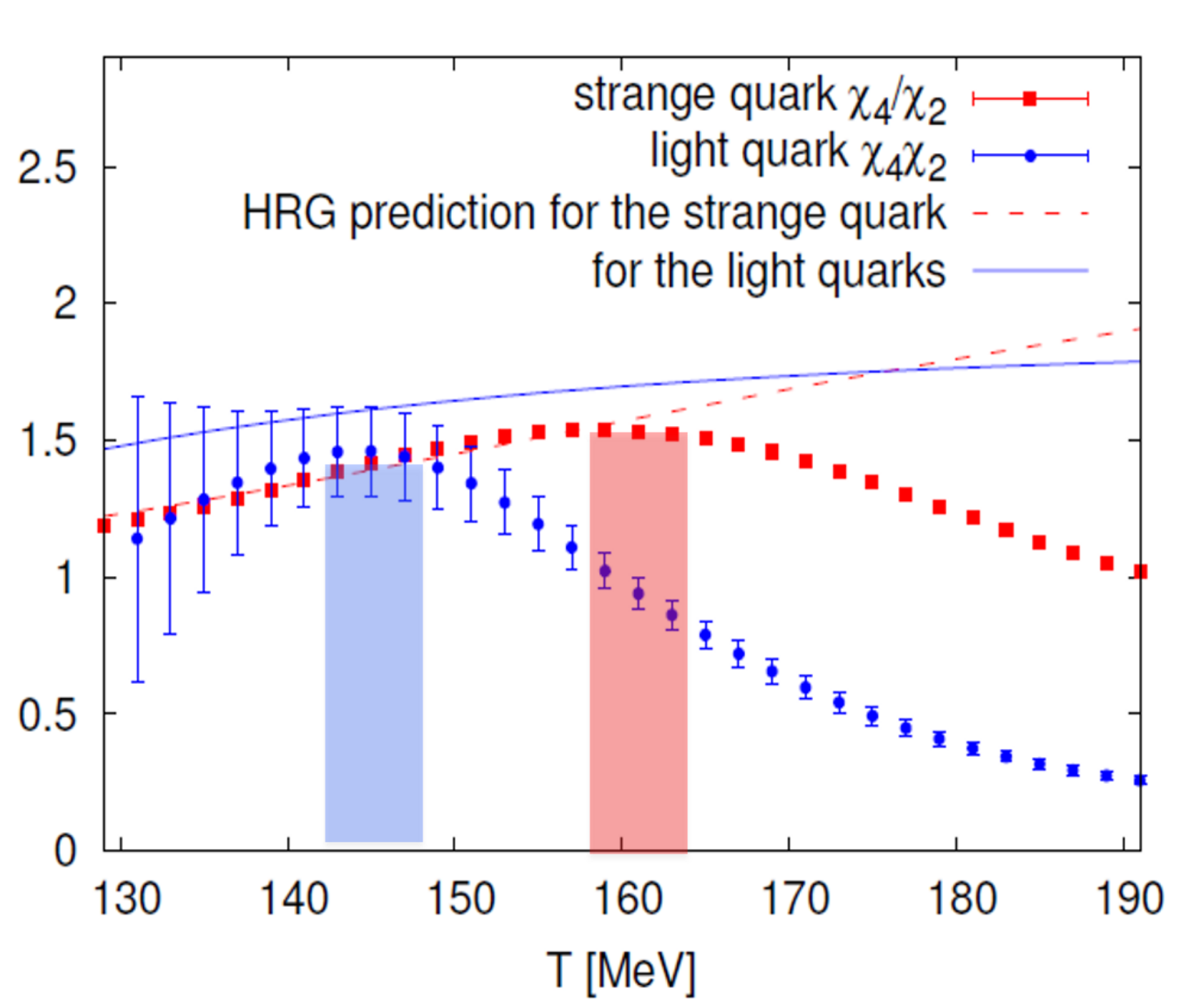}
\end{center}

\caption{\noindent Continuum extrapolated lattice QCD results on the
        difference between light and strange quark susceptibility ratios, in
        this case $\chi$$_{4}$/$\chi$$_{2}$~\protect\cite{Bellwied:2013ctar}.}
        \label{form1D}
\end{figure}
Recent lattice QCD calculations in the QCD crossover transition region 
between a deconned phase of quark and gluons and a hadronic resonance gas 
have revealed a potentially interesting sub-structure related to the 
hadronization process. Studies of avor dependent susceptibilities, which 
can be equated to experimental measurements of conserved quantum number 
uctuations, seem to indicate a slight avor hierarchy in the three quark 
sector (u,d,s) in thermalized systems. Specically, the ratios of higher 
order susceptibilities in the strange sector show a higher transition
temperature than in the light sector~\cite{Bellwied:2013ctar}. Both
pseudo-critical temperatures are still within the error bars of the
quoted transition temperature based on all lattice QCD order 
parameters~\cite{Borsanyi:2010bpr,Bazavov:2014pvzr} which is 154$\pm$9~MeV,
but the difference of the specific susceptibilities is around 18~MeV
and well outside their individual uncertainties, as shown in Fig.~\ref{form1D}.

This difference might also be confirmed by statistical thermal model
calculations that try to describe the yields of emitted hadrons from
a QGP based on a common chemical freeze-out temperature. Although the
yields measured by ALICE at the LHC in 2.76~TeV PbPb collisions can be
described by a common temperature of 156$\pm$2~MeV, with a reasonable
$\chi$$^{2}$, the fit improves markedly if one allows the light quark
baryons to have a lower temperature than the strange quark baryons,
which is shown in Fig.~\ref{form2D}~\cite{Floris:2014ptar}.
\begin{figure}[tbh]
\begin{center}
\includegraphics[width=3.5in]{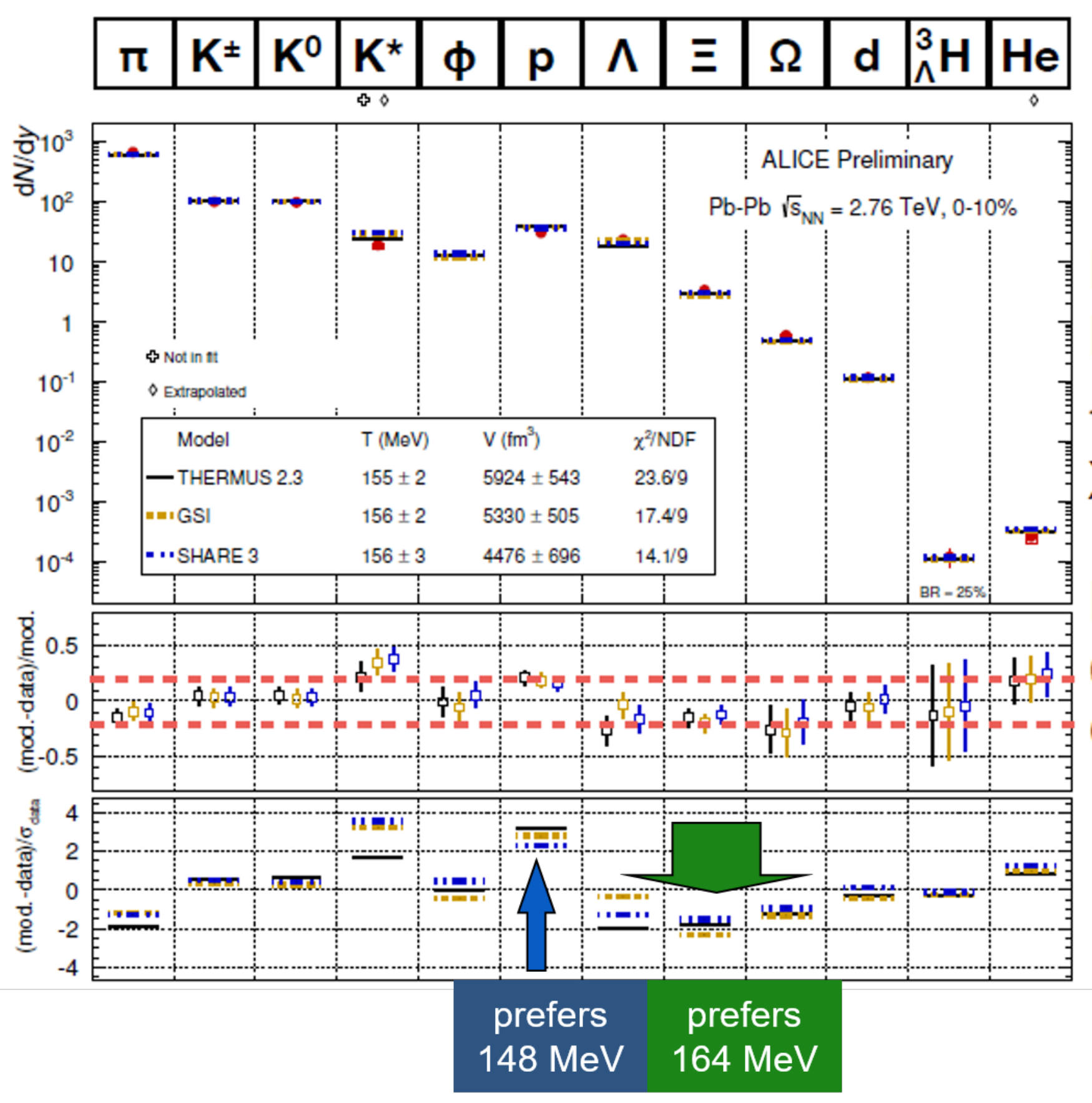}
\end{center}

\caption{\noindent Statistical Hadronization Model fits to all measured
	particle yields in ALICE~\protect\cite{Floris:2014ptar}. Although 
	all three models converge on a common temperature, the main 
	deviations seem to be flavor dependent. }
	\label{form2D}
\end{figure}

A similar result has been found when the thermal fluctuations of particle
yields as measured by STAR~\cite{Adamczyk:2013dalr,Adamczyk:2014fiar},
which can be related to the light quark dominated susceptibilities of the
electric charge and the baryon number on the lattice, have been compared to
statistical model calculations~\cite{Alba:2014ebar}. Fig.~\ref{form3D} shows that the
deduced chemical freeze-out temperature is well below a common fit based on
particle yields which  takes into account all particle species, including
strange particles.
\begin{figure}[tbh]
\begin{center}
\includegraphics[width=3.5in]{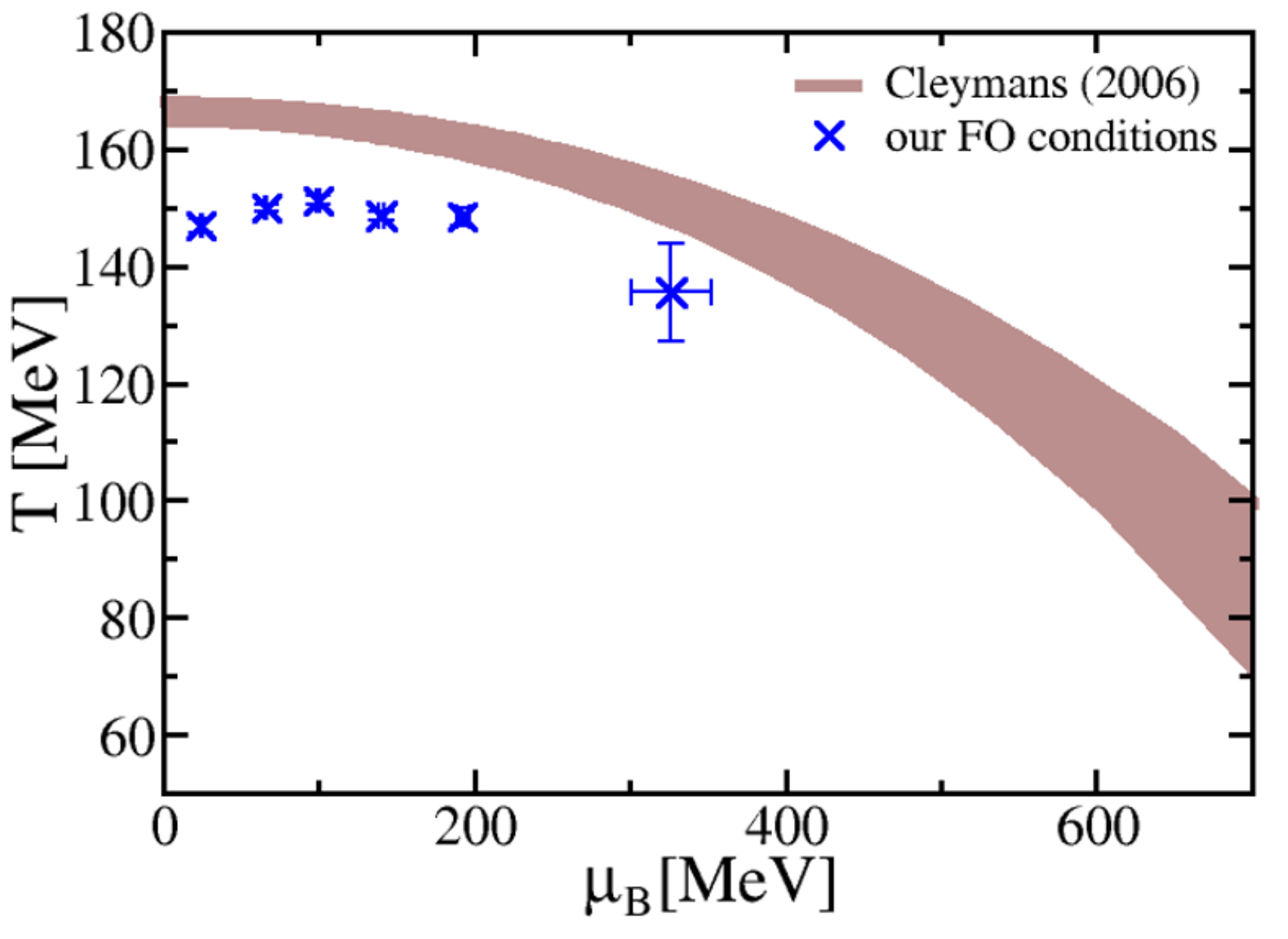}
\end{center}

\caption{\noindent Chemical freeze-out temperature as a function of 
	the baryo-chemical potential from a Hadron Resonance Gas model 
	analysis of STAR net-proton and net-charge fluctuations, compared 
	to a common freeze-out curve based on all measured particle 
	yields~\protect\cite{Alba:2014ebar}.}
	\label{form3D}
\end{figure}

\item \textbf{Possible Measurements in the Strange Sector}

If one assumes that strange and light quarks indeed prefer different
freeze-out temperatures then the question arises how this could impact
the hadronization mechanism and abundance of specific hadronic species.
In other words, is the production of strange particles enhanced in a
particular temperature range in the crossover region ?

This question is rather complex since most of the accessible species
are quark mixtures of light and strange quarks, and the only pure strange
baryon, the $\Omega$, has a low cross section, even in relativistic heavy
ion collisions. Furthermore besides the production of ground state baryons,
one also has to account for the formation of resonant states and potentially
even exotic strange multi-quark states and hypernuclei. Although hypernuclei
are likely formed through coalescence at the very end of the evolution of
the hadronic system, i.e., between chemical and kinetic freeze-out of the
thermal hadronic system, studies comparing measured yields of molecular
matter and anti-matter produced at the LHC with predictions from statistical
hadronization models have shown that the actual yield of triton and helium
states is well described with models assuming a chemical freeze-out near
156~MeV~\cite{Andronic:2010qur}. Although any state formed at these
temperatures with a binding energy in the keV range will almost certainly
dissolve, it seems its regeneration cross section is fixed by the chemical
freeze-out from the deconfined state. This is likely due to the fact that
the entropy to baryon ratio is frozen in at that time~\cite{Siemens:1979dzr}.
Therefore, with respect to determining a likely hypernucleus yield one can
employ the SHM predictions as long as the assumed temperature is within the
range determined by lattice QCD susceptibility calculations. Recent 
measurements performed at RHIC~\cite{Abelev:2010rvr} and the 
LHC~\cite{Adam:2015ytar} reveal cross sections which are in agreement with 
thermal model predictions at a
freeze-out temperature in between the two extremes of pure light and strange
quark states, although the error bars are such that no strong conclusion can
be drawn from these statistics limited measurements.

SHM predictions for exotic states, in particular the H-Dibaryon and the strange
pentaquark, postulated early on based on measurements by NA49 at the 
CERN-SPS~\cite{Alt:2003vbr}, have been compared to measurements in PbPb and 
pp collisions
at the LHC, and no evidence for these exotic states has been found, see 
Figs.~\ref{form4D} and \ref{form5D}~\cite{Adam:2015ncar,Abelev:2014qqar} and 
Dr.~Doenigus presentation
at this workshop. At the same time Fig.5 shows that signals of hadronic 
resonant
states, in this case the $\Xi$(1530) are strong at LHC energies and can be well
reconstructed through their hadronic decay channels.
\begin{figure}[htb]
\begin{center}
\includegraphics[width=4.in]{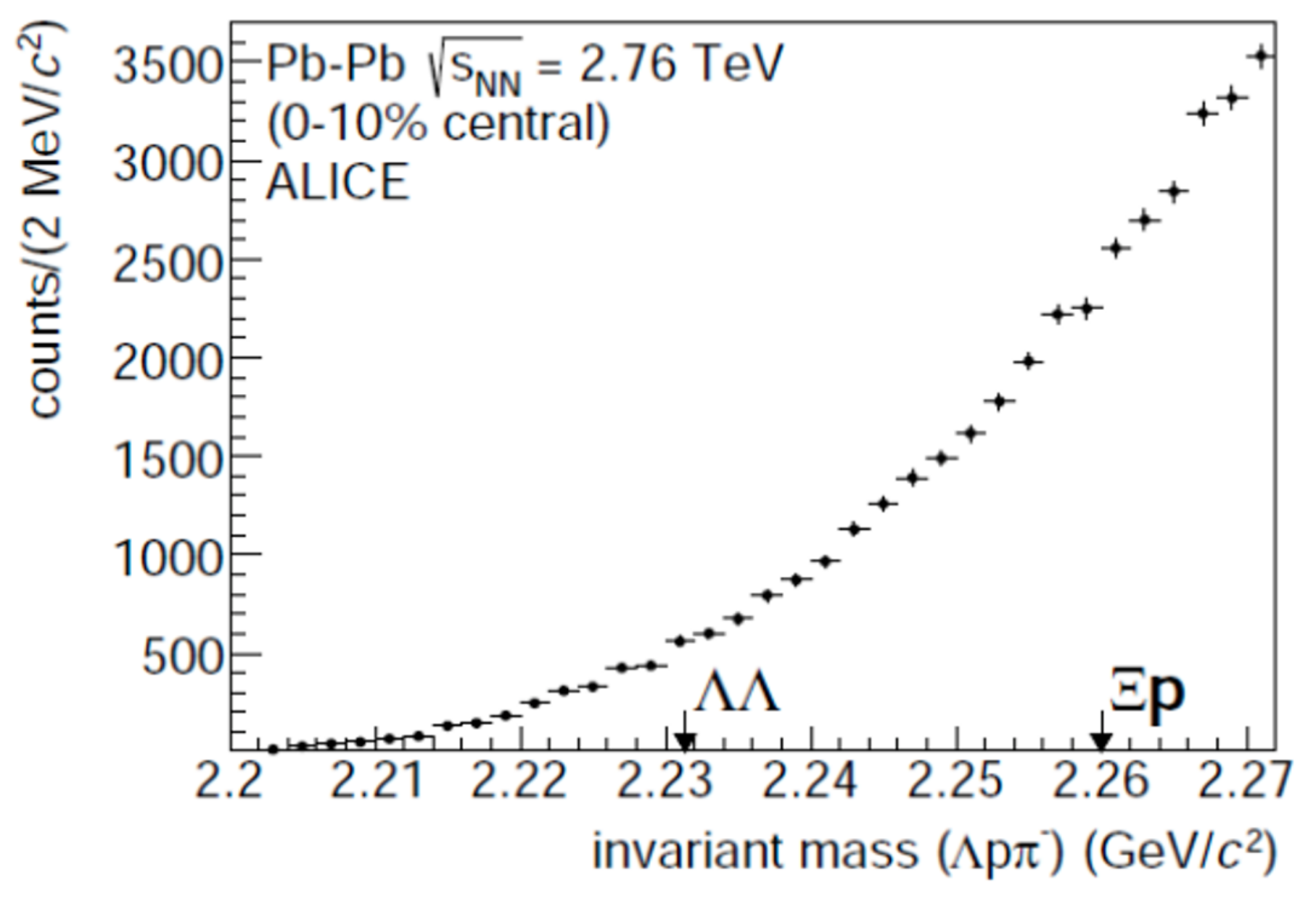}
\end{center}

\caption{\noindent Results from searches for $\Lambda\Lambda$ and $\Xi$p bound
	states in 2.76~TeV PbPb collisions in ALICE~\protect\cite{Adam:2015ncar}. No evidence
	for Di-baryon states has been found at limits well below the thermal model
	predictions.}
	\label{form4D}
\end{figure}
\begin{figure}[tbh]
\begin{center}
\includegraphics[width=4.in]{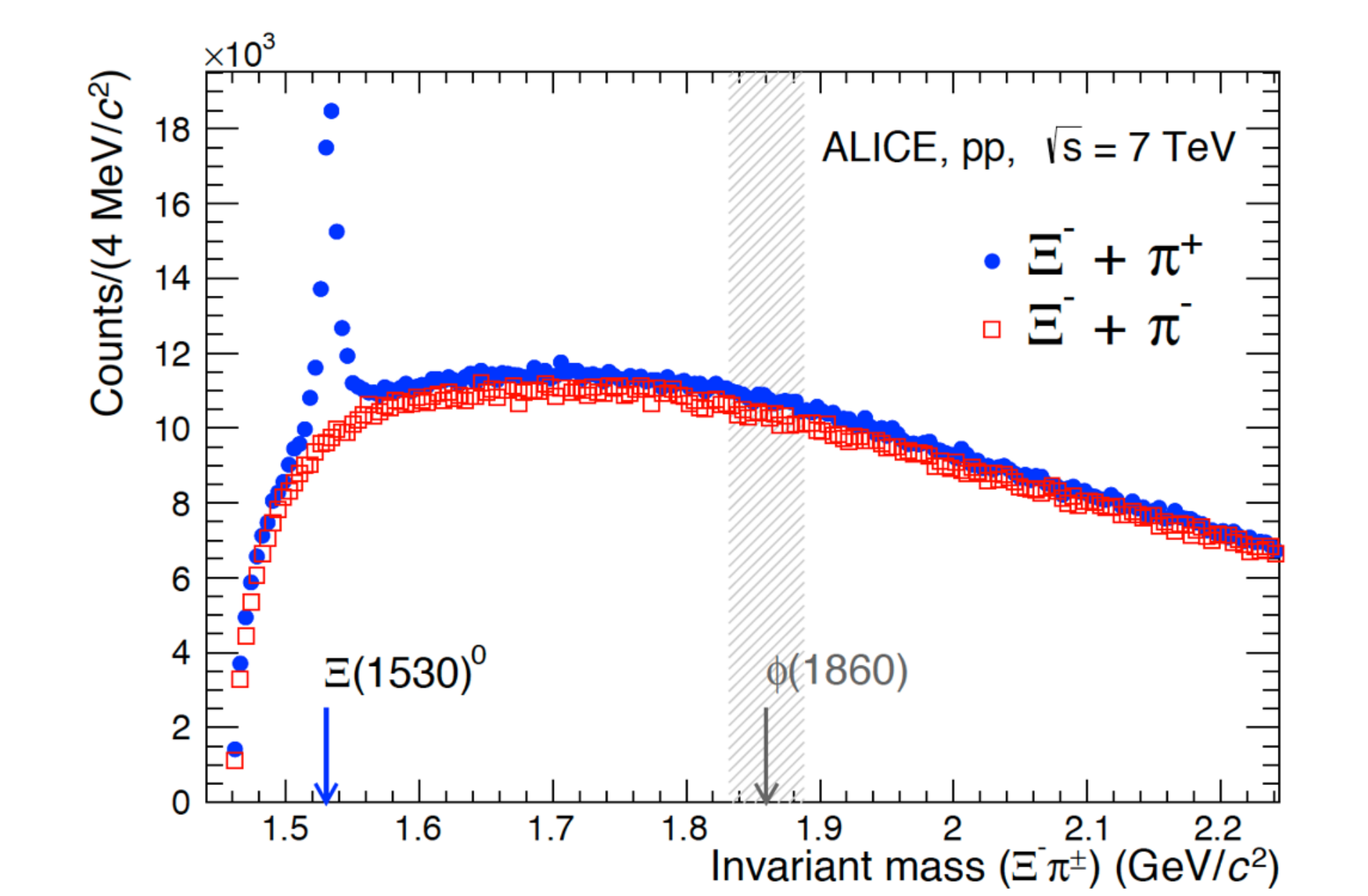}
\end{center}

\caption{\noindent Results from searches for the $\Xi$(1530) resonance and
	the $\phi$(1860) pentaquark in 7~TeV pp collisions in 
	ALICE~\protect\cite{Adam:2015ncar}.  No evidence for the strange 
	pentaquark state has been found at limits well below the thermal 
	model predictions.} \label{form5D}
\end{figure}

Since the evidence for exotic states in the strange sector is small and
the hypernuclei seem to be not particularly enhanced, the emphasis of any
future experimental program trying to understand hadron production should
be shifting towards strange baryonic resonance production. Elementary
collisions at RHIC and LHC energies might well provide the link between
future measurements in the YSTAR experiment at JLab and the future analysis
of strange resonance enhancements in heavy ion collisions at RHIC and the
LHC. Both the $\Lambda$(1405) and the $\Lambda$(1520) have been determined
in relativistic heavy ion 
experiments~\cite{Agakishiev:2012xkr,Markert:2002xir}, so has the 
$\Sigma$(1385)~\cite{{Abelev:2014qqar}}. Figs.~\ref{form6D}
show the latest results in pPb collisions from ALICE for the $\Sigma$ and 
$\Xi$ resonant states.
\begin{figure}[h]
\begin{center}
\begin{tabular}{c c}
\includegraphics[width=3.in]{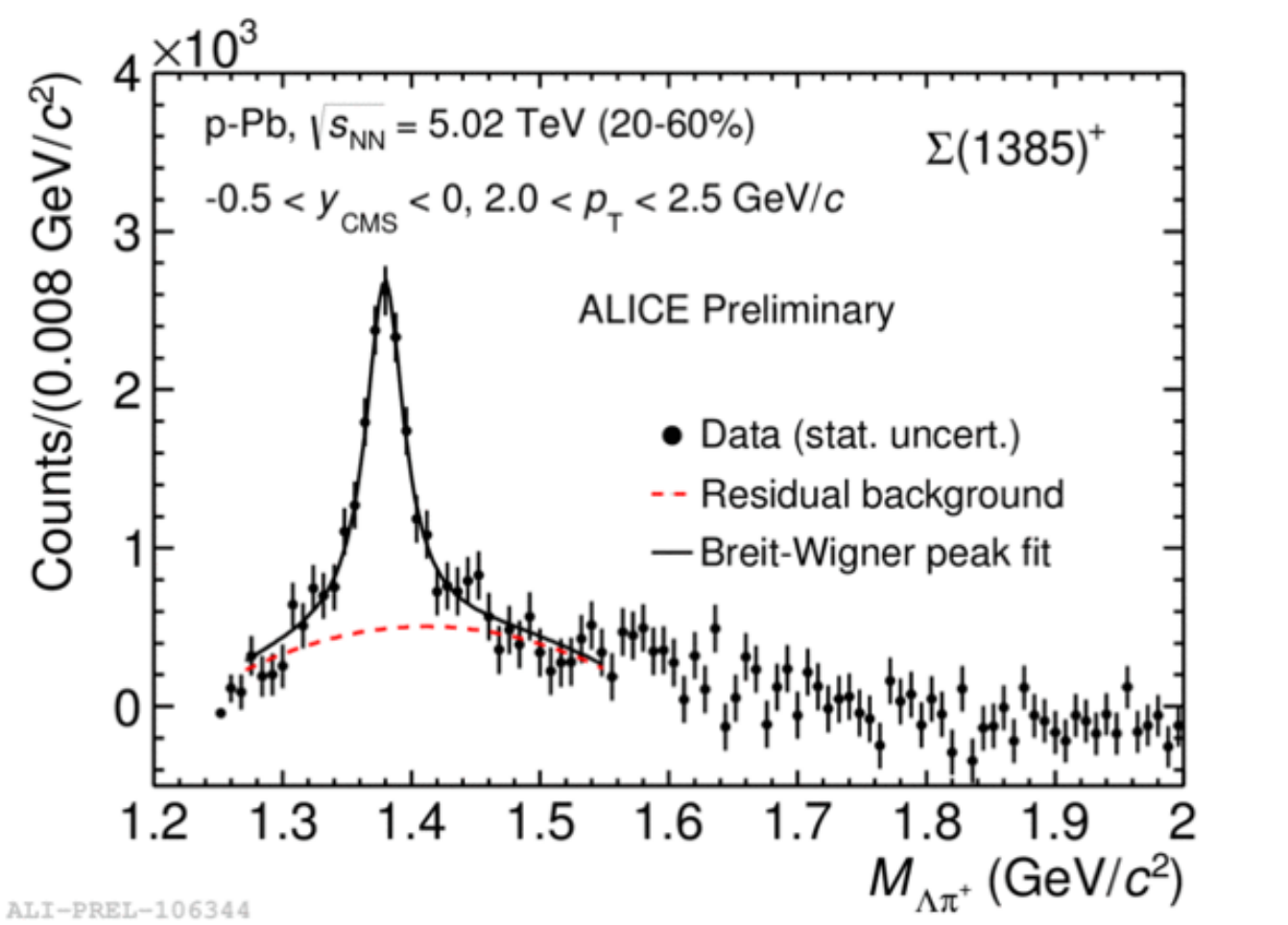} &
\includegraphics[width=3.in]{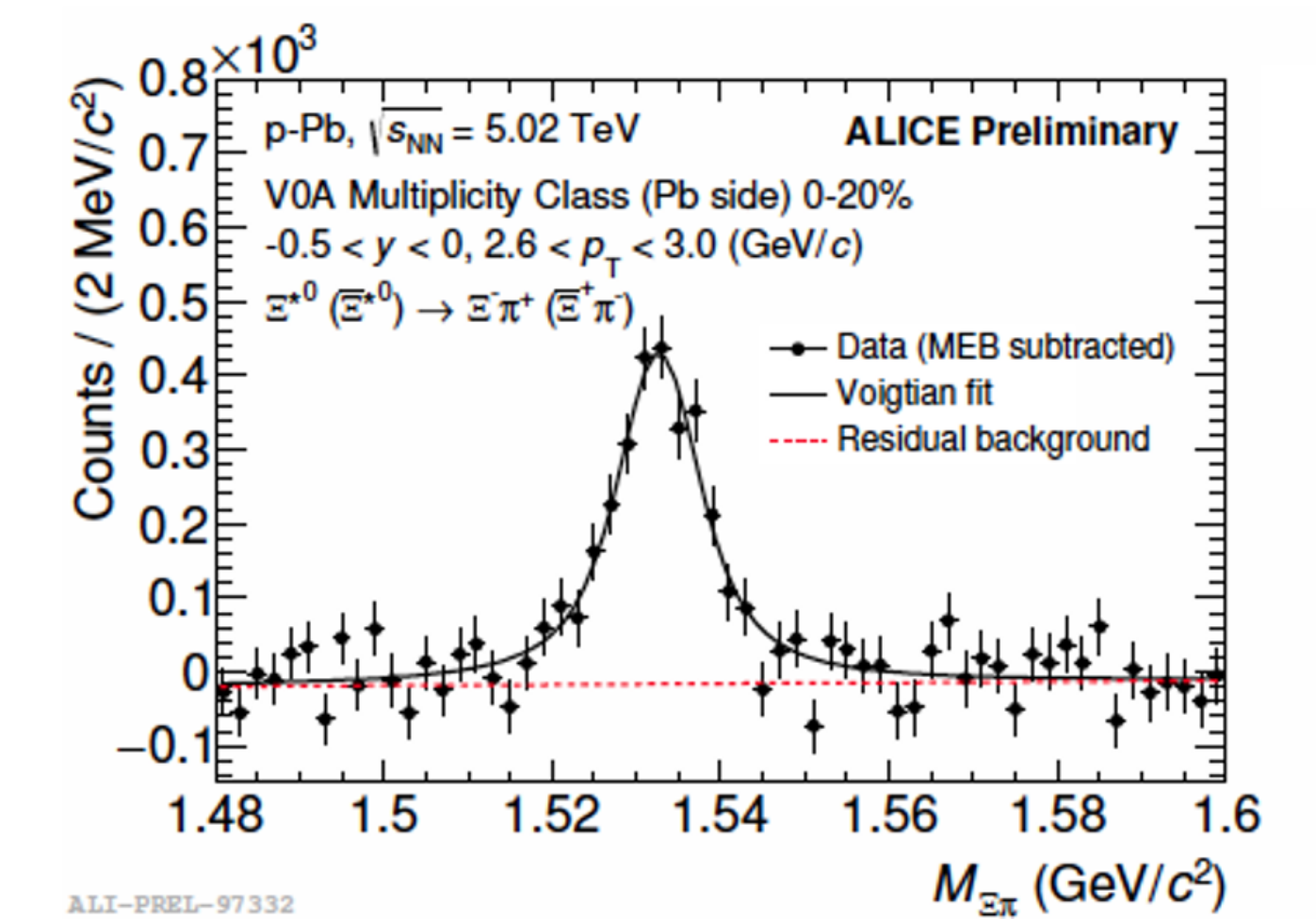}
\end{tabular}
\end{center}

\caption{\noindent Preliminary results for $\Sigma$(1385)and $\Xi$(1530)
	resonant states in 5.02~TeV pPb collisions in ALICE.}
	\label{form6D}
\end{figure}

\item \textbf{Input from Theoretical Calculations}

In addition to the experimental determination of potential enhancements
of strange ground and resonant states, the topic can also be addressed by
comparing lattice QCD calculations with model predictions based on a
non-interacting hadronic resonance gas (HRG). I will here only briefly
describe the basics, since the contribution by Prof.~Ratti to these
proceedings gives much more detail. The HRG model has shown to be very
effective in describing the hadronic phase of lattice QCD thermodynamics
very well up to the pseudo-critical temperature. The early model by
Hagedorn, which effectively describes interactions of hadrons by simply
increasing the number of possible resonant states exponentially as a
function of temperature, captures the behavior of the hadronic gas very
well as long as the temperature is low enough~\cite{Hagedorn:1976efr}. In
case of a flavor hierarchy in the transition region one would expect the
number of strange resonances to higher temperature, than the number of
light quark resonances. Such an approach has been proposed in the recent
past~\cite{Bazavov:2014xyar}, based on predictions which highlight the
difference between strange resonances in the PDG 2008 
listings~\cite{Amsler:2008zzbr} and the simplest non-relativistic Quark 
Model~\cite{Capstick:1986bmr}. Indeed adding states to the PDG listing 
seemed to initially improve the agreement between HRG and lattice QCD 
below the transition temperature. More detailed studies have followed, 
which are
presented in Prof.~Ratti's ntribution. An interesting conclusion that
arises out of these studies is that the improvement in the listing of
strange resonances between PDG-2008~\cite{Amsler:2008zzbr} and 
PDG-2014~\cite{Agashe:2014kdar} definitely brought the HRG calculations 
closer to
the lattice QCD data. By looking at details in the remaining discrepancy
it seems that the effect is more carried by singly strange resonances
rather than multi-strange resonances. This is good news for the experiments
since the $\Lambda$ and $\Sigma$ resonances below 2~GeV/c$^{2}$ are will
within reach of both the JLab and the RHIC/LHC experiments.

\item \textbf{Conclusion}

It is my opinion that comparisons between quark model predictions and 
lattice QCD, in addition to future measurements in the strange baryonic 
resonance sector, will shed light on a multitude of interesting questions:
\begin{itemize}
\item   Which quark model gives the best description of the experimental
	situation in a hadron gas near the QCD crossover? It seems the simple
	non-relativistic quark model considerably over-estimates the 
	number of possible states, but models based on di-quark 
	structures~\cite{Santopinto:2014opar} or enhanced quark interactions 
	in the baryon (hypercentral models~\cite{Giannini:2015ziar}) give a 
	better description of the thermal hadronic system calculated by 
	lattice QCD. 
\item Can one go as far as using model comparisons to  postulate evidence 
	or lack of evidence for di-quark structures? Can one 
	specify the necessary interactions between quarks in the baryons? 
\item Do we understand the hadronization mechanism of light-strange quark 
	baryons by focusing on potential yield enhancements of specific 
	resonant states? 
\item Can we determine exotic state probabilities in the strange 
	sector from measuring new strange resonances? 
\item Can we determine a flavor hierarchy in the QCD transition based on 
	specific strange resonance yields? 
\end{itemize}

There has been up to now relatively little interest in expanding the 
list of hyperon resonances in heavy ion erence in yield between pp and 
heavy ion reactions has been primarily used to determine the lifetime 
of the hadronic phase between chemical and kinetic freeze-out rather 
than determining potential enhancement patterns in specic channels. But 
the advent of the detailed lattice QCD studies, as well as the interest 
in an expanding YSTAR program at JLab in order to better understand the 
hadronization mechanism in the non-perturbative regime, denitely generates 
a bridge towards more detailed measurements of resonance production
out of the deconned phase at RHIC and LHC. Together with the YSTAR 
program these measurements might well solve the mystery of hadron 
formation and the interactions of the contributing quarks and gluons.
\end{enumerate}


\newpage
\subsection{Light Flavour Hadron Production at the LHC:
	Equilibrium Thermodynamics at Work}
\addtocontents{toc}{\hspace{2cm}{\sl B.~D\"onigus}\par}
\setcounter{figure}{0}
\setcounter{table}{0}
\setcounter{equation}{0}
\setcounter{footnote}{0}
\halign{#\hfil&\quad#\hfil\cr
\large{Benjamin D\"onigus}\cr
\textit{Institut f\"ur Kernphysik}\cr
\textit{Goethe-Universit\"at Frankfurt}\cr
\textit{D-60438 Frankfurt am Main, Germany}\cr}

\begin{abstract}
The measurement of the production of light flavoured hadrons in
high-energy collisions can be used to determine the chemical
freeze-out temperature by assuming a (statistical) thermal model.
In particular, we focus here on the production of (multi-)strange
hadrons, resonances, (hyper-)nuclei and the search for strange
exotica with ALICE at the LHC in view of the equilibrium thermal
model. Recent measurements of (multi-)strange hadrons and resonances 
performed in high-multiplicity proton-proton (pp) and proton-lead 
(p-Pb) collisions have shown features that are similar to those 
observed in lead-lead (Pb-Pb) collisions. The values of particle 
ratios of strange to non-strange hadrons in high-multiplicity pp 
and p--Pb collisions approach the ones in Pb--Pb. A similar trend 
is observed for light nuclei, namely the deuteron-to-proton ratio 
is rising in pp and p--Pb and reaches the thermal model expectation 
for Pb--Pb multiplicities. In addition, the thermal model can be 
used to predict particle yields of possible states and thus can be 
used to test the existence of these particles. This is shown 
exemplarily for a search for two exotic dibaryons whose upper 
limits on the production yields are significantly below the 
thermal model expectations whereas the known states are quite 
well described within an equilibrium thermal model approach.
\end{abstract}

\begin{enumerate}
\item \textbf{Introduction}

Ultra-relativistic heavy-ion collisions at the LHC offer a unique 
way to study QCD matter at very high temperatures. Lattice QCD
calculations~\cite{hotqcdb,bwgb} suggest that a transition from
confined hadrons to a state of deconfined matter, the so-called
quark-gluon plasma, is happening at temperatures above $T_c =
(154 \pm 9)$~MeV and/or energy densities above $\epsilon_c = 0.34
\pm 0.16~\mathrm{GeV/fm}^3$. These are clearly reached in these
collisions as one can deduce for instance from the measurement of
direct photon transverse momentum spectra. These lead to effective
temperatures of $T_{eff} = (297 \pm 12 (stat) \pm 41 (syst))$~MeV
averaged over the collision, extracted through the slope of the
spectra~\cite{dir_photonsb}. Comparisons to models lead to initial
temperatures of up to 740~MeV~\cite{dir_photons_modelb}.

The evolution of the collisions themselves are imagined usually as
a sequence of the following stages: two Lorentz contracted nuclei
approach each other, they collide, and after a short time (less
than 1~fm/$c$) a quark-gluon plasma is formed which eventually
expands, cools down and hadronises. Finally the hadrons rescatter 
and freeze out. The temperature where the hadrons stop being produced 
is called chemical freeze-out temperature $T_{ch}$ and the temperature 
when the hadrons stop scattering is denoted as kinetic freeze-out 
temperature $T_{fo}$.

If one uses an analogy between a light source and a particle source
one can extract also a temperature from the measured multiplicities
of the different particle species. This is possible using an approach
based on a grand canonical ensemble with the main ingredients: 
(chemical freeze-out) temperature $T_{ch}$, volume $V$ and 
baryo-chemical potential $\mu_B$. The latter is basically zero at the 
LHC, namely baryons and anti-baryons are produced with equal 
amounts~\cite{sqm2013b}.  This approach is called (statistical) thermal 
model and the measurement of the production yield of different particle 
species, such as $\pi$, K, p, etc. can be used to extract a temperature 
of about 156~MeV at the LHC~\cite{qm2014b}. If this is done for the 
different available energies at different laboratories one sees that 
there is an increase up to energies reached at the SPS at CERN and then 
the temperature stays constant.  This is another hint that the 
quark-gluon plasma is formed around a temperature of 159~MeV, which is 
the average of the extracted temperatures in the aforementioned plateau 
($T_{lim} $). This leads to the assumption $T_{lim} \approx T_{ch} 
\approx T_c$~\cite{gsi_modelb}.
\begin{figure}[h]
\begin{center}
 \includegraphics[width=0.9\textwidth]{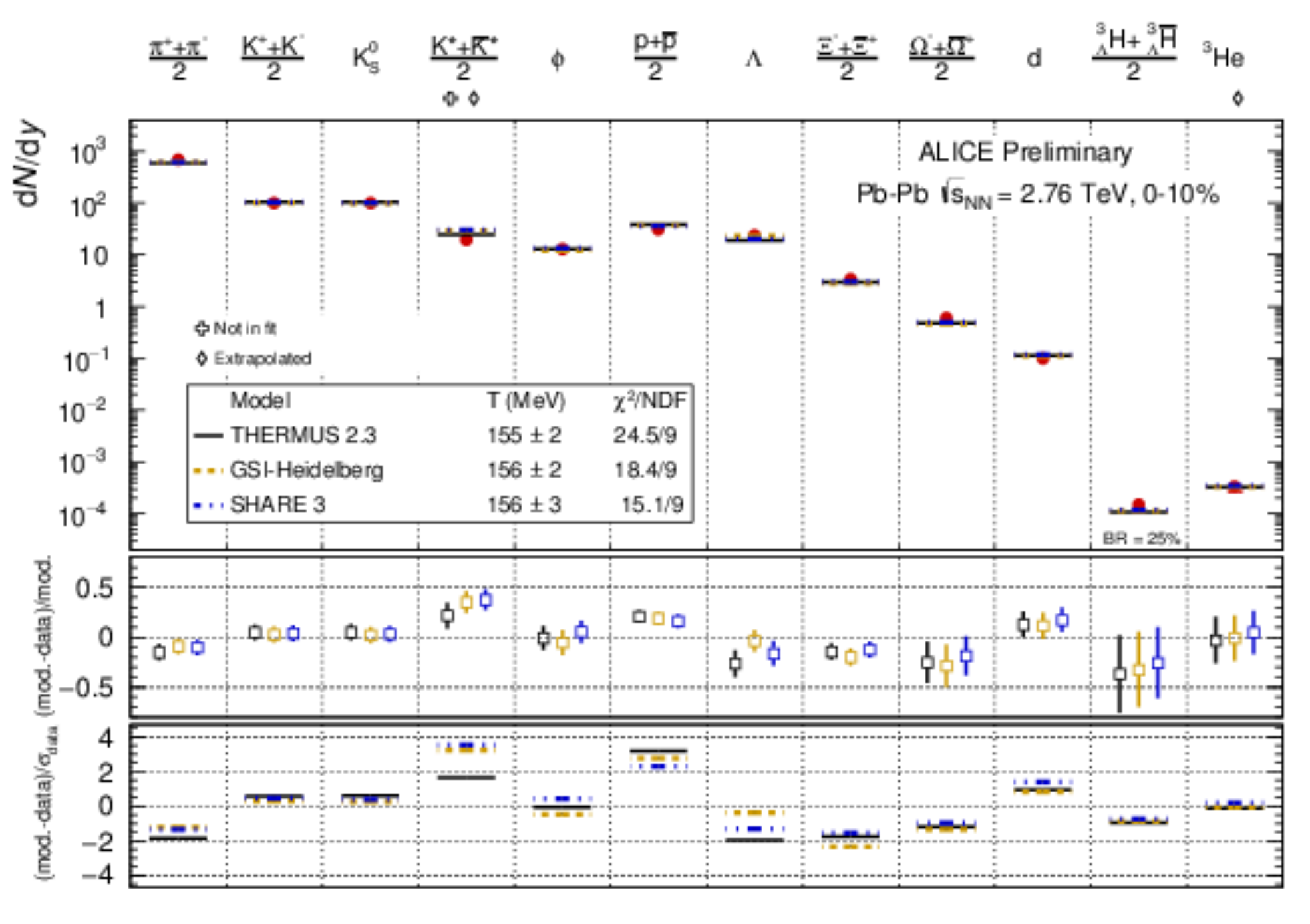}

\caption{\label{therm_fit} Comparison of three thermal model
	implementations, used to fit light flavoured hadron 
	production yields measured in Pb--Pb collisions at 
	$\sqrt{s_{\mathrm{NN}}}=2.76$~TeV with the ALICE detector. 
	The models THERMUS~\protect\cite{thermusb}, 
	GSI-Heidelberg~\protect\cite{gsi_modelb} and 
	SHARE~\protect\cite{shareb} agree well, as visible in the 
	lower two panels showing the difference between model and 
	data once as ratio to the model and once as ratio to the
	experimental uncertaint. All model fits lead to a 
	temperature of about $T_{ch}=156$~MeV.}
\end{center}
\end{figure}

An example thermal model fit using production yield measurements of
the ALICE Collaboration is shown in Fig.~\ref{therm_fit}. The fit
compares three different implementations of thermal models
(THERMUS~\cite{thermusb}, GSI-Heidelberg~\cite{gsi_modelb} and
SHARE~\cite{shareb}). The three different thermal model 
implementations agree well and result in a temperature of about 
$T_{ch}=156$~MeV.  The thermal model can also be used to predict 
the production of particles as for instance described in~\cite{horstb}.

\item \textbf{Strangeness}

Recently, pp and p--Pb collisions have received large attention from
the heavy-ion community. Normally, these are only seen as baseline
measurements compared to the quark-gluon plasma formed in Pb--Pb
collisions, for instance quantified in the nuclear modification
factor $R_\textrm{AA}$. Nowadays, studies have been carried out
to compare these small systems more directly to heavy-ion collisions.
To achieve this, measurements of the production yields of different
particle species have been performed as a function of the charged
particle multiplicity d$N_\textrm{ch}$/d$\eta$.
\begin{figure}[h]
\begin{center}
 \includegraphics[width=0.6\textwidth]{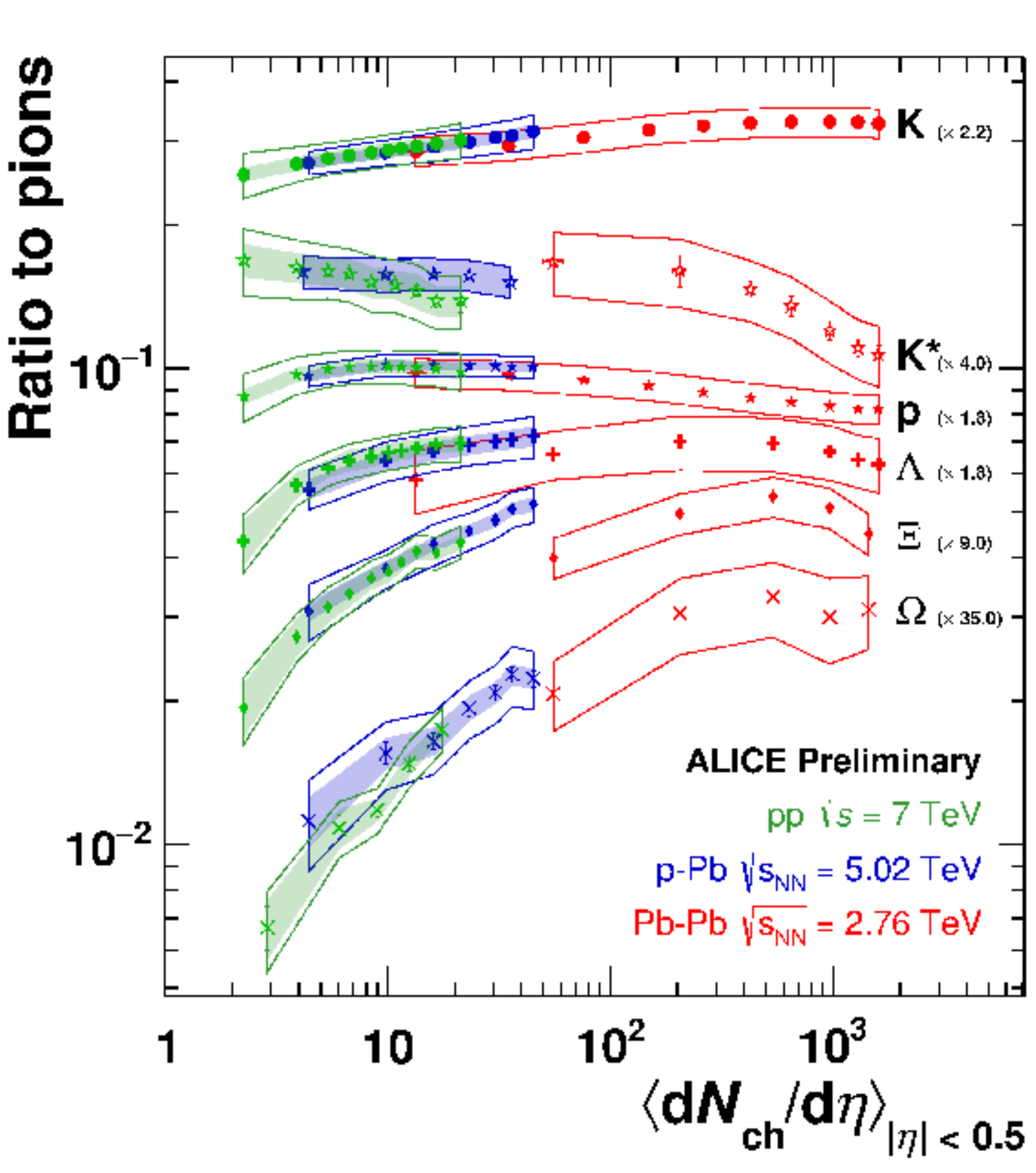}

\caption{\label{ratios} Ratios of the production yield to the yield
	of pions for K$^{-}$, K(892)$^\ast$, p, $\Lambda$, $\Xi^{-}$ 
	and $\Omega^{-}$ as a function of the mean charged particle 
	multiplicity $\langle\textrm{d}N_\textrm{ch}/\textrm{d}
	\eta\rangle$ measured at mid-rapidity ($|\eta|<0.5$).}
\end{center}
\end{figure}

The ratios of the production yield of K$^{-}$, K(892)$^\ast$, p,
$\Lambda$, $\Xi^{-}$ and $\Omega^{-}$ compared to the yield of pions
as a function of the mean charged particle multiplicity
$\langle\textrm{d}N_\textrm{ch}/\textrm{d}\eta\rangle$ measured at
mid-rapidity ($|\eta|<0.5$) are displayed in Fig.~\ref{ratios}. One
can see that the K/$\pi$ and the p/$\pi$ ratios are basically flat
for all collision systems, wheras the K(892)$^{*}$/$\pi$ ratio is
nearly constant in pp and p--Pb collisions but starts to drop for
increasing multiplicities in Pb--Pb collisions. The hyperon-to-pion
ratios, namely $\Lambda$/$\pi$, $\Xi$/$\pi$ and $\Omega$/$\pi$ show
a clear rise already at low multiplicities and seem to reach a
saturation for Pb-Pb multiplicities, where the thermal model
expectation value is reached. More details on the measurement
can be found in~\cite{aliceb,alice1b}.

Before its observation, many signatures of the quark-gluon plasma
have been proposed. One of the earliest ones was the possible
enhancement of strangeness in ultra-relativistic heavy-ion collisions,
relative to the production in elementary collisions~\cite{rafelskib,
mullerb}. This was observed at the SPS~\cite{antinorib} but is nowadays 
often interpreted in an opposite direction, as a lifting of the 
suppression when moving from elementary collisions such as pp to 
Pb--Pb for instance~\cite{blumeb,hamiehb}. This can be understood 
also as the necessity to treat strangeness canonically and not as 
introduced before in a grand canonical ensemble~\cite{redlichb,tounsib}.

\item \textbf{Resonances}

Resonances are of particular interest in the study of heavy-ion
collisions, mainly because their lifetimes are typically shorter
than the lifetime of the fireball, i.e., 10~fm/$c$. Therefore,
resonances can decay inside the fireball and can either be lost
or regenerated. As such, resonances can be used as probes for the
lifetime of the hadronic phase, because of their different lifetimes.
The suppression of K(892)$^\ast$ only visible in Pb--Pb collisions
relative to pions as shown in Fig.~\ref{ratios} is clearly supporting
this fact. The lifetime of the K(892)$^\ast$ is 4.2~fm/$c$ and it is
clearly stronger suppressed with increasing multiplicity of the
collision. In contrary, the $\phi(1020)$ has a lifetime of 45~fm/$c$
and seems to be unsuppressed~\cite{phi_kstarb,phi_kstar1b}. The
suppression effect can be semi-quantitatively described by
EPOS~\cite{epos0b,eposb} with an UrQMD~\cite{urqmdb,urqmd1b}
afterburner for the hadronic phase~\cite{andersb}.
\begin{figure}[hb]
\begin{center}
 \includegraphics[width=0.7\textwidth]{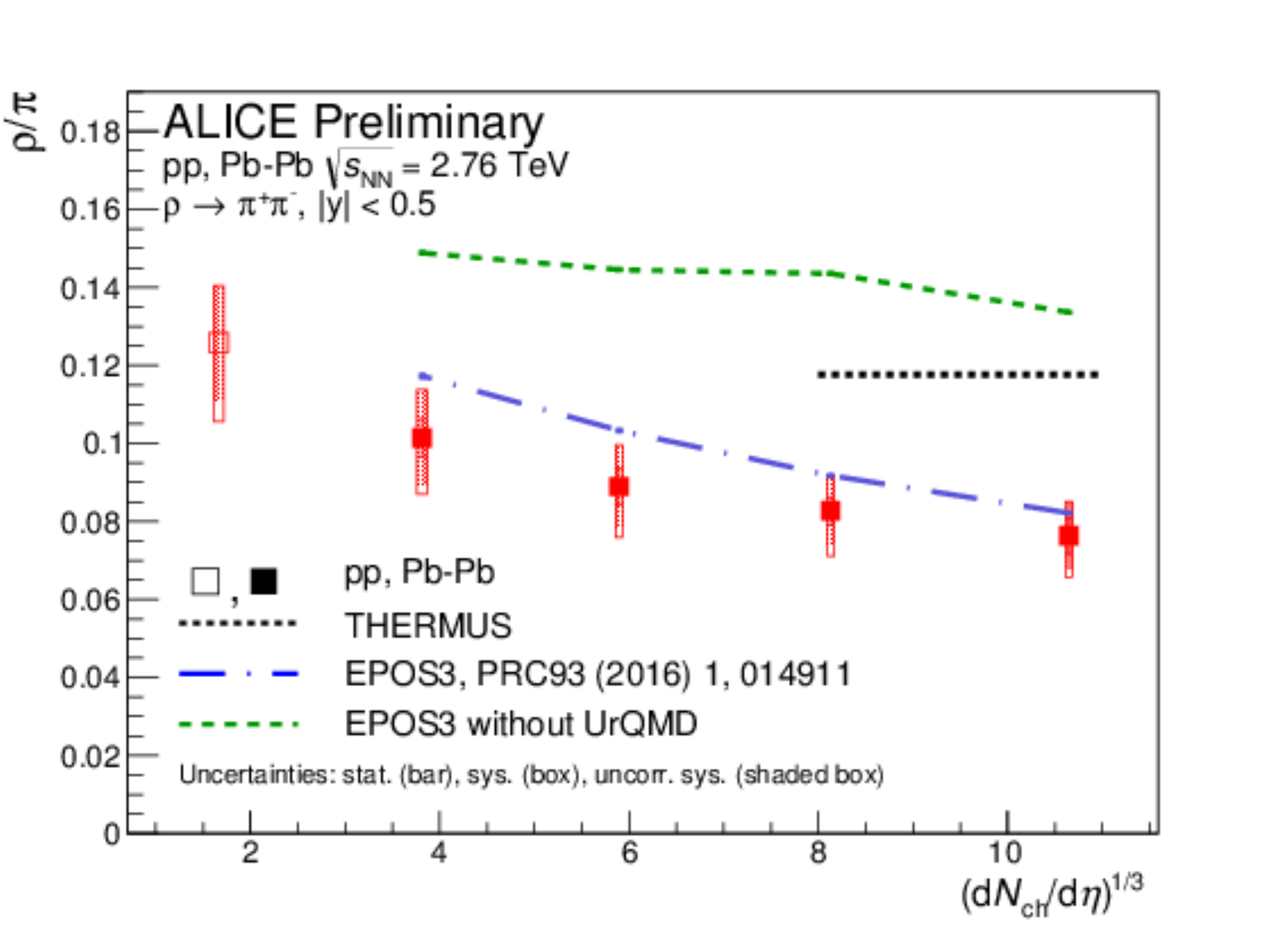}

\caption{\label{rho} $\rho$/$\pi$ ratio as a function of the scaled 
	charged particle multiplicity ($\langle\textrm{d}N_\textrm{ch}/
	\textrm{d}\eta\rangle^{1/3}$), measured in $|y|<0.5$ in pp and 
	Pb--Pb collisions. In addition, a thermal model prediction and 
	two EPOS predictions are depicted. From Ref.~\protect\cite{anders1b}.}
\end{center}
\end{figure}

Recently, the ALICE Collaboration has released new data including
also the study of the $\rho(770)$~\cite{anders1b}, with a lifetime
of only 1.3~fm/$c$. Also this shows a clear suppression with
increasing multiplicity as visible from Fig.~\ref{rho}. The figure
shows in addition the comparison to a thermal model expectation and
the trend expected from EPOS for two cases, once with and once without
a UrQMD afterburner to simulate the hadronic phase. The trend of the
data matches the case of EPOS with the afterburner.

Clearly, more data and especially the study of more resonances is
needed to get a complete picture here.

\item \textbf{(Hyper-)Nuclei}

Another set of interesting results comes from the measurement of
light (anti-)(hyper-)nuclei, which seem to contradict the observations
discussed before. When one looks at the deuteron-to-proton ratio as a
function of the charged particle multiplicity as visualised in
Fig.~\ref{dtop} a similar trend is observed as for the hyperon-to-pion
ratios discussed before, namely an increase in the small systems with
increasing multiplicity which stops when the thermal model expectation
is reached for the Pb--Pb values~\cite{nucleib}.
\begin{figure}[h]
\begin{center}
 \includegraphics[width=0.7\textwidth]{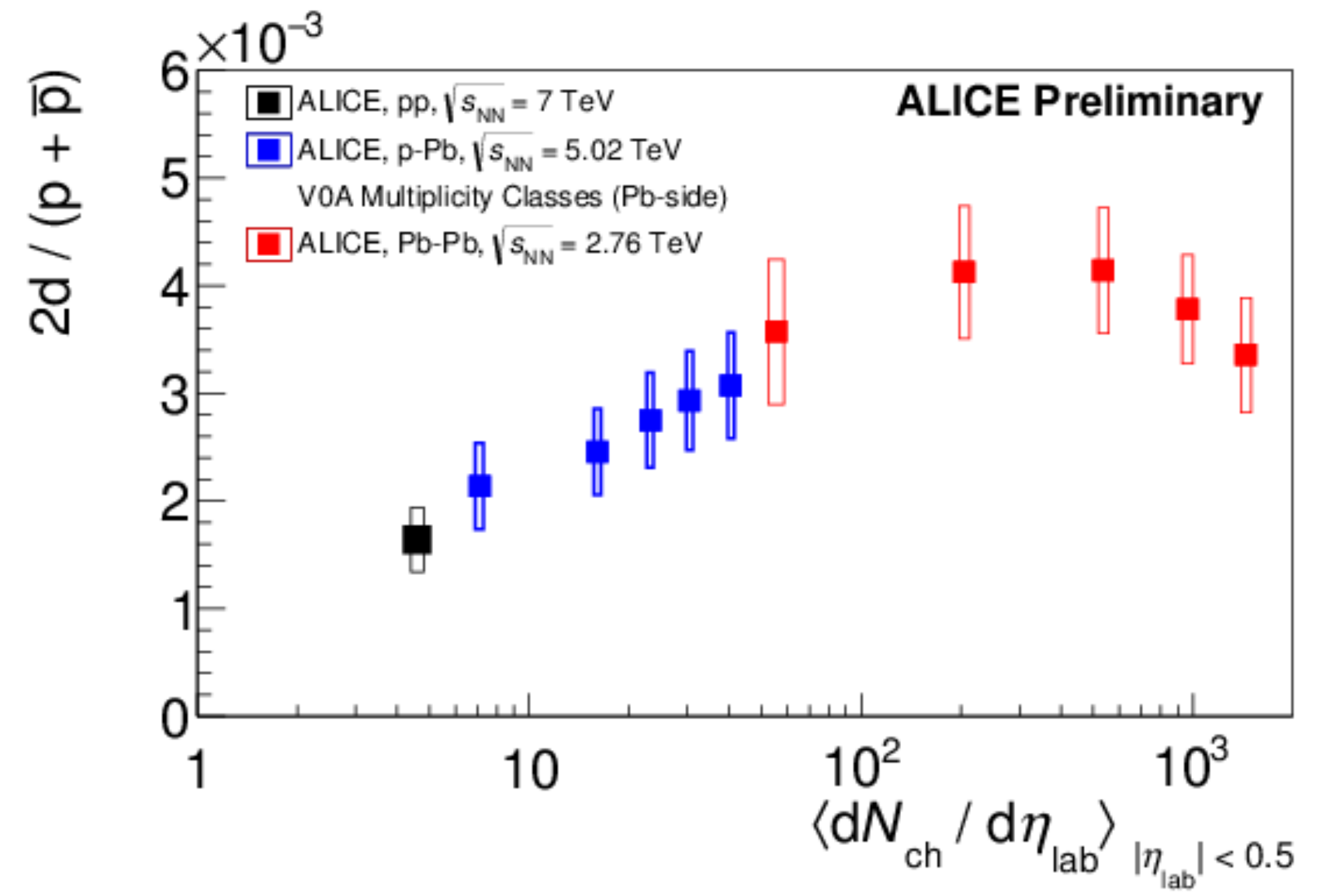}

\caption{\label{dtop} Deuteron-over-proton ratio as a function of the
	charged particle multiplicity for pp, p--Pb and Pb--Pb. Taken
	from Ref.~\protect\cite{natashab}.}
\end{center}
\end{figure}

The aforementioned contradiction comes from the comparison with the
resonance results discussed before, where the existence of a hadronic
phase is observed from the suppressed production yields of short lived
resonances due to rescattering. A similar rescattering should also occur
for light nuclei, especially because the nuclei are expected to be
produced at the temperature of 156~MeV (as all other particles) which
is 2-3 orders of magnitude above the binding energies of the nucleons
in the (hyper-)nuclei. The most extreme example is the hypertriton,
which has a separation energy of the $\Lambda$ of only 130~keV but
its production is very well described by the equilibrium thermal
model~\cite{hypertritonb}.

\item \textbf{Exotica}

Since also the production of light (anti-)(hyper-)nuclei is well
described by the thermal model it is expected that it can be used
to make predictions for unobserved exotic states, as for instance
strange baryons as $\Lambda\Lambda$ and $\Lambda$n bound states. A
search was performed and is reported in~\cite{exoticab}. The estimated
upper limits of these states are well below the expectation of the
thermal model, clearly depending on the lifetime and the branching
ratio of these states. If reasonable values are picked for the
lifetimes and the branching ratios of these states the upper limits
are about 25 times below the thermal model expectation as shown in
Fig.~\ref{comp}.
\begin{figure}[h]
\begin{center}
 \includegraphics[width=0.8\textwidth]{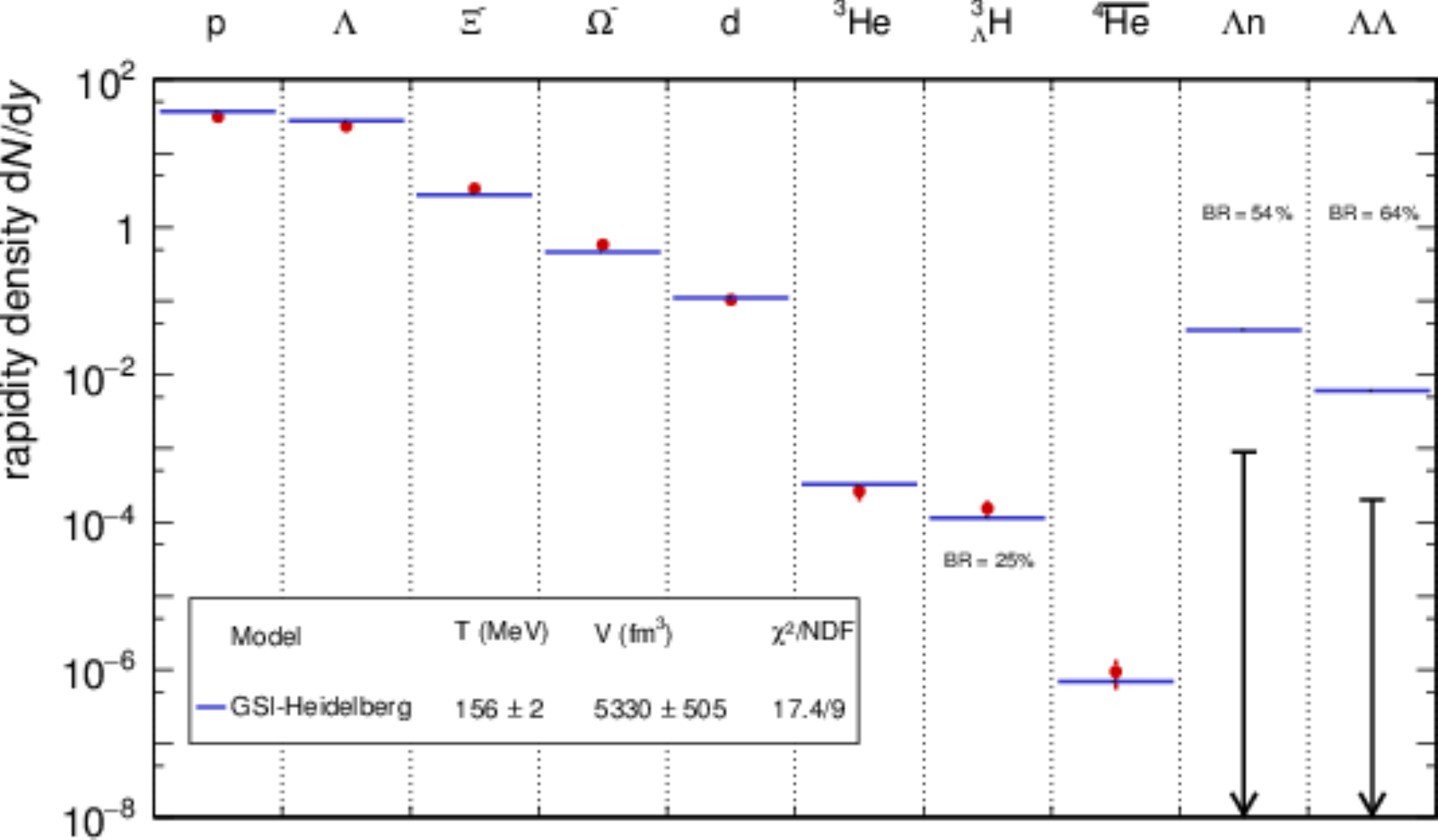}

\caption{\label{comp} Comparison of data to thermal-model
	predictions. The production yields measured with ALICE 
	are shown as red dots and the results of the calculations 
	are shown as blue lines. The upper limits at 99\% CL 
	determined by ALICE are indicated as black arrows. 
	Figure taken from Ref.~\protect\cite{courierb}.}
\end{center}
\end{figure}

\item \textbf{Conclusion}

Interesting physics has been observed in the strangeness sector
in all collision systems at the LHC. The trend seen for the
hyperon-to-pion ratio matching pp, p--Pb and Pb--Pb collisions
can be understood as a lifting of the strangeness suppression
in elementary collisions.   Short lived resonances show a
clear suppression with increasing multiplicity, which is due
to the rescattering in the hadronic phase. Here new resonances
could clearly help to map the lifetime of the hadronic phase
further down. A copious production of loosely bound objects
is measured by ALICE as predicted by the thermal model. This
is creating a riddle, because the clearly observed hadronic
phase and the rescattering for resonances is contradicting
the measured yields of light (hyper-)nuclei which should also
be destroyed and thus reduced in the hadronic phase. The
estimated upper limits for the searched exotica are 25 times
below the thermal model expectation which makes their existence
questionable. More data and more studies are needed in order to
establish a scenario which seamlessly describes all observations.
Therefore, the observation of new hyperon resonances can clearly
help to complete the picture.

\item \textbf{Acknowledgments}

The author gratefully acknowledges the support from the German 
Bundesministerium fur Bildung, Wissenschaft, Forschung und 
Technologie (BMBF) though the project grant 05P2015 - ALICE at 
High Rate (BMBF-FSP202).
\end{enumerate}


\newpage
\subsection{Constraining the Hadronic Spectrum from Lattice QCD 
	Thermodynamics}
\addtocontents{toc}{\hspace{2cm}{\sl P.~Alba, R.~Bellwied, S.~Borsanyi,
	Z.~Fodor, J.~G\"unther, S.~Katz, V.~Mantovani~Sarti,}\par}
\addtocontents{toc}{\hspace{2cm}{\sl
	J.~Noronha-Hostler, P.~Parotto, A.~Pasztor, I.~Portillo~Vazquez, 
	and C.~Ratti}\par}
\setcounter{figure}{0}
\setcounter{table}{0}
\setcounter{equation}{0}
\setcounter{footnote}{0}
\halign{#\hfil&\quad#\hfil\cr
\large{Paolo Alba}\cr
\textit{Frankfurt Institute for Advanced Studies}\cr
\textit{Goethe Universit\"at Frankfurt}\cr
\textit{D-60438 Frankfurt am Main, Germany}\cr\cr
\large{Rene Bellwied}\cr
\textit{Physics Department}\cr
\textit{University of Houston}\cr
\textit{617 SR1 Building}\cr
\textit{Houston, TX 77204, U.S.A.}\cr\cr
\large{Szabolcs Borsanyi}\cr
\textit{Department of Physics}\cr
\textit{Wuppertal University}\cr
\textit{D-42119 Wuppertal, Germany}\cr\cr
\large{Zoltan Fodor}\cr
\textit{Department of Physics}\cr
\textit{Wuppertal University}\cr
\textit{D-42119 Wuppertal, Germany \&}\cr
\textit{Inst. for Theoretical Physics}\cr
\textit{E\"otv\"os University}\cr
\textit{H-1117 Budapest, Hungary \&}\cr
\textit{J\"ulich Supercomputing Centre}\cr
\textit{Forschungszentrum J\"ulich}\cr
\textit{D-52425 J\"ulich, Germany}\cr\cr
\large{Jana G\"unther}\cr
\textit{Department of Physics}\cr
\textit{Wuppertal University}\cr
\textit{D-42119 Wuppertal, Germany}\cr\cr
\large{Sandor Katz}\cr
\textit{Inst. for Theoretical Physics}\cr
\textit{E\"otv\"os University}\cr
\textit{H-1117 Budapest, Hungary \&}\cr
\textit{TA-ELTE "Lend\"ulet" Lattice Gauge Theory Research Group}\cr
\textit{H-1117 Budapest, Hungary}\cr\cr
\large{Valentina Mantovani Sarti}\cr
\textit{Department of Physics}\cr
\textit{Torino University and INFN}\cr
\textit{10125 Torino, Italy}\cr\cr
\large{Jacquelyn Noronha-Hostler}\cr
\textit{Department of Physics}\cr
\textit{University of Houston}\cr
\textit{617 SR1 Building}\cr
\textit{Houston TX 77204, U.S.A.}\cr\cr
\large{Paolo Parotto}\cr
\textit{Physics Department}\cr
\textit{University of Houston}\cr
\textit{617 SR1 Building}\cr
\textit{Houston, TX 77204, U.S.A.}\cr\cr
\large{Attila Pasztor}\cr
\textit{Department of Physics}\cr
\textit{Wuppertal University}\cr
\textit{D-42119 Wuppertal, Germany}\cr\cr
\large{Israel Portillo Vazquez}\cr
\textit{Physics Department}\cr
\textit{University of Houston}\cr
\textit{617 SR1 Building}\cr
\textit{Houston, TX 77204, U.S.A.}\cr\cr
\large{Claudia Ratti}\cr
\textit{Department of Physics}\cr
\textit{University of Houston}\cr
\textit{617 SR1 Building}\cr
\textit{Houston TX 77204, U.S.A.}\cr}

\begin{abstract}
We study the effect of not-yet-detected or poorly known hadronic states 
on several thermodynamic observables, which we calculate in the Hadron 
Resonance Gas (HRG) model and compare to lattice QCD results. Our 
analysis shows that the inclusion of such states is crucial to reproduce 
the lattice data.
\end{abstract}

\begin{enumerate}
\item \textbf{Introduction}

The precision reached by lattice QCD simulations in recent years allows 
unprecedented quantitative calculations of thermodynamic quantities. In 
particular, the transition temperature of QCD~\cite{Aoki:2006brc,
Aoki:2009scc,Borsanyi:2010bpc,Bazavov:2011nkc}, the QCD equation of 
state at zero~\cite{Borsanyi:2010cjc,Borsanyi:2013biac,Bazavov:2014pvzc,
Borsanyi:2016kswc} and small chemical potential~\cite{Borsanyi:2012crc,
Gunther:2016vcpc} and fluctuations of quark flavors and/or conserved 
charges near the QCD transition~\cite{Borsanyi:2011swc,Bazavov:2012jqc,
Bellwied:2015lbac} are now known with high accuracy. The latter are 
particularly useful because they carry information on the chemical 
composition of the system and can be used to extract information on 
the degrees of freedom which populate the system in the vicinity of 
the phase transition~\cite{Koch:2005vgc,Bazavov:2013dtac,Bellwied:2013ctac}. 
Besides, they can be related to the experimental measurements of moments 
of net-charge distributions in order to extract the chemical freeze-out 
temperature and chemical potential of a heavy ion 
collision~\cite{Karsch:2012wmc,Bazavov:2012vgc,Borsanyi:2013hzac,
Borsanyi:2014ewac,Mukherjee:2015mxcc,Noronha-Hostler:2016rpdc}.

The Hadron Resonance Gas (HRG) model~\cite{Dashen:1969epc,Venugopalan:1992hyc,
Karsch:2003vdc,Karsch:2003zqc,Tawfik:2004swc} yields a good description of 
most thermodynamic quantities in the hadronic phase. It is based on the idea 
that strongly interacting matter in the ground state can be described in 
terms of a non-interacting gas of hadrons and resonances. The only input of 
the model is the hadronic spectrum: usually it includes all well-known 
hadrons in the Particle Data Book, namely the ones rated with at least two 
stars. Recently, it has been noticed that some more differential observables 
present a discrepancy between lattice and HRG model results. The inclusion of 
not-yet-detected states, such as the ones predicted by the Quark Model 
(QM)~\cite{Capstick:1986bmc,Ebert:2009ubc} has been proposed in order to 
improve the agreement~\cite{Majumder:2010ikc,Bazavov:2014xyac}.

\item \textbf{Approach and Results}

The HRG model provides a satisfactory description of several thermodynamic 
quantities below the phase transition temperature. However, it was recently 
noticed that discrepancies are starting to appear when more differential 
observables are being calculated. These observables are often expressed in 
terms of fluctuations of conserved charges, defined as
\begin{equation}
	\chi_{lmn} ^{BQS} =\left( \dfrac{\partial ^{l+m+n}P(T,\mu_B,\mu_Q,
	\mu_S)/T^4}{\partial
	(\mu_B/T)^l \partial (\mu_Q/T)^m \partial (\mu_S/T)^n}
	\right)_{\mu_i=0}.
\end{equation}\label{eq2}
Such quantities can be evaluated on the lattice, but also in the HRG model 
approach. The starting point for the model is the total pressure, written as 
a sum over all known baryons and mesons of the pressure of a non interacting 
bosonic and fermionic gas:
\begin{eqnarray}
	&&P_{tot}(T,{\mu_k}) = \sum _k P_k(T,\mu_k)
	= \sum _k (-1)^{B_k+1} \dfrac{d_k T}{(2\pi) ^3}\int d^3\vec{p} \nonumber\\   
	&&\ln  \left(1+ (-1)^{B_k+1} \exp \left[ -\dfrac{(\sqrt{\vec{p} 
	^2+m_k^2}-\mu_k)}{T} \right] \right).
\end{eqnarray}\label{eq1}
Here the single particle chemical potential is defined with respect to the 
global conserved charges (baryonic $B$, electric $Q$ and strangeness $S$) 
as $\mu_k=B_k\mu_B+Q_k\mu_Q+S_k\mu_S$. More details on the HRG model used 
here can be found in Ref.~\cite{Alba:2014ebac}.

In principle, the chemical potentials for baryon number, electric charge and
strangeness can be chosen independently. However, to match the experimental
situation in a heavy-ion collision, a common choice is to fix them by imposing
the following conditions:
\bea
	\langle n_S\rangle =0~~~~~~~~~~~~\langle n_Q\rangle=\tfrac{Z}{A} 
	\langle n_B\rangle\simeq\langle0.4n_B\rangle.
\eea
In lattice calculations to leading order in $\mu_B$, the ratio $\mu_S/\mu_B$
which satisfies the above conditions reads~\cite{Borsanyi:2013hzac, 
Bazavov:2012vgc}:
\begin{equation}
	\left( \dfrac{\mu_S}{\mu_B}\right) _{LO}=-\dfrac{\chi_{11} ^{BS}}
	{\chi_2 ^{S}}-\dfrac{\chi_{11} ^{QS}}{\chi_2^{S}}\dfrac{\mu_Q}{\mu_B};
	\end{equation}\label{eq3}
\begin{figure}[h!]
\centering
\includegraphics[width=0.45\textwidth]{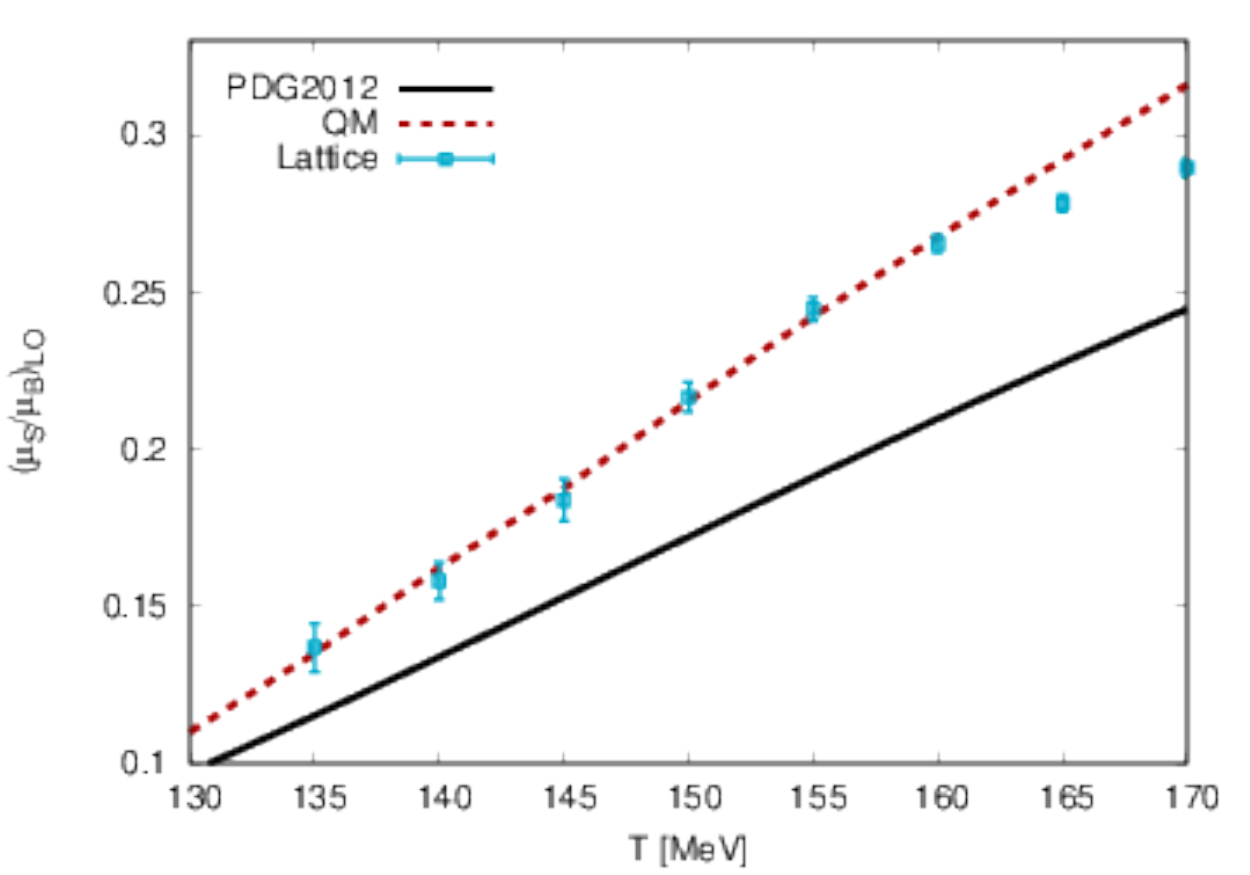}

	\caption[]{\label{fig2}Ratio $\mu_S/\mu_B$ at leading order as 
	a function of the temperature. The two curves are the HRG model 
	result obtained with the well-known states from the PDG2012 
	(full black line) and Quark Model (dotted red line).}
\end{figure}

The lattice results for this observable are shown in Fig.~\ref{fig2}, in 
comparison to the HRG model results based on the well-established states 
from the PDG2012 (full, black line) and the Quark Model (dotted, red line). 
The improvement due to the inclusion of the QM states is evident.
However, for other observables such as $\chi_4^S/\chi_2^S$ and 
$\chi_{11}^{us}$, the agreement between HRG model and lattice gets worse 
when the QM states are included (see the two panels of Fig.~\ref{fig4}).
\begin{figure}[h!]
\centering
\includegraphics[width=0.45\textwidth]{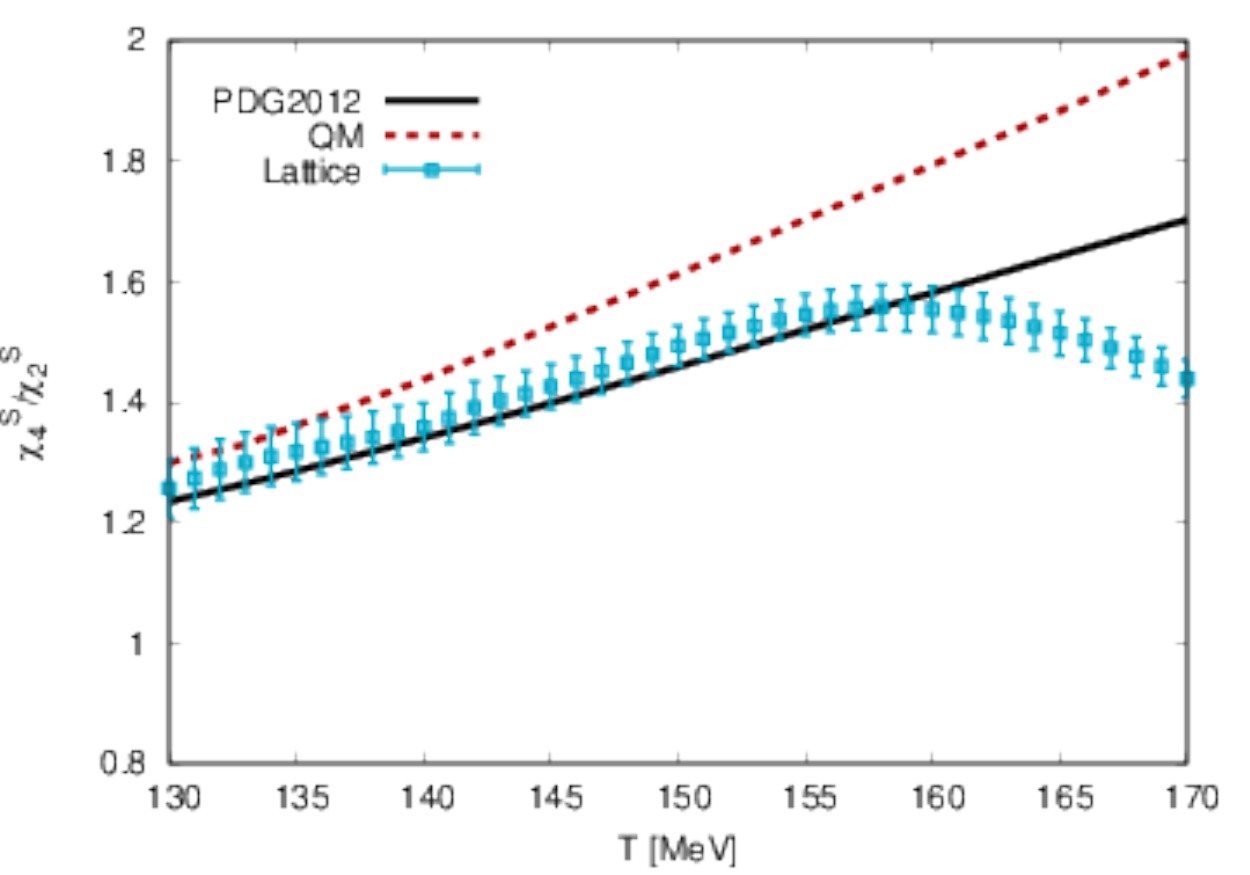}\\
\includegraphics[width=0.46\textwidth]{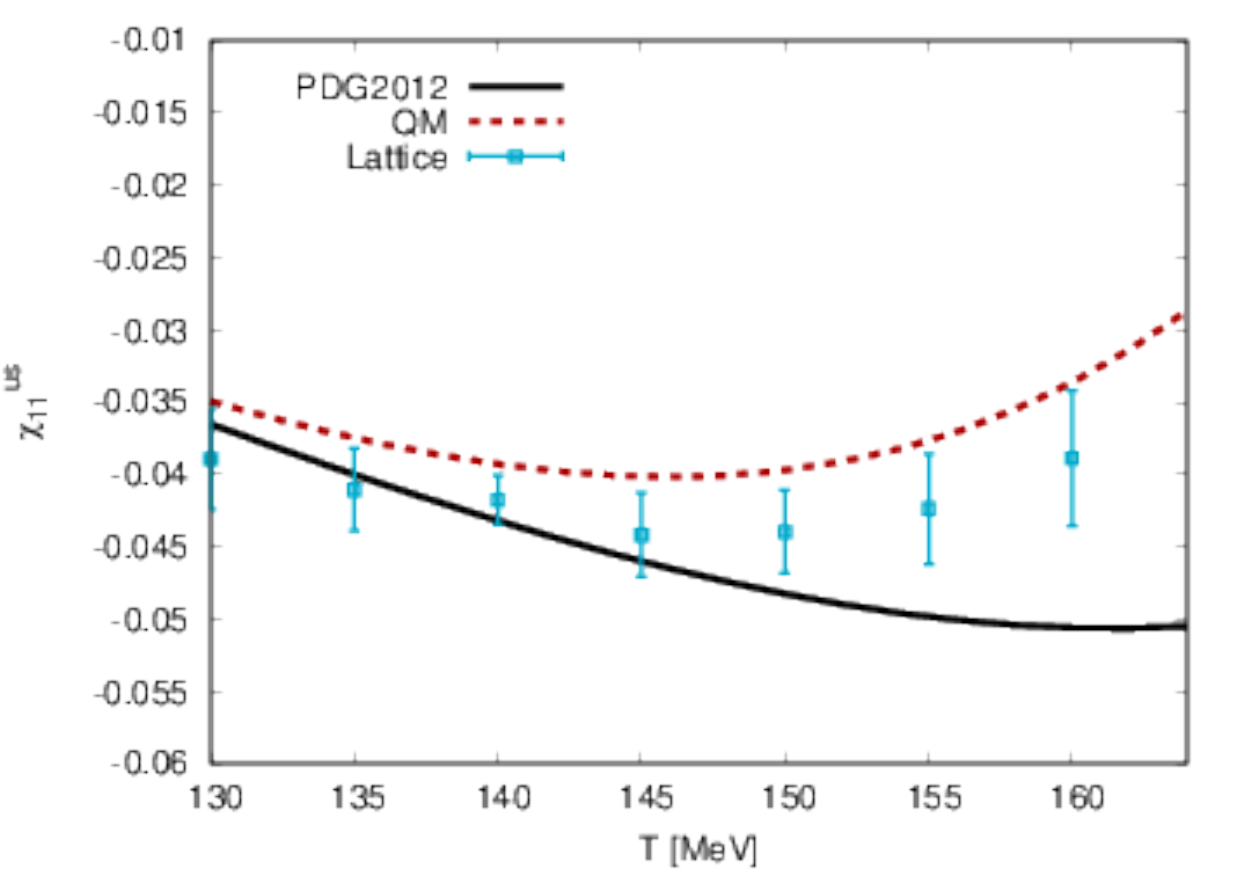}

\caption[]{\label{fig4}Upper panel: $\chi_4 ^{S}/\chi_2 ^{S}$ 
	as a function of the temperature. The lattice results 
	from Ref.~\protect\cite{Bellwied:2013ctac} are compared 
	to the HRG model calculations based on the PDG$2012$ 
	(black solid line) and the QM (red dashed line). Lower 
	panel: comparison of up-strange correlator $\chi_{11}^{us}$ 
	simulated on the lattice~\protect\cite{Bellwied:2015lbac} 
	and calculated in the HRG model using the PDG$2012$ (solid 
	black line) and the QM (dashed red line) spectra.}
\end{figure}

By observing these plots we can already try to understand what the issue 
could be. The ratio $\chi_4^S/\chi_2^S$ is proportional to the average 
strangeness squared in the system. The fact that the QM overestimates 
the data means that it either predicts too many multi-strange states or 
not enough $S=1$ states. Analogously, $\chi_{11}^{us}$ measures the 
correlation between $u$ and $s$ quarks: it is positive for baryons and 
negative for mesons. The fact that the QM overestimates the data means 
that it either predicts too many strange baryons or not enough strange 
mesons.

In order to solve this ambiguity, we decided to look at each particle 
family separately, dividing them according to their baryonic and 
strangeness content. In order to do so, we defined the partial pressures 
for each family in the hadronic phase, according to the following 
equations~\cite{Bazavov:2013dtac}:
\begin{eqnarray}
	P(\hmu_B,\hmu_S) &=& P_{00}+P_{10}\cosh(\hmu_B)+P_{0|1|} \cosh(\hmu_S)
	\nonumber \\
	&+& P_{1|1|} \cosh(\hmu_B-\hmu_S)
	\nonumber \\
	&+& P_{1|2|} \cosh(\hmu_B-2\hmu_S)
	\nonumber \\
	&+& P_{1|3|} \cosh(\hmu_B-3\hmu_S) \;,
\end{eqnarray}
where $\hmu_i=\mu_i/T$, and the indexes are $P_{B|S|}$.
\begin{figure*}[ht]
\centering
\includegraphics[width=0.32\textwidth]{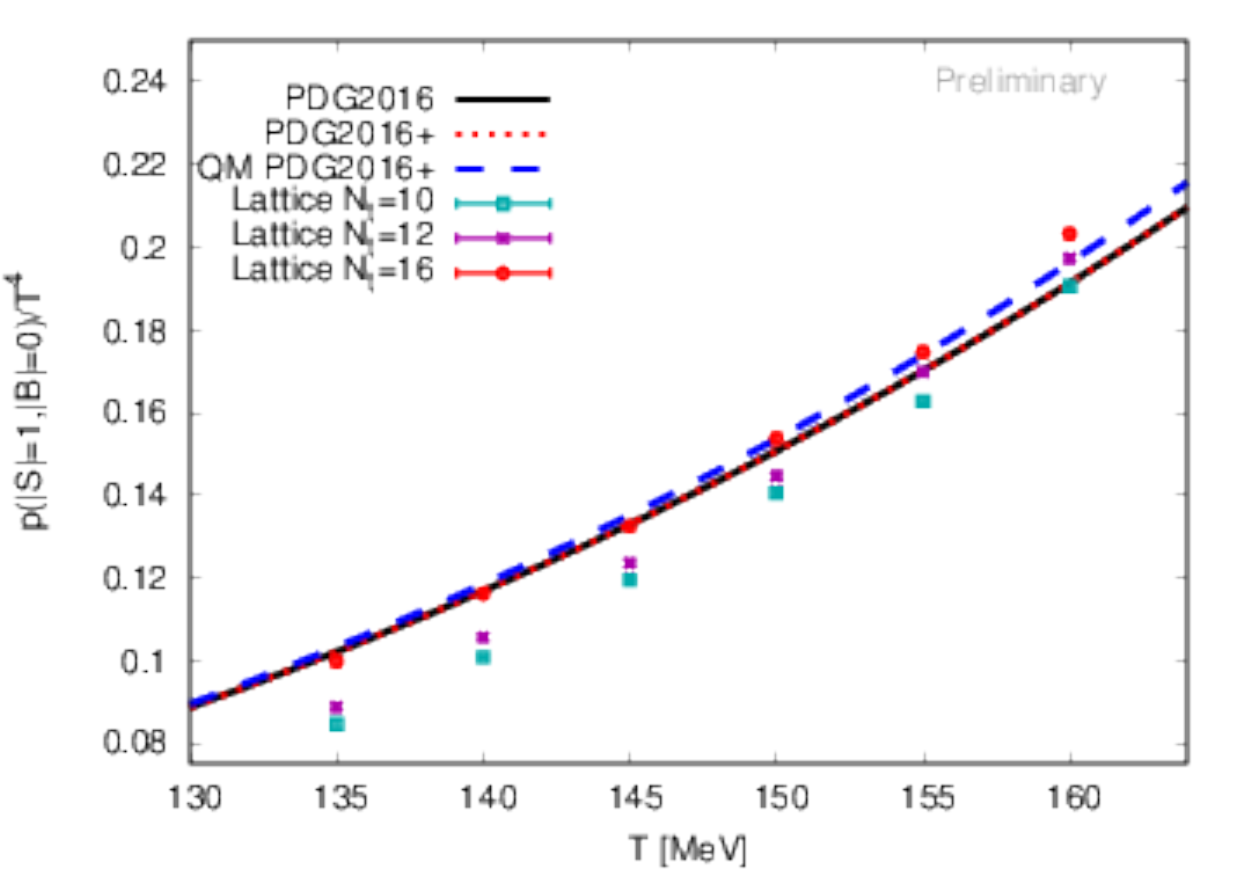}
\includegraphics[width=0.32\textwidth]{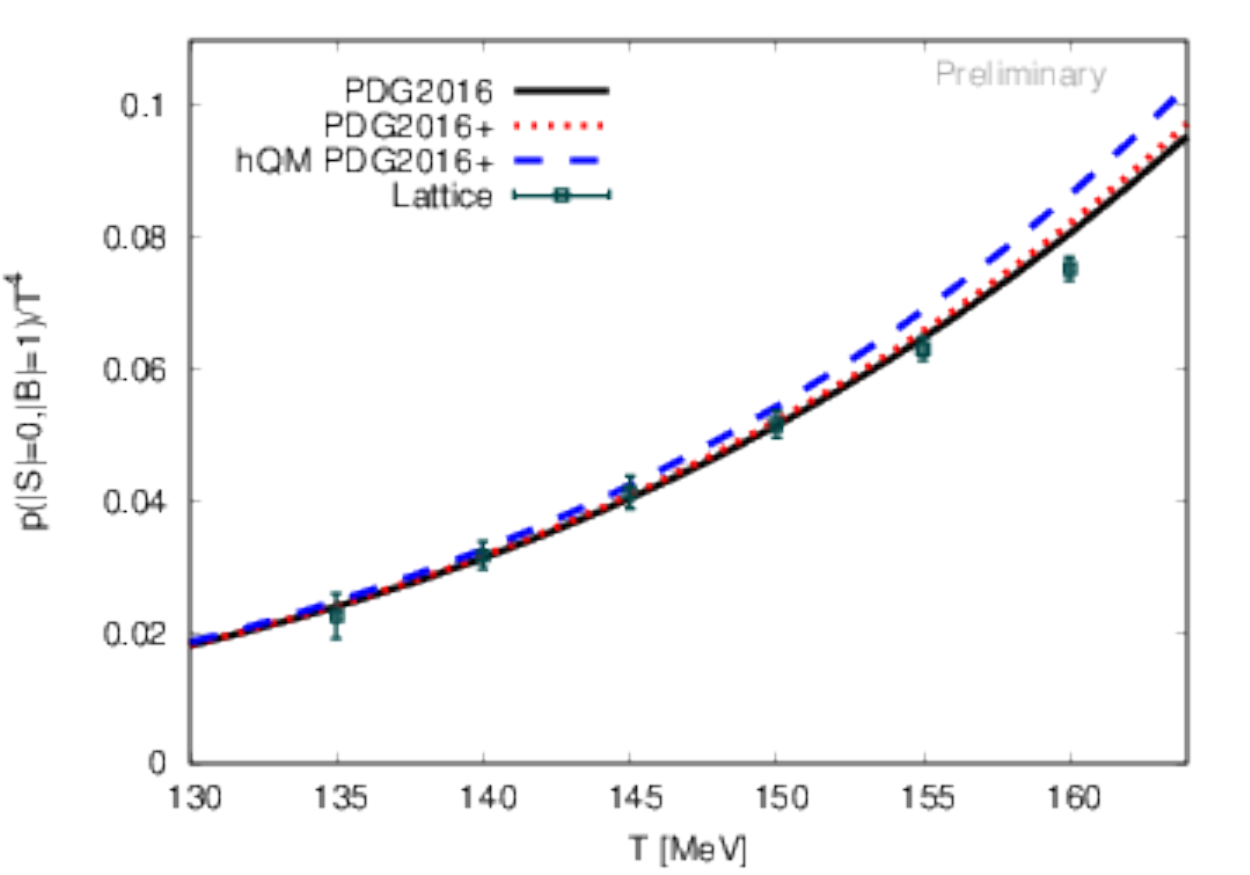}
\includegraphics[width=0.32\textwidth]{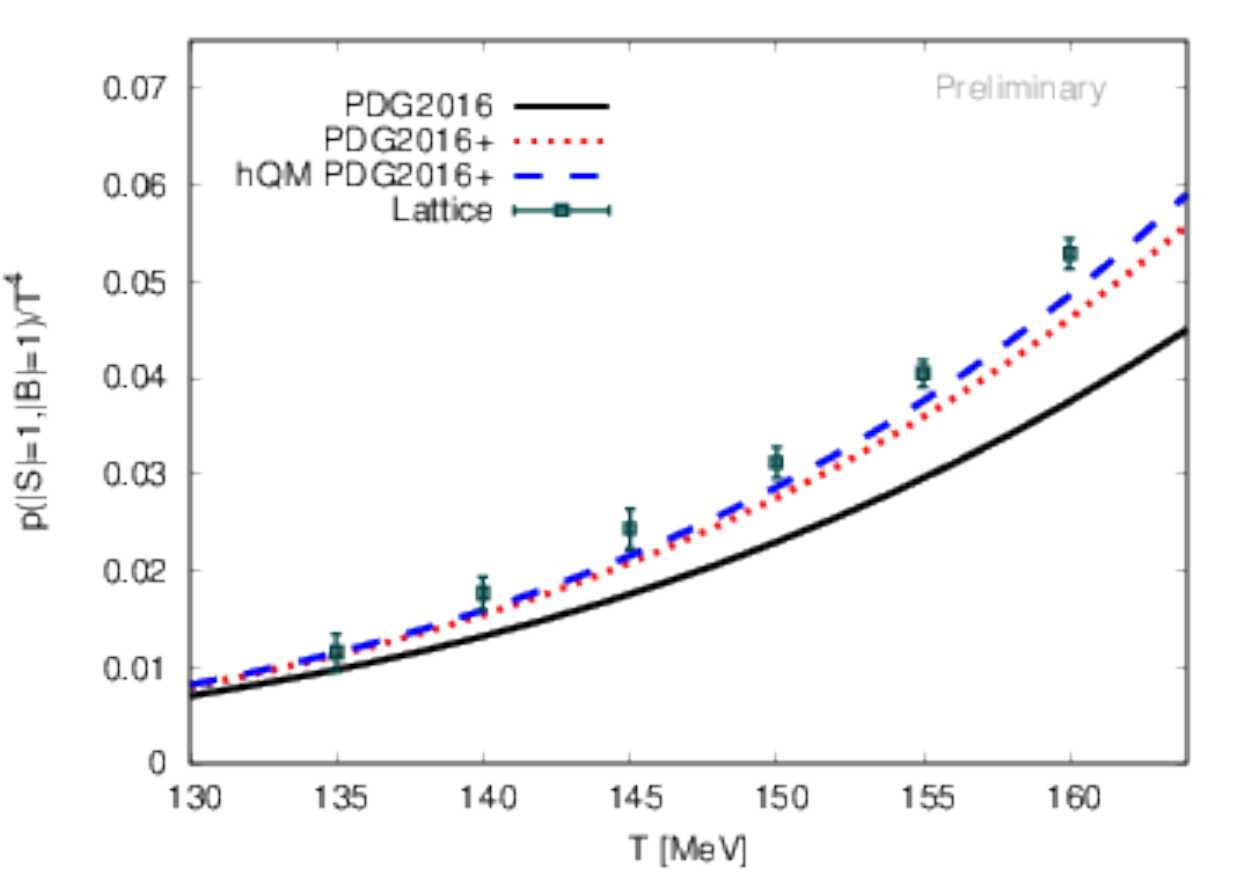}
\includegraphics[width=0.32\textwidth]{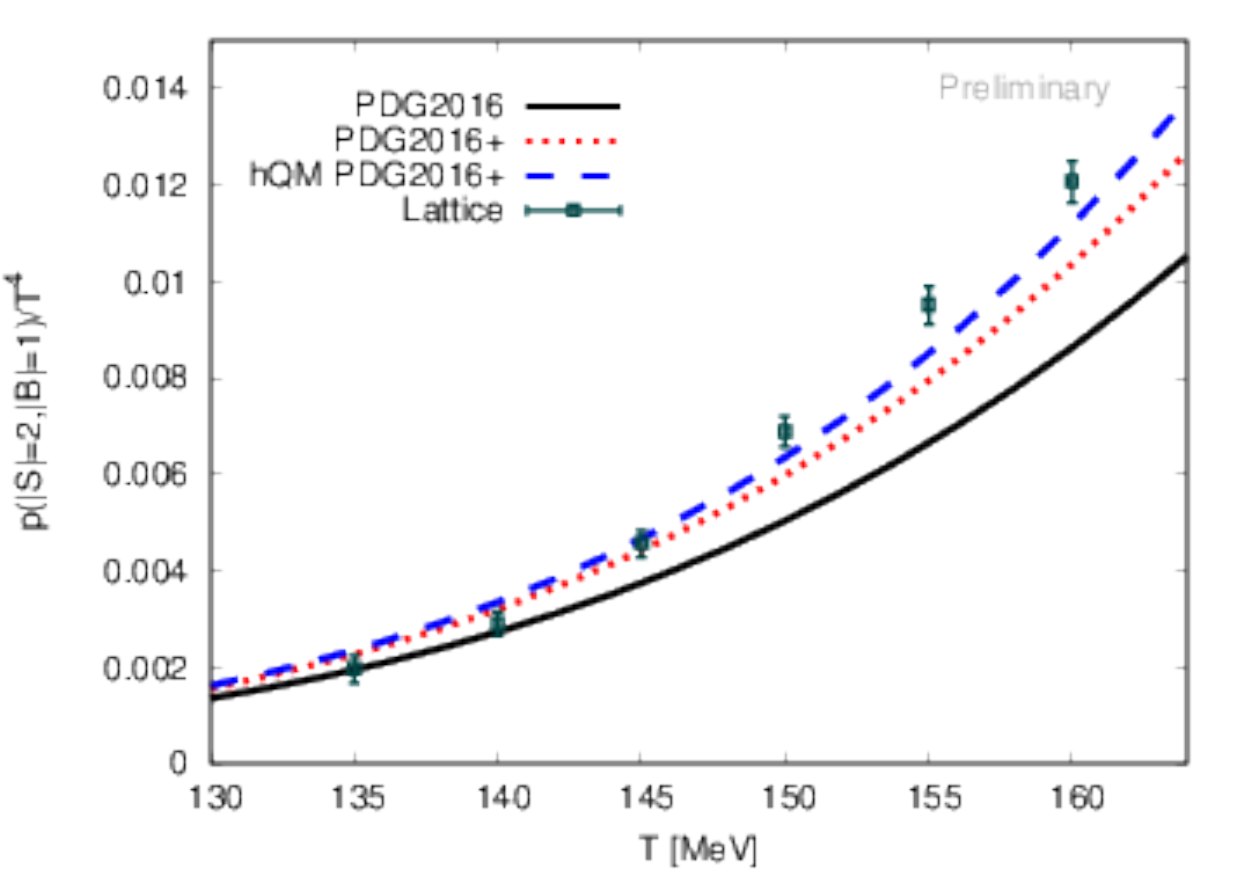}
\includegraphics[width=0.32\textwidth]{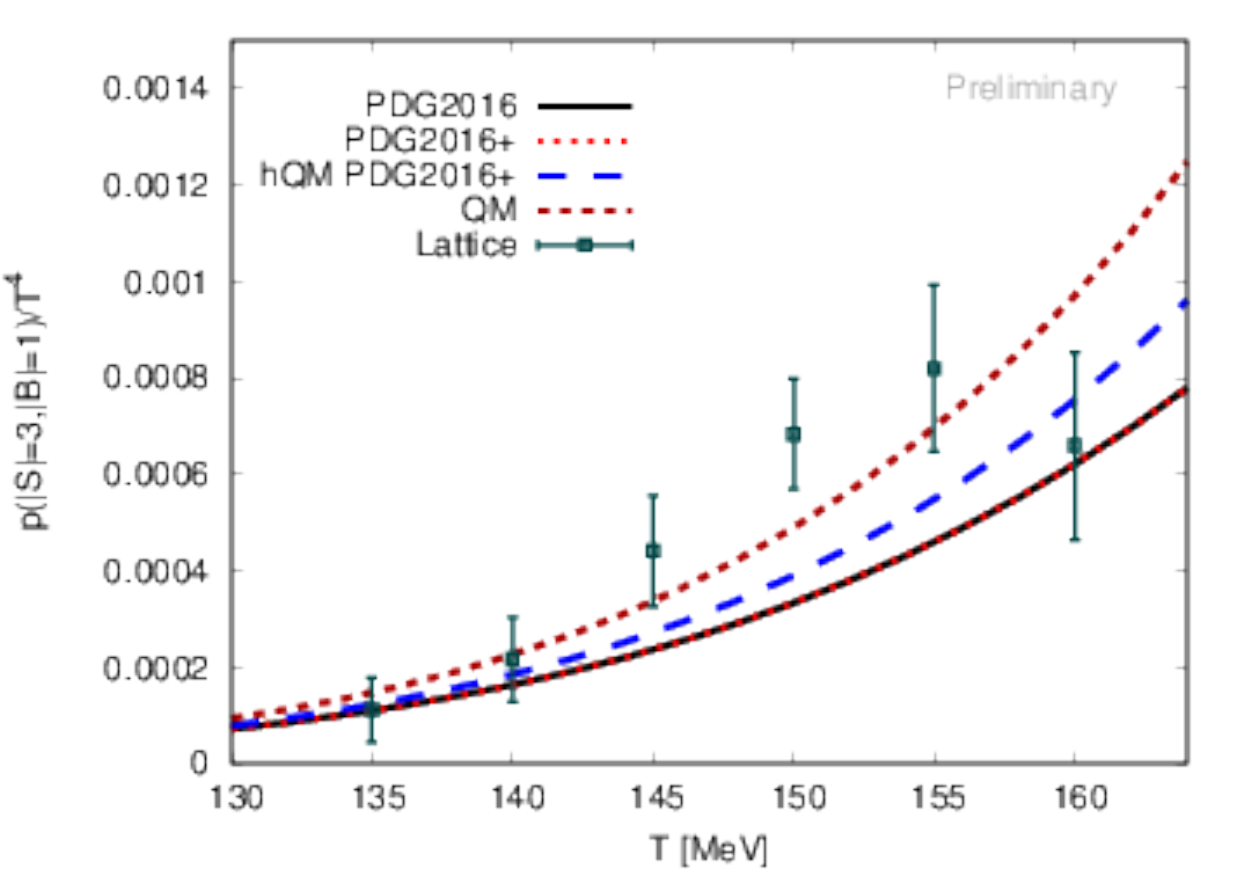}
\includegraphics[width=0.32\textwidth]{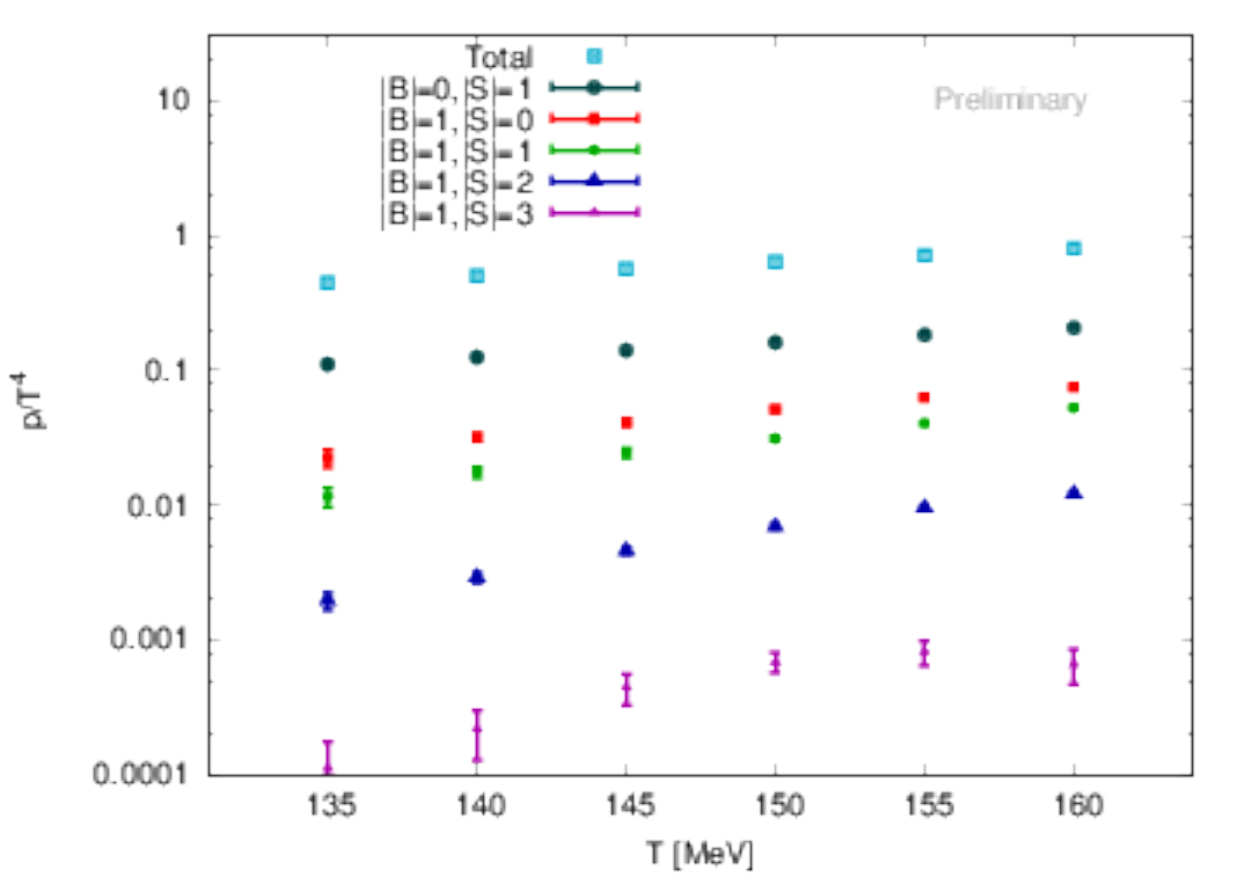}

\caption{\label{fig5} Partial pressures for the different hadronic 
	families as functions of the temperature. The upper panels 
	show the contributions due to strange mesons (left), 
	non-strange baryons (center) and baryons with $|S|=1$ 
	(right). The lower panels show the contributions due to 
	$|S|=2$ baryons (left), $|S|=3$ baryons (center) and the 
	relative contributions of the single families with respect 
	to the total (right). In the first five panels, the points 
	are the lattice results, while the curves are  PDG$2016$ 
	(solid black), PDG$2016+$ (with inclusion of one star 
	states, red dotted), PDG$2016+$ and additional states from 
	the hQM (blue, dashed)~\protect\cite{Ferraris:1995uic,
	Giannini:2001kbc,Giannini:2005ksc,Ferretti:2015adac,
	Giannini:2015ziac}. In the lower center panel, the dark 
	red short-dashed curve corresponds to the PDG$2016+$ with 
	inclusion of the $\Omega$ resonances predicted by the 
	QM~\protect\cite{Capstick:1986bmc,Ebert:2009ubc}.}
\end{figure*}

Our results are shown in Fig.~\ref{fig5}. The first five panels show the 
contributions of strange mesons, non-strange baryons and baryons with 
$|S|=1,2,3$ respectively. All lattice results are continuum-extrapolated, 
with the exception of the strange mesons, for which the finite-$N_t$ 
lattice data do not scale. The last panel shows the relative contribution 
of the single families to the total pressure, from which it is evident 
that there are more than three orders of magnitude between the total 
pressure and the smallest contribution due to the $|S|=3$ baryons. The 
method we used for this analysis, namely simulations at imaginary $\mu_B$ 
and $\mu_S$, was crucial in order to extract a signal for the 
multi-strange baryons. From all panels except the non-strange baryons, 
it is evident that the HRG model based on the well-established states 
listed in the PDG2016 is not enough to reproduce the lattice results. 
The additional one-star states are the minimal choice which allows to 
reproduce most of the data. In some cases (strange mesons and $|S|=3$ 
baryons), even more states are required to reproduce the data: for the 
$|S|=3$ baryons a better agreement is obtained when including the 
$\Omega$ resonances predicted by the QM, while for the strange mesons, 
the continuum extrapolation (once available) will lie above all curves, 
even the ones including the QM states. Therefore, our analysis shows 
that additional states are needed in order to reproduce the lattice 
results, with respect to the the well-established ones listed in the 
most updated version of the PDG. 

The minimal choice which reproduces almost all the observables is the 
PDG2016+. Besides reproducing most of the partial pressures, this 
choice also allows to correctly describe $(\mu_S/\mu_B)_{LO}$ and 
$\chi_4^S/\chi_2^S$, as shown in Fig.~\ref{fig6}. Inclusion of 
additional states predicted by the hQM pushes the agreement with the 
lattice to higher temperature for most observables.
\begin{figure*}[ht]
\includegraphics[width=0.45\textwidth]{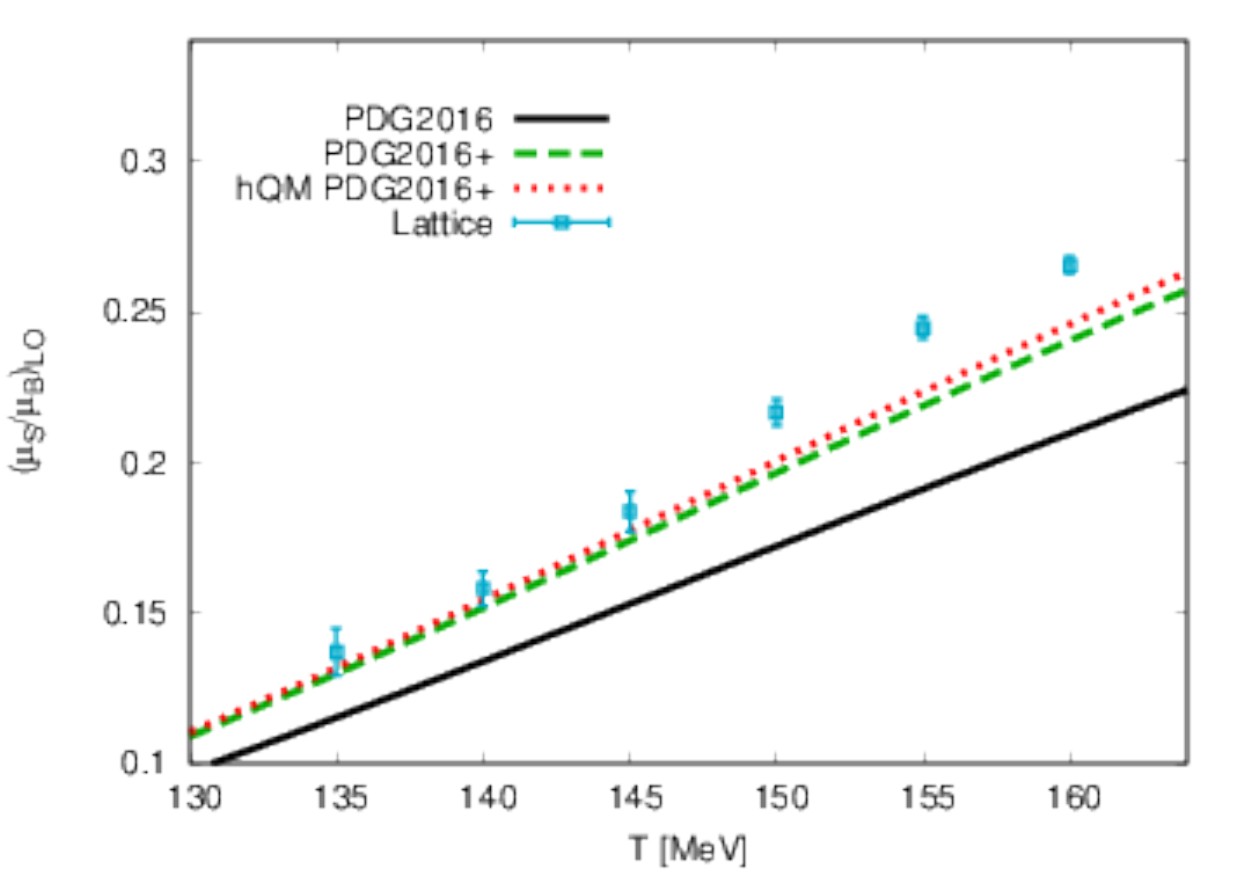}
\includegraphics[width=0.46\textwidth]{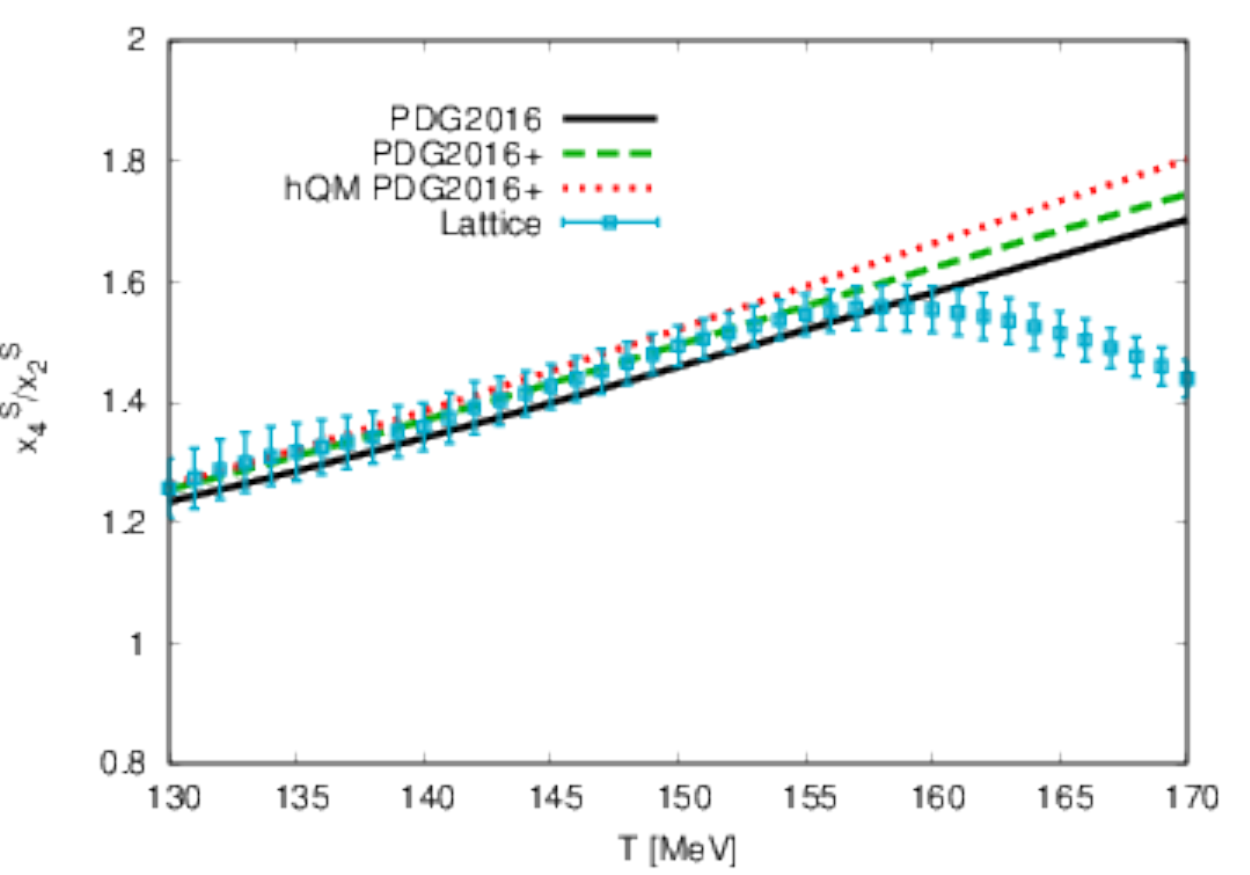}

\caption[]{\label{fig6} Left: $(\mu_S/\mu_B)_{LO}$ as a function of the 
	temperature. Right: $\chi_4 ^{S}/\chi_2 ^{S}$ as a function of 
	the temperature. In both panels, the lattice data are shown in 
	comparison to the HRG model curves based on the PDG 2016 (black, 
	solid line), the PDG2016+ (green, dashed line) and the PDG2016+ 
	with additional states from the hQM (red, dotted line).}
\end{figure*}

\newpage
\item \textbf{Conclusions}

We have calculated several thermodynamic quantities on the lattice and
compared them to the predictions of the HRG model based on the PDG2016,
PDG2016+ and PDG2016+ with the addition of states predicted by the hQM.
Besides $\chi_4^S/\chi_2^S$ and $(\mu_S/\mu_B)_{LO}$, we have extracted 
the contribution to the QCD pressure in the hadronic phase due to 
several hadronic families, grouped according to their baryon and 
strangeness content. Our analysis clearly shows that the 
well-established states listed in the most updated version of the PDG 
are not enough to correctly reproduce the lattice results. The minimal 
choice which reproduces most of the data is the PDG2016 with the
inclusion of one-star states which are not well-established. In some
cases, such as the strange mesons and $|S|=3$ baryons, there is room 
to add even more states. These results will potentially affect all
phenomenological applications of the HRG model, including thermal fits
and other calculations relevant to heavy-ion physics.

\item \textbf{Acknowledgments}

We acknowledge fruitful discussions with Elena~Santopinto, Igor~Strakovsky, 
Moskov~Amaryan and Mark~Manley. This material is based upon work supported 
by the National Science Foundation under grant no. PHY--1513864 and by the 
U.S. Department of Energy, Office of Science, Office of Nuclear Physics, 
within the framework of the Beam Energy Scan Theory (BEST) Topical 
Collaboration. The work of R.~Bellwied is supported through DOE grant
DE--FG02--07ER41521. An award of computer time was provided by the 
INCITE program.  This research used resources of the Argonne Leadership 
Computing Facility, which is a DOE Office of Science User Facility 
supported under Contract DE--AC02--06CH11357. The authors gratefully 
acknowledge the Gauss Centre for Supercomputing (GCS) for providing 
computing time for a GCS Large-Scale Project on the GCS share of the 
supercomputer JUQUEEN~\cite{juqueenc} at J\"ulich Supercomputing Centre 
(JSC). The authors gratefully acknowledge computing time on the Hazel 
Hen machine at HLRS.
\end{enumerate}


\newpage
\subsection{Strange Hadrons from the Lattice}
\addtocontents{toc}{\hspace{2cm}{\sl R.~G.~Edwards}\par}
\setcounter{figure}{0}
\setcounter{table}{0}
\setcounter{equation}{0}
\setcounter{footnote}{0}
\halign{#\hfil&\quad#\hfil\cr
\large{Robert~G.~Edwards}\cr
\textit{Theory Center}\cr
\textit{Jefferson Lab}\cr
\textit{12000 Jefferson Ave.}\cr
\textit{Newport News, VA 23606, U.S.A.}\cr}

\begin{abstract}
Lattice QCD has recently made signicant progress in determining the excited
state spectrum of Quantum Chromodynamics. We describe in these proceedings 
a program for these calculations and indicate future directions.
\end{abstract}

\begin{enumerate}
\item \textbf{The End of the Beginning}

One of the enigmatic features of Quantum Chromodynamics (QCD) is its
excited state spectrum. The presence of such states was inferred, at
least initially, by their characteristic bumps in scattering cross
sections. The plethora of such states drove the development of an
explanation via effective models of the spectrum -- such as the Quark
Model composed of constituent particles called quarks . A feature of
these models is that there is a global symmetry of the spectrum under
global rotations of a ``flavor'' under the Lie group $SU(3)$. The
three constituents particle are the up, down and strange. The Quark
Model works reasonable well at describing the spectrum of baryon
states, but there is a catch. By construction in the model, the
overall wave function for baryons are symmetric which is clearly at
 odds with their fermionic nature. It was postulated that the quarks
carry a new kind of ``charge'' that transformed under a {\em local}
gauge symmetry called ``color''. With three kinds of color charge,
baryon wave-functions can be antisymmetric as required by Fermi-Dirac
statistics.

The existence of the $\Omega$ baryon provided strong support for the
existence of color. Combined with insight gained from deep inelastic
scattering experiments, we now believe that the underlying theory of
the strong interactions is, in fact, QCD which is described in terms
of the interactions of quarks and gluons. These ``quarks'', though,
are thought to have a fairly small amount of mass. How do we reconcile
this nice, simple, and elegant theory of quarks with the massive hadrons
seen in experiments? What is the origin of spin? In particular, there
are spin $\tfrac{5}{2}$ hadrons. How is that possible?

The existence of the $\Omega$ is only a small sample of the implications
implied by the existence of three color charges. The color gauge fields
themselves must carry color, thus implying the glue interacts with itself.
One picture of the ``constituent'' quarks in the Quark Model is that they
gain their mass through screening, or ``dressing'' of the ``current''
quarks by the glue as well as other virtual fluctuations from the vacuum.
It seems a preferred pattern of these constituent quarks is groups of
three to form a baryon, or in a quark and antiquark pair configuration
to form a meson. But this is only an effective picture. Is it correct
in any level of detail? Why only these simple quark configurations?
There are new indications that tetraquark and even pentaquark 
configurations might exist. What about the glue itself -- does the 
existence of three color charges have more implications? 

Spectroscopy has been an active field of study over many years. The
unexplained mysteries arising from experiment challenges our understanding
of QCD. To address these fundamental questions we need theoretical guidance
and predictions directly from QCD. Thus, we also need a new generation of
experiments to find if there are yet even more unexpected configurations,
and certainly unexplained, properties of hadrons as well as measuring the
production amplitudes for these states. A burning question is what is the
spectrum of QCD featuring strange quarks? The presence of these states
might help explain properties of the QCD thermodynamic phase transition
at finite temperature, and is a motivation for the Ystar program.
\begin{figure*}[tbh!]
\begin{center}
\includegraphics[width = 0.6\textwidth]{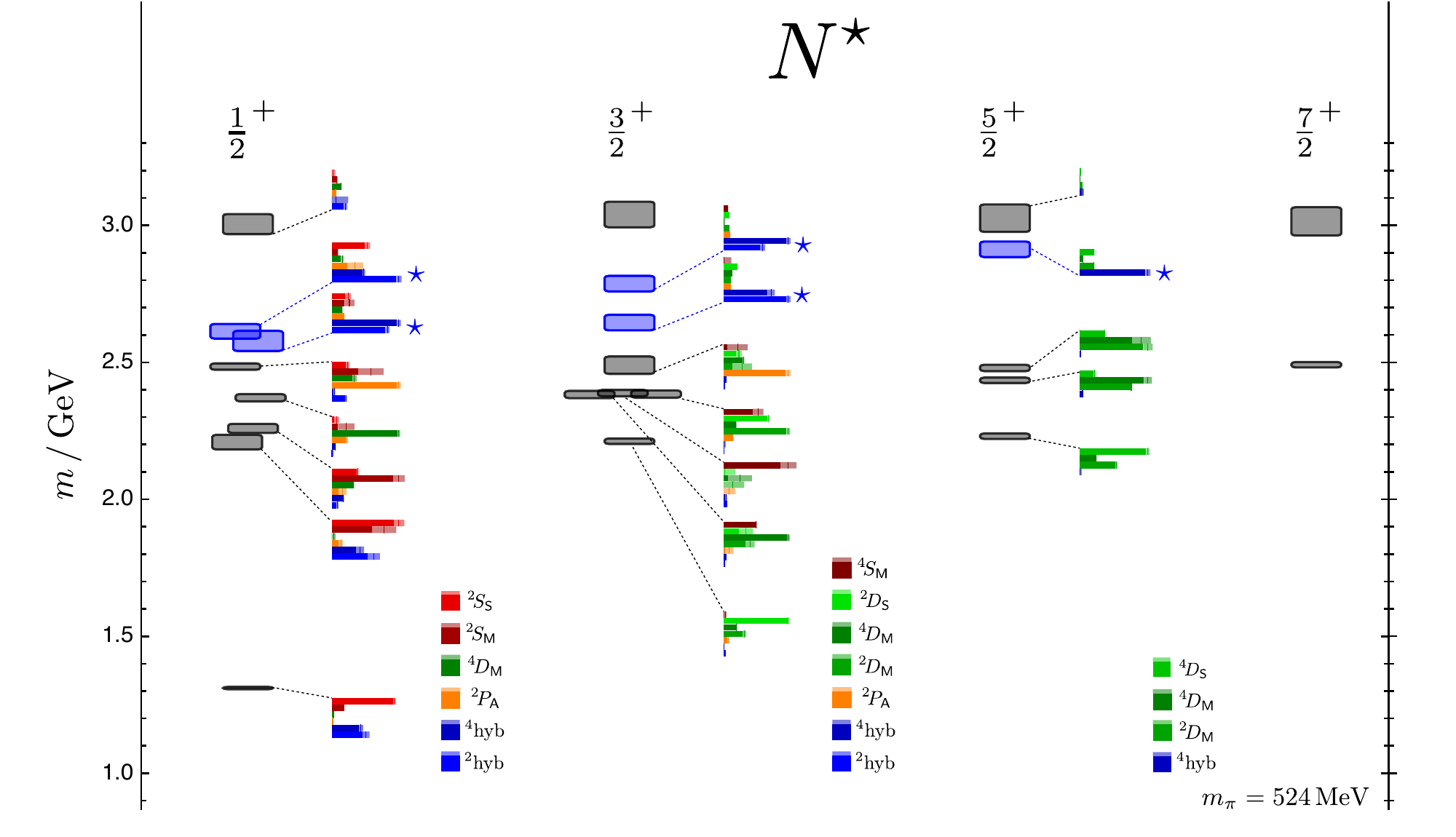}
\vspace{0.3cm}
\includegraphics[width = 0.6\textwidth]{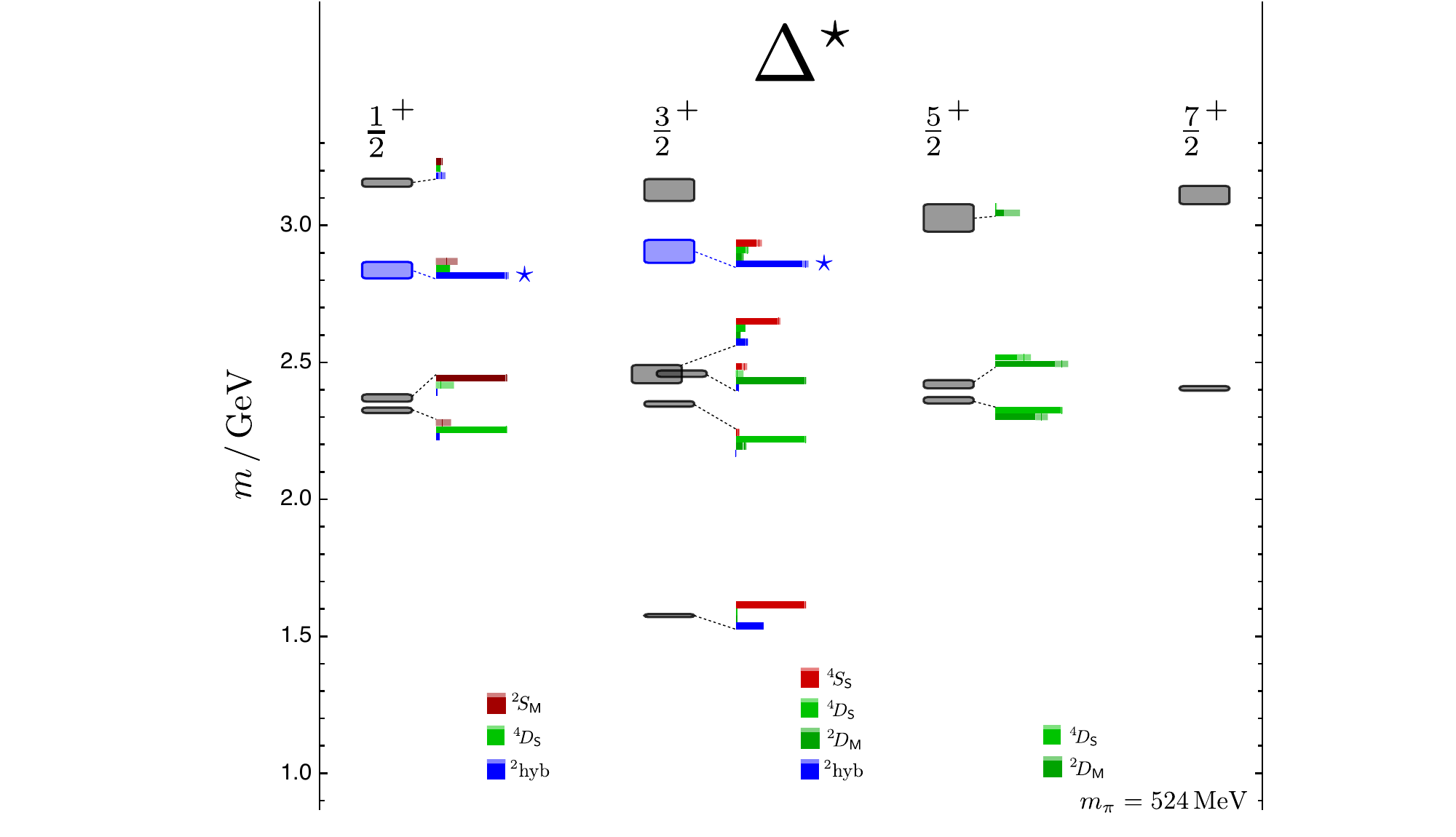}
\end{center}

\caption{ \label{fig:808histo} Results from Ref.~\protect\cite{Dudek:2012agg}.
	Extracted spectrum of Nucleon and Delta states indicated by 
	$J^P$ at a pion mass of 524 MeV. Rectangle height indicates the 
	statistical uncertainty in mass determination. Histograms 
	indicate the relative size of matrix elements $\big\langle 
	\mathfrak{n} \big| \mathcal{O}^\dag_i(0) \big| 0\big\rangle$ 
	for a ``non-relativistic" subset of the operators used with 
	the lighter area at the head of each bar being the statistical 
	uncertainty. Operator labeling follows a spectroscopic convention 
	as described in~\protect\cite{Edwards:2011jjg} with
	${}^{2S+1}L_\pi$ and $\pi$ indicating the permutation structure 
	of the quark fields. The  operator constructions include those 
	featuring only upper Dirac-spin components as well as lower 
	components, and also hybrid-like constructions featuring 
	chromomagnetic gluon fields.  Asterisks indicate states having 
	dominant overlap onto hybrid operators. The counting of the 
	levels is consistent with the $SU(6)\times O(3)$ predictions of 
	the non-relativistic quark model, but with the important
	addition of states that are hybrid-like.}
\end{figure*}
\begin{figure*}[tbh]
\begin{center}
\includegraphics[width=0.8\textwidth] {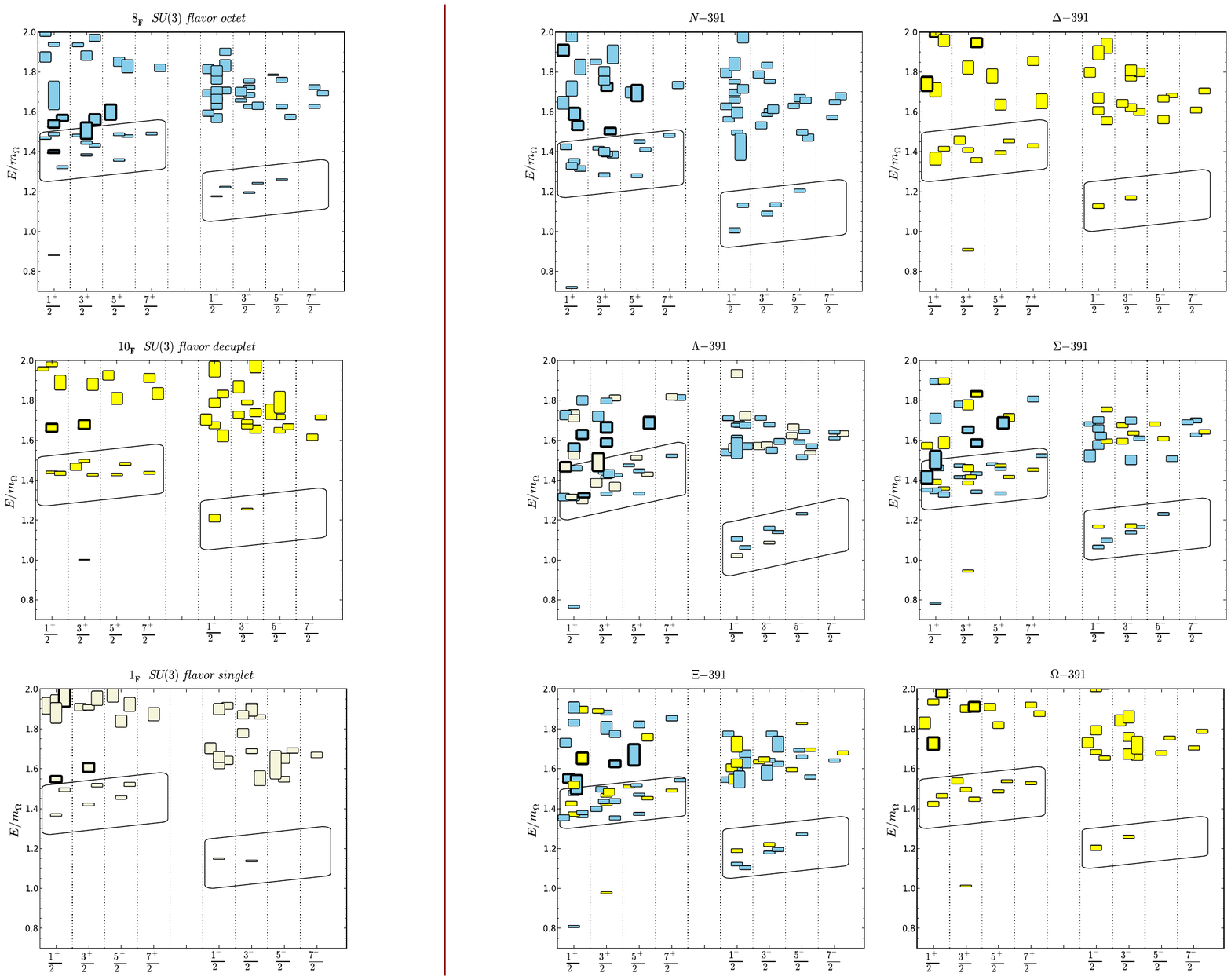}
\end{center}

\caption{Results from Ref.~\protect\cite{Edwards:2012fxg}.
	Left panel: baryon excited states in flavor irreps
	${\bf 8_F}$, ${\bf 10_F}$ and $\bf{1_F}$ obtained using the 
	flavor-symmetric point, with $m_{\pi}$ = 702~MeV are shown 
	versus $J^P$.
	Right panel: results for baryon excited states using the 
	ensemble with $m_{\pi}$ = 391~MeV are shown versus $J^P$.
	Colors are used to display the flavor symmetry of dominant 
	operators as follows:
	blue for ${\bf 8_F}$; beige for  ${\bf 1_F}$; yellow for 
	${\bf 10_F}$.  Symbols with thick border lines indicate 
	states with strong hybrid content. The lowest bands
	of positive- and negative-parity states are highlighted 
	within slanted boxes. \label{fig:840_SU6xO3_spectra}}
\end{figure*}

There has been significant theoretical progress  towards answering some
of these fundamental questions. Lattice QCD is the only known fully
 non-perturbative, systematically improvable method we have available
to answer these questions. Recent progress is highly suggestive that
there is a rich spectrum of hadrons beyond what is predicated by the
Quark Model. What has been missing from these initial studies is the
information on branching fractions of decays. There has been very recent
progress in this area as well.

This proceedings summarizes some of the results from the Hadron Spectrum
Collaboration pertaining to the calculation of the light and strange
quark baryon spectrum of QCD.
\begin{figure*}[tbh]
\begin{center}
\includegraphics[width=0.6\columnwidth]{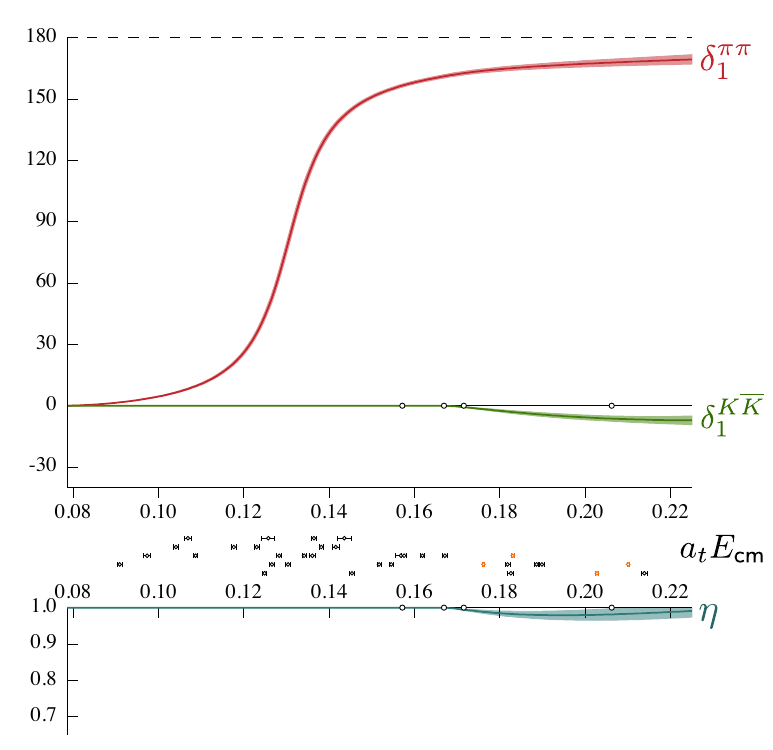}
\end{center}

\caption{Results from Ref.~\protect\cite{Wilson:2015dqag} with a pion mass of
        236~MeV. Shown are the coupled $P$-wave $\pi\pi$ and $K\overline{K}$
        isospin-1 phase shifts $\delta$ and inelasticity $\eta$ from a
        single $K$-matrix fit. Statistical uncertainty shown by the shaded
        band. The central points show the energy levels constraining the
        amplitude extraction with those dominated by $\pi\pi$-like and
        $\overline{q}q$-like operators shown in black and those with
        significant $K\overline{K}$ contributions shown in orange. On axis
        circles show the opening of the $4\pi$, $\bar{K}K$, $\eta \pi\pi$
        and $\pi \bar{K}K$ thresholds.}\label{coupled_default_K}
\end{figure*}

\item \textbf{Predictions from QCD}

\begin{enumerate}
\item \textbf{Glimpse of the Baryon Spectrum}

The fundamental theory of the strong interactions -- QCD -- is simple
and elegant, but it is not amenable to perturbative methods to extract
some observable, and it is precisely the strongly interacting properties
of the three color charges that make it so complicated -- and interesting!
We must {\em regulate} the theory to make a calculation, and a
finite-volume lattice is a straightforward method to do so. The basic
quantity that we compute are correlation functions of quark and gluon
fields. These functions in Euclidean space can be written as an infinite
sum of exponentials decaying in time. The coefficients in the exponents
is the discrete energy of the finite-volume versions of states, and the
coefficient in front of the exponential is the matrix element of the
overlap of the creation (or annihilation) operator with the states going
into the vacuum. These overlap factors are thus a probe of the
wave-function of the state. A large basis of operators thus provides
multiple probes of the state, and also provides a powerful method to
extract the towering of finite-volume energies.

The Hadron Spectrum collaboration, centered at Jefferson Lab, has been
pursuing a campaign to compute the spectrum of QCD. In
Figure~\ref{fig:808histo} is shown the determination of the discrete
energy of states in a finite-volume where the pion mass in this world,
with two light and one strange dynamical set of quarks, is 524 MeV.
Shown is the positive parity spectrum of the Nucleon and Delta. The
horizontal bars show the magnitude of the overlap factor for a particular
operator across each of the states.  The counting of the levels and the
structure of the operator constructions is consistent with the $SU(6)\times
O(3)$ prediction of the non-relativistic quark model, but with the important
addition of states that have a hybrid-like nature, and are indicated by a
thick black border for the boxes denoting the mass.

In the left panel of Figure~\ref{fig:840_SU6xO3_spectra} is shown the
positive and negative parity spectrum in the flavor symmetric limit of
QCD, where the $u$, $d$ and $s$ quarks are degenerate. In this limit,
there is an exact $SU(3)_F$ symmetry. The overlap factors and energies
suggest again a pattern of states consistent with $SU(6)\times O(3)$
with the number of states determined by the flavor representation.
\begin{figure*}[tbh]
\begin{center}
\includegraphics[width = 0.6\textwidth]{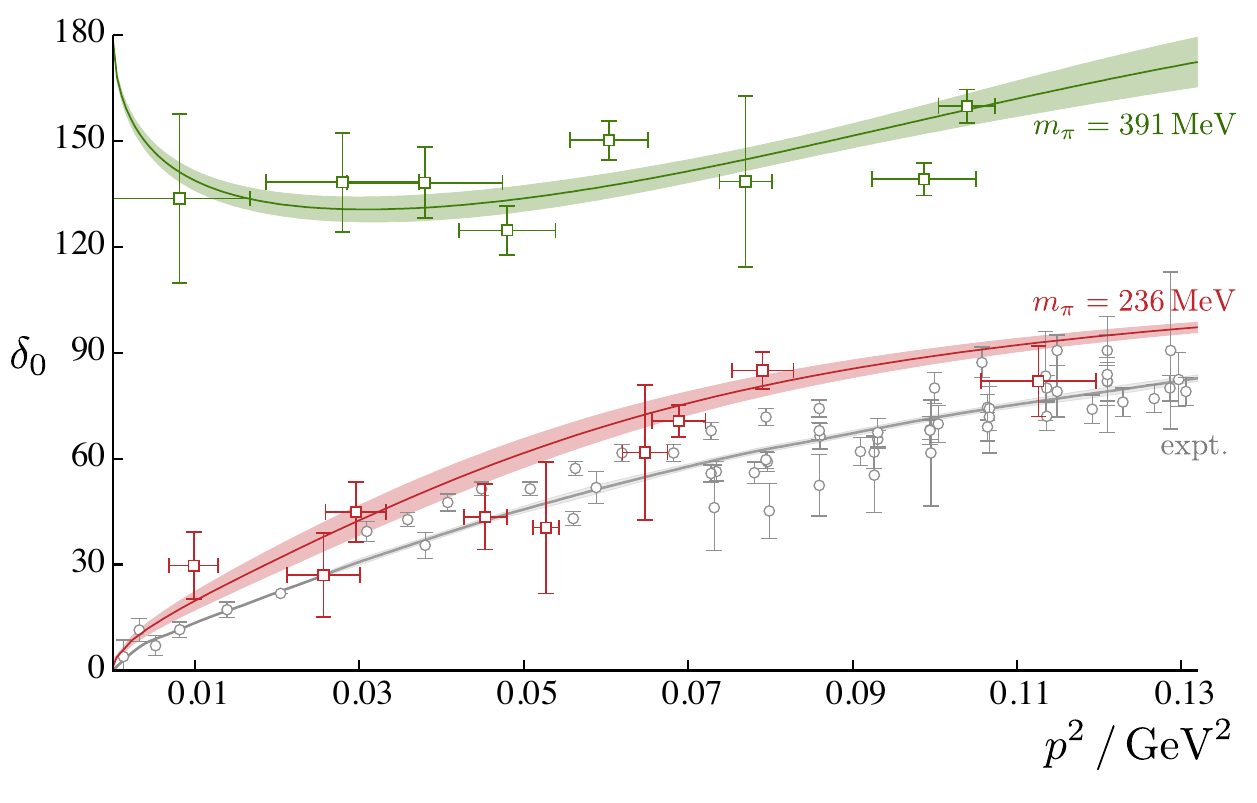}
\end{center}

\caption{Results from Ref.~\protect\cite{Briceno:2016mjcg}.
       Isoscalar $S$-wave $\pi\pi$ elastic scattering phase-shift,
       $\delta_0$, plotted against the scattering momentum, ${p^2 =
       ( E_\mathsf{cm}/2)^2 - m_\pi^2}$, computed at a pion mass of
       391 and 236~MeV. The colored curves are the result of a
       $K$-matrix ``pole plus constant'' with Chew-Mandelstam
       phase-space parameterization. The gray points show
       experimental data~\protect\cite{Protopopescu:1973shg,
       Hyams:1973zfg,Grayer:1974crg,Estabrooks:1974vug} and the
       gray curve shows the constrained dispersive description of
       these data presented in Ref.~\cite{GarciaMartin:2011jxg}.}
\end{figure*}

As we decrease the light-quark masses, the $SU(3)$ flavor structure is
broken suggesting there should be mixing of flavor representations within
a fixed isospin and total flavor number. Shown in the right panel of
Figure~\ref{fig:840_SU6xO3_spectra} is the spectrum with a pion mass of
391~MeV. The color of the boxes denotes the flavor structure of the
dominant overlap factors for the states. As the light quark mass is not 
appreciably smaller than the strange quark mass, we expect the mixing 
to be small, and indeed, all the states could be cleanly separated into 
fixed flavor representations. In this quark mass limit, the isospin=0 
with strangeness=-1 components of the $1_F$ and $8_F$ operator 
constructions are allowed to mix, and as seen in the panel for the 
$\Lambda$ in the figure, we see a pattern of both $\bf 8_F$ and $\bf 1_F$  
states superimposed on each other, but with negligible flavor mixing. A 
similar observation is found for mixing of the $\Sigma$ and $\Xi$ states 
in $\bf 8_F$ and $\bf 10_F$. We expect that is the quark mass is 
decreased that there should be more mixing among the appropriate flavor 
representations.

The operators are constructed with flavor and spin transforming according
to the appropriate flavor representation and spin construction. This
operator flavor and spin part is multiplied by a spatial part that has
some number of units of spatial gauge covariant derivatives which resemble
orbital angular momentum. In the spectroscopic notation, we group operators
according to the number of units of orbital angular momentum. The bands
around the states indicate their grouping according to their dominant
spectral overlaps which we find to be consistent with this grouping of
orbital angular momentum.

That the grouping follows such a pattern moving up in energy has a
plausible explanation. Each unit of orbital angular momentum flips the
parity of the state. In analogy to the Bohr model, each unit of orbital
angular momentum brings in some unit of energy, roughly on the scale of
$\Lambda_{QCD}$. Thus, we might expect a band of states alternating as
we go up in energy. Of course, the states become more numerous as we go
up in units of orbital angular momentum, so this simple picture will not
be so neat and clean.

This pattern of states also happens to be a prediction of the Quark Model.
It should be emphasized that the Quark Model are not baked into the
lattice calculations. Rather, the operator constructions happen to be
all that are allowed by continuum symmetries. They are easily classified
according to the spectroscopic notation which happens to be the preferred
way of constructing the Quark Model. That the two happen to agree in
overall structure (but certainly not in detail) is suggestive that their
are some effective degrees of freedom at play that are captured in the
Quark Model.

However, the addition of the hybrid states is something new again. The
overall picture so far from the lattice calculations suggest the spectrum
roughly resembles a collection of constituent quarks and a constituent
glue, with the quarks transforming as $S=\tfrac{1}{2}$ and $\bf 3_C$ in
color, and the glue transforming according to $J_{PC}=1^{+-}$ in an $\bf
8_C$ in color. The overall state must be a singlet in color. The pattern
of energies suggest that the constituent glue has an effective mass of
roughly 1.3~GeV.

The lattice results so far did not attempt to classify resonances
according to their structure. The calculations shown have been at
pion masses at 391 MeV and above. If these states are indeed resonances,
they must decay. That the lattice results so far appear so clean and
consistent is possibly that the pion mass is large and the phase space
for decays is small. To make this picture more rigorous requires extending
the calculations to allow for decays, and that program is described next.

\item \textbf{Resonances and Decays}

What is observably measurable are scattering amplitudes as a function of
energy. A resonance could appear as a bump in these amplitudes, but more
generally, we ascertain a resonance as a pole in the scattering amplitude
when continued into the complex plane of energy. Of course, very few
resonances are expected to be a pole in a purely elastic decay into a
single channel two particles. More generally, we expect decays of a
resonance into many channels. Using a partial wave picture, a pole can
appear in many partial waves. It is thus critical that the lattice methods
be extended to provide such decay information.
\begin{figure*}[tbh]
\begin{center}
\includegraphics[width = 0.4\textwidth]{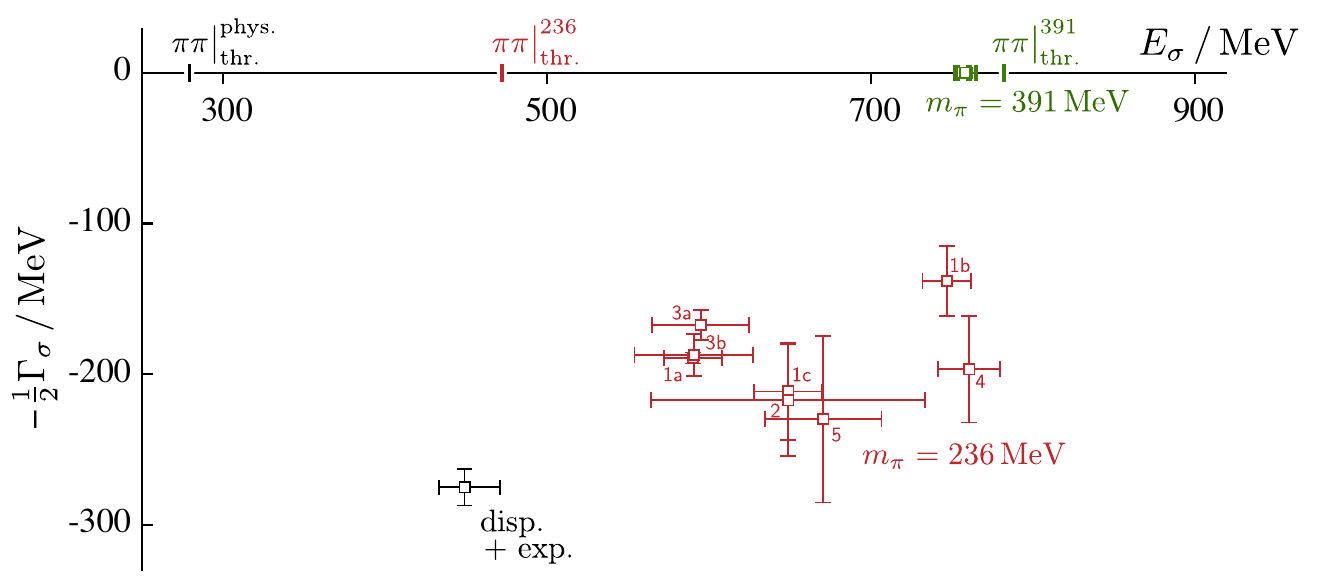}
\includegraphics[width = 0.4\textwidth]{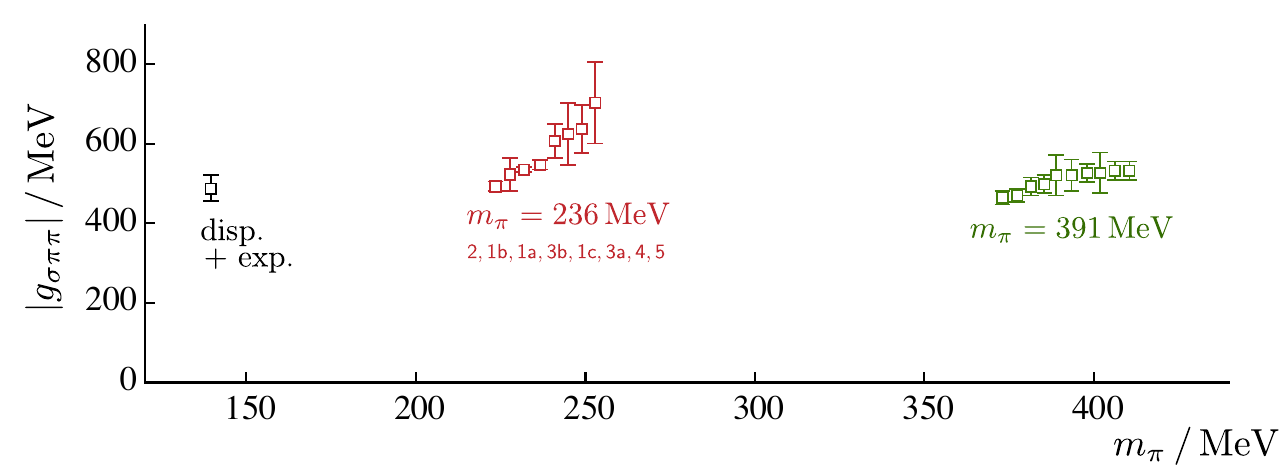}
\end{center}

\caption{Results from Ref.~\protect\cite{Briceno:2016mjcg} for
        isoscalar elastic $\pi\pi$ scattering.
        Left panel: $t$-matrix pole positions for a variety of
        parameterizations [$K$-matrix ``pole plus polynomial''
        forms with ($\mathsf{1a-c}$) and without Chew-Mandelstam
        phase-space ($\mathsf{2}$) and/or Adler
        zero~\protect\cite{Adler:1964umg}($\mathsf{3a,b}$),
        relativistic Breit-Wigner($\mathsf{4}$) and effective
        range expansion($\mathsf{5}$)].
        Red points: resonant pole (unphysical sheet) at $m_\pi =
        236$~MeV.
        Black point: Resonant pole from dispersive analysis of
        experimental data (conservative average presented in
        Ref.~\protect\cite{Pelaez:2015qbag}).
        Right panel: Coupling $g_{\sigma\pi\pi}$ from $t$-matrix
        residue at the pole - points from various parameterizations
        are shifted horizontally for clarity. The phenomenological
        determinations of the experimental results employ Roy-Steiner
        constraints on the scattering amplitudes while the lattice
        results do not (yet). Nevertheless, there is consistency
        of the extracted coupling to the $\pi\pi$ channel across
        between the phenomenological and lattice results.}
        \label{fig:poles}
\end{figure*}

Exploiting the finite-volume nature of the lattice calculations, an elegant
formalism has been developed, generically called the {\em L\"uscher}
formalism, that relates the infinite-volume Minkowski scattering amplitude
to the discrete energy spectrum encoded within a set of known, finite-volume
functions. These L\"uscher functions are also functions of the lattice cubic
representation and total momenta of the system. The discrete spectrum of
energies changes significantly according to the total momenta and cubic
group representation. Thus, we can map out scattering amplitudes by
determining the energy spectrum in different moving frames. The
finite-volume formalism has also been extended to coupled channel
systems of two particles.

To illustrate the method, shown in Figure~\ref{coupled_default_K} is the
first calculation of the $P$-wave scattering amplitude(s) for the simplest
resonance in QCD -- the $\rho$ meson. The sets of dots between the upper
and lower panel show the discrete energies measured in different frames.
Solving the relation for the energies, the form of the amplitude, including
any parameters within it, must be adjusted to agree with the ``measured''
lattice energies. As the lattice energies can be determined in many moving
frames, there are many constraints on the details of the amplitude. In the
$\rho$ system, one can see that as the CM energy of the system is increased,
the amplitude opens at the threshold for the $\bar{K}K$ channel and the
amplitude (here encoded as a phase shift) for that decay mode becomes
small, but non-zero. This is the first demonstration from a lattice
calculation of such behavior, and opens the door to the calculation of
amplitudes in more complicated systems.

One such system that has been a mystery for decades is the existence of
the $\sigma$ meson in isospin=0 $\pi\pi$ scattering. In
Figure~4 is the first such calculation of the elastic
$S$-wave isoscalar scattering amplitude, as a function of the pion
mass. We see that the form of the amplitude channels significantly
as the phase space for the decay becomes more open, and it approaches
the phenomenologically determined amplitude.
Shown in Figure~\ref{fig:poles} are the pole locations and the decay
couplings determined from the residue of the pole in the complex energy
plance of the scattering amplitude. While the pole location changes
significantly as the pion mass decreases, we see the coupling appears
to be fairly consistent with a constant.

These results suggest that it is possible to determine scattering
amplitude and resonance properties directly from QCD. Determining
amplitudes for mesons is technically simpler for lattice QCD, but
it is possible to extend these calculations to the baryon sectors.
First results have appeared from other groups.
\end{enumerate}

\item \textbf{Outlook}

The Hadron Spectrum Collaboration has embarked on a ambitious
program to determine resonance decays for mesons with an emphasis
on exotic mesons -- relevant for the GlueX program -- and also
the baryon spectrum relevant for both GlueX and CLAS12, as well
as other facilities around the world. There is a great deal of
work to do, but the results to date are encouraging. It is possible
to determine scattering amplitudes and extract resonance properties
directly from QCD. As the pion mass in the calculations is decreased,
the phase space for decays can become large, thus the scattering
formalism will be essential to extract resonance properties,
reliably, in complicated channels.

The lattice QCD efforts do not exist in a vacuum (sorry, bad
pun), but go hand in hand with the experimental analysis efforts.
The scattering amplitudes determined from the lattice must agree
with those from experiment. The formalism for scattering amplitudes
in complicated channels can be complicated mathematically. Thus a
close interaction between the communities is needed for a successful
spectroscopy program.

\end{enumerate}


\newpage
\subsection{Implications of Missing Resonances in Heavy Ions Collisions}
\addtocontents{toc}{\hspace{2cm}{\sl J.~Noronha-Hostler}\par}
\setcounter{figure}{0}
\setcounter{table}{0}
\setcounter{equation}{0}
\setcounter{footnote}{0}
\halign{#\hfil&\quad#\hfil\cr
\large{Jacquelyn Noronha-Hostler}\cr
\textit{Department of Physics}\cr
\textit{University of Houston}\cr
\textit{617 SR1 Building}\cr
\textit{Houston TX 77204, U.S.A.}\cr}

\begin{abstract}
Between 2004 until today many new resonances in the QCD spectrum
have been added by the Particle Data Group.  However, it is still
possible that there are further states that have yet to be measured,
as predicted by Lattice QCD calculations and Quark Models. In this
paper, I review how these missing states would influence the field
of heavy-ion collisions.  Additionally, the quantum numbers that
characterize these missing states are also discussed.
\end{abstract}

\begin{enumerate}
\item \textbf{Introduction}

Efforts at Jefferson Laboratory have resulted in a significant
increase in the number of known QCD resonances since 2004, which
are now listed in the Particle Data Group (PDG). Furthermore,
various theory groups have predicted even further states using
Lattice QCD~\cite{Dudek:2009qfh} and quark 
models~\cite{Capstick:1986bmh,Ebert:2009ubh}, giving marginal support 
to an exponentially increasing mass spectrum~\cite{Hagedorn:1965sth}
up to a given value of mass.  Even amongst the resonances listed
within the PDG some are more confidentially measured than others.
Thus, a star system rating was developed where states with the
highest confidence level are listed as **** states and those with
the least are $\ast$ states.

It is natural to wonder how many more states are still missing and
what is the top possible maximum mass of these resonances.  The
implications of these missing resonances have been studied
extensively over the last 15 years in the field of heavy-ion
collisions. Results that have garnered the most attention include
the effect of missing resonances on the transport coefficients in
the hadron gas phase~\cite{NoronhaHostler:2008juh} as well as the
possibility that there are missing strange baryons found using
Lattice QCD~\cite{Bazavov:2014xyah}.  This proceedings reviews
some of the main findings in the field of heavy-ion collisions
and also discusses the mass range and the quantum numbers within
which further missing states may be measured.

\item \textbf{Comparisons with Lattice QCD}

At the transition region between the hadron gas phase into the Quark
Gluon Plasma, Lattice Quantum Chromodynamics (QCD) has been enormously
successful in calculating thermodynamic quantities. In the early 2000's
the hadron resonance gas model (where it is assumed that an interacting
gas of hadrons can be described by a gas of non-interacting hadrons and
their resonances) was used to calculate thermodynamic quantities. For
instance, the pressure of the hadron resonance gas model is
\begin{equation}
	p/T^4= \sum_{i}^{PDG}\frac{d_i}{2\pi^2}\left(\frac{m_i}{T}\right)^2
	\sum_{k=1}^{\infty}\frac{(\pm1)^{k+1}}{k^2}K_2\left(\frac{km_i}{T}
	\right)
	\cosh\left[\frac{k\left(B_i\mu_B+S_i\mu_S+Q_i\mu_Q\right)}
	{T}\right]
\end{equation}
where $d_i$ is the degeneracy, $m_i$ is the mass, $B_i$ is the baryon
number, $S_i$ is the strangeness, and $Q_i$ is the electric charge of
each individual hadron taken into account.  Thus, the only real variable
in this model is the total number of resonances as dictated by the PDG.

Due to having fewer experimentally measured resonances in the early
2000's, the hadron gas model was unable to match Lattice QCD calculations
of the pressure, energy density and entropy~\cite{Karsch:2003vdh} (an
unphysical pion mass was also an issue on the lattice side but that has
now been resolved).   Using the PDG lists from 2004 predictions were made
that missing states would improve the fits to Lattice QCD 
results~\cite{NoronhaHostler:2008juh,Majumder:2010ikh,NoronhaHostler:2012ugh},
which were later confirmed~\cite{Borsanyi:2013biah} when the PDG's 2014
list~\cite{Agashe:2014kdah} was released. However, more massive resonances
should not contribute strongly to the total pressure and, thus, if further
missing states are measured it is very unlikely that they would affect the
comparisons with Lattice QCD thermodynamic quantities.  Additionally, such
comparisons make it very difficult to determine the quantum numbers of
missing resonances and, thus, more differential observables from Lattice
QCD are currently being introduced. As a final note, it has also been
postulated that missing states can assist in understanding phase changes
from hadronic to deconfined matter and the order of the phase 
transition~\cite{Moretto:2005izh,Begun:2009anh,Zakout:2006zjh,
Ferroni:2008ejh,Bugaev:2008iuh,Ivanytskyi:2012yxh}.

\item \textbf{Searching for Missing Resonances Assuming an Exponential 
	Mass Spectrum}

In the 1960's Hagedorn~\cite{Hagedorn:1965sth} suggested that QCD resonances
followed an exponentially increasing mass spectrum.   In his picture, there
was a limiting temperature $T_H$ above which as one adds in more  energy to
the system the temperature would not increase but rather it would open up
new degrees of freedom.
\begin{figure} [h]
\begin{center}
  \includegraphics[width=0.6\textwidth]{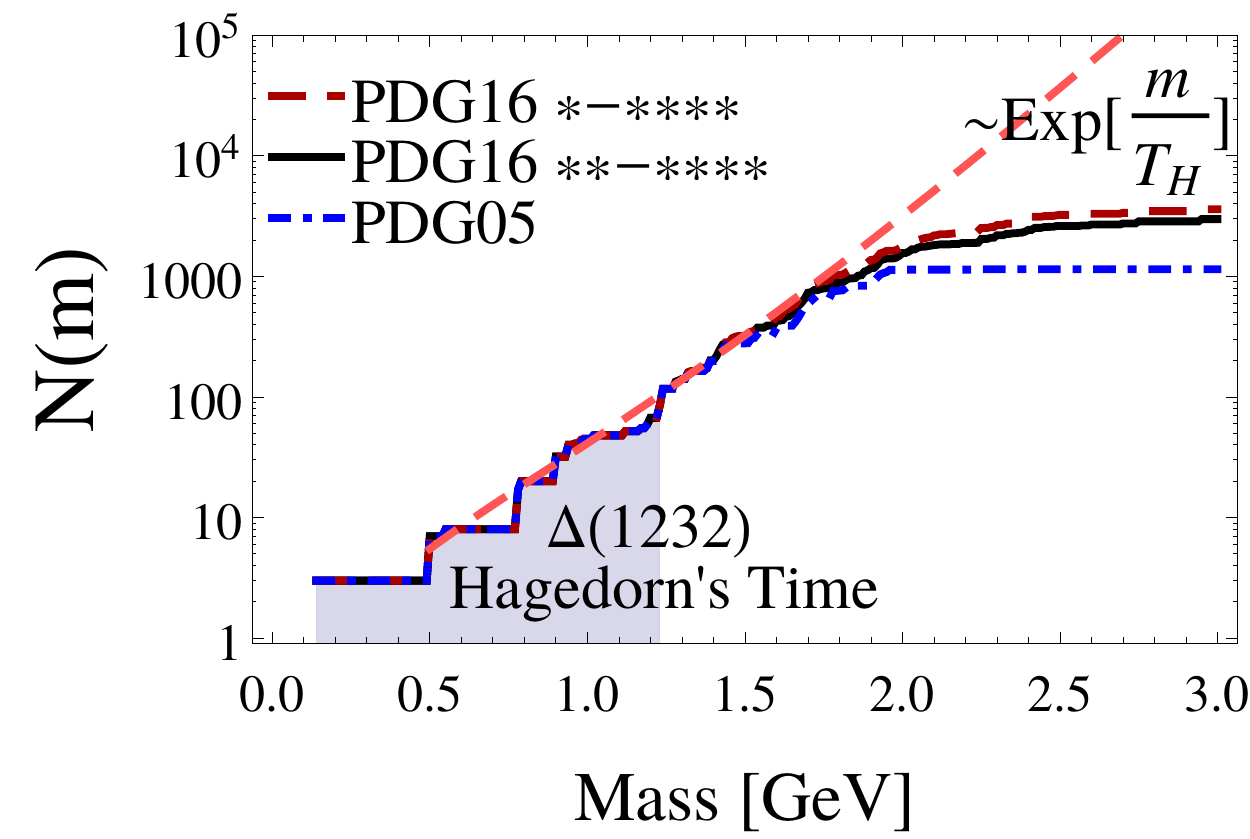}

  \caption{Mass spectrum of all known (light and 
	strange) hadrons as listed in the PDG from 
	2004~\protect\cite{Eidelman:2004wyh} and 
	2016~\protect\cite{Olive:2016xmwh} either for more 
	well-measured states (**-****) or all possibly measured 
	states (*-****). } \label{fig:all}
\end{center}
\end{figure}

In Hagedorn's time resonances were only measured up to $\Delta(1232)$.
However, in the most recent release from the Particle Data Group 
(PDG)~\cite{Olive:2016xmwh} (light and strange) hadrons were measured up 
to $m\sim 2.5$~GeV.  One can see in Fig.~\ref{fig:all} that the mass 
range that demonstrates an exponentially increasing mass spectrum has 
expanded significantly from the 1960's and, in fact, even between 2004 
to 2016 many new states have been measured (although reservations 
regarding an exponential mass spectrum were also discussed 
here~\cite{Cohen:2011crh}).

Returning to  Fig.~\ref{fig:all}, the mass spectrum of all light and
strange hadrons are plotted versus their mass such that:
\begin{equation}
	N(m)=\sum_i d_i \Theta(m-m_i)
\end{equation}
where $d_i$ is their degeneracy and $m_i$ is the mass of the $i^{th}$
resonance.  The exponential fit made in Fig.~\ref{fig:all} is based
upon Hagedorn's Ansatz for the degeneracy of ``Hagedorn States" (i.e.,
the massive states missing from the exponentially increase mass spectrum):
\begin{equation}
	\rho(M)=\int_{M_{0}}^{M}\frac{A}{\left[m^2 +(m_{0})^2
	\right]^{a}}e^{\frac{m}{T_{H}}}dm
\end{equation}
where $m_0$ is the mass of the lowest stable particle of the spectrum,
$T_{H}=165$~MeV is the Hagedorn temperature such that $T_c \leq T_H$ (a
discussion on the choice of $T_H$ follows below), $M_0$ is the minimum
mass where Hagedorn states begin, and $A$ is a free parameter. The power
of $a$ has a non-trivial affect such that when $a=3/2$ Hagedorn states
are more likely to decay into two body decays and when $a=5/4$ multi-body
decays are more likely~\cite{Frautschi:1971ijh}.  I assume that multi-body
decays are more likely with heavier resonances (and a brief glance through
the PDG booklet confirms this assumption) so $a=5/4$ is taken.

In Refs.~\cite{Broniowski:2004yhh,Chatterjee:2009kmh,Lo:2015ccah}, the 
known hadrons were separated by families (light mesons, strange mesons 
etc.) in order to extract their individual mass spectra.  Here I return 
to this approach to determine:
\begin{itemize}
\item if all families see an exponentially increasing mass spectrum,
\item if there are any gaps in the mass spectrum that could possibly be
	filled with missing states,
\item if there are variations in the needed Hagedorn temperature by
	species (following the idea in~\cite{Bellwied:2013ctah,
	Noronha-Hostler:2016rpdh}).
\end{itemize}
In Fig.~\ref{fig:light}, the mass spectra of the light mesons and
baryons are shown where both are fitted to a temperature of $T_H=155$~MeV,
which is almost exactly what one would expect from Lattice QCD.  Light
mesons begin to deviate from an exponential mass spectrum at around
$M\sim 1.5$~GeV whereas light baryons deviate around $M\sim 2$~GeV.
In both cases a visible improvement was seen going from the PDG 2004 to 
2016.  For light hadrons the difference between *-**** states and
**-**** states are negligible.
\begin{figure} [h]
\begin{center}
\begin{tabular}{c c}
  \includegraphics[width=0.45\textwidth]{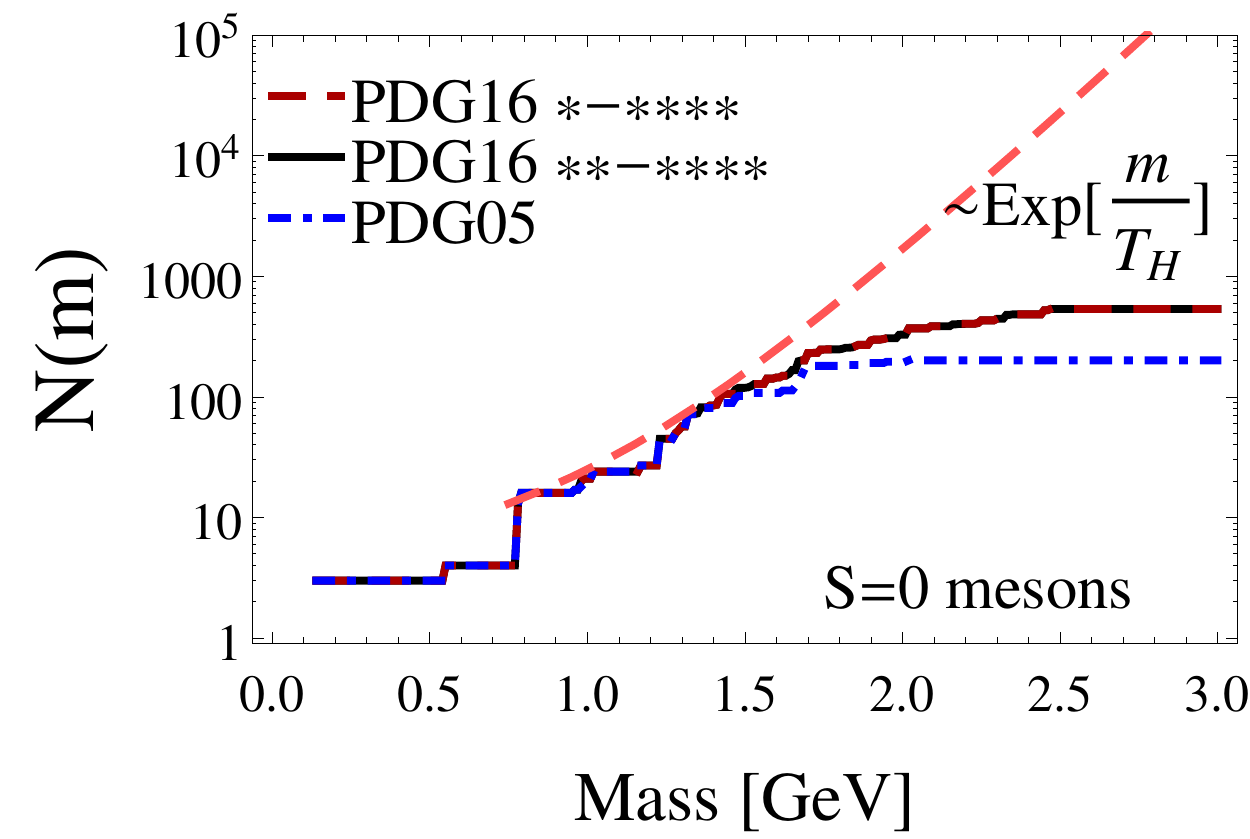} &
  \includegraphics[width=0.45\textwidth]{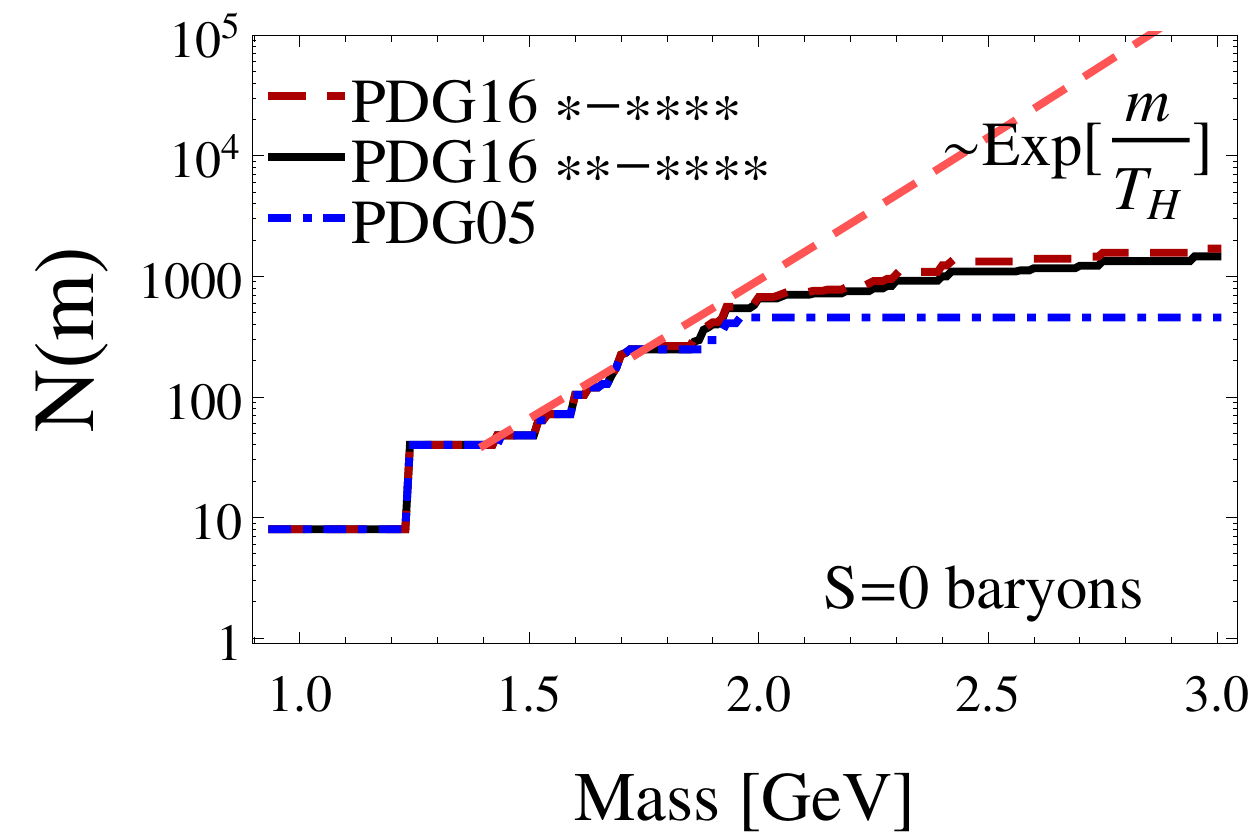}
\end{tabular}

\caption{Mass spectrum of all known light mesons
	(left) and light baryons (right) as listed in the PDG from 
	2004~\protect\cite{Eidelman:2004wyh} and 
	2016~\protect\cite{Olive:2016xmwh} either for more 
	well-measured states (**-****) or all possibly measured
	states (*-****). }  \label{fig:light}
\end{center}
\end{figure}

In Fig.~\ref{fig:strange}, the strange mesons and baryons are shown by
their strangeness content.  Unlike for the light hadrons, a higher
Hagedorn temperature of $T_H=175$~MeV is needed in order to fit the
mass spectra (the one exception being Omega baryons), which is in-line
with the possibility that strange particles have a higher transition
temperature than light ones.  While $|S|=1$ and  $|S|=2$ baryons work
well with an exponential mass spectrum up to $M\sim 2$~GeV and $M\sim
1.7$~GeV, respectively, the strange mesons and Omega baryons do not
fit well to exponential mass spectra.  In the case of strange mesons,
it appears that there is a large gap in the $M=1-1.5$~GeV region.  For
Omega baryons there are so few states that the only possibility is to
fit the spectrum with a $T_H=300$~MeV because the $T_H=175$~MeV was
significantly too stiff, which is well above the expected values.
Thus, looking for mid-mass strange mesons and $|S|=3$ baryons are
likely candidates for missing states.  Additionally, $|S|=1$ and
$|S|=2$ may have missing states but most likely only in the more
massive region $M\gtrsim 2$~GeV.
\begin{figure} [h]
\begin{center}
\begin{tabular}{c c}
  \includegraphics[width=0.45\textwidth]{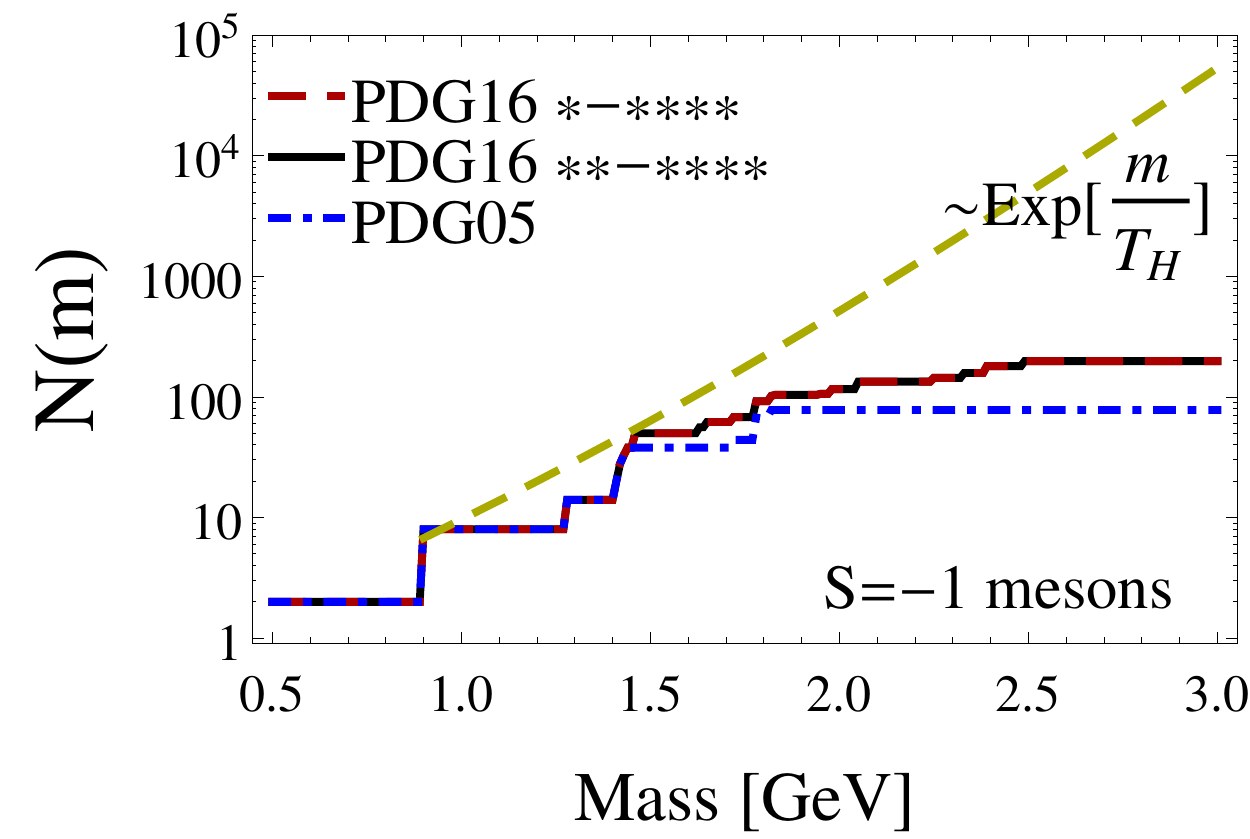} &
  \includegraphics[width=0.45\textwidth]{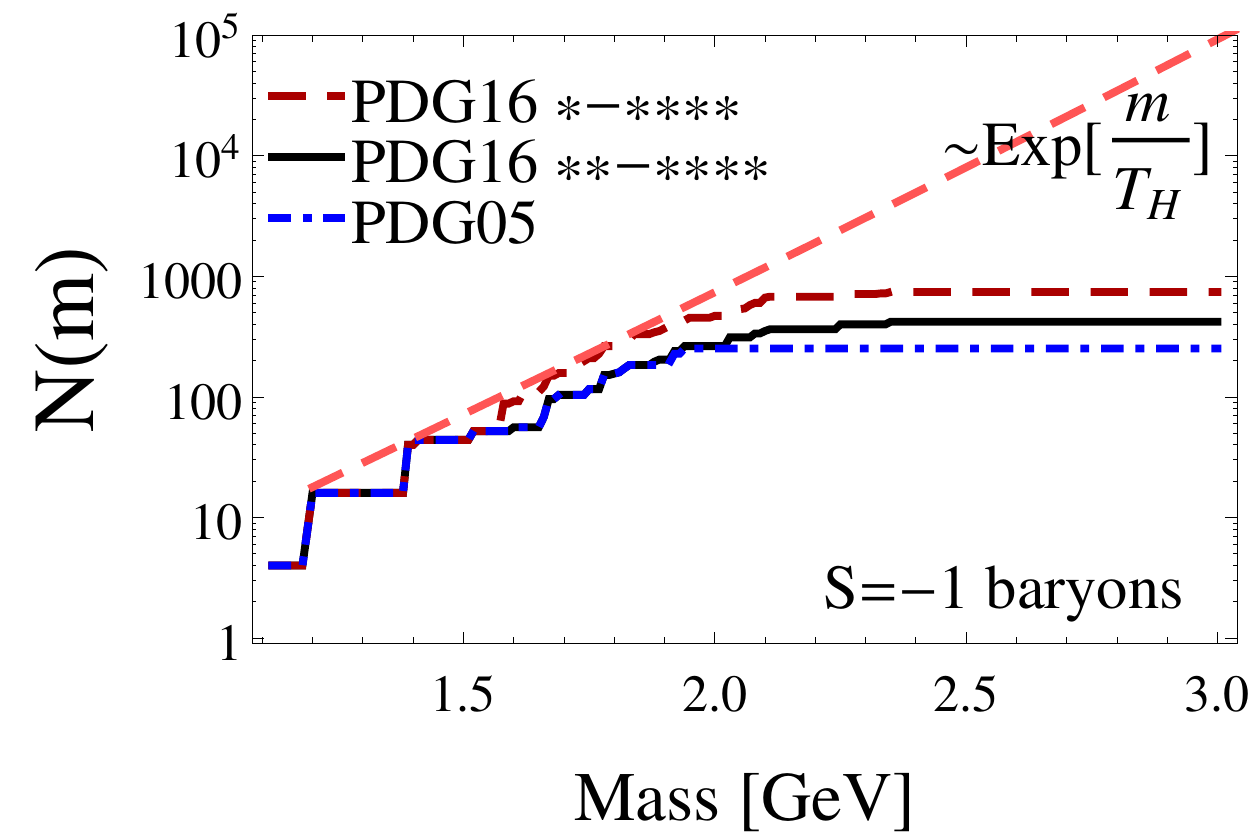} \\
  \includegraphics[width=0.45\textwidth]{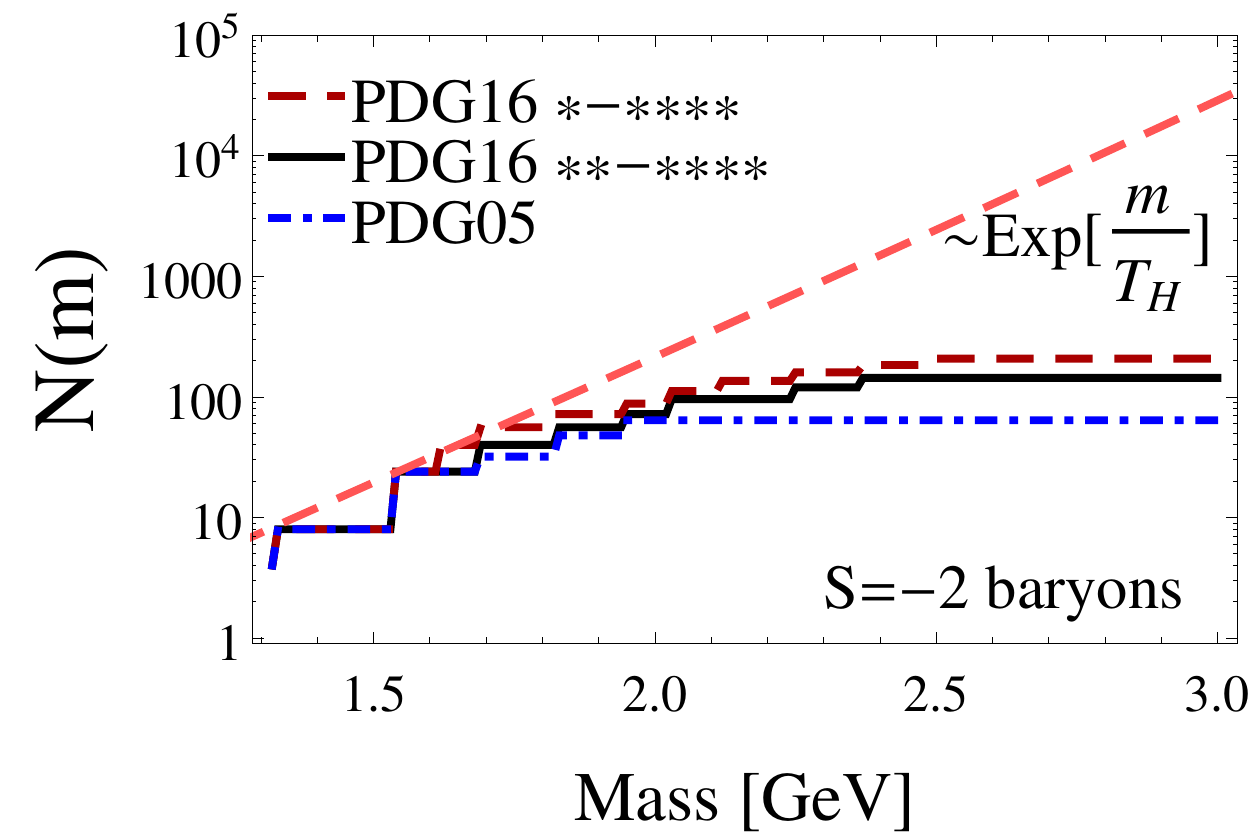} &
  \includegraphics[width=0.45\textwidth]{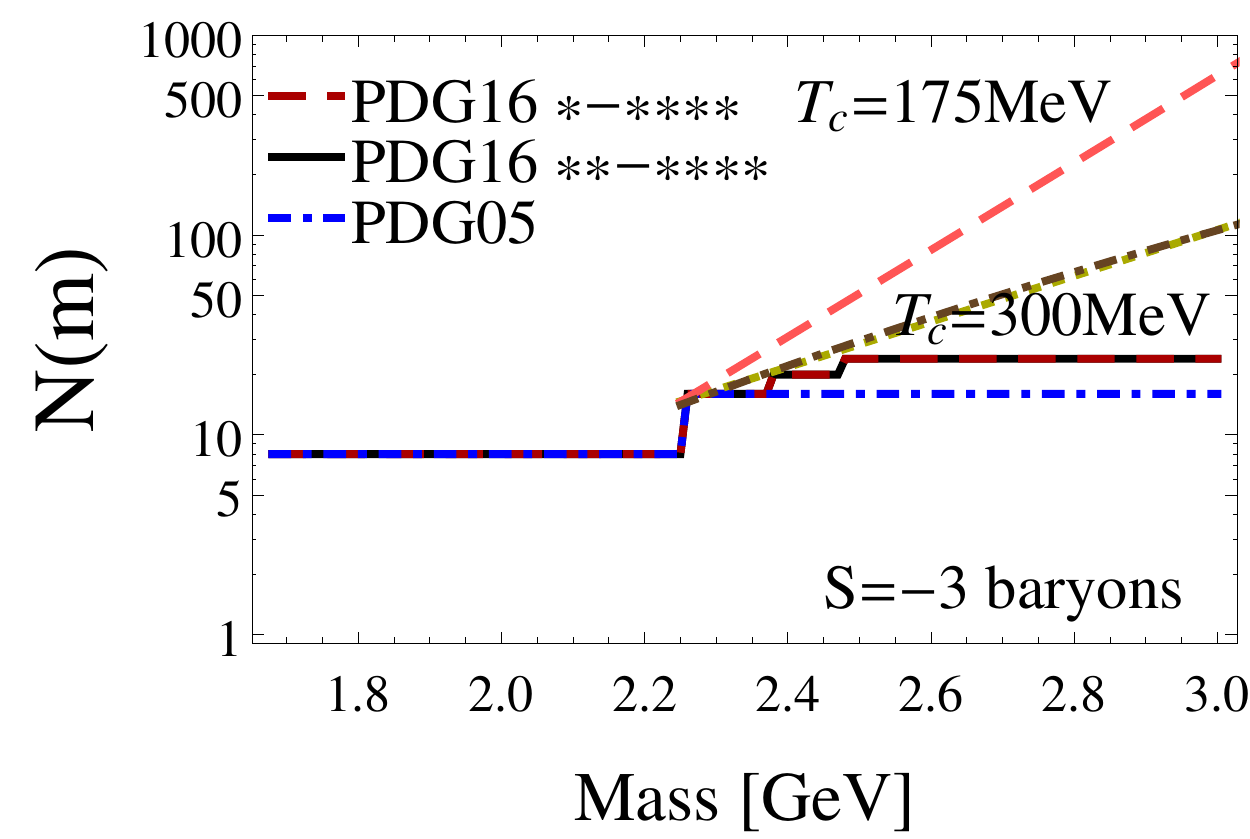} \\
\end{tabular}

  \caption{Mass spectrum of all known strange 
	hadrons (mesons and baryons $|S|=1-3$) as listed in the 
	PDG from 2004~\protect\cite{Eidelman:2004wyh} and 
	2016~\protect\cite{Olive:2016xmwh} either for more
	well-measured states (**-****) or all possibly measured 
	states (*-****). } \label{fig:strange}
\end{center}
\end{figure}

\item \textbf{Freeze-out Parameters and Chemical Equilibration Time}

In heavy-ion collisions, the system size is extremely small and is
constantly expanding (at almost the speed of light) and cooling 
down. Thus, due to dynamical effects, hadrons may not manage to 
reach chemical equilibrium on such short time scales.  However, due 
to the success of thermal fits~\cite{Andronic:2005yph}, many believe
that even on such short time scales chemical equilibrium can be
reached.  Studies found that $2\leftrightarrow 2$ body reactions
were not able to reach chemical equilibrium on such short time
scales, especially at SPS energies and higher.  It was then
suggested that multi-mesonic reactions could speed up chemical
equilibration times~\cite{Rapp:2000gyh,Kapusta:2002pgh} to explain
SPS energies, however, with the known resonances it was still not
enough to explain RHIC data~\cite{Heinz:2006urh}. To describe RHIC
energies extra, massive resonances were needed that had large decay
widths and could act as a catalyst to speed up hadronic 
reactions~\cite{Greiner:2004vmh,Pal:2005rbh,NoronhaHostler:2007fgh,
NoronhaHostler:2007jfh,NoronhaHostler:2009cfh,Pal:2013ohah,
Beitel:2014kzah,Beitel:2016ghwh}.
\begin{figure} [h]
\begin{center}
\begin{tabular}{c c}
  \includegraphics[width=0.4\textwidth]{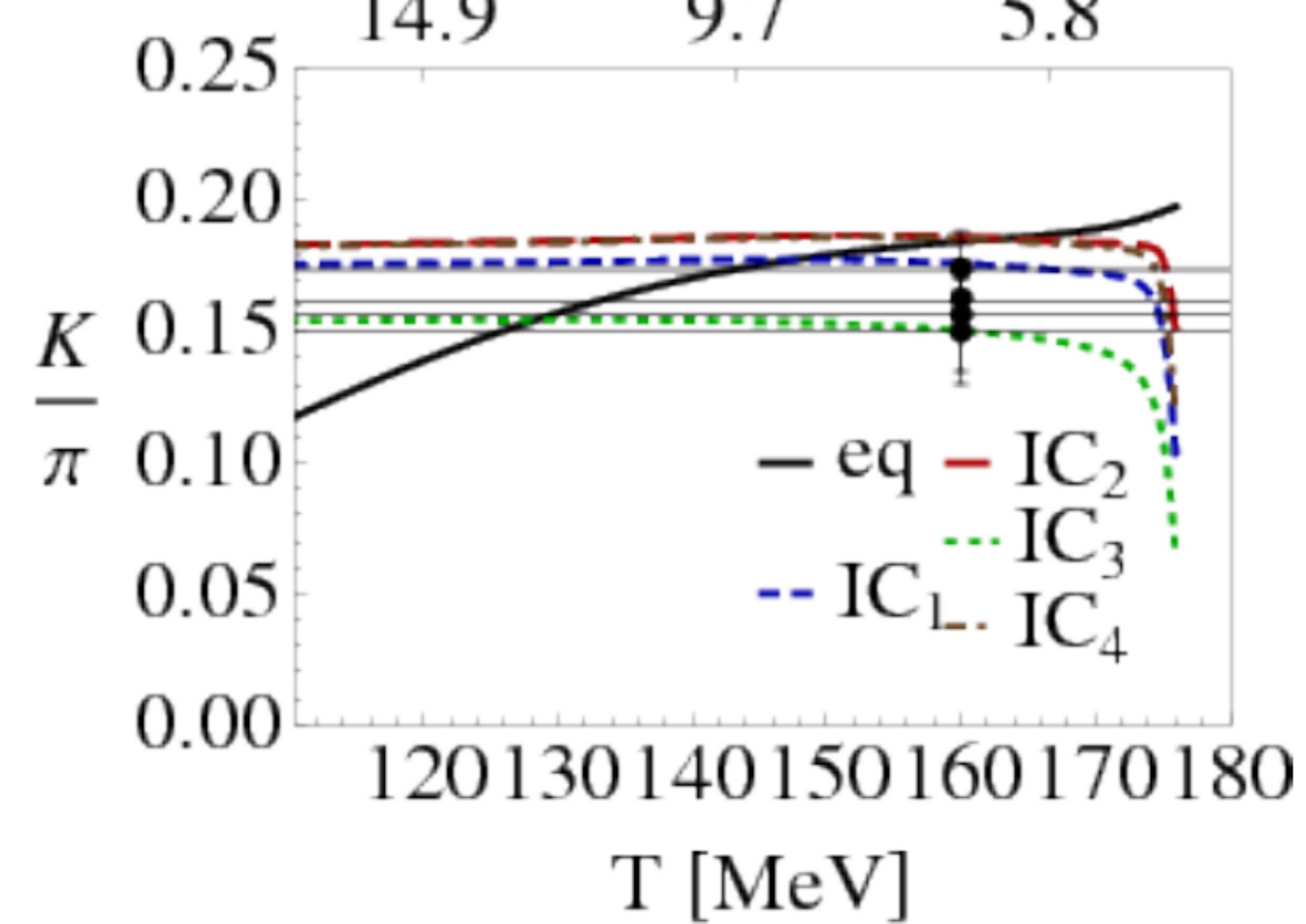} &
  \includegraphics[width=0.5\textwidth]{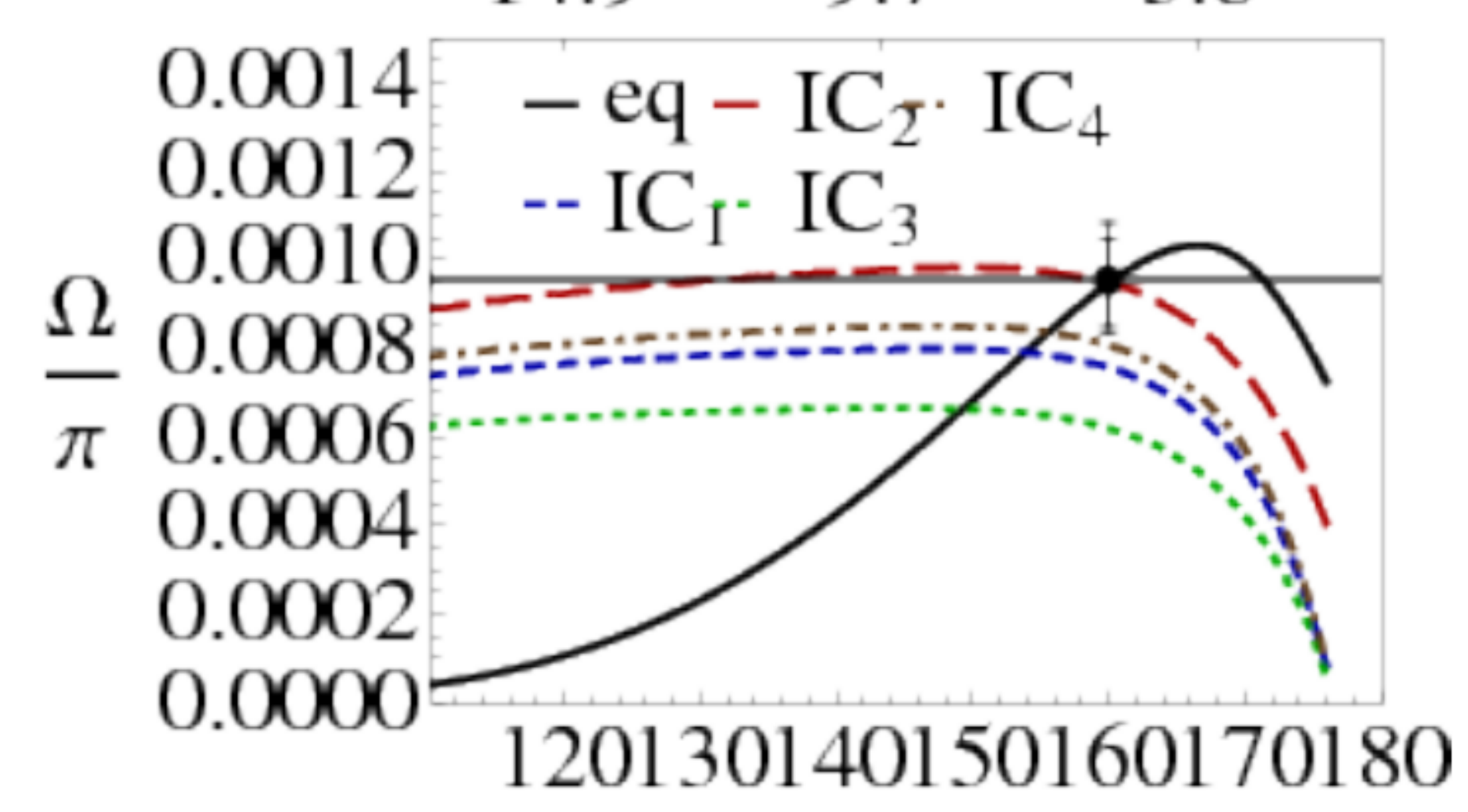}
\end{tabular}

  \caption{$K/\pi$ ratio and $\Omega/\pi$ ratio
	in a cooling, expanding fireball where the reactions are 
	driven by Hagedorn states~\protect\cite{NoronhaHostler:2007jfh,
	NoronhaHostler:2009cfh} compared to RHIC data. }
	\label{fig:dyn}
\end{center}
\end{figure}

In Fig.~\ref{fig:dyn}, the description of these missing states 
from~\cite{NoronhaHostler:2007jfh,NoronhaHostler:2009cfh} is shown 
within a cooling, expanding fireball compared to experimental data 
from RHIC. Because all reactions are dynamical, there is not a set
``freeze-out temperature" but each species is allowed to reach
chemical equilibrium on its own, which varies according to the
description of the Hagedorn Spectrum, branching ratios, and
initial conditions. A summary of the final particle ratios is
shown in Fig.~\ref{fig:sum} compared to LHC experimental particle
ratios.
\begin{figure} [h]
\begin{center}
  \includegraphics[width=0.7\textwidth]{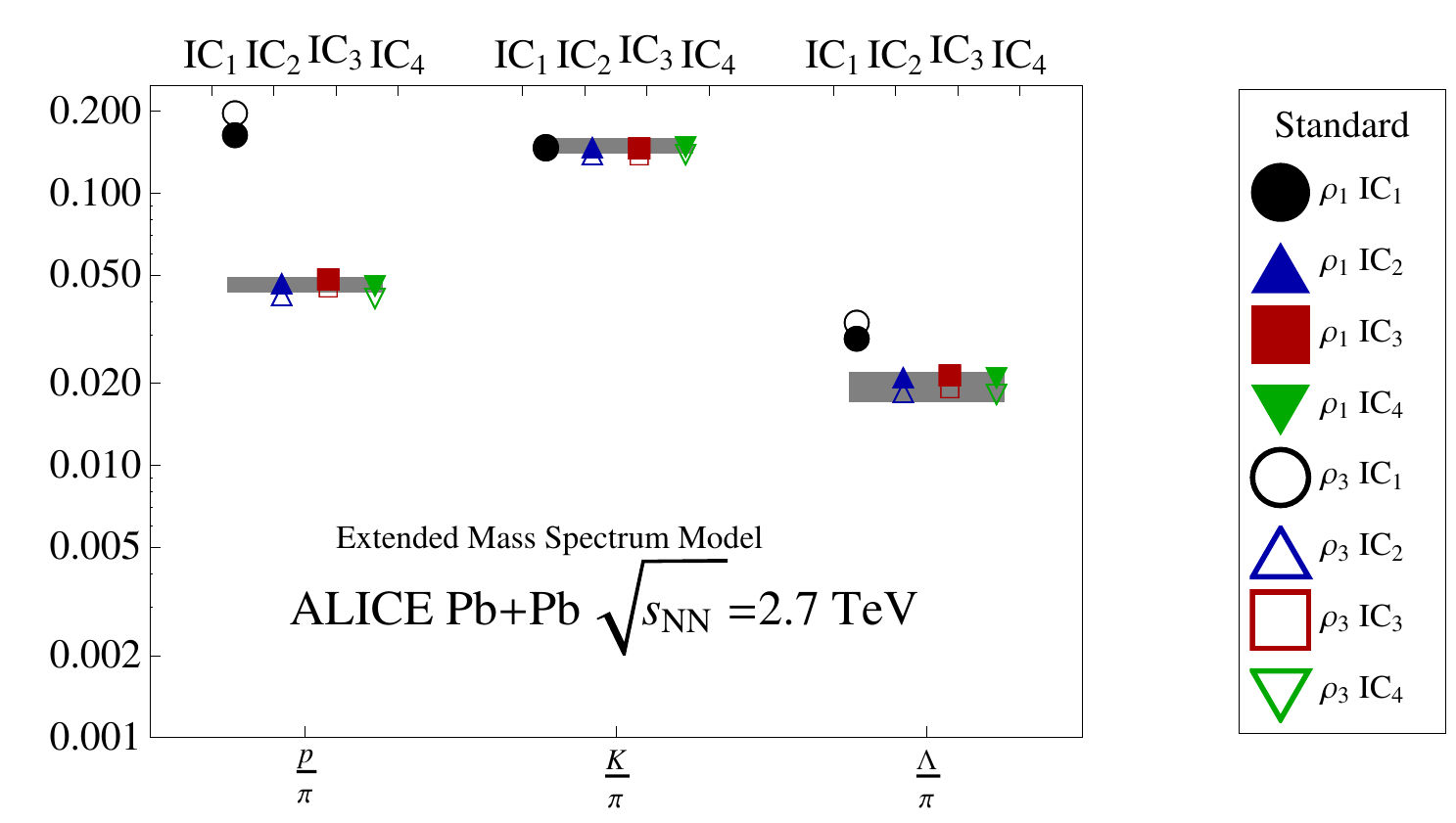}

  \caption{Particle ratios calculated within
	a dynamical multi-body interaction model using Hagedorn 
	states~\protect\cite{NoronhaHostler:2007jfh,
	NoronhaHostler:2009cfh} compared to ALICE data.}
	\label{fig:sum}
\end{center}
\end{figure}

Further studies have investigated if these missing resonances
could affect the freeze-out temperature of 
hadrons~\cite{NoronhaHostler:2009tzh,Bazavov:2014xyah}, i.e., at 
what temperature hadrons reach chemical equilibrium. However, it
is unlikely that additional states beyond the PDG 2004/2016
will strongly affect the freeze-out 
temperature~\cite{Noronha-Hostler:2016sjeh}.  Additionally, they 
were proposed as a solution for the tension between strange and
light particles yields in theoretical models~\cite{Bazavov:2014xyah,
Noronha-Hostler:2014usah,Noronha-Hostler:2014aiah} though another
likely solution may be a difference between the light and strange
freeze-out temperatures~\cite{Bellwied:2013ctah,
Noronha-Hostler:2016rpdh}.

\item \textbf{Transport Coefficients and Hydrodynamics}

One of the most significant findings in heavy-ion collisions is
that the Quark Gluon Plasma is a nearly perfect fluid (extremely
small viscous effects), which can be well described by relativistic
viscous hydrodynamics (for a review, see~\cite{Heinz:2013thh}).
Adding in missing resonances helped to explain the extremely
low shear viscosity to entropy density ratio expected at the
phase transition between the hadron gas phase and the Quark
Gluon Plasma~\cite{NoronhaHostler:2008juh,NoronhaHostler:2012ugh}
as shown  on the left in Fig.~\ref{fig:trans}. This calculation
also gave support to the presence of a peak in the bulk 
viscosity~\cite{Karsch:2007jch} at the phase transition region, 
as shown on the right in Fig.~\ref{fig:trans}.
\begin{figure} [h]
\begin{center}
\begin{tabular}{c c}
  \includegraphics[width=0.45\textwidth]{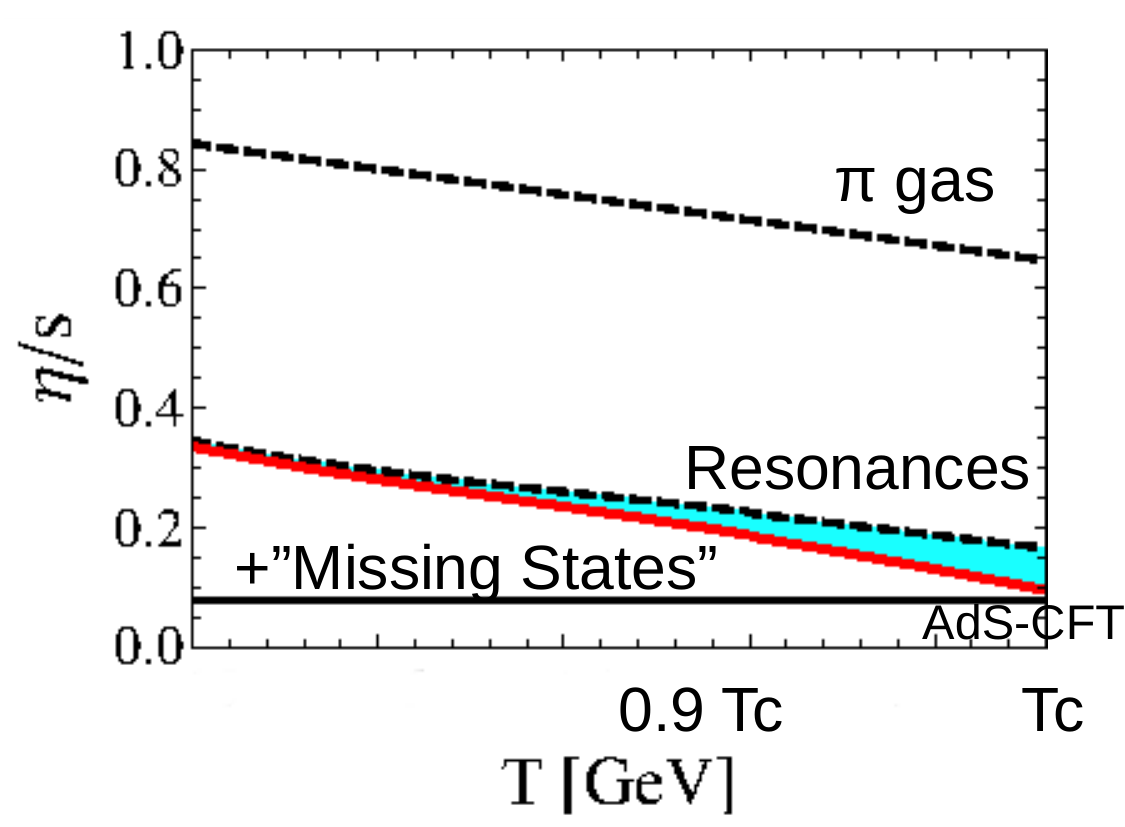} &
  \includegraphics[width=0.45\textwidth]{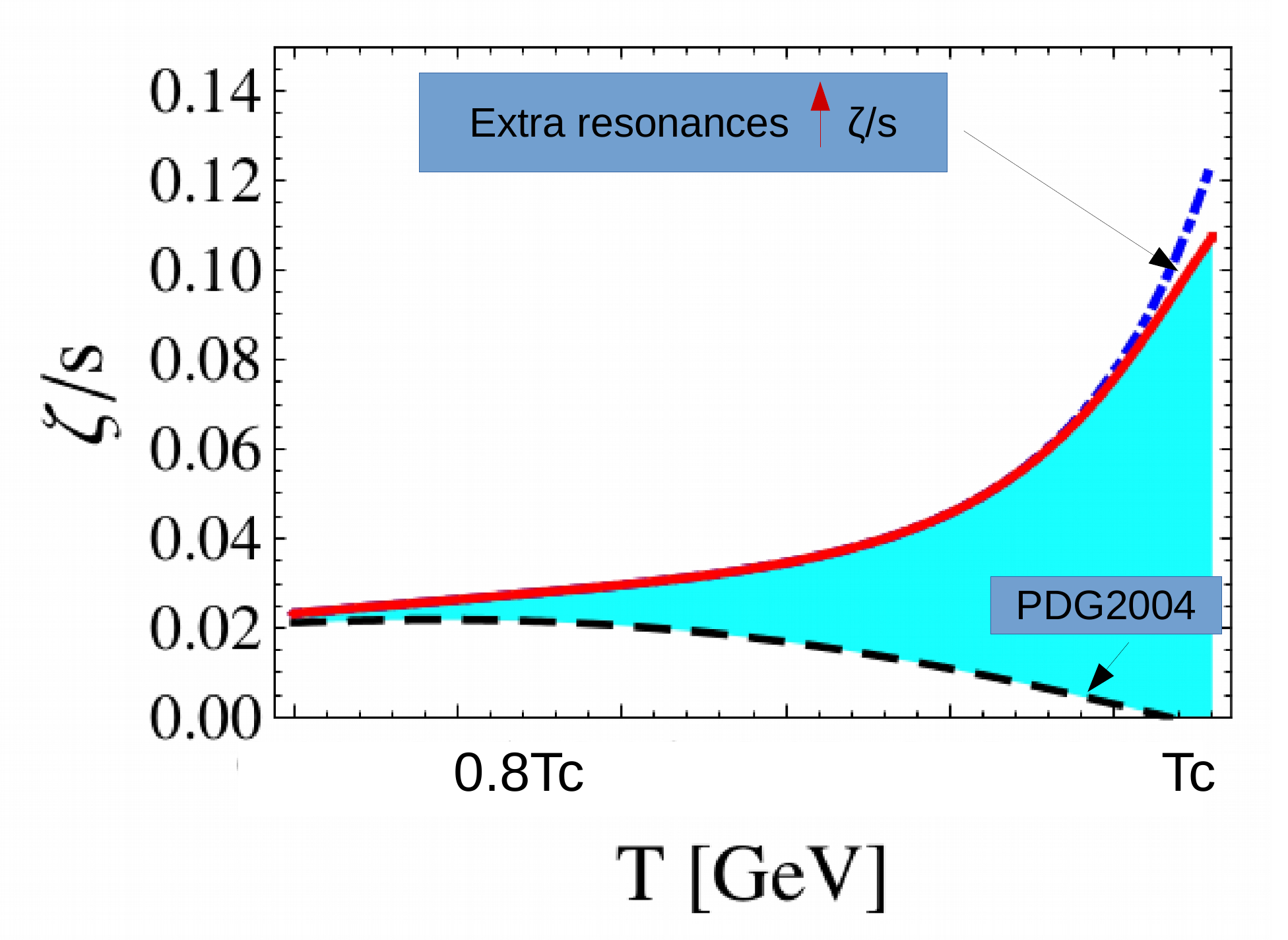}
\end{tabular}

  \caption{The effect of adding in missing
	resonances to the shear viscosity to entropy density 
	ratio (left) and bulk viscosity to entropy density 
	ratio (right) within the hadron gas 
	phase~\protect\cite{NoronhaHostler:2008juh}.}
	\label{fig:trans}
\end{center}
\end{figure}

The major signature of perfect fluidity in heavy-ion collisions
is known as elliptical flow, $v_2$.  Elliptical flow is the
second Fourier coefficient of the particle spectra and indicates
that there is a strong (spatial) elliptical shape in the initial
conditions, which is turned into an elliptical shape in momentum
space due to the Quark Gluon Plasma's nearly perfect fluid-like
nature.  If the Quark Gluon Plasma did not act as a fluid or if
it was very viscous than $v_2\sim 0$.
\begin{figure} [h]
\begin{center}
\begin{tabular}{c c}
  \includegraphics[width=0.4\textwidth]{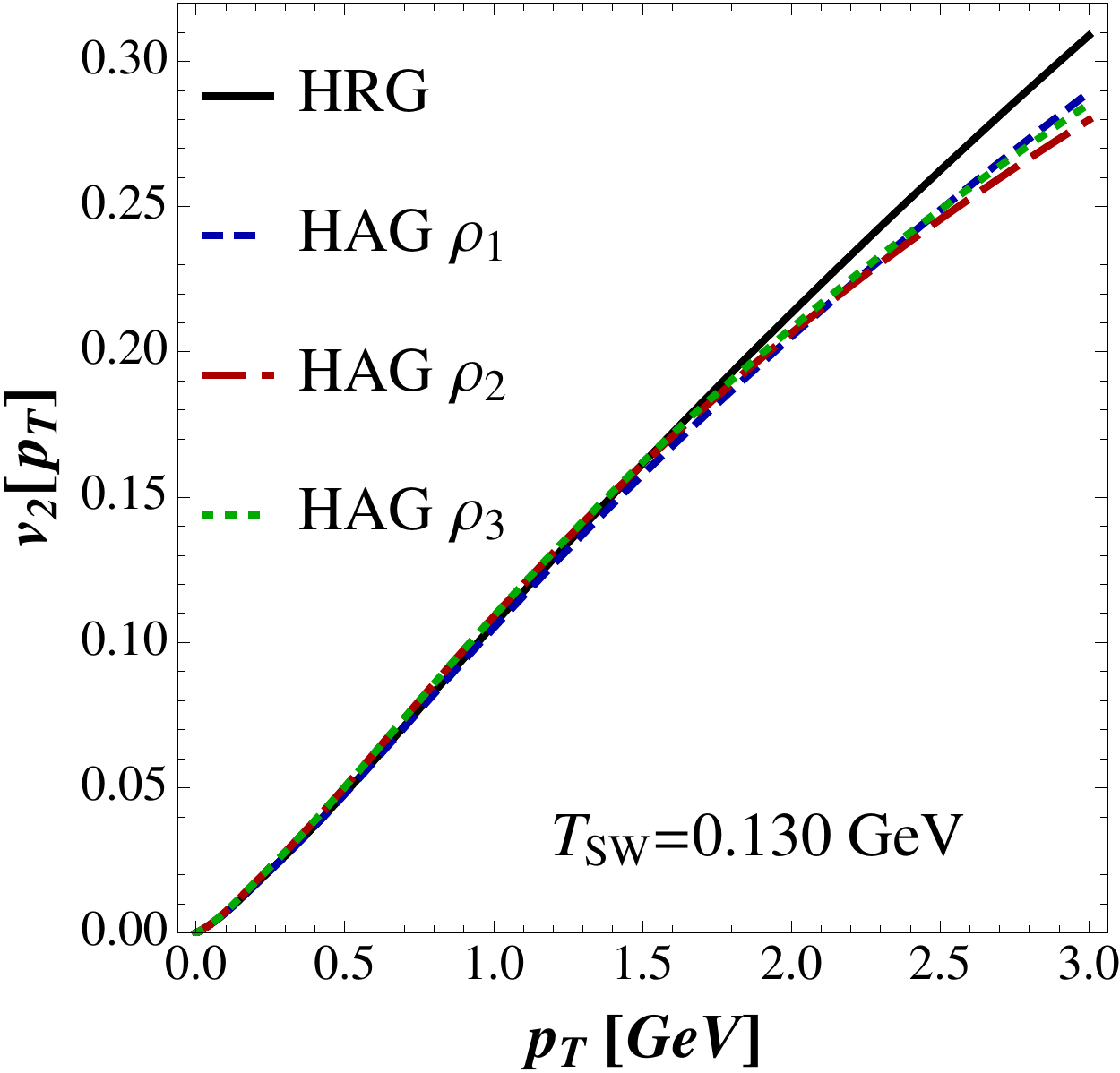} &
  \includegraphics[width=0.4\textwidth]{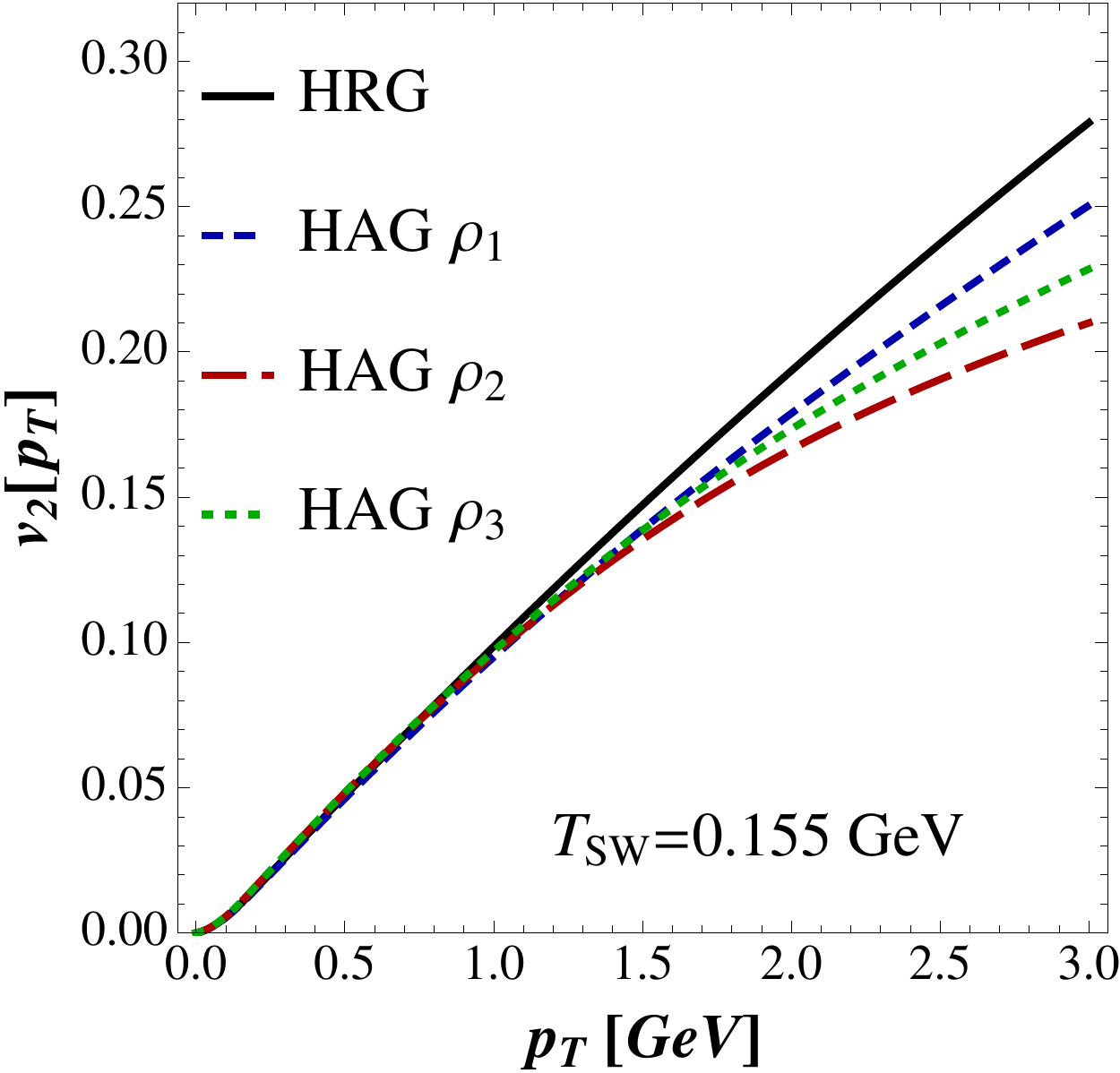}
\end{tabular}

  \caption{Effect of different descriptions
	of Hagedorn states on the elliptical flow at 
	$T_{SW}=130$~MeV (left) and $T_{SW}=155$~MeV (right) 
	from~\protect\cite{Noronha-Hostler:2013rcwh}. Solid 
	line includes all PDG2014 resonances.} \label{fig:v2}
\end{center}
\end{figure}

In Ref.~\cite{Noronha-Hostler:2013rcwh}, the effects of adding in
missing states were studied on $v_2$. For high momenta and higher 
freeze-out temperatures extra resonances suppressed elliptical flow. 
Thus, missing states not included in hydrodynamical models introduce 
a systematic error in comparisons of $v_2$ to experimental data.  
Recently, the hadron gas phase was shown to play a significant
role in photon production~\cite{Paquet:2015ltah} though theoretical 
calculations still underpredict experimental photon yields. It would 
be interesting to see if effects from missing resonances play a role 
in solving the photon ``puzzle".

\item \textbf{Conclusions}

In conclusion, missing resonances can affect various aspects of
heavy-ion collisions.  The dynamics of the hadron gas phase plays
a crucial role in comparisons between theoretical models and
experimental data. If there are missing resonances, they contribute
as a source of systematic error in all theoretical calculations.
Using an exponential mass spectra, there may be a gap in the strange
mesons between $m=1-1.5$~GeV, which is a potential region to search
for new resonances.  Therefore, strange hadron observables may be
especially sensitive to missing states.

Due to the sensitivity of heavy-ion collisions to missing resonances,
a strong collaboration between the fields of hadron spectroscopy and
heavy-ion collisions could be extremely fruitful.  For instance, the
creation of a database that included all the known PDG resonances
(and eventually also states predicted by Lattice QCD or Quark Models)
with all their characteristic information (degeneracy, mass, quantum
numbers, decay channels, etc) sorted by their star rating in a format
that is compatible with heavy-ion codes, would significantly speed up
comparisons between the fields and allow for systematic checks on
which resonances influence heavy-ion observables the most.

\item \textbf{Acknowledgments}

This  work is supported by the National Science Foundation under
grant no. PHY-1513864.
\end{enumerate}


\newpage
\subsection{Thermal Shifts, Fluctuations, and Missing States}
\addtocontents{toc}{\hspace{2cm}{\sl E.~Ruiz~Arriola, W.~Broniowski,
	E.~Meg\'{\i}as, and L.~L.~Salcedo}\par}
\setcounter{figure}{0}
\setcounter{table}{0}
\setcounter{equation}{0}
\setcounter{footnote}{0}
\halign{#\hfil&\quad#\hfil\cr
\large{Enrique~Ruiz~Arriola}\cr
\textit{Departamento de F\'{\i}sica At\'omica, Molecular y Nuclear and}\cr
\textit{Instituto Carlos I de F\'{\i}sica Te\'orica y Computacional}\cr
\textit{Universidad de Granada}\cr
\textit{E-18071 Granada, Spain}\cr\cr
\large{Wojciech~Broniowski}\cr
\textit{Institute of Physics, Jan Kochanowski University}\cr
\textit{25-406 Kielce, Poland \&}\cr
\textit{The H. Niewodnicza\'nski Institute of Nuclear Physics}\cr
\textit{Polish Academy of Sciences}\cr
\textit{PL-31342 Cracow, Poland}\cr\cr
\large{Eugenio~Meg\'{\i}as}\cr
\textit{Max-Planck-Institut f\"ur Physik (Werner-Heisenberg-Institut)}\cr
\textit{D-80805 Munich, Germany \&}\cr
\textit{Departamento de F\'{\i}sica Te\'orica}\cr
\textit{Universidad del Pa\'{\i}s Vasco UPV/EHU, Apartado 644}\cr
\textit{48080 Bilbao, Spain}\cr\cr
\large{Lorenzo~L.~Salcedo}\cr
\textit{Departamento de F\'{\i}sica At\'omica, Molecular y Nuclear and}\cr
\textit{Instituto Carlos I de F\'{\i}sica Te\'orica y Computacional}\cr
\textit{Universidad de Granada}\cr
\textit{E-18071 Granada, Spain}\cr}

\begin{abstract}
Thermal shifts and fluctuations at finite temperature below the
deconfinement crossover from hadronic matter to the quark-gluon plasma
provide a viable way to search for missing states with given quantum
numbers in the hadronic spectrum. We analyze three realizations of the
hadron resonance gas model in the light quark ($uds$) sector: the
states from the Particle Data Group tables with or without width and
from the Relativized Quark Model. We elaborate on the meaning of
hadronic completeness and thermodynamical equivalence on the light of
lattice QCD trace anomaly, heavy quark entropy shift and
baryon,charge and strangeness susceptibilities.
\end{abstract}

\begin{enumerate}
\item \textbf{Introduction}

The concept of missing states in QCD is intimately related to the
completeness of the hadronic spectrum.  The issue was anticipated by
Hagedorn in the mid 60's~\cite{Hagedorn:1965stz} when analyzing the
mass-level density $\rho(M)$ and {\it predicting} the bulk of states
at higher masses, which later on were experimentally confirmed.  This
also implies that the states may be {\it counted} one by one (and
hence ordered) by, say, the cumulative number of states function,
\begin{eqnarray}
	N(M) = \sum_n g_n \theta (M-M_n) \, ,
	\label{eq.count}
\end{eqnarray}
with $g_n$ the total degeneracy and $\rho(M)=dN(M)/dM$.  Updated
analyses of the Hagedorn hypothesis may be found
in~\cite{Broniowski:2000bjz,Broniowski:2004yhz}. The function $N(M)$
assumes integer values, and the best mass resolution is $\Delta M =
\min_n (M_{n+1}-M_n)$. For bound states, where the spectrum is
discrete, this is a well defined procedure. In the continuum, this can
only be done by putting the system in a box with finite but
sufficiently large volume which acts as an infrared cut-off $ V^{1/3}
\Delta M \gg 1$. The ultraviolet cut-off is the maximum mass $M_{\rm
max}$ in Eq.~(\ref{eq.count}).

The commonly accepted reference for hadronic states is the Particle
Data Group (PDG) table~\cite{Olive:2016xmwz}, a compilation reflecting
a consensus in the particle physics community and which grades states
*,**,***, and ****, according to the growing confidence in their
existence, respectively.  Global features of the hadronic spectrum may
depend on whether we decide to promote or demote their significance,
according to some theoretical prejudice. Of course, we expect the PDG
hadronic states to have a one-to-one correspondence with colour
neutral eigenstates of the QCD Hamiltonian; indeed, ground and some
excited states have been determined on the
lattice~\cite{Edwards:2012fxz}. For hadronic states with only light
($uds$) quarks the maximum mass, $M_{\rm max}$, recorded by PDG is
around $2.5~{\rm GeV}$ for mesons and baryons, hence currently
$N(M_{\rm max}) \sim 2.5 \times 10^3$. So far, the states listed by
PDG echo the standard quark model classification for mesons ($\bar q
q$) and baryons ($qqq$). Because of this feature, it will be pertinent
to consider also the Relativized Quark Model (RQM) for
mesons~\cite{Godfrey:1985xjz} and Baryons~\cite{Capstick:1986bmz}, as
first done for $N(M)$ in~\cite{Broniowski:2000hdz}~(Fig.~9).  The
remarkable coincidence $N_{\rm PDG} (M) \sim N_{\rm RQM} (M)$ up to
$M_{\rm max}$ for {\it bot} mesons and baryons has been shown in
Ref.~\cite{Arriola:2014bfaz}. The so-called ``further states'' may or
may not be confirmed or expected and have not been clearly regarded by
the PDG as identified, although they could be exotic tetraquarks,
$\bar q q \bar q q $, pentaquarks, $\bar q qqqq$, glueballs $gg$,
$ggg$ or hybrids $\bar q qg$~\cite{Briceno:2015rltz}.

In this contribution we analyze thermodynamic measures (various
susceptibilities) which are sensitive to missing states.  The setup
corresponds to heating up the vacuum without dissolving its
constituent hadrons into quarks and gluons and testing quark-hadron
duality at temperatures below $T_c \sim 150~{\rm MeV}$.  Obviously,
such a framework is inefficient for individual states, but becomes
competitive if globally a relatively large number of states are
missing.  As reported in~\cite{Majumder:2010ikz,Broniowski:2016hvtz} the
Hagedorn conjectured behaviour of $N(M) \sim A e^{M/T_H}$ for $M >
M_{\rm max}$ may influence the results close to $T_c$, at temperatures
above $T>140$~MeV.  According to~\cite{Broniowski:2016hvtz}, there is
not much room for such states in the one-body observables, where they
would spoil the agreement with the lattice data, unless suitable
repulsion between states is simultaneously incorporated.  Here will
make no attempt to complete the spectrum beyond $M_{\rm max}$.

\item \textbf{Prehistory of Missing States}

The use of thermodynamical arguments to characterize the existence of
missing states is a rather old subject which goes back to the early
beginnings of the kinetic theory of gases and the equipartition
theorem. In its most general form it states that every degree of
freedom contributes to the mean energy with $\frac12 k_B
T$.\footnote{The story around this principle illustrates many of the
	issues under discussion, including the contribution of an 
	anonymous referee~\cite{brush1967foundationsz} 
	D.~Bernoulli~\cite{bernoulli1738hydrodynamicaz} was the first 
	who found in 1738 that the Boyle-Mariotte, Gay-Lussac, and 
	Charles equations could be unified and understood by means 
	of Newton's equations and in statistical terms. His work was 
	forgotten, and only in 1845 J.~J.~Waterston submitted a paper 
	to the Philosophical Transactions of the Royal Society (PTRS) 
	which was rejected with a remark ``The paper is nothing but 
	nonsense, unfit even for reading before the Society''. Hence 
	this work was also ignored. Maxwell, in 1859, managed to 
	publish the case of rigid molecules, and Boltzmann
	generalized it in 1868 to its modern form including rotational 
	and vibrational degrees of freedom. Lord Rayleigh in 1895, 
	found by chance Waterston's paper in the archives and decided 
	to publish it in PTRS twelve years after Waterston's death 
	with a commentary: ``had he put forward his investigation as 
	a development of Bernoulli a referee migh have hesitated to 
	call it nonsense. It is probable that Waterston was 
	unaquainted with his work.''.}.  
Therefore $\bar E = N \nu k_B T/2$, where $\nu$ is the total number 
of degrees of freedom.  Generally, $\nu = \nu_{\rm translation}+ 
\nu_{\rm vibration}+ \nu_{\rm rotation} $ and the molar specific 
heat is $c_V/R= \nu/2 $. A major obstacle at the time was pointed 
out by J.~C.~Maxwell in 1860 in connection to the specific heat of the
diatomic gas such as, e.g., $H_2$, where a priori the total number of
degrees of freedom is $\nu= 3_{\rm trans}+ 2_{\rm vib}+ 2_{\rm rot} =
7$. This would imply $c_V/R = 7/2$, whereas experimentally at room
temperature one has $c_V/R \approx 5/2 $.  This is because the
vibrational degrees of freedom are not active due to high excitation
energy, and become visible only as the dissociation temperature of
$\sim 3200$~K is approached. Likewise, as $T$ is decreased, the
rotational degrees of freedom are also frozen and below $\sim 70$~K,
$c_V/R = 3/2$, as for the monoatomic molecules.

In modern terms the ``freezing'' of degrees of freedom is related to
the quantization of energy levels for the Hamiltonian $H \Psi_n = E_n
\Psi_n$ with energy eigenvalues above the temperature, $ E_n > T$,
contributing negligibly to the partition function
\begin{eqnarray}
	Z= {\rm Tr} e^{-H/T}= \sum_n e^{-E_n/T}. \label{eq:Z}
\end{eqnarray}
In QCD, the quantized energy levels are the masses of the existing
hadronic states and, like in the Maxwell argument, the states which 
are not activated when $M_n > T$ do not contribute.

\item \textbf{Completeness of the Hadron Spectrum}

Completeness of the listed PDG states~\cite{Olive:2016xmwz} is a
subtle issue.  On the one hand they are mapped into the $\bar q q $
and $qqq$ quark model states. On the other hand, most reported states
are not stable particles but resonances produced
as intermediate steps in a scattering process.

With a finite lifetime $\tau_R$, they are characterized by a mass
distribution $\rho_R (M)$, with a central value $M_R$ and a width
$\Gamma_R \sim \hbar/\tau_R$.  From a rigorous point of view
resonances are poles of the {\it exact} amplitude in the second
Riemann sheet in the complex $s$ plane at $s = M^2 - i M \Gamma$. For
multichannel scattering with $N$ channels one has $2^N$ Riemann
sheets, depending on which cuts have been crossed (see, 
e.g.,~\cite{Nieves:2001wtz,GarciaRecio:2002tdz} for discussions in the
meson-baryon $S=0,-1$ sectors). Despite the rigor of these
definitions, complex energies are not directly measured. An analytic
continuation of a phenomenological and approximate scattering
amplitude, taking into account a process dependent background, is
needed and the arbitrariness grows with the width of the
resonance~\cite{Ciulli:1975smz} (see, e.g., for the specific $0^{++}$
case~\cite{Caprini:2016uxyz}). On average, most of the resonances
listed by PDG~\cite{Olive:2016xmwz} can be regarded as narrow, since
one finds $ \langle \Gamma_R/M_R \rangle = 0.12(8) $ {\it both} for
mesons and baryons~\cite{Arriola:2011enz,Masjuan:2012gcz}, a fact
numerically consistent with the large $N_c$ theoretical expectation
$\Gamma_R/M_R = {\cal O} (N_c^{-1})$~\cite{Witten:1979khz}. In the
Hamiltonian picture, resonances are identified as the so-called Gamow
states and are not normalizable in the usual Hilbert space, as they
are not conventional irreducible representations of the Poincar\`e
group~\cite{Bohm:2004ziz}. The completeness relation involves bound
states and the continuum, which can be rewritten as a discrete sum of
the Gamow states and a remainder~\cite{Berggren:1968zzz}.

The meaning of completeness is fairly clear within a given Hilbert
space ${\cal H}$ with specified degrees of freedom when only
bound states are possible. For instance, if we restrict ourselves 
to the meson $(\bar q q)$ or baryon $(qqq)$ sectors, such as in
RQM~\cite{Godfrey:1985xjz,Capstick:1986bmz}, we can diagonalize the
$\bar q q$ and $qqq$ Hamiltonians with confining potentials in a 
given already complete basis, which is truncated but large enough 
that states with $M_n \le M_{\rm max}$ converge. Thus we write
\begin{eqnarray}
	{\cal H}_{\rm RQM} = {\cal H}_{\bar q q } \oplus {\cal H}_{qqq} \oplus
	{\cal H}_{\bar q \bar q \bar q}
\end{eqnarray}
Within this framework, hadrons are stable, extended, and composite
particles. This is explicitly illustrated by the virial relations in
the massless quark limit~\cite{Arriola:2014bfaz} $ M_{\bar q q} = 2
\sigma \langle r \rangle_{\bar q q} $ and $ M_{qqq} = N_c \sigma
\langle r \rangle_{qq} $, which shows that hadrons are larger the
heavier they become. Many of these states may decay by strong
processes, such as $\rho \to 2 \pi$ or $\Delta \to N \pi$, where
a coupling to the continuum is needed by incorporating the ${\cal
  H}_{\bar q q \bar q q } $ and ${\cal H}_{\bar q q qqq}$ Fock state.
As a result, the pole mass is shifted into the complex plane $ M \to M +
\Delta M - i \Gamma/2 $. The mass-shift $\Delta M \sim \Gamma$ depends
parametrically on the coupling to the continuum $\Delta M \sim \Gamma$
so that in the large $N_c$ limit, $\Delta M /M = {\cal O}
(N_c^{-1})$~\cite{Masjuan:2012skz}.

On the lattice, hadrons are constructed as interpolating fields in a
finite-volume box. Completeness proceeds along similar lines, with the
important modification that resonances are characterized by 
volume-independent and real mass shifts. The connection to physical
resonances in the complex energy plane requires also analytical
extrapolation (for a review see, e.g.,~\cite{Briceno:2014tqaz}).

\item \textbf{Thermodynamic Equivalence}

Be it the PDG~\cite{Olive:2016xmwz}, RQM~\cite{Godfrey:1985xjz,
Capstick:1986bmz}, or the lattice excited QCD~\cite{Edwards:2012fxz}, 
the partition function can be constructed from the (complete) energy 
localized colour neutral eigenstates, Eq.~(\ref{eq:Z}). The lattice 
at finite temperatures, or the ultrarelativistic heavy ions collisions, 
generate global colour neutral configurations which along the crossover 
are expected to delocalize. Most of the emerging physical quark-hadron 
duality picture has to do with the thermodynamical equivalence of 
different approaches.

According to the quantum virial expansion~\cite{Dashen:1969epz} one can
compute the partition function from the knowledge of the $S$-matrix in
the complete Hilbert space, i.e., involving all possible processes
with any number of elementary particles in both the initial and final
states, $n \to m$. In practice, hadrons have been taken as the
building blocks in this approach, which for obvious practical reasons
has never been taken beyond the $2 \to 2$ reactions, where the
corresponding phase shifts are involved. In the case of narrow
resonances one can replace the total contribution entering in terms of
phase shifts by the resonance itself~\cite{Dashen:1974jwz}, whereby the
resonance can be assumed to be elementary and
point-like~\cite{Dashen:1974yyz}. The result conforms to the Hadron
Resonance Gas (HRG) as initially proposed by
Hagedorn~\cite{Hagedorn:1965stz}.  This provides the formal basis for
modern HRG calculations using the PDG compilation. As mentioned above,
most states entering the HRG are resonances with a given width,
$\Gamma$. Therefore we will also consider the effect of smearing the
mass distribution according to the replacement
\begin{eqnarray}
	\sum_R F(m_R^2) \to \int d\mu^2 F(\mu^2) \Delta_\Gamma ( \mu^2
	-m_R^2)
\end{eqnarray}
for an observable $F(\mu^2)$.
\footnote{Ideally the profile function should be determined from the 
	scattering phase-shift~\cite{Dashen:1974jwz}, which displays 
	cancellations~\cite{Venugopalan:1992hyz,Broniowski:2015ohaz} 
	and irrelevance of some weakly bound 
	states~\cite{Arriola:2015graz} but it is not always available. 
	Here we take a simple normalized Gaussian profile distribution. 
	A Breit-Wigner representation works well around the resonance, 
	but it has very long tails which do not faithfully represent 
	the background. An upper bound for the error is to use the 
	half-width rule~\cite{Arriola:2011enz,Arriola:2012vkz} according 
	to which PDG masses are varied within half the width, i.e., 
	taking $M_R \pm\Gamma_R/2$. We do not use this large $N_c$ 
	motivated prescription here as we feel that it largely 
	overestimates the uncertainties for {\it all resonances} in 
	the $N_c=3$ world.}

However, the elementary constituents are both quarks and gluons. A
different derivation proceeds along chiral quark-gluon models with a
quantum and local Polyakov loop~\cite{Megias:2004hjz,Megias:2006bnz}.
\footnote{This is unlike the more popular PNJL model~\cite{Fukushima:2003fwz,
	Ratti:2005jhz}, where the quantum and local nature of $\Omega 
	(\vec x) $ is ignored, thus introducing an undesirable group 
	coordinates dependence. In addition, in PNJL the Polyakov loop 
	in the adjoint representation is not quenched, contradicting 
	lattice calculations.} 
The action corresponds to creating, e.g., a quark at location $\vec x$ 
and momentum $\vec p$ in the medium
\begin{eqnarray}
	e^{-E(\vec p)/T} \Omega(\vec x)^\dagger,
\end{eqnarray}
where in the static gauge $\Omega(\vec x)= e^{i g A_4 (\vec x)/T}$. 
Consequently, the total action can be separated into different quark 
and gluon sectors according to the low temperature partonic expansion 
around the vacuum~\cite{RuizArriola:2012wdz,Arriola:2014bfaz}
\begin{eqnarray}
	Z= Z_0+ Z_{\bar q q }+ Z_{qqq} + Z_{\bar q \bar q \bar q} + \dots\sim
	Z_{\rm RQM}.
\end{eqnarray}
Subsequent hadronization of $\bar q q$ and $qqq$ states uses the
cluster properties of the Polyakov loop correlator and group
properties of the Haar measure, as well as the quantum, composite and
extended nature of hadronic states. One appealing feature of this
``microscopic'' derivation of the HRG is the counting of states
according to the quark model for the lowest Fock state components, but
ambiguities arise when a given colour neutral multiquark state admits
a separation into colour neutral irreducible subsystems~\cite{Arriola:2014bfaz,
Megias:2013xaaz}. We take this result as our justification to use RQM.
\footnote{Bound state masses are shifted when coupled to the continuum, 
	so if we take a simple average estimate $\langle \Delta M/M 
	\rangle_{\rm RQM}\sim\langle\Gamma/M \rangle_{\rm PDG}\sim 0.12(8)$.  
	This roughly corresponds to take $5\%-20\%$ uncertainty in $T$.}

The fact that we use thermodynamic quantities to make a quantitative
comparison does not sidestep the problem of discriminating different
spectra. The best example is provided by a direct comparison of
HRG using either PDG or RQM~\cite{Godfrey:1985xjz,Capstick:1986bmz} in 
terms of the trace anomaly, ${\cal A}(T) \equiv (\epsilon-3 P)/T^4$ 
which are hardly indistinguishable within the lattice QCD uncertainties 
from the WB~\cite{Borsanyi:2013biaz} and HotQCD~\cite{Bazavov:2014pvzz}
collaborations (see Fig.~1 of Ref.~\cite{Arriola:2014bfaz}). As already 
mentioned the states with $M > M_{\rm max}$ with an exponential Hagedorn 
distribution are relevant below $T_c$~\cite{Broniowski:2016hvtz} only at 
$T >140$~MeV, and their contribution may be overcome with repulsive 
effects. Actually, the volume effects are expected to play a significant 
role; the excluded volume exceeds the total volume around $T 
\lesssim T_c$ (see Fig.~9 of Ref.~\cite{Arriola:2014bfaz}).

\item \textbf{Thermal Shifts}

The idea of thermal shifts is to study the change of thermodynamic
quantities under the presence of local external sources. This looks
very much like adding an impurity to a macroscopic system or adding a
grain of salt to a bunch of snow. By looking a these thermal shift we
may also assess a possible existence of missing states.  An 
interesting hadronic example is provided by the free energy shift
caused by a heavy quark placed in a hot medium with vacuum quantum
numbers, which corresponds to a ratio of partition functions which 
can be identified with the Polyakov loop expectation value. This free
energy shift is ambiguous and hence it is better to deal with the 
corresponding entropy shift and the specific heat, which are directly 
measurable quantities. A hadronic representation of Polyakov loop and 
its entropy has been analyzed~\cite{Megias:2012kbz,Megias:2012hkz}. The 
implications of thermal shifts due to a heavy source or a heavy $Q 
\bar Q$ pair located at a fixed distance $r$ at the hadronic level 
has recently been considered in~\cite{Megias:2016bhkz,
RuizArriola:2016iftz,Megias:2016onbz}.

\begin{figure}[tbc]
\begin{center}
 \epsfig{figure=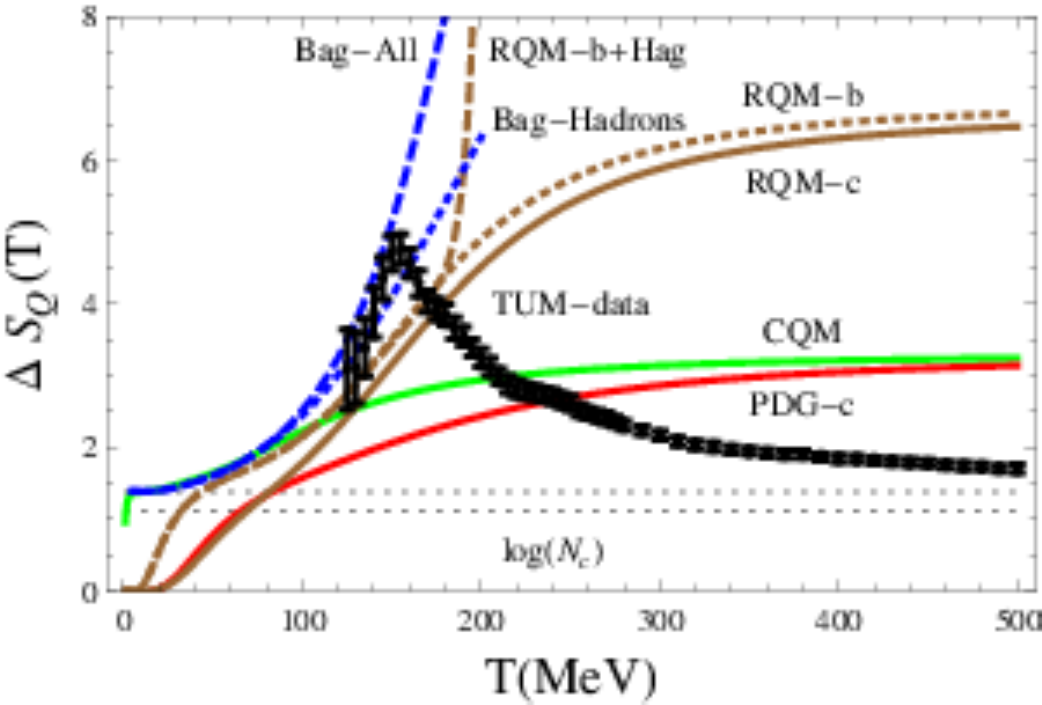,height=7cm,width=9cm}

\caption{The entropy as a function of the temperature.  We show
	results from various hadronic models: the bag model 
	including all ($Q\bar q$, $Q qq$ and $Q\bar qg$) states 
	and just hadrons, the RQM with one $c$- or $b$-quark and 
	the PDG states with one $c$-quark. The Hagedorn 
	extrapolation of the $b$-spectrum is also displayed. We 
	also plot the CQM with $uds$ quarks and constituent
	mass $M=300~MeV$ and the bare $m_u=2.5~MeV $, $m_d= 
	5~MeV$, $m_s= 95~MeV$ masses. Horizontal lines mark 
	$\Delta S_Q(0) = \log 2 N_f$, with $N_f=2$ the number 
	of light degenerate flavours, and $\Delta S_Q(\infty)= 
	\log(N_c)$. Lattice data for 2+1 flavours are taken 
	from Ref.~\protect\cite{Bazavov:2016uvmz}.}
	\label{fig:entropy-models}
\end{center}
\end{figure}

Fig.~\ref{fig:entropy-models} from Ref.~\cite{Megias:2016bhkz} makes a
good case for different categories of missing charm or bottom
states. On the one hand the PDG is clearly insufficient to describe
the entropy shift. So, we clearly miss higher mass states. Guided by
the thermodynamic equivalence of PDG and RQM in the $uds$
sector~\cite{Arriola:2014bfaz}, we may complete the PDG spectrum
using the RQM in the c- or b- sectors. As we see there is a big
improvement and, moreover, the change when going from c to b is
sufficiently small. Nonetheless, we have still missing states, a
feature that is not mended when extending the spectrum a la
Hagedorn. When a Bag model with the heavy source located at the center
is considered for singly heavy hadrons $Q \bar q$, $Q q q$, and a
hybrid $Q \bar q g$ the TUM lattice data are well reproduced.

\item \textbf{Fluctuations}

The connection between fluctuations and the abundance of hadronic
resonances was pointed out by Jeon and Koch~\cite{Jeon:1999grz}, who
later~\cite{Jeon:2000wgz} proposed it as a signal for the Quark-Gluon
Plasma formation  from the partition function (for pedagogical reviews 
see, e.g.,~\cite{Koch:2008iaz,Asakawa:2015ybtz}). Implications for heavy
ion collisions are reviewed in~\cite{Braun-Munzinger:2015hbaz}. 
In Ref.~\cite{Bazavov:2014xyaz}, the event-by-event statistical analysis 
of ultrarelativistic heavy ions-collisions was compared to the HRG with 
a given chemical potential. Of course, any mismatch in this kind of 
analyses suggests missing resonances. Here we are concerned with the 
simplest vacuum zero density case. Actually, some authors have 
understood the significance of fluctuations as a possible hint of 
missing states~\cite{ManLo:2016pgdz}.

Fluctuations of conserved charges, i.e., fulfilling $[Q_A,H]=0$, are a
way of selecting given quantum numbers~\cite{Asakawa:2015ybtz} and
become particularly simple in terms of the grand-canonical partition
function which is given by
\begin{eqnarray}
	Z= {\rm Tr} e^{- (H - \sum_A \mu_A  Q_A )/T} \qquad \Omega 
	= - T \log Z.
\end{eqnarray}
with $\Omega$ the corresponding potential. One then gets
\begin{equation}
	-\frac{\partial \Omega}{\partial \mu_A} = \langle Q_A \rangle_T ,
	\quad
	- T \frac{\partial^2 \Omega}{\partial \mu_A \partial \mu_B} =
	\langle \Delta Q_A \Delta Q_B \rangle_T ,
\end{equation}
where $\Delta Q_A = Q_A - \langle Q_A \rangle_T$. In the $uds$ sector 
the only conserved charges are the electric charge $Q$, the baryon 
charge $B$ and the strangeness $S$, which is equivalent to the number 
of $u$, $d$, and $s$ quarks. We consider the hot vacuum (no chemical 
potential) $ \langle B\rangle_T = \langle Q \rangle_T = \langle S 
\rangle_T = 0$.

For $N_f=2+1$, fluctuations have been computed on the lattice by the
WB~\cite{Borsanyi:2011swz} and HotQCD~\cite{Bazavov:2012jqz}
collaborations with the high temperature asymptotic limits
\begin{eqnarray}
	\chi_{BB}(T) &=& V^{-1}\langle B^2 \rangle_T \to \frac1{N_c} \\
	\chi_{QQ} (T) &=&  V^{-1} \langle Q^2 \rangle_T \to \sum_{i=1}^{N_f}
	q_i^2 \\
	\chi_{SS}(T) &=& V^{-1} \langle S^2 \rangle_T \to 1,
\end{eqnarray}
where $(q_u,q_d,q_s, \dots ) = (2/3,-1/3,-1/3, \dots)$. Higher order
cumulants, such as skewness and kurtosis originally analyzed in
Ref.~\cite{Ejiri:2005wqz}, have also recently been computed more
accurately~\cite{Bellwied:2015lbaz}, but we do not discuss them here.

In the hadron resonance model, the charges are carried by various
species of hadrons, $Q_A = \sum_i q^{(i)}_A N_i$, where $N_i$ is 
the number of hadrons of type $i$, hence
\begin{equation}
	\langle \Delta Q_A  \Delta Q_B \rangle_T
	= \sum_{i,j}  q^{(i)}_A q^{(j)}_B \langle \Delta N_i 
	\Delta N_j \rangle_T .
\end{equation}
The average number of hadrons is
\begin{equation}
	\begin{split}
	\langle N_i \rangle_T &= V \int \frac{d^3 k}{(2\pi)^3} 
	\frac{g_i}{e^{E_{k,i} /T} + \eta_i} \nonumber \\
	&= \frac{VT^3}{2\pi^2} \sum_{n=1}^\infty g_i 
	\frac{(-\eta_i)^{n+1}}{n}
	\left( \frac{M_i}{T} \right)^2 K_2(n M_i/T)
	\end{split}
\end{equation}
where $E_{k,i}= \sqrt{M_i^2 + k^2}$, $g_i$ is the degeneracy and
$\eta_i=\mp 1$ for bosons/fermions respectively. In practice the 
Boltzmann approximation (i.e., just keeping $n=1$) is sufficient. 
Regarding the fluctuations, since the different species are uncorrelated
$\langle \Delta n_\alpha \Delta n_\beta \rangle_T = \delta_{\alpha\beta}
\langle n_\alpha \rangle_T (1 - \eta_\alpha \langle n_\alpha \rangle_T )$,
for the occupation numbers. Since $\langle n_\alpha \rangle_T \ll 1$,
\begin{eqnarray}
	\langle \Delta Q_A \Delta Q_B \rangle_T \approx
	\sum_{i} q_A^{(i)} q_B^{(i)} \langle N_i \rangle_T.
\end{eqnarray}
\begin{figure*}
\begin{center}
\epsfig{figure=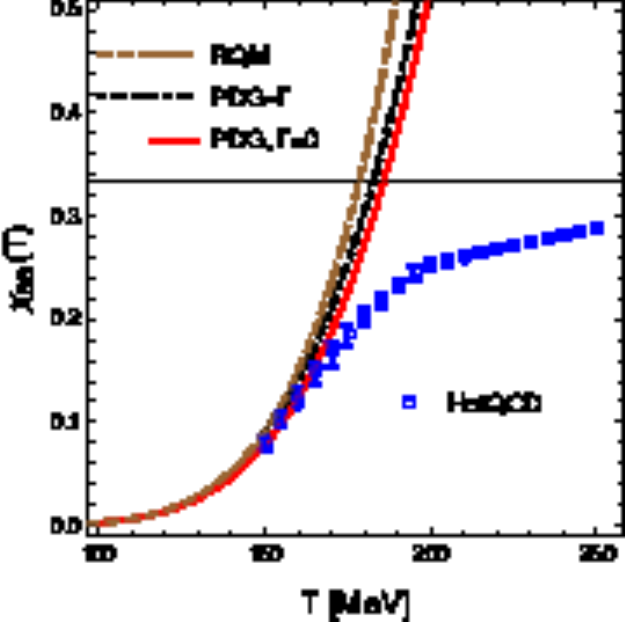,width=5cm,angle=0}
\epsfig{figure=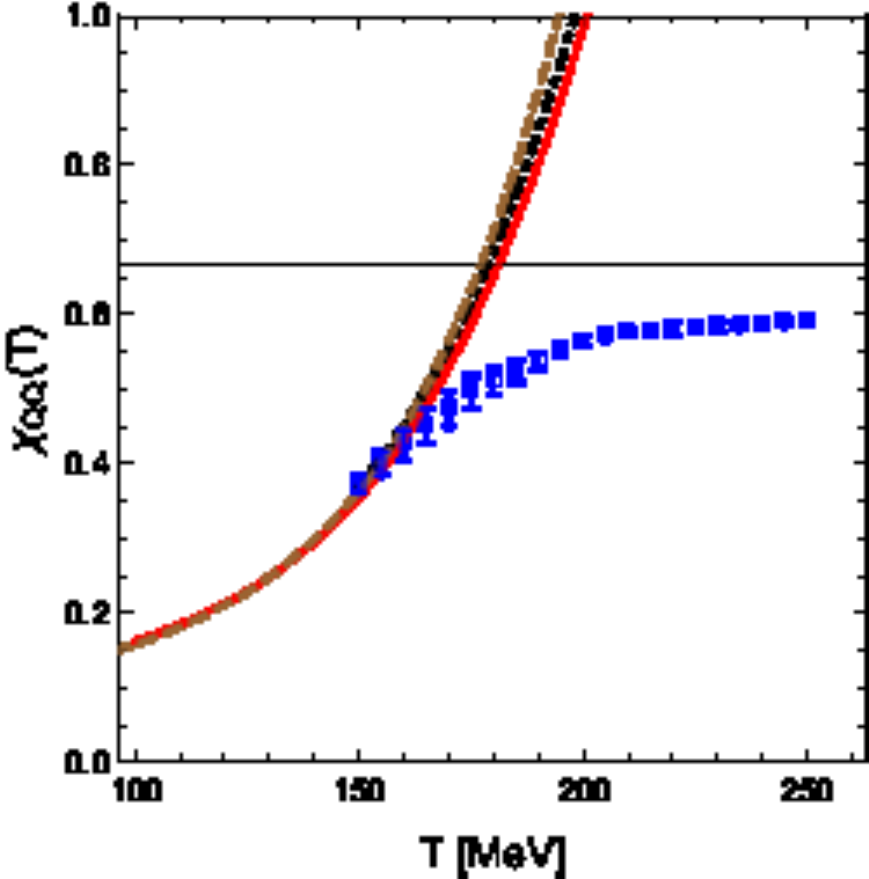,width=5cm,angle=0}
\epsfig{figure=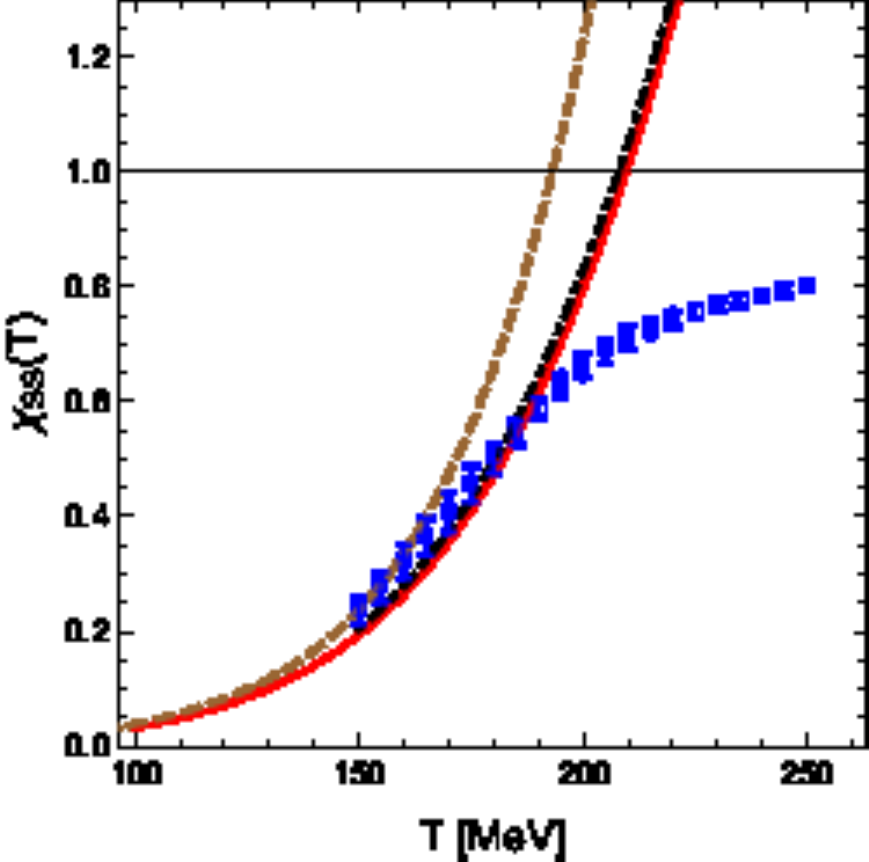,width=5cm,angle=0}\\

\epsfig{figure=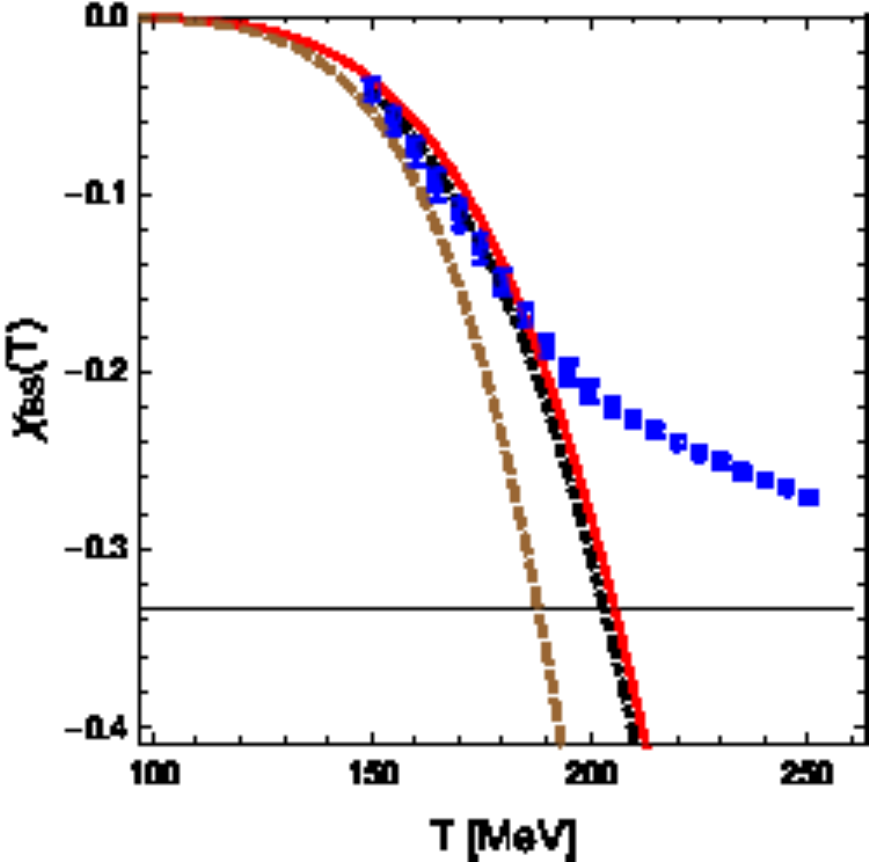,width=5cm,angle=0}
\epsfig{figure=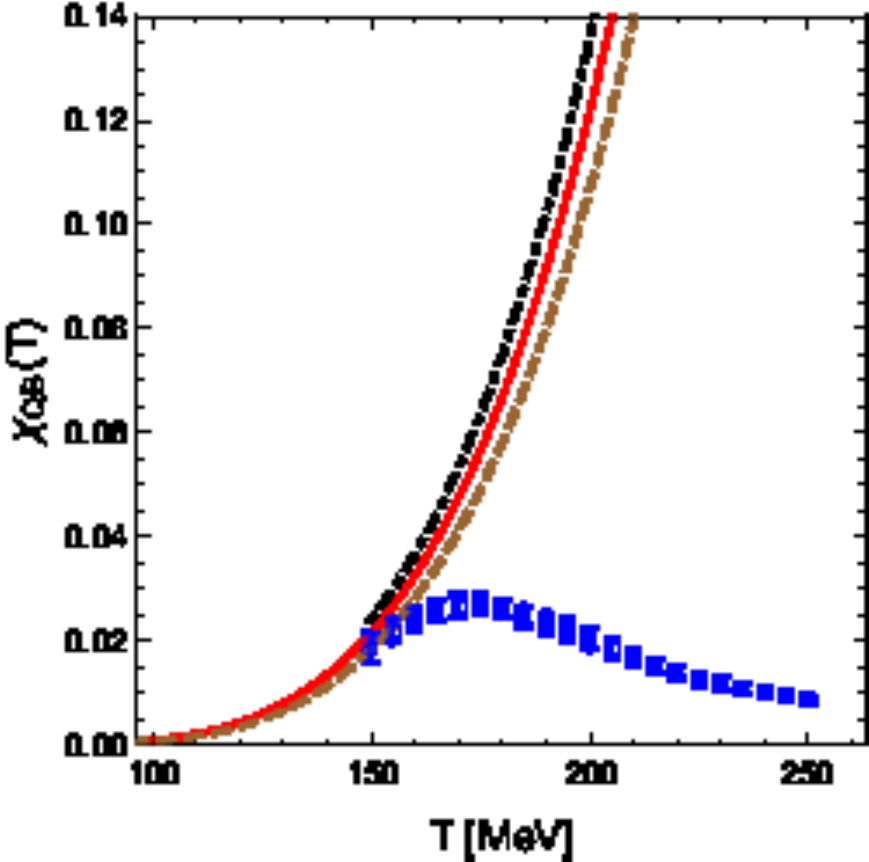,width=5cm,angle=0}
\epsfig{figure=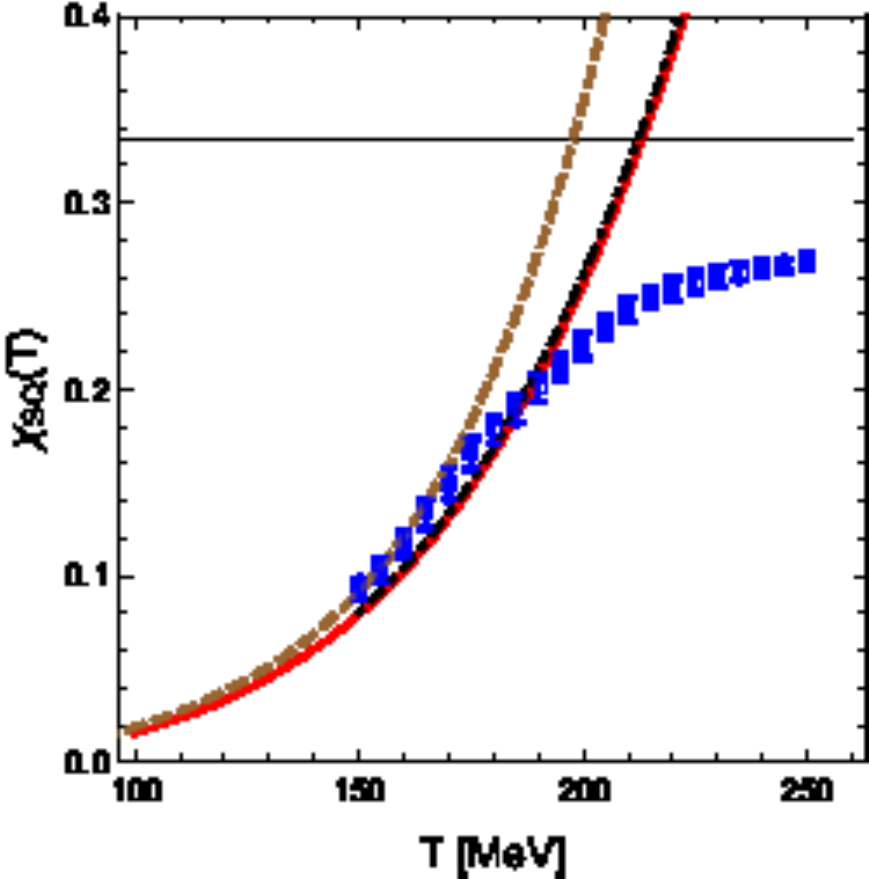,width=5cm,angle=0}

\caption{Baryon, charge and strangeness susceptibilities from HRG 
	with the PDG, PDG($\Gamma$) and RQM spectra, compared to 
	the lattice HotQCD~\protect\cite{Bazavov:2012jqz} data. WB 
	data~\protect\cite{Borsanyi:2011swz} are compatible with 
	them so they are not plotted.} \label{fig:HRG-PDG}
\end{center}
\end{figure*}

Our results for the susceptibilities are depicted in
Fig.~\ref{fig:HRG-PDG} where we show the HotQCD lattice
data~\cite{Bazavov:2012jqz} (the earlier WB data~\cite{Borsanyi:2011swz}
are compatible with them so they are not included in the figure to
avoid cluttering.). We compare with the standard HRG model, denoted as
PDG, the HRG including a Gaussian width profile, which we denote as
PDG ($\Gamma$), and the RQM.

Our scheme here is to include {\it all} states from PDG, which as
mentioned are mapped into the standard quark model classification of
mesons as $\bar q q$ and baryons as $qqq$ as the only hadronic
states. This choice of states provides a visible effect in the $SB$
correlator bringing it closer to the lattice data as compared
to~\cite{ManLo:2016pgdz} where only $****$ PDG states are
considered. The inclusion of width effects is also generally quite
sizeable and cannot be ignored, as it has routinely been done in many
HRG comparisons in the past (see however~\cite{Arriola:2012vkz,
Broniowski:2016hvtz}). Nevertheless, there are other ways to include 
the width profile which will somehow blur the PDG($\Gamma$) result, 
and a more systematic study, perhaps including also volume effects, 
would be most helpful.

The remarkable good agreement of the trace anomaly found between PDG
and RQM~\cite{Arriola:2014bfaz} or the
PDG($\Gamma$)~\cite{Broniowski:2016hvtz} compared with lattice QCD
results from WB~\cite{Borsanyi:2013biaz} and
HotQCD~\cite{Bazavov:2014pvzz} collaborations gets a bit spoiled in
terms of the considered fluctuations, where these spectra may 
feature missing or exceeding states. For instance, a look at the $BB$
correlation in Fig.~\ref{fig:HRG-PDG} suggests that the RQM has too
many baryonic states but not too many charged states. Therefore, the
thermodynamic equivalence will depend on the quantum numbers,
enhancing the relevance of a fluctuation analysis, as done here, in
the discussion of quark-hadron duality.

\item \textbf{Conclusions}

In the present contribution we have revised the thermodynamical
equivalence between the PDG, RQM, and lattice QCD for temperatures
below the hadron-gas---quark-gluon-plasma crossover for the case of 
an entropy shift due to a heavy quark and fluctuations via Baryon, 
Charge and Strangeness susceptibilities as diagnostic tools for 
missing states.

The analysis of the entropy shift due to a heavy quark suggests that
there are conventional (high mass) missing states in single charm, 
or bottom hadrons ($Q \bar q$ and $Q qq$) and it looks likely that a
large number of hybrids ($Q \bar q g$) is also missing.

In the pure light $uds$ sector, our perception on the missing states 
may change when finite width effects are placed into the calculation.
This effectively corresponds to redistribute the mass spectrum
weighted with an asymmetric Boltzmann factor. From that point of view
the missing states effect could also be regarded as a missing mass
effect. At this level the highest temperature of agreement for the
trace anomaly seems to be $T \lesssim 150~{\rm MeV}$ between either the
HRG based on PDG, PDG ($\Gamma$) or RQM spectra and current QCD finite
temperature calculations. However, the separate analysis in terms of
$B,Q,S$ fluctuations reveals a less obvious pattern regarding the
verification of quark-hadron duality. While the HRG has arbitrated the
lattice QCD discrepancies for the trace anomaly in the past, in the
case of fluctuations we are now confronted with the opposite
situation. Lattice data agree but are not universally reproduced by
any of the three HRG realizations considered here. This may offer a
unique opportunity to refine these models including other effects 
and which deserves further studies.

\item \textbf{Acknowledgments}

We thank Pok Man Lo and Michal Marczenko for useful communications.
This work is supported by Spanish Ministerio de Econom\'{\i}a y
Competitividad and European FEDER funds under contracts
FIS2014-59386-P and FPA2015-64041-C2-1-P, Junta de Andaluc\'{\i}a
grant FQM-225, and Spanish Consolider Ingenio 2010 Programme CPAN
(CSD2007-00042). W.B. is supported by the Polish National Science
Center grant 2015/19/B/ST2/00937.  The research of E.M. is supported
by the European Union under a Marie Curie Intra-European fellowship
(FP7-PEOPLE-2013-IEF) with project number PIEF-GA-2013-623006, and by
the Universidad del Pa\'{\i}s Vasco UPV/EHU, Bilbao, Spain, as a
Visiting Professor.
\end{enumerate}


\newpage
\subsection{Interplay of Repulsive Interactions and Extra Strange Resonances}
\addtocontents{toc}{\hspace{2cm}{\sl P.~Alba}\par}
\setcounter{figure}{0}
\setcounter{table}{0}
\setcounter{equation}{0}
\setcounter{footnote}{0}
\halign{#\hfil&\quad#\hfil\cr
\large{Paolo Alba}\cr
\textit{Frankfurt Institute for Advanced Studies}\cr
\textit{Goethe Universit\"at Frankfurt}\cr
\textit{D-60438 Frankfurt am Main, Germany}\cr}

\begin{abstract}
From simulations on lattice it is suggested that there exist missing
states in the strange sector of the hadronic spectrum. Since those
predictions mainly rely on the Hadron-Resonance Gas model assumption,
it is important to check the influence of all the known possible
implementation of such a model, in particular the effect of repulsive
forces among hadrons. I explore the interplay between the inclusion
of extra states predicted by the Quark Model, which I expect to be
in line with possible future discoveries, and the corresponding
repulsive forces parametrized trough the Excluded Volume. I find
that the inclusion of Quark Model states improves the description of 
some observables, but in order to have an overall improvement for 
most of the available observables, repulsive interactions are needed. 
I check experimental measured yields and results from lattice
simulations as well. I find that there is a better description of 
the data when including both effects, within a reasonable temperature 
range.
\end{abstract}

\begin{enumerate}
\item \textbf{Introduction}

In recent years the HIC program was very successful in providing
results on strong interactions. Among these it is remarkable the
success of the statistical hadronization model, which assumes that
immediately after the collision the system thermalizes into a 
fireball from which hadrons are emitted. This result has been 
furtherly tested against hadron production~\cite{BraunMunzinger:1994xry}, 
and with hydrodynamical simulations~\cite{Schnedermann:1993wsy}. 
However it is worth to mention that other options exist, e.g., a 
microscopical description through transport models, from which it 
is possible to access directly the partonic nature of those 
interactions.

Lattice simulations provided excellent results, and in recent years 
it was possible to improve the accuracy and precision of those, in 
order to be able to analyze experimental data~\cite{Borsanyi:2013hzay}, 
and being able to do important step forward in the understanding of 
the QCD transition; this is extremely important because allows to 
directly connect the experiment with first principle calculations. 
From these, we know that the confinement transition is a 
crossover~\cite{Aoki:2006wey}, i.e., does not allow to clearly asses 
when there is the passage between the two phases, but only to estimate 
a (pseudo-)critical temperature of about 150~MeV~\cite{Borsanyi:2010bpy}. 
This has important consequences, e.g., the fact that the crossover 
acts differently for different constituent, hinting for a flavor 
hierarchy in the critical temperature~\cite{Bellwied:2013ctay}.

Fluctuations of conserved charges have been proposed to be able to 
test these assumptions, and have been proven to be very sensitive 
observables from lattice simulations~\cite{Bellwied:2013ctay}, and 
from experimental measurements~\cite{Adamczyk:2013daly,Alba:2015ivay}.

One important tool used to overcome the inner differences between
experimental measurements and lattice is the Hadron-Resonance Gas
(HRG) model, which has proven to be successful in the description
of particle yields~\cite{Andronic:2008guy}, and shows a good 
agreement with lattice simulations~\cite{Borsanyi:2011swy} and 
experimental measurements for the fluctuations of conserved 
charges~\cite{Alba:2014ebay}.

\item \textbf{Quark Model}

From the discrepancy between lattice calculations and HRG predictions
for a specific observable, it has been proposed that the actual 
measured hadronic spectrum is lacking in some states in the strange 
sector~\cite{Bazavov:2014xyay}, and that Quark-Model (QM) predictions 
could fill the gap restoring the agreement between the two frameworks.
However it is easy to check that the wild inclusion of all those
states can ruin the agreement with other observables, due in particular
to multi-strange baryons.

Indeed, this is connected with the uncertainty coming from the 
particle list used as an input in the HRG model, which relies 
essentially on the measured experimental states listed by the Particle 
Data Group~\cite{Olive:2016xmwy}.

In order to check the reliability of those states, I show in 
Fig.~\ref{fit_to_yields}, panel (a), the $\chi^2$ profile in temperature
for the fit to particle yields from PbPb collisions at 2.76~TeV measured
by the ALICE collaboration~\cite{Abelev:2013veay,Abelev:2013xaay,
ABELEV:2013zaay} (see~\cite{Alba:2016hwxy} for more details), with
$\chi^2$ given by:
\begin{equation}\label{xi}
	\chi ^2~=~\frac{1}{N_{\rm dof}}\sum_{h=1}^N \frac{\left( \langle
	N_h^{\rm exp}\rangle~-~
	\langle N_h \rangle \right)^2}{\sigma_h^2}~.
\end{equation}
\begin{figure}
\begin{centering}
\includegraphics[width=0.5\textwidth]{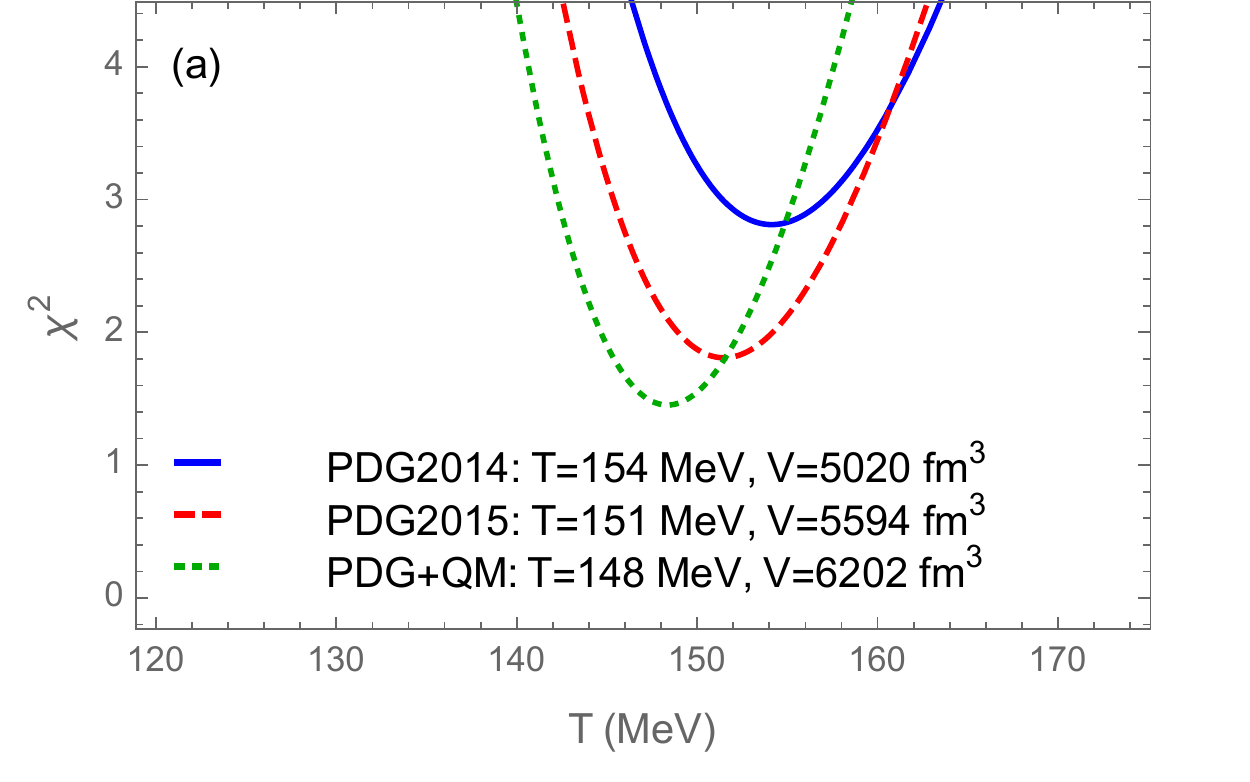}
~\hspace{-2em}~
\includegraphics[width=0.5\textwidth]{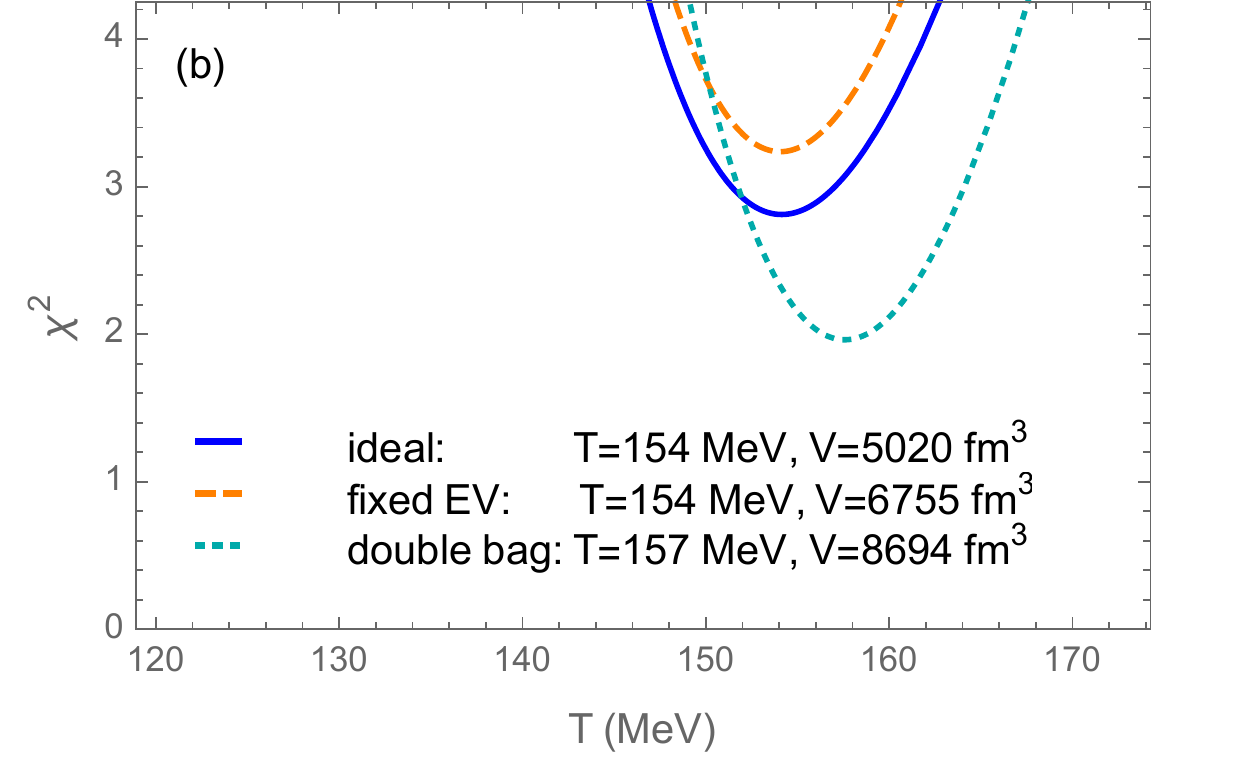}

\caption{\label{fit_to_yields} $\chi^2$ profile with temperature from
        the fit to particle yields measured by the ALICE collaboration
        for PbPb collision at 2.76~TeV~\protect\cite{Abelev:2013veay,
        Abelev:2013xaay,ABELEV:2013zaay}. For every temperature the
        volume is minimized, but the baryon chemical potential is
        fixed to zero. On panel (a) is showed the effect of the
        inclusion of higher-mass states, while on panel (b) is showed
        the effect of EV effects.}
\end{centering}
\end{figure}

It comes out that the updates from the PDG improve the description of
the data (see also Table~\ref{tabella}); extracting the informations
on the branching ratios from the PDG 2015 and applying them to the QM
states, the $\chi^2$ is further decreased, leaving the freeze-out
parameters almost unaffected.

By the way there are different options available for QM calculations
with respect to the one employed here, which are essentially the most
crude and most abundant.

A similar improvement can be achieved accounting for repulsive
interactions (see Fig.~\ref{fit_to_yields} panel (b)), as will be
explained in the following section.
\begin{table}
\begin{center}

\caption{\label{tabella} Freeze-out parameters, and corresponding 
	$\chi^2$ from the fit to particle yields measured by the 
	ALICE collaboration for PbPb collision at 
	2.76~TeV~\protect\cite{Abelev:2013veay,Abelev:2013xaay,
	ABELEV:2013zaay} for different particle lists and EV 
	parameterizations. In the last column is shown the $\chi^2$ 
	for observables calculated on the lattice for temperatures 
	below 164~MeV; the parameters are the ones obtained from
	the fit to particle yields.}
\begin{tabular}{clllll}
\hline\noalign{\smallskip}
list    & EV &$\chi^2_{yields}$ & $T$ (MeV) & $V$ (fm$^3$)&$\chi^2_{lattice}$\\
\hline
PDG2014 & \emph{id} &22.49/8 $\cong$ 2.81 & 154.19 $\pm$ 2.29& 5047 $\pm$ 663 & 9.49\\
PDG2015 & \emph{id} &14.47/8 $\cong$ 1.8  & 151.53 $\pm$ 2.12& 5620 $\pm$ 705 &8.65\\
QM      & \emph{id} &11.62/8 $\cong$ 1.45 & 148.39 $\pm$ 1.18& 6227 $\pm$ 722 &15.905\\
PDG2014 & \emph{fix}&22.65/7 $\cong$ 3.23 & 154.11 $\pm$ 2.28& 5934 $\pm$ 701 &10.95\\
QM      & \emph{fix}&11.74/7 $\cong$ 1.67 & 148.33 $\pm$ 1.18& 7131 $\pm$ 760 &6.98\\
PDG2014 & \emph{2b} &11.77/6 $\cong$ 1.96 & 157.64 $\pm$ 2.46& 5734 $\pm$ 620 &14.07\\
QM      & \emph{2b} &13.47/6 $\cong$ 2.24 & 149.27 $\pm$ 1.8&  7483 $\pm$ 704 &1.705\\
\hline
\end{tabular}
\end{center}
\end{table}

\item \textbf{Repulsive Forces}

The inclusion of resonance formation mediates the attractive interactions 
among hadrons, neglecting the repulsive ones which however are present in 
the experimental scattering measurements. The last can be implemented 
within the HRG model with the so called Excluded Volume 
(EV)~\cite{Rischke:1991key}; hadrons are considered as hard spheres, and 
are assumed to repel each other when their effective radii $r_i$ overlap.
In this picture hadrons posses an eigenvolume given by $v_i=\frac{16}{3}\pi 
r_i^3$, which must be subtracted to the total volume of the system. This 
implies a transcendental equation for the system pressure $p$ with a 
shifted single particle chemical potential $\bar\mu_i=\mu_i-v_ip$. The 
others thermodynamical quantities are obtained from usual relations, 
e.g., the particle densities are:
\begin{equation}
\label{densities}
	n_i(T,\mu_B)~=~\frac{n_i^{\rm id}(T,\mu_i^\ast)}{1
	+\sum_j\,v_j\,n_j^{\rm id}(T,\mu_j^\ast)}\,,
\end{equation}
where it is clear the double fold suppression, coming from the shifted 
chemical potential and the overall denominator.

It has been pointed out that the proper inclusion of repulsive forces 
is relevant for resonances like the $\sigma$ and the 
$\kappa$~\cite{Broniowski:2015ohay,Friman:2015zuay}, and in general can 
influence other resonances.

Since data are not available for all the hadronic species present in 
our lists, I employ different parameterizations for the particle 
eigenvolumes: fixed for all species, directly proportional to the 
particle mass (as one could expect for radial excitations from QM 
calculations), and inversely proportional. The last case, even if 
may look counterintuitive, has been explored 
theoretically~\cite{Friedmann:2009mzy}, and can derive from the 
assumption of diquarks as constituents building blocks together with 
quarks. It has been pointed out how this assumption can improve the 
description of particle yields~\cite{Alba:2016hwxy}, and of lattice 
simulations in the pure gauge sector~\cite{Alba:2016fkuy}.

It is worth to note that for a fixed radius the relative densities 
stay constant (see Eq.\ref{densities}), essentially leaving unaffected 
results based on particle yields. The situation is different when 
looking at higher order cumulants, as can be seen in Fig.~\ref{lattice}.
\begin{figure}
\begin{centering}
\includegraphics[width=0.5\textwidth]{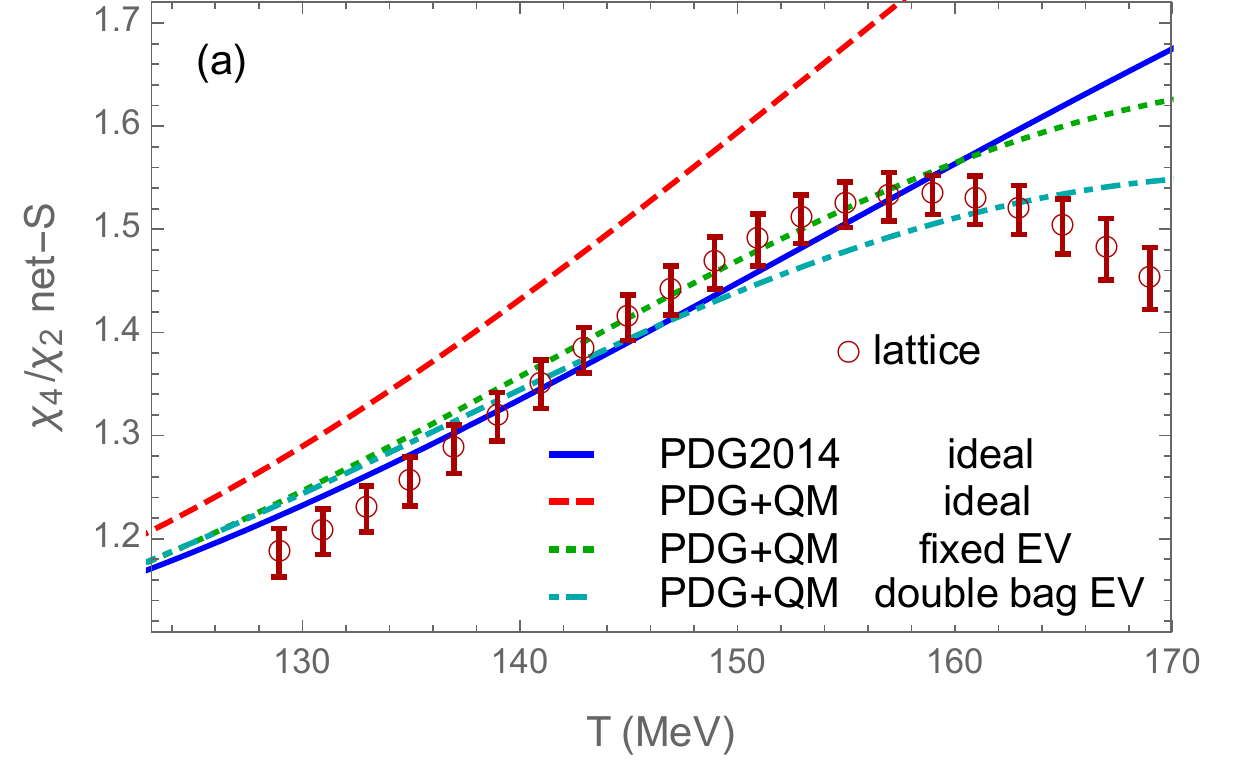}
~\hspace{-2em}~
\includegraphics[width=0.5\textwidth]{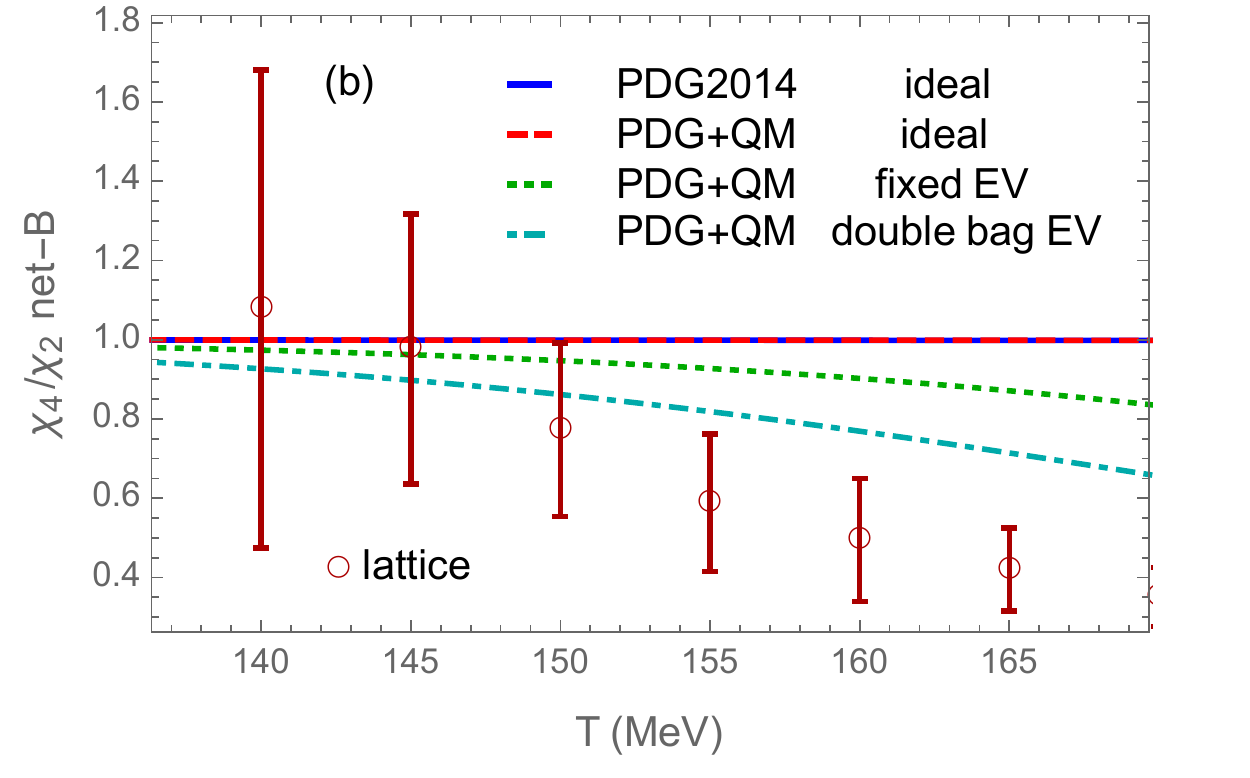}\\
\includegraphics[width=0.5\textwidth]{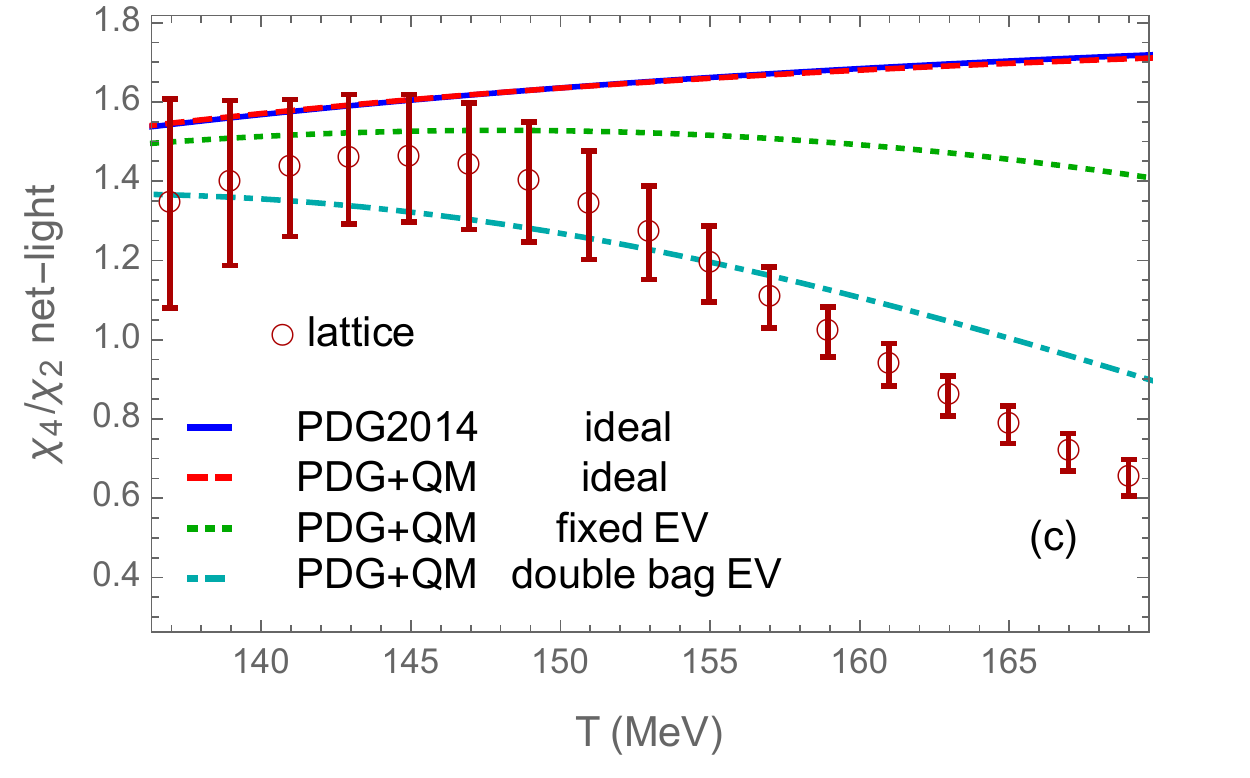}
~\hspace{-2em}~
\includegraphics[width=0.5\textwidth]{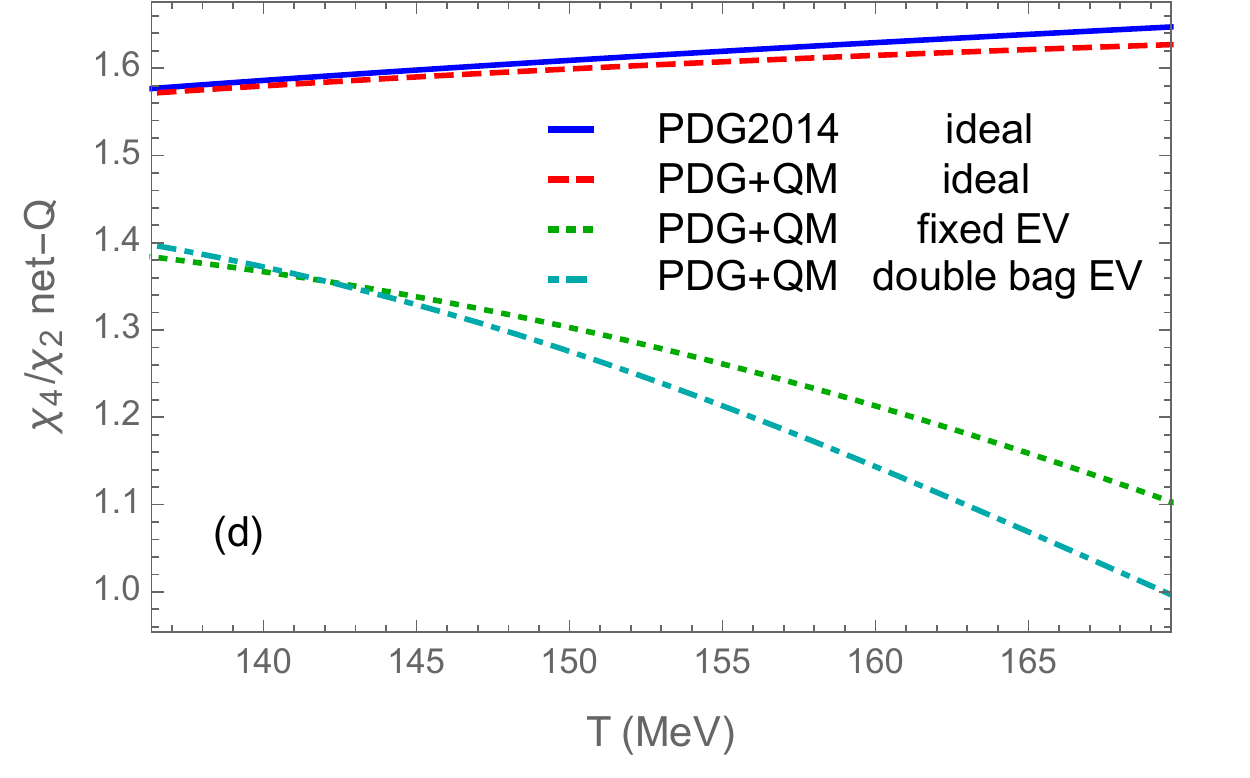}

 \caption{\label{lattice} Comparison between lattice
        data~\protect\cite{Bellwied:2013ctay,Bellwied:2015lbay} and
        HRG calculations with different particle lists and particle
        eigenvolumes, for $\chi_4/\chi_2$ observables for different
        quantum numbers. Currently there are no lattice data
        available for net-electric charge.}
\end{centering}
\end{figure}

\item \textbf{Comparison with Lattice}

In Table~\ref{tabella}, I list the results from the fit to particle yields 
using the standard HRG (\emph{id}), an EV with fixed radii (\emph{fix}) 
and with radii proportional to the particle mass with proportionality 
constant dependent on the flavor (\emph{2b}). The same EV parameters are 
applied for PDG2014, PDG2015 and QM lists. Here I also calculate the 
$\chi^2$ for the different HRG options against the lattice data for 
pressure, interaction measure, $\chi^{us}_{11}$, $\chi^{ud}_{11}$, 
$\chi^{s}_{2}$, $\mu_S/\mu_B|_{LO}$, $\chi_4/\chi_2|_{net-B}$, 
$\chi_4/\chi_2|_{net-S}$ and $\chi_4/\chi_2|_{net-light}$ for 
temperatures below 164~MeV~\cite{Borsanyi:2013biay,Borsanyi:2013hzay,
Bellwied:2015lbay}, which should be a reasonable range of temperatures 
near the crossover. I show that with the inclusion of QM states and with 
a \emph{2b} EV with $r_p=0.36$~fm and $r_{\Lambda}=0.27$~fm (it is 
convenient to parametrize the $r_i$ with the ground state hadrons in 
order to have an immediate comparison), I can systematically improve 
the description of both experimental and lattice data, with hints from 
both sectors for smaller strange hadrons with respect to the light ones 
with the same mass.

In Fig.~\ref{lattice}, I show a comparison between lattice and HRG 
predictions for the $\chi_4/\chi_2$ for net-light, net-B and net-S 
quantum numbers; the improvement due to the EV could be understood in 
terms of the statistical suppression due to the finite sizes of 
resonances; in particular in the strange sector the EV suppression 
balances the effect of the inclusion of multi-strange baryons (leaving 
unaffected the results for the $\mu_S\mu_B|_{LO}$). There are no data
for the net-electric charge, and here I show my prediction for this 
quantity.

It is worth to note at this point that the EV parameters employed are 
solely obtained from a fit to particle yields, which result in 
freeze-out parameters compatible with the \emph{id} case.

\item \textbf{Conclusion}

In conclusion, I studied the balance between attractive and repulsive 
forces within the HRG framework. I find that the simultaneous inclusion 
of higher-mass states and EV interactions improves the description of 
a large set of observables both from experimental measurements and 
lattice simulations, hinting in particular to a flavor dependent size. 
This could depend on the QM calculations employed, and can be critical 
the inclusion of exotic resonances like the $\kappa$(800), which need 
still to be confirmed but which would have a large influence on 
observables related to strangeness.

A systematic study of all the different versions of the Quark Model 
against repulsive interactions is mandatory.

\item \textbf{Acknowledgments}

This work was supported by HIC for FAIR within the LOEWE program of 
the State of Hesse. I acknowledge the University of Houston for 
hosting me during the period in which this work was completed.
\end{enumerate}


\newpage
\subsection{Missing Resonance Decays in Thermal Models}
\addtocontents{toc}{\hspace{2cm}{\sl V.~V.~Begun, V.~Yu.~Vovchenko, and
	M.~I.~Gorenstein}\par}
\setcounter{figure}{0}
\setcounter{table}{0}
\setcounter{equation}{0}
\setcounter{footnote}{0}
\halign{#\hfil&\quad#\hfil\cr
\large{Viktor~V.~Begun}\cr
\textit{Faculty of Physics}\cr
\textit{Warsaw University of Technology}\cr
\textit{00-662 Warsaw, Poland}\cr\cr
\large{Volodymyr~Yu.~Vovchenko}\cr
\textit{Institute f\"ur Theoretishe Physik}\cr
\textit{Goethe Universit\"at Frankfurt}\cr
\textit{D-60438 Frankfurt am Main, Germany \&}\cr
\textit{Frankfurt Institute for Advanced Studies}\cr
\textit{Goethe Universit\"at Frankfurt}\cr
\textit{D-60438 Frankfurt am Main, Germany \&}\cr
\textit{Department of Physics}\cr
\textit{Taras Shevchenko National University of Kiev}\cr
\textit{03022 Kiev, Ukraine}\cr\cr
\large{Mark~I.~Gorenstein}\cr
\textit{Frankfurt Institute for Advanced Studies}\cr
\textit{Goethe Universit\"at Frankfurt}\cr
\textit{D-60438 Frankfurt am Main, Germany \&}\cr
\textit{Bogolyubov Institute for Theoretical Physics}\cr
\textit{03680 Kiev, Ukraine}\cr}

\begin{abstract}
Detailed information on decay channel probabilities is absent for
many high mass resonances, which are typically included in thermal
models. In these cases, the sum over all known decay branching
probabilities is smaller than 1. Due to this systematic
uncertainty of the model, the exact charge conservation may appear
to be violated. We estimate the corresponding number of missing
charge states in the canonical ensemble formulation of the hadron
resonance gas for p+p reactions at the SPS energy $E_{\rm lab}=
158$~GeV: $\Delta B \simeq 0.16$ for baryon charge, $\Delta
Q\simeq 0.12$ for electric charge, and $\Delta S=-0.01$ for
strangeness.
The value of the considered effect is 5-8\%, which seems to be
important enough to include it as a systematic error in the
calculations within a hadron gas.
\end{abstract}

\begin{enumerate}
\item 

The hadron-resonance gas model (HRG) allows to obtain particle
multiplicities at different collision energies with a relatively
good accuracy. In a simplest HRG hadrons and resonances are
assumed to be non-interacting, and full chemical equilibrium is
imposed. This model has just two free parameters - the
temperature, $T$, and the baryon chemical potential, $\mu_B$,
which follow a simple analytic dependence as the function of
collision energy~\cite{Cleymans:2005xvq,Vovchenko:2015idtq}.
The HRG model works well for nucleus-nucleus (A+A) collisions, and
even for elementary particle reactions $p+p$, $p+\bar{p}$ and
$e^++e^-$~\cite{Becattini:1995ifq,Becattini:1997rvq}
An analytic assumption about the form of the fireball hypersurface
at freeze-out allows further to obtain the $p_T$ spectra of
measured particles for the cost of just one more parameter - the
ratio of the freeze-out radius and
time~\cite{Broniowski:2001weq,Begun:2013ngaq,Begun:2014rsaq}.

Many modifications to the standard HRG exist, which improve
agreement with experimental data, in particular, the one at the
LHC. These include sequential
freeze-out~\cite{Chatterjee:2013ygaq}, a mechanism of proton
suppression due to re-scattering during the
freeze-out~\cite{Becattini:2012xbq}, introducing different proper
volumes for different particles~\cite{Alba:2016hwxq}, or
considering the possibility of pion
condensation~\cite{Begun:2013ngaq,Begun:2015ifaq}.

A list of stable particles and resonances is a key ingredient of a HRG 
model. Taking into account more/missing hadron resonances helped with 
data description at the SPS~\cite{Schnedermann:1993wsq}
and the LHC~\cite{Noronha-Hostler:2014aiaq}.
The problem is that the list of resonances and their properties is
known well only up to $m\sim 1.5$~GeV.
In particular, many decay channels for measured heavy resonances
are unknown. Therefore, the amount of charge that is calculated in
a HRG after including only the known decays of these resonances
listed in Particle Data Tables~\cite{pdgq} is different from the
one before the decays~\footnote{Another aspects of missing charges
	in p+p reactions were recently reported
	in~\cite{Stroebele:2016lklq}.}.
The size of the latter effect can be estimated. For the case of a
proton-proton reaction the charge of the system is known exactly,
$B=Q=2$, $S=0$. If several total $4\pi$ multiplicities are
available experimentally, then their fit in the HRG within the
canonical ensemble gives the thermal parameters: temperature $T$,
system volume $V$, and strangeness undersaturation parameter
$\gamma_S$~(see Ref.~\cite{Vovchenko:2015idtq} for details).
The resulting missing charge for the p+p reactions at laboratory
momentum $158$~GeV/c is $\Delta B/B\simeq 8\%$ for the baryon
charge, $\Delta Q/Q\simeq 6\%$ for the electric charge, and
$\Delta S=-0.01$ for strangeness~\cite{Vovchenko:2015idtq}.
These numbers reflect a strongest effect of the missing decay
channels probabilities for heavy positively charged baryons.

There are alternative ways to deal with the problem of missing
branching ratios. One option is to additionally normalize all the
branching ratios to 100\%~\cite{Torrieri:2004zzq,Wheaton:2004qbq}.
However, such a normalization produces some error: it artificially
enhances the known channels, and, therefore, suppresses yet
undiscovered channels.
Another option is to assign same/similar branching ratios based on
analogies to the nearest states with the
same quantum numbers and known branching ratios~\cite{Andronic:2008guq}.

Therefore, the value of the considered effect is 5-8\%, which
seems to be important enough to include it as a systematic error
in the calculations within a HRG.

\item \textbf{Acknowledgments}

We thank H.~Stroebele for a fruitful discussion. V.~Y.~V. acknowledges 
the support from HGS-HIRe for FAIR. The work of M.~I.~G. was supported 
by the Program of Fundamental Research of the Department of Physics and 
Astronomy of National Academy of Sciences of Ukraine.
\end{enumerate}


\newpage
\subsection{Discussion and Summary}
\addtocontents{toc}{\hspace{2cm}{\sl J.~L.~Goity, P.~Huovinen, J.~Ritman, 
	and A.~H.~Tang}\par}
\setcounter{figure}{0}
\setcounter{table}{0}
\setcounter{equation}{0}
\setcounter{footnote}{0}
\halign{#\hfil&\quad#\hfil\cr
\large{Jose Goity}\cr
\textit{Department of Physics}\cr
\textit{Hampton University}\cr
\textit{Hampton, VA 23668, U.S.A. \&}\cr
\textit{Thomas Jefferson National Accelerator Facility}\cr
\textit{Newport News, VA 23606, U.S.A.}\cr\cr
\large{Pasi Huovinen}\cr
\textit{Institute of Theoretical Physics}\cr
\textit{University of Wroclaw}\cr
\textit{PL-50204 Wroclaw, Poland}\cr\cr
\large{James Ritman}\cr
\textit{Institut f\"r Kernphysik}\cr
\textit{Forschungszentrum J\"ulich}\cr
\textit{D-52424, J\"ulich, Germany}\cr\cr
\large{Aihong Tang}\cr
\textit{Brookhaven National Laboratory}\cr
\textit{Upton, NY 11973, U.S.A.}\cr}

\vspace{1cm}
After the scientific presentations were complete, and directly 
before the summary and concluding remarks, a session was held 
to discuss the common issues presented by multiple talks, in 
particular those items that are particularly relevant to the 
physics motivation and for a K0Long facility and what needs to 
be done to realize this facility. The very intense and lively 
discussions had a wide range of participants.  This discussion 
was structured into several themes as presented below:

\begin{enumerate}
\item \textbf{What are the Synergies between Heavy Ion Collision 
	Physics and the K0Long Facility?}

Hadron production yields have been analyzed in heavy ion collisions 
with incident beam energies ranging from several GeV per nucleon up 
to Pb+Pb  interactions at LHC energies. Statistical model 
calculations using specific parameters such as thermal and chemical 
freezeout temperatures, baryon-, strangeness- and charm- chemical 
potentials, give surprisingly accurate descriptions of the measured 
results over many orders of magnitude of yield. These models give a 
very important baseline, upon which one can gauge, to which extent 
other observations are "simple" many-body effects, or can be 
attributed to, e.g., consequences of the deconfined phase.

Despite the tremendous success of these calculations, some specific 
aspects require dedicated attention. In particular the calculation 
of yields for many longer lived resonances require knowledge of the 
higher lying states, that feed down to the lower lying states, which 
are finally observed by the detectors. The importance for Charmonium 
and Bottomonium states is very apparent. In the baryon sector the 
calculations take into account the resonances listed in the Particle 
Data Book. For light quark states this approach seems to work well. 
For the hyperon sector, the quality of the results depends strongly 
upon which hyperon states are included in the calculation. The 
calculations are far off the data if only well-established resonances 
are included. If however those states with low confidence ("zero-star"
or "1-star" resonances) are also included, then the level of agreement 
between the calculations and the data improves. Although the heavy 
ion data can not be used to prove the exact parameters of specific 
narrow resonances, they do however imply that a large number of narrow 
hyperon states, in a narrow mass interval must exist.

Therefore, it is evident that the spectroscopic results from a K0Long 
facility will greatly improve the reliability of the thermal model 
predictions in the baryon/hyperon sector. Concurrently, the heavy ion 
results provide a strong indication of how many narrow hyperon states 
are to be expected with masses up to roughly $2~GeV/c^2$.

\item \textbf{Should these Experiments Rather Measure PP ?}

The K0L facility is based on using a tertiary beam (electrons, 
photons, K0Long), with the corresponding reduction of rate at each 
step of the beam generation process. In contrast, proton-proton 
interactions can be operated at many orders of magnitude higher 
luminosity. Consequently, excited hyperons are produced at higher 
rates in the primary proton-proton collisions. These resonances 
however need to be cleanly identified in these high flux 
environments. Kinematic constraints that exploit exclusive event 
reconstruction tend to favor beam energies close to the production 
threshold, where the signal cross section drops rapidly compared 
to the total reaction cross section. Additional complications 
arise since either projectile or target (or both) can be excited. 
Thus, the multi-baryon final state has a large, irreducible 
physics background. 

The yield of exotics in proton-proton collisions and/or heavy ion 
collisions might be enhanced due to the high temperature and high 
density environment, however, due to the background mentioned 
above, attempts of search in those collisions would be better 
viewed as opportunistic play.

\item \textbf{How to go beyond "Bump-Hunting"?}

In the relevant kinematic range it is expected that there are many 
overlapping resonances of various spin-parity quantum numbers. As 
clearly shown in, e.g., photon, pion and kaon induced reactions, it 
is essential to apply partial wave analyses to the data to reliably 
extract resonance information from the ensemble of final states. 
This is well understood in the community, and powerful techniques 
are being further developed. This development and the use of these 
programs requires continued, very close interaction between theory 
and experiment, as for instance already practiced at JPAC.

\item \textbf{From Theory: How Many Y$^\ast$ States do we need to 
	Measure?}

Of course this general formulation of the question begs the answer 
"as many as possible". Instead it is more quantitative to ask how 
many states are needed for a specific purpose. For instance, how 
many states are needed to improve the statistical model calculations 
for heavy ion collisions to a given level of precision? Similarly, 
one could ask how many states are needed to corroborate/confirm 
Lattice QCD predictions for this sector?  In either case we are 
talking about on the order of 10 to tens of states. 

\item \textbf{From Experiment: How Many States are Practically 
	Feasible?}

The experimental effort connected to measuring on the order of 10 
states indicates that a dedicated measurement program is needed, 
and not just a single measurement. This is very important, since 
it means that the success of the K0L facility does not depend upon 
the existence or measurability of any single state. 

In order to achieve these stated scientific goals, the scope of 
the program will comprise the instrumental setup, and will require 
several months of beamtime.  Specific details on the effort 
involved to setup the K0L facility as well as the amount of 
required beamtime have been presented during this workshop and 
will be included in the proposal for the facility.  

\item \textbf{Which Decays Should be Measured?}

An integral part of the K0L facility is the GlueX spectrometer, 
which will surround the hyperon production target. This provides 
excellent charged particle tracking capability for particles from 
both the primary and also delayed vertices. Furthermore, the 
calorimeters are well suited to measure photons from neutral pion 
decays. 

Kinematic over-constraints do allow events to be reconstructed if 
one particle is unmeasured. However, the resolution of the beam 
momentum worsens significantly for systems with increasing hyperon 
masses. Consequently, final states where the daughters are charged 
hadrons, or include at most one neutral pion, are ideal. This 
explicitly includes delayed decays of neutral particles, such as 
Lambda into proton $\pi^-$. The use of Cherenkov detectors in the 
GlueX spectrometer will significantly enhance the selectivity to 
charged kaons in the final state, which is vital for the S=-2 and 
S=-3 studies.

\item \textbf{Rates at JLab -- Comparison to Other Facilities?}

In the useful momentum range of 0.3 to 10~GeV/c there will be up to 
$10^4$ K0L per second incident on the target inside GlueX. This flux 
will result in about $2 \times 10^5$ excited cascades and $4 \times 
10^3$ excited Omegas per month.

A similar program for KN scattering is under development at J-PARC 
with charged kaon beams. The current maximum momentum of secondary 
beamline of 2~GeV/c is available at the K1.8 beamline. This is 
however not sufficient to produce the first excited cascade states 
within the quark model.  However, there are plans to create a high 
energy beamline in the momentum range 5 -- 15~GeV/c to be used with 
the spectrometer commonly used with the J-PARC P50 experiment. The 
statistical power of proposed experiment with a K0L beam at JLab 
will be of the same order as that in J-PARC with a charged kaon 
beam. In BES-III and Belle, the hyperons are produced in charmonium 
or bottomium decays. The branching ratios of heavy quarkonia into 
multi-strange baryon-antibaryon states are however not very large. 
In LHCb, inclusive strange hyperons are produced in abundance, but 
PWA requires a well-defined initial state, which makes only hyperons 
from known states (e.g., bottomium) suitable. The production rates 
of hyperons from such decays are low. PANDA can access the full 
double- and triple-strange hyperon spectrum even with relatively 
low luminosities, and the hyperons and the antihyperons can be 
measured exclusively. Those measurements are however on a longer 
time scale compared to when the K0L facility can be realized.

\item \textbf{Which Other Physics Potential is there with a K0L Beam?}

A very interesting further opportunity for the K0L facility is to 
investigate K0L reactions on complex nuclei. By selecting events 
with the appropriate beam momentum together with a fast forward 
going pion, events can be identified, in which a hyperon is produced 
at low relative momentum to the target nucleus or into a bound state. 
Baryons with strangeness embedded in the nuclear environment, 
hypernuclei or hyperatoms, are the only available tool to approach 
the many-body aspect of the three-flavor strong interaction.

Further potential exists to search for - or exclude - possible 
exotic baryonic states that can not easily be described by the 
usual 3 valence quark structure. Recent results from  LHCb provide 
tantalizing hints for the existence of so-called pentaquarks that 
include a charm valence quark, however the interpretation of those 
results is under discussion. In contrast, elastic scattering of K0L 
with a hydrogen target gives unambiguous information on the 
potential existence of such states. With the given flux of K0L at 
the proposed facility, a clear proof of existence or proof of 
absence will be obtained within the integrated luminosity required 
for the excited hyperon spectroscopy program that forms the basis 
of this proposal.
\end{enumerate}

\newpage
\section{List of Participants of YSTAR2016 Workshop}

\begin{itemize}
\item Paolo Alba, Frankfurt Inst. for Advanced Studies <alba@fias.uni-frankfurt.de>
\item Hisham Albataineh, Texas A\&M U.-Kingsville <hisham@jlab.org>
\item Moskov Amaryan, ODU       		  <mamaryan@odu.edu>
\item Mikhail Bashkanov, Edinburgh U.		  <mikhail.bashkanov@ed.ac.uk>
\item Viktor Begun, Jan Kochanowski U.		  <viktor.begun@gmail.com>
\item Rene Bellwied, Houston U.			  <bellwied@uh.edu>
\item Jay Benesch, JLab				  <benesch@jlab.org>
\item William J.~Briscoe, GWU                     <briscoe@gwu.edu>
\item Dilini Bulumulla, ODU			  <dilinib@jlab.org>
\item Simon Capstick, FSU			  <capstick@fsu.edu>
\item Eugene Chudakov, JLab	  		  <gen@jlab.org>
\item Volker Crede, FSU			  	  <crede@fsu.edu>
\item Pavel Degtyarenko, JLab	  		  <pavel@jlab.org>
\item Benjamin D\"onigus,			  <b.doenigus@gsi.de>
\item Robert G.~Edwards, JLab			  <edwards@jlab.org>
\item Hovanes Egiyan, JLab 			  <hovanes@jlab.org>
\item Mathieu Ehrhart, 				  <mehrh002@odu.edu>
\item Ashley Ernst, FSU				  <ace15f@my.fsu.edu>
\item Paul Eugenio, FSU				  <peugenio@fsu.edu>
\item Ishara Fernando, Hampton U./JLab		  <ishara@jlab.org>
\item Igor Filikhin, NCCU			  <ifilikhin@nccu.edu>
\item Humberto Garcilazo, IPN, Mexico		  <humberto@esfm.ipn.mx>
\item Jos\'e~L.~Goity, Hampton U./JLab 		  <goity@jlab.org>
\item Helmut Haberzettl, GWU 			  <helmut@gwu.edu>
\item Avetik Hayrapetyan, JLU Giessen		  <Avetik.Hayrapetyan@uni-giessen.de>
\item Pasi Huovinen, Wroclaw U.			  <huovinen@ift.uni.wroc.pl>
\item Charles Hyde, ODU			  	  <chyde@odu.edu>
\item David Jenkins, Virginia Tech		  <jenkins@jlab.org>
\item Kanchan Khemchandani, Rio de Janeiro U.	  <kanchan.khemchandani@uerj.br>
\item Chan Kim, GWU				  <kimchanwook@gwu.edu>
\item Franz Klein, GWU			  	  <fklein@jlab.org>
\item Valery Kubarovsky, JLab			  <vpk@jlab.org>
\item Victoria Lagerquist, ODU			  <vlage001@odu.edu>
\item Ilya Larin, Umass, Amherst		  <ilarin@jlab.org>
\item Ziwei Lin, East Carolina U.		  <linz@ecu.edu>
\item Maxim Mai, GWU 			  	  <maximmai@gwu.edu>
\item Mark Manley, KSU 			  	  <manley@kent.edu>
\item Valentina Mantovani Sarti, Torino U.-INFN	  <mantovan@to.infn.it>
\item Alberto Martinez Torres, San Paulo U.	  <amartine@if.usp.br>
\item Georgie Mbianda Njencheu, ODU 		  <gmbia001@odu.edu>
\item Eric Moffat, ODU				  <emoffat1@hotmail.com>
\item Wayne Morris, ODU				  <morris3wh@gmail.com>
\item Hugh Montgomery, JLab			  <mont@jlab.org>
\item Fred Myhrer, USC			  	  <myhrer@sc.edu>
\item James Napolitano, Temple U.		  <tuf43817@temple.edu>
\item Joseph Newton, ODU			  <jnewt018@odu.edu>
\item Jacquelyn Noronha-Hostler, Houston U.	  <jakinoronhahostler@gmail.com>
\item Hiroyuki Noumi, RCNP, Osaka U.		  <noumi@rcnp.osaka-u.ac.jp>
\item Hiroaki Ohnishi, RIKEN/RCNP Osaka U. 	  <h-ohnishi@riken.jp>
\item Gabriel Palacios Serrano, ODU		  <gpala001@odu.edu>
\item Kijun Park, JLab				  <parkkj@jlab.org>
\item Eugene Pasyuk, JLab			  <pasyuk@jlab.org>
\item David Payette, ODU 			  <payette@jlab.org>
\item William Phelps, FIU 			  <wphelps@jlab.org>
\item Jiwan Poudel, ODU				  <jpoud001@odu.edu>
\item Jianwei Qiu , JLab			  <gxiaofeng2001@yahoo.com>
\item Claudia Ratti, Houston U.			  <cratti@uh.edu>
\item James Ritman, FZJ  			  <j.ritman@fz-juelich.de>
\item Torri Roark, ODU 			  	  <troar001@odu.edu>
\item Enrique Ruiz Arriola, Granada U. Spain	  <earriola@ugr.es>
\item Daniel Sadasivan, GWU			  <dansadasivan@gmail.com>
\item Sergey Kuleshov, Federico Santa Maria U.    <sergey.kuleshov@usm.cl>
\item Igor Strakovsky, GWU                        <igor@gwu.edu>
\item Joachim Stroth, Goethe U./ GSI		  <j.stroth@gsi.de>
\item Aihong Tang, BNL				  <aihong@bnl.gov>
\item Simon Taylor, JLab 			  <staylor@jlab.org>
\item Yusuke Tsuchikawa, Nagoya U.		  <tsuchikawa@phi.phys.nagoya-u.ac.jp>
\item Daniel Watts, Edinburgh U.		  <dwatts1@ph.ed.ac.uk>
\item Nilanga Wickramaarachchi, ODU		  <nwick001@odu.edu>
\item Nicholas Zachariou, Edinburgh U.		  <nicholas@jlab.org>
\end{itemize}
\end{document}